\documentclass[iop]{emulateapj}
\usepackage{epstopdf}

\submitted{Draft version: \today}

\def\msun{M$_\odot$}
 
\shortauthors{Gon\c{c}alves et al.}
\shorttitle{IFU Spectroscopy of LBAs}

\begin{document}

\title{The Kinematics of Ionized Gas in Lyman-Break Analogs at z $\sim$ 0.2} \author{\sc Thiago S. Gon\c{c}alves,\altaffilmark{1} Antara Basu-Zych,\altaffilmark{2} Roderik Overzier,\altaffilmark{3}  D. Christopher Martin,\altaffilmark{1} David R. Law,\altaffilmark{4} David Schiminovich,\altaffilmark{5} Ted K. Wyder,\altaffilmark{1} Ryan Mallery,\altaffilmark{4} R. Michael Rich,\altaffilmark{4} Timothy H. Heckman\altaffilmark{6}}

\altaffiltext{1}{California Institute of Technology, MC 278-17, Pasadena, CA 91125}

\altaffiltext{2}{NASA Goddard Space Flight Center}

\altaffiltext{3}{MPA}

\altaffiltext{4}{UCLA}

\altaffiltext{5}{Columbia University}

\altaffiltext{6}{Johns Hopkins}

\email{tsg@astro.caltech.edu}

\begin{abstract}

We present results for 19 ``Lyman Break Analogs'' (LBAs) observed with Keck/OSIRIS with an AO-assisted spatial resolution of less than 200 pc. We detect satellites/companions, diffuse emission and velocity shear, all with high signal-to-noise ratios. These galaxies present remarkably high velocity dispersion along the line of sight($\sim$ 70 km s$^{-1}$), much higher than standard star-forming spirals in the low-redshift universe. We artificially redshift our data to $z \sim 2.2$ to allow for a direct comparison with observations of high-z LBGs and find striking similarities between both samples. This suggests that either similar physical processes are responsible for their observed properties, or, alternatively, that it is very difficult to distinguish between different mechanisms operating in the low versus high redshift starburst galaxies based on the available data. The comparison between morphologies in the UV/optical continuum and our kinemetry analysis often shows that neither is by itself sufficient to confirm or completely rule out the contribution from recent merger events. We find a correlation between the kinematic properties and stellar mass, in that more massive galaxies show stronger evidence for a disk-like structure. This suggests a co-evolutionary process between the stellar mass build-up and the formation of morphological and dynamical sub-structure within the galaxy.

\end{abstract}


%
%

\section{Introduction}\label{section:intro}

Our understanding of galaxy formation has changed considerably over the course of the last two decades. Remarkable progress has been made with numerical simulations that reproduce the growth of the large-scale structure in the universe, and the results from these simulations agree well with studies of galaxy clusters and the cosmic microwave background \citep[e.g.][]{Springel2005, Benson2010}. However, the small-scale, nonlinear baryonic physics that goes into forming the galactic structure remains an open question. The so-called ``gastrophysics", comprising AGN feedback and supernova winds among other processes, is still poorly understood. Simulations rely on {\em ad hoc} recipes, which are in turn based on observational results and are purely phenomenological; the underlying physical processes are not yet known.

The traditional paradigm of galaxies forming from slowly cooling shock-heated gas (e.g. \citet{White1978, Mo1998, Baugh2006} and references therein) does not seem to apply in many cases. An elevated fraction of galaxies at high redshift display clumpy structures \citep{Elmegreen2008}, which might form from internal instabilities \citep{Noguchi1999, Immeli2004, Bournaud2007} or, alternatively, from mergers of subgalactic gas clumps \citep{Taniguchi2001}, in agreement with the idea of hierarchical galaxy formation in LCDM models. Furthermore, recent numerical simulations indicate that star formation at high redshift might be fed through cooling flows supplying the centers of dark matter haloes directly with gas at just below the virial temperature \citep{Dekel2006, Dekel2009, Keres2009}. In this context, it becomes important to analyze the kinematics in these galaxies, and to confront the relative contributions from ordered rotation, random motions and merger-induced features with predictions from the aforementioned models. Because {\it stellar} kinematics at high redshift are largely beyond reach of current instruments and telescopes, the bright nebular emission line gas is often used as a tracer for the underlying kinematics.

In an early attempt to study kinematics of star-forming galaxies at $z \sim 2-3$, \citet{Erb2006} analyzed long-slit spectra of H$\alpha$ emission in UV-selected galaxies \citep{Steidel2003, Steidel2004}, detecting significant velocity shears in 12\% of the objects in their sample. In all cases, velocity dispersion in the ionized gas was high in comparison with the observed velocity shears, with $v_{\rm c}/\sigma \sim 1$. These observations are challenging, since they are seeing-limited and slit-alignment plays an important role in actually detecting any shears \citep{Erb2006, Law2006}.

More recently, \citet{Law2009} improved on this result, with spatially resolved kinematics of the gas from adaptive-optics (AO) assisted integral-field spectroscopy of 12 star-forming galaxies at redshift $z \sim 2.5$. This technique has the advantage of not depending on alignment choice, detecting velocity shears all across the extent of the galaxy, while the AO system resolves features at sub-kpc scales. The authors detect, again, high velocity dispersion values of $\sigma \simeq 60-70$ km s$^{-1}$. In most cases there is no evidence for ordered rotation across the galaxy, and in general the gas dynamics appear to be dominated by random motions. The authors also find a mild trend of rotational properties with stellar mass, with massive galaxies typically displaying more pronounced velocity shears.

In a similar study, \citet{ForsterSchreiber2009} studied a large sample of 62 star-forming galaxies at similar redshifts with the SINFONI instrument. This work differs from \citet{Law2009} in that most observations are seeing-limited, with spatial resolution elements of approximately 4 kpc. In addition, most galaxies in this sample were drawn from the BzK sample of \citet{Daddi2004}, and are typically two times as massive as the UV-selected galaxies. The authors found that their sample can be subdivided into three groups: rotation-dominated objects, with pronounced velocity shears and $v_{\rm c}/\sigma$ values of up to 4; dispersion-dominated objects, with little to no velocity shear across the major axis; and mergers, with multiple components or peculiar velocity profiles. In addition, \citet{ForsterSchreiber2009} also found a trend of properties with stellar mass, with more massive galaxies presenting higher $v_c/\sigma$ ratios and larger sizes. A number of observations at intermediate and high redshifts also support the hypothesis of extreme starbursts being protodisks resulting either from minor mergers or smooth accretion from the intergalactic medium \citep{Bouche2007, Cresci2009, Wright2009}.

\citet{Jones2009} also studied the kinematics of the ionized gas in high-redshift star-forming galaxies, but a sample of strongly lensed objects was used instead. The authors were then able to reconstruct the kinematic structure by applying models of the gravitational lens, achieving much higher spatial resolution ($\sim$100 pc) along one spatial dimension. Out of a sample of 6 objects, 5 display characteristics of rotating gas disks, again with trends in velocities as a function of size and dynamical mass. Although the results help us understand the dynamical structures of such galaxies, it is challenging to construct a statistically significant sample of lensed objects. Additionally, in many cases the major axis is not aligned with the lens shear, in which case the velocity shear comprises few resolution elements in the data.

Studies to date explore complementary regions of parameter space.ÊThe difference in the prevalence of different kinematics observed is probably a function primarily of parent sample, compounded with differences in the sensitivity regime of different techniques. The benefit of AO is that it obtains greater spatial resolution but is not sensitive to low surface brightness features (if present), while non-AO probes lower surface brightnesses and larger radii but with less fidelity. In both cases observations are technically challenging, due to the distance to the galaxies, which results in low intrinsic spatial resolution and cosmic surface brightness dimming. Therefore, it is advantageous to observe similar galaxies at lower redshifts in order to assess whether certain features derived from observations at high redshift are intrinsic or biased due to observational effects.

\citet{Heckman2005} have selected a sample of UV-bright galaxies in the low-redshift universe ($z \sim 0.2$) from GALEX data \citep{Martin2005}, referred to as Ultraviolet Luminous Galaxies (UVLGs). The authors found that these galaxies could be subdivided into two main groups, one consisting of massive spirals, and the other consisting of compact objects undergoing intense starbursts. \citet{Hoopes2007} further expanded this analysis, subdividing UVLGs into three categories with respect to their FUV surface brightness.

The UV characteristics for the most compact of these galaxies, the ``supercompact UVLGs'', were chosen to match those of typical LBGs (e.g., $L\sim L^*_{z=3}$, where $L^*_{z=3}$ is the characteristic luminosity of LBGs at $z\sim3$). \citet{Hoopes2007} found that these objects indeed present similar properties to star forming galaxies at higher redshift, with comparable star formation rates, colors and metallicities, as inferred from their SDSS spectra. \citet{BasuZych2007} also determined, from radio continuum and mid-infrared observations, that these objects have significantly less dust attenuation when compared to galaxies of similar star formation rates in the local universe, as is the case for LBGs. We have therefore previously referred to these galaxies as ``Lyman Break Analogs'' (LBAs), and will do so for the remainder of this paper.

To study the morphologies of LBAs, \citet{Overzier2009, Overzier2010} obtained HST ultraviolet and optical imaging of 30 galaxies. In general, their ultraviolet morphologies are dominated by clumpy features indicative of massive and compact star forming regions, while many furthermore show clear signs of recent merger events. Interestingly, when the data is redshifted to $z \sim 2-4$, their morphologies are remarkably similar to LBGs at these epochs \citep[e.g.][]{Giavalisco1996, Papovich2005, Lotz2006,Law2007a}, while the subtle, low surface brightness merger features tend to disappear even in the deepest rest-frame UV or optical imaging data. This implies that on the basis of morphologies alone, it cannot be ruled out that LBGs grow through clumpy accretion and mergers, perhaps together with rapid gas accretion through other means \citep{Overzier2010}.

Furthermore, strong hydrogen lines and compact sizes make them ideal candidates for IFU spectroscopy. In \citet{Basu-Zych2009}, we presented preliminary results of the IFU survey discussed here for three LBAs, showing how these galaxies resemble the kinematic structures of high-redshift star-forming galaxies. In this work, we expand the sample to investigate the ionized gas kinematics of 19 LBAs, observed with spatial resolution down to $\sim$200 pc. The paper is divided as follows: in section (2), we describe the data acquisition and analysis, including target selection and how we artificially redshift our data to $z=2.2$ in order to make direct comparisons with LBGs; in Section (3), we describe properties of individual objects; in Section (4) we describe our results, including general trends for these galaxies; in section (5) we discuss and analyze the results described in the previous section, and in section (6) we summarize our findings.

Throughout this paper, we assume standard cosmology, with $H_0 = 70$ km s$^{-1}$ Mpc$^{-1}$, $\Omega_{\rm m} = 0.30$ and $\Omega_\Lambda = 0.70$.

\section{Observations and Data Reduction}\label{section:obsred}

\subsection{Sample Selection}\label{section:sample}

We investigate a subsample of the ultraviolet-luminous galaxies (UVLGs). These objects were first defined by \citet{Heckman2005} to have far-ultraviolet (FUV) luminosities $\geq 2\times 10^{10} L_\odot$, which is roughly halfway between the characteristic luminosity of present-day galaxies and that of higher redshift Lyman-Break Galaxies (LBGs).

As described in the previous section, \citet{Hoopes2007} later expanded the analysis of these objects and subdivided the sample in terms of average FUV surface brightness ($I_{1530}$), using the SDSS $u$-band half-light radius as proxy for the UV size of the galaxies. The sample was divided in three categories: large UVLGs ($I_{1530} \leq 10^8 L_\odot$ kpc$^{-2}$), compact UVLGs ($I_{1530} > 10^8 L_\odot$ kpc$^{-2}$) and supercompact UVLGs ($I_{1530} > 10^9 L_\odot$ kpc$^{-2}$). The latter represents the aforementioned LBAs.

The LBAs are compact systems undergoing intense star formation; in fact, they are among the most star-forming galaxies in the low-redshift universe. The observed physical properties, such as metallicity, dust attenuation, UV/optical morphologies and star formation rates, are remarkably similar to those of high-redshift LBGs. We further discuss the analogy between low- and high-redshift objects in subsequent sections.

\subsection{Observations and Data Reduction}\label{section:obs}

LBAs are selected to have high surface brightness values, which translate into small physical sizes, ranging from ~0.4 to 1.9 kpc half-light radii in the ultraviolet \citep{Overzier2010}. Together with the high star formation rates up $\sim100$ \msun yr$^{-1}$ \citep{Hoopes2007}, which translates into extremely bright nebular hydrogen emission lines, LBAs are highly suitable targets for adaptive optics (AO) assisted integral field spectrography. 

We have used OSIRIS in the Keck II telescope \citep{Larkin2006}. OSIRIS is an integral field unit (IFU) available solely for use with AO. It provides a spectral resolution of $R\sim 3800$ and a field of view (FOV) of a few arcseconds, depending on the configuration utilized. Its design is based on a lenslet array, with variable spatial pixel scales (spaxels) depending on the need for better PSF sampling or a larger FOV. In good weather conditions, we are able to achieve near diffraction limited resolution, or approximately 70 milli-arcseconds (mas) FWHM in angular size. 

We have targeted the Pa-$\alpha$ emission line (rest wavelength $\lambda = 1875.1$ nm), which is expected to be $\sim$ 8 times fainter than the H-$\alpha$ line, depending on gas temperature \citep{OsterbrockDonaldE.2006}. In all cases this is redshifted into the redder half of the K-band, with observed wavelength varying between 2055 nm $\lesssim \lambda_{\rm obs} \lesssim$ 2350 nm for the objects in our sample.

The objects observed for this work were selected from the original 30 objects observed with HST presented in \citet{Overzier2010}. Due to a lack of bright nearby guide stars, we have used the Laser Guide Star Adaptive Optics (LGS-AO) system for all objects presented here \citep{vanDam2006, Wizinowich2006}. The selection of galaxies for each observing run was based purely on availability during a given night, proximity of prominent sky lines to the wavelength of the Pa-$\alpha$ line at each redshift and lesser impact of space command closures (when observers are prevented from using the laser due to possible collisions with artificial satellites). Therefore, no biases were introduced in the data beyond the original LBA selection.

The properties of individual objects are shown in Table \ref{table:obstab} along with observing information. The stellar masses were taken from the SDSS/DR7 MPA-JHU value-added catalog \footnote[1]{http://www.mpa-garching.mpg.de/SDSS/DR7/}. These masses were calculated by fitting a large grid of spectral synthesis models from Bruzual \& Charlot (BC03,2003) to the SDSS $u^\prime$,$g^\prime$,$r^\prime$,$i^\prime$,$z^\prime$ photometry. The lack of near-infrared data and TP-AGB stars from the synthesis library should introduce an uncertainty of $\sim$0.3 dex \citep[see][]{Overzier2009}, small enough that our results, spanning two orders of magnitude in stellar mass, are unaffected. The BC03 models assume a Chabrier (2003) initial mass function, and were chosen to span a large range in star formation histories and ages. Prior to the fitting, the magnitudes were corrected for emission line flux by assuming that the relative contribution of the lines to the broad-band photometry is the same inside the fibre as outside. Because most of our objects are not or barely resolved in SDSS, this is a reasonable assumption. The final mass estimate is taken to be the mean value of the mass likelihood distribution constructed from all models. A more detailed analysis of the masses, ages and star formation histories of LBAs based on rest-frame UV to far-IR photometry and resolved emission line spectroscopy is currently under investigation. Star formation rates presented here are measured from combined H$\alpha$ and MIPS-24$\mu$m data; they typically present an uncertainty up to 0.3 dex \citep{Overzier2009}. For an in-depth discussion of properties of LBAs and comparison with high-z galaxies, see \citet{Hoopes2007} and \citet{Overzier2009, Overzier2010}. 

Given the limited physical size of the detector, there is a trade-off between spatial coverage and wavelength coverage; since we are interested in a single emission line, we have chosen to use the narrowband mode for most galaxies in order to maximize the spatial coverage of the data. In most cases we observed with the 50 mas spaxel scale; the UV sizes of the remaining objects were larger and we chose to use the 100 mas scale with double the FOV.

In many cases, the object occupies a significant portion of the FOV of the instrument. Because appropriate sky subtraction is crucial for a reliable detection of emission lines in the data, we have ensured an exclusive sky frame was taken in conjunction with each science frame. The best strategy to maximize on-target telescope time was to observe in 45 minute blocks of science-sky-science frames, with 15 minute exposures in each case. Weather ranged from acceptable to excellent in all cases, with uncorrected seeing (in V band) varying from $\sim$ 1" in moderate conditions to 0.5" in the best cases. Weather conditions directly affect spatial resolution in our data, since the quality of AO corrections depend on the stability and brightness of the laser guide star and the tip-tilt star.

Data were reduced with the OSIRIS pipeline, which subtracts the sky frames and translates the two-dimensional detector image into a 3D datacube, composed of two spatial dimensions and one wavelength dimension \citep[for details, see][]{Wright2009}. In addition, we have written custom IDL code to further subtract sky emission residuals still present in the datacube. This is done for each galaxy simply by fitting the 1D spectrum at spaxels where we believe no signal from the observed galaxy exists; this is then subtracted from all spaxels in the datacube.

\begin{deluxetable*}{lcccccccc}
\tablewidth\linewidth
\tabletypesize{\scriptsize}
\tablecaption{Summary of LBA Observations} 
\tablehead{
  \colhead{Name} & \colhead{$z$} & \colhead{Observing} & \colhead{Spaxel} &  \colhead{Exposure} & AO FWHM & \colhead{SFR (H$\alpha$ + 24$\mu$m)} & \colhead{$R_{\rm l}$} \altaffilmark{a} & \colhead{log M$_*$}\\
  & & \colhead{date (UT)} & \colhead{scale (mas)} & \colhead{(s)} & \colhead{(mas)} & \colhead{($M_\odot$ yr$^{-1}$)} & \colhead{(kpc)} & \colhead{(\msun)}
  \label{table:obstab}
  }
\startdata
$005527$ & 0.167 & Oct 01, 2007 & 50 & 900 & 90 & 55.4 & 0.36 & 9.7\\
$015028$ & 0.147 & Oct 20, 2008 & 50 & 2400 & 82 & 50.7 & 1.34 & 10.3\\
$021348$ & 0.219 & Oct 19, 2008 & 100 & 2100 & 177 & 35.1 & 0.38 & 10.5\\
$032845$ & 0.142 & Jan 24, 2010 & 50 & 1800 & 103 & 8.7 & 0.86 & 9.8\\
$035733$ & 0.204 & Oct 20, 2008 & 100 & 1800 & 116 & 12.7 & 1.00 & 10.0\\
$040208$ & 0.139 & Sep 13, 2009 & 50 & 2100 & 80 & 2.5 & 0.80 & 9.5\\
$080232$ & 0.267 & Jan 24, 2010 & 100 & 1800 & 115 & 30.4 & 3.01 & 10.7\\
$080844$ & 0.096 & Feb 06, 2010 & 100 & 1200 & 187 & 16.1 & 0.08 & 9.8\\
$082001$ & 0.218 & Jan 25, 2010 & 50 & 2400 & 69 & 40.0 & 2.78 & 9.8\\
$083803$ & 0.143 & Feb 05, 2010 & 50 & 1800 & 105 & 6.2 & 1.02 & 9.5\\
$092600$ & 0.181 & Feb 26, 2008 & 50 & 1800 & 101 & 17.0 & 0.68 & 9.1\\
$093813$ & 0.107 & Feb 06, 2010 & 50 & 1800 & 77 & 19.8 & 0.65 & 9.4\\
$101211$ & 0.246 & Feb 06, 2010 & 50 & 1200 & 96 & 6.2 & N/A & 9.8\\
$113303$ & 0.241 & Jan 24, 2010 & 50 & 2400 & 76 & 7.7 & 1.36 & 9.1\\
$135355$ & 0.199 & Feb 05, 2010 & 50 & 2100 & 68 & 19.4 & 1.45 & 9.9\\
$143417$ & 0.180 & Feb 26, 2008 & 50 & 2700 & 98 & 20.0 & 0.90 & 10.7\\
$210358$ & 0.137 & Oct 20, 2008 & 50 & 1500 & 65 & 108.3 & 0.44 & 10.9\\
$214500$ & 0.204 & Sep 13, 2009 & 50 & 2400 & 70 & 16.4 & 1.13 & 9.9\\
$231812$ & 0.252 & Sep 13, 2009 & 100 & 1800 & 130 & 63.1 & 2.77 & 10.0
\enddata
\tablenotetext{a}{UV half-light radius from HST data}
\end{deluxetable*}

\subsection{Kinematic Maps}\label{section:maps}

In order to produce velocity moment maps, we fit gaussian functions to the emission lines detected at each spaxel. In most cases, our LBA spectra do not show any continuum, only the Pa-$\alpha$ line emission. The zero-point of the fit is the center of a gaussian fit to the integrated one-dimensional spectrum of the collapsed datacube.

We smooth every datacube spatially with a kernel of $1.5-2$ pixels, depending on the data quality and seeing in each case. While this results in a slight loss of spatial resolution, it also reduces noise, allowing detection of line emission at regions with lower surface brightness, especially at the outskirts of the galaxies, where gas velocity offsets from the center will likely be higher and thus can strongly affect our kinematic measurements. In addition, to produce the images shown in Fig. \ref{fig:vdmaps}, we oversample the image by a factor of 2, so that features are smoother. This is simply a visualization technique and has not been used in any of the quantitative analyses discussed in the following sections.

The signal-to-noise ($S/N$) ratios shown are obtained by dividing the area of the gaussian fit to the emission line in each spaxel by the sum of the noise fluctuation over the same wavelength range. The noise is determined from a region of the sky with no emission line detection. We introduce a minimum threshold of $S/N=6$ for a fit to be deemed acceptable; anything smaller is discarded. This minimizes the presence of artifacts in the final maps. This $S/N$ threshold represents a detection limit in star formation surface density of order $\Sigma_{\rm SFR}\sim 0.1$ M$_\odot$ yr$^{-1}$ kpc$^{-2}$, comparable to surface brightness limits determined in \citet{ForsterSchreiber2009} and an order of magnitude deeper than the data presented in \citet{Law2009}. The velocity-dispersion ($\sigma$) maps, corrected for instrumental broadening, always show values greater than the intrinsic instrumental resolution of $\sim$35 km s$^{-1}$, with the exception of some low surface brightness spaxels.

Figure \ref{fig:vdmaps} shows the recovered kinematic maps for each of the objects in our sample. In each case, the two left panels show the HST images of the galaxy, with line emission contours overlaid. The third panel shows the zero-th moment of the fit, which is simply the total intensity in each spaxel, shown as the signal-to-noise of the fit in each spaxel. The fourth panel shows the velocity maps, and the final panel shows the velocity dispersion maps. We also show the resolution element, given by the FWHM of a star observed before the galaxy, in the exact same configuration (band filter and pixel scale). Also shown is a horizontal bar indicating a physical size of one kpc at the redshift of the galaxy. Three of these galaxies (092600, 143417 and 210358) have been previously analyzed in \citet{Basu-Zych2009}.

\subsection{Comparison with HST morphologies}\label{section:hst}

Figure \ref{fig:vdmaps} shows the HST images for each galaxy in rest-frame optical (left panel) and UV (second-to-left). Images are scaled at logarithmic (black) and linear(blue) stretch, to distinguish between low surface brightness structures and more compact ones. Pa-$\alpha$ flux contours, in red, typically enclose 1/3 of the rest-frame optical flux, and above 60\% of the UV flux. In general we are able to detect emission where the bulk of the stellar mass is present, unless no significant star formation is present (e.g. the southeast components in 080844 and 210358).

Comparison between both bands in HST shows more extended structures in the rest-frame optical, in particular at low surface brightness (black). This might indicate an underlying older stellar population in which star-forming regions exist. A complete discussion of LBA morphologies in both bands can be found in \citet{Overzier2009, Overzier2010}.

\begin{figure*}[ht]

\includegraphics[width=.19\linewidth]{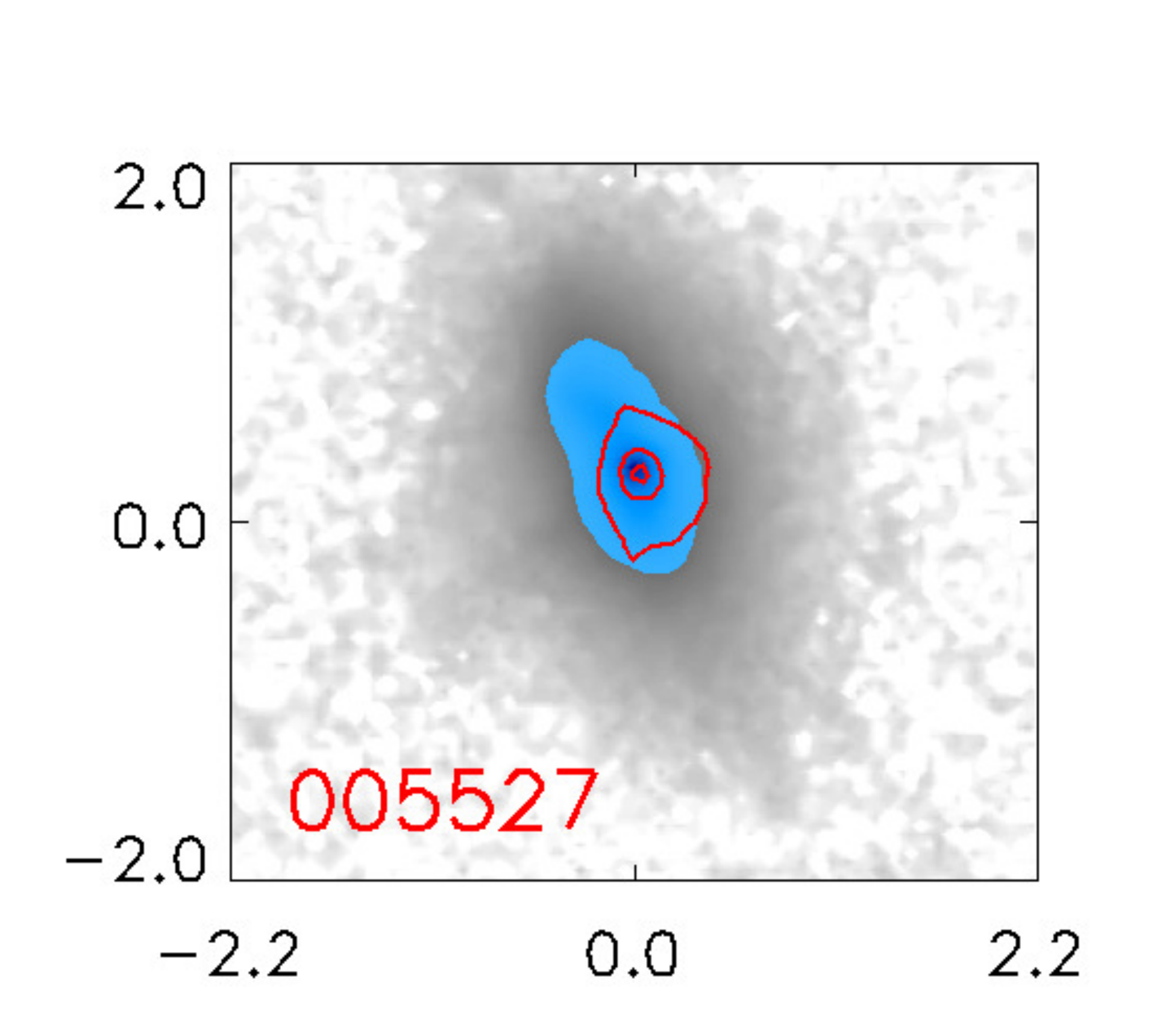} \hskip.05in
\includegraphics[width=.19\linewidth]{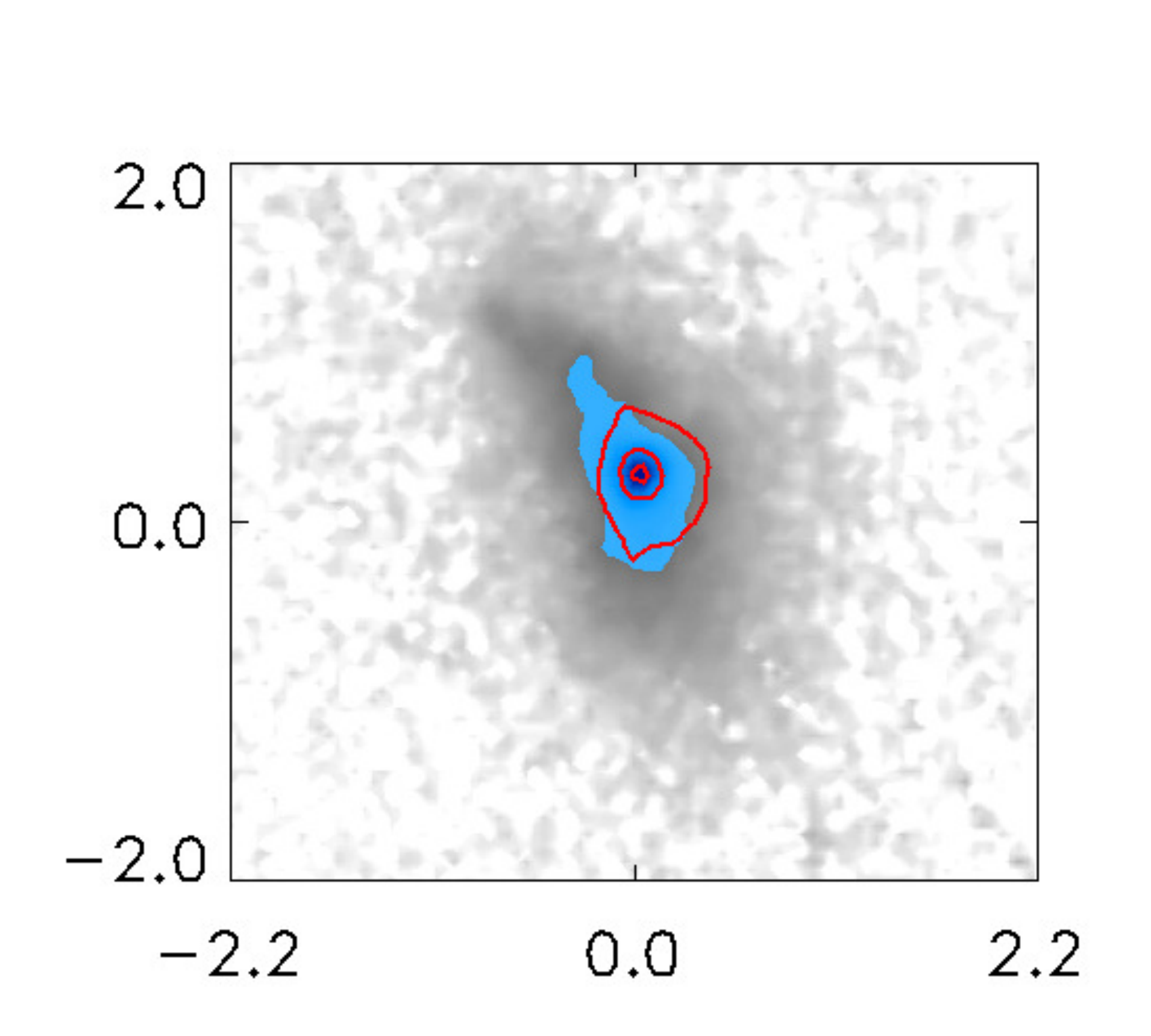} \hskip.05in
\includegraphics[width=.19\linewidth]{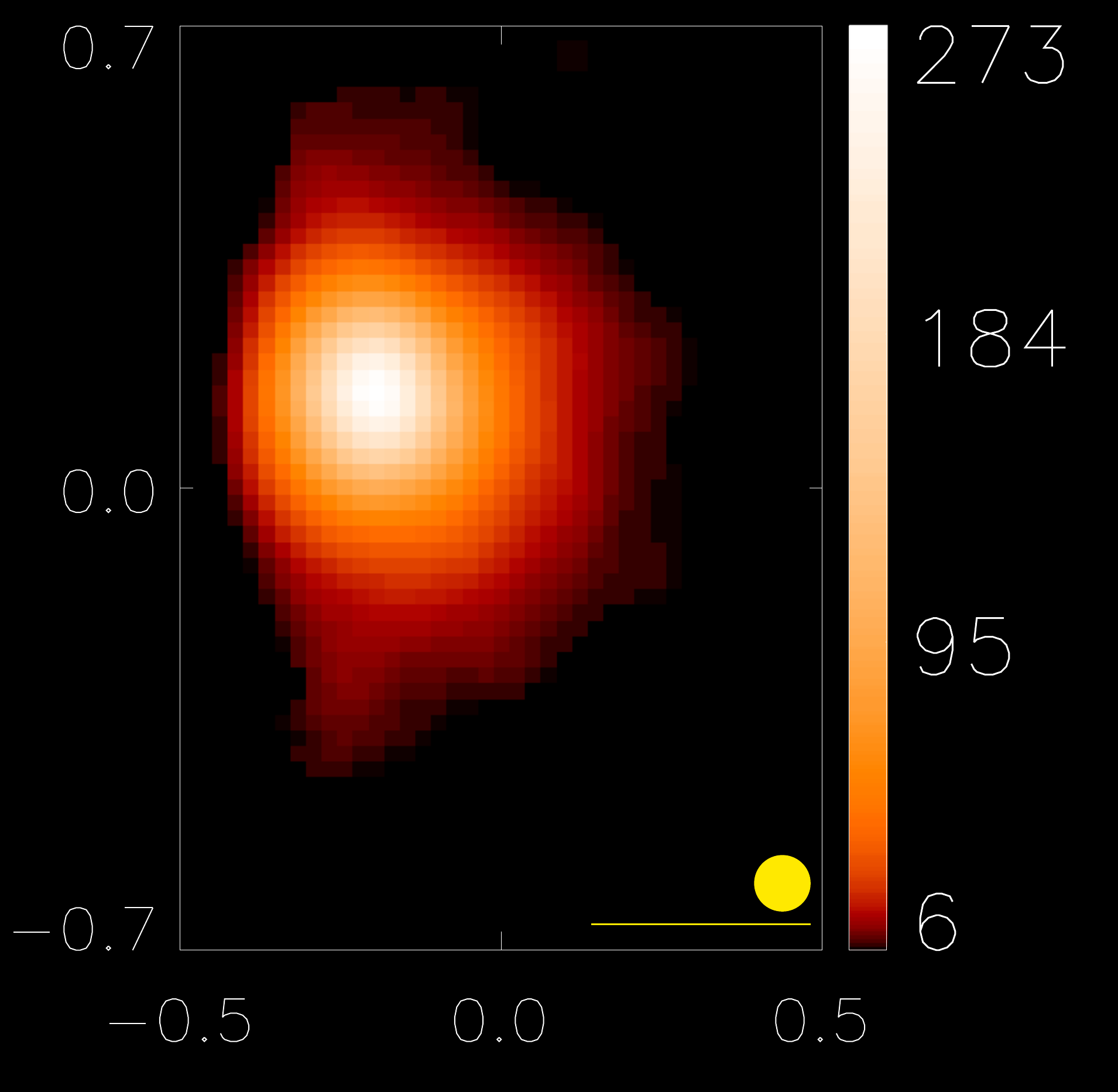} \hskip.05in
\includegraphics[width=.19\linewidth]{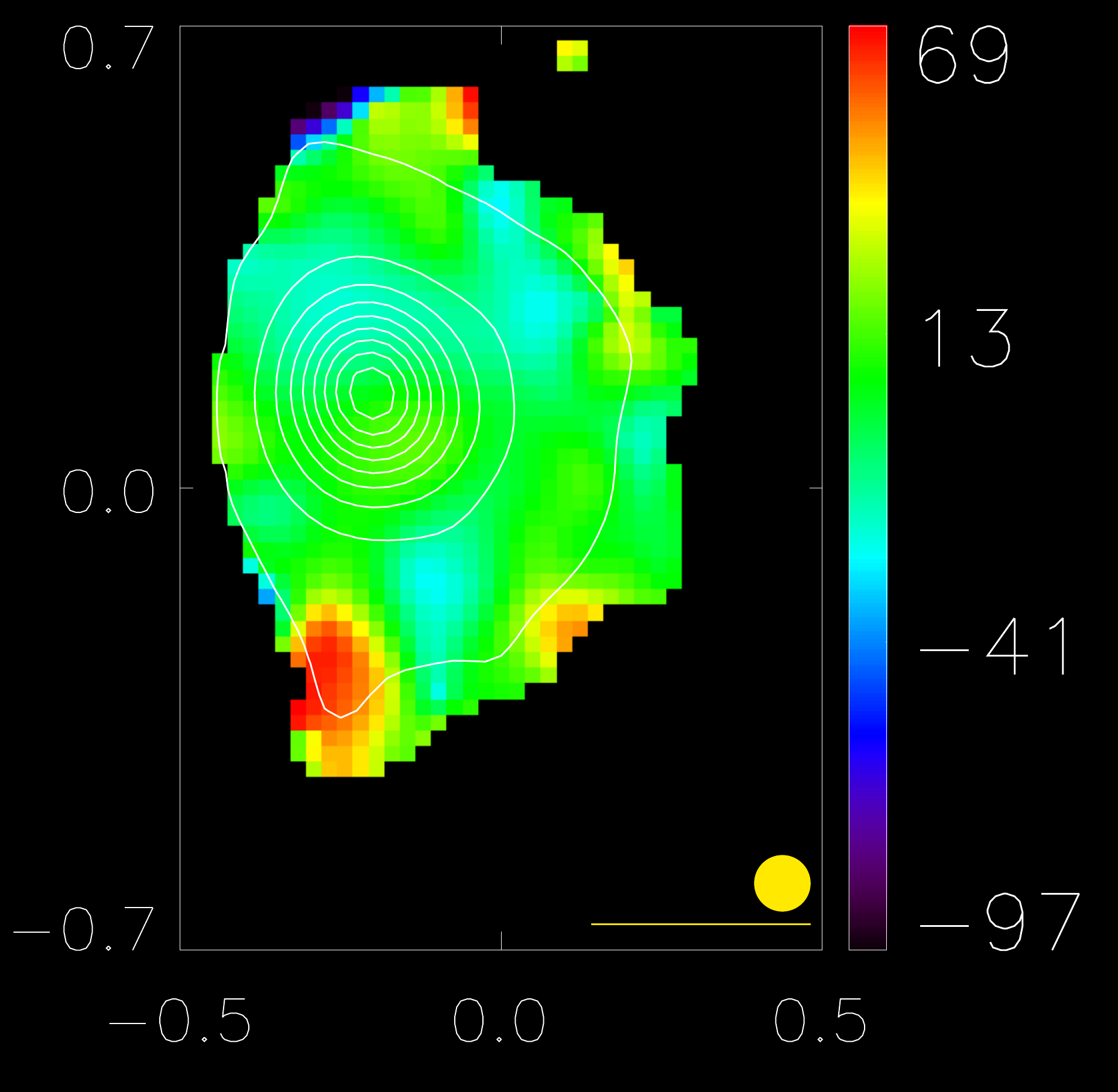} \hskip.05in
\includegraphics[width=.19\linewidth]{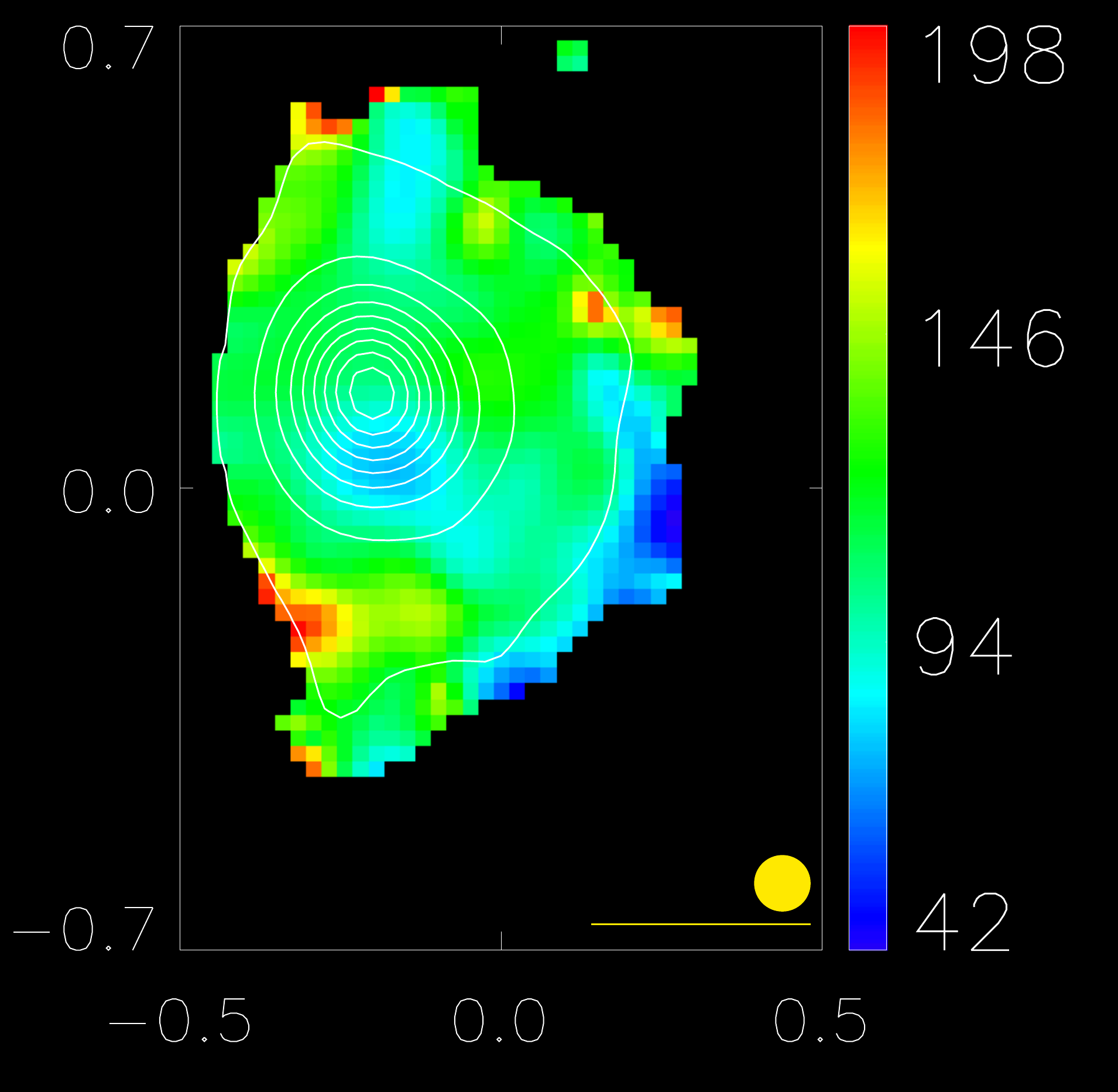} \hskip.05in
\vskip .1 in
\includegraphics[width=.19\linewidth]{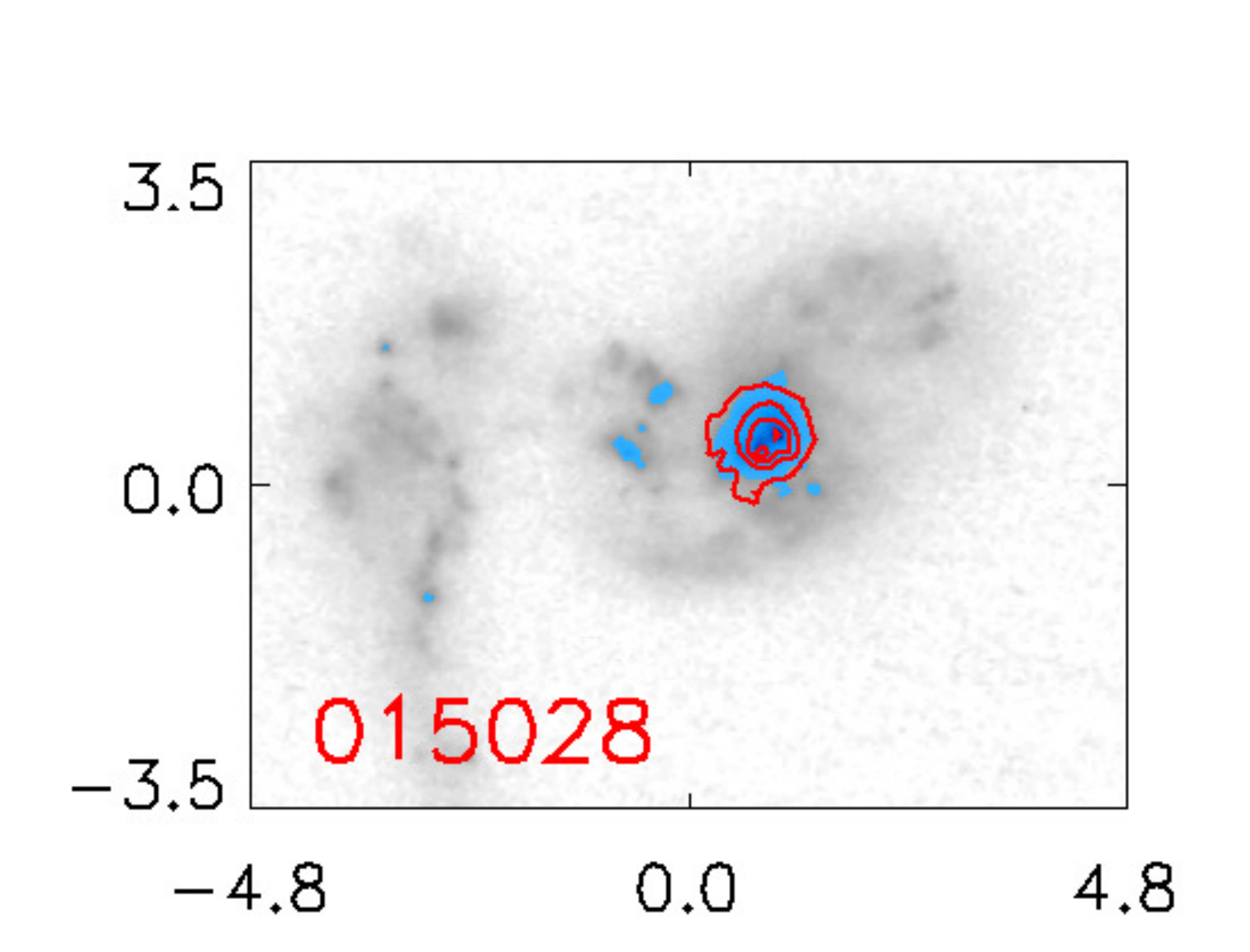} \hskip.05in
\includegraphics[width=.19\linewidth]{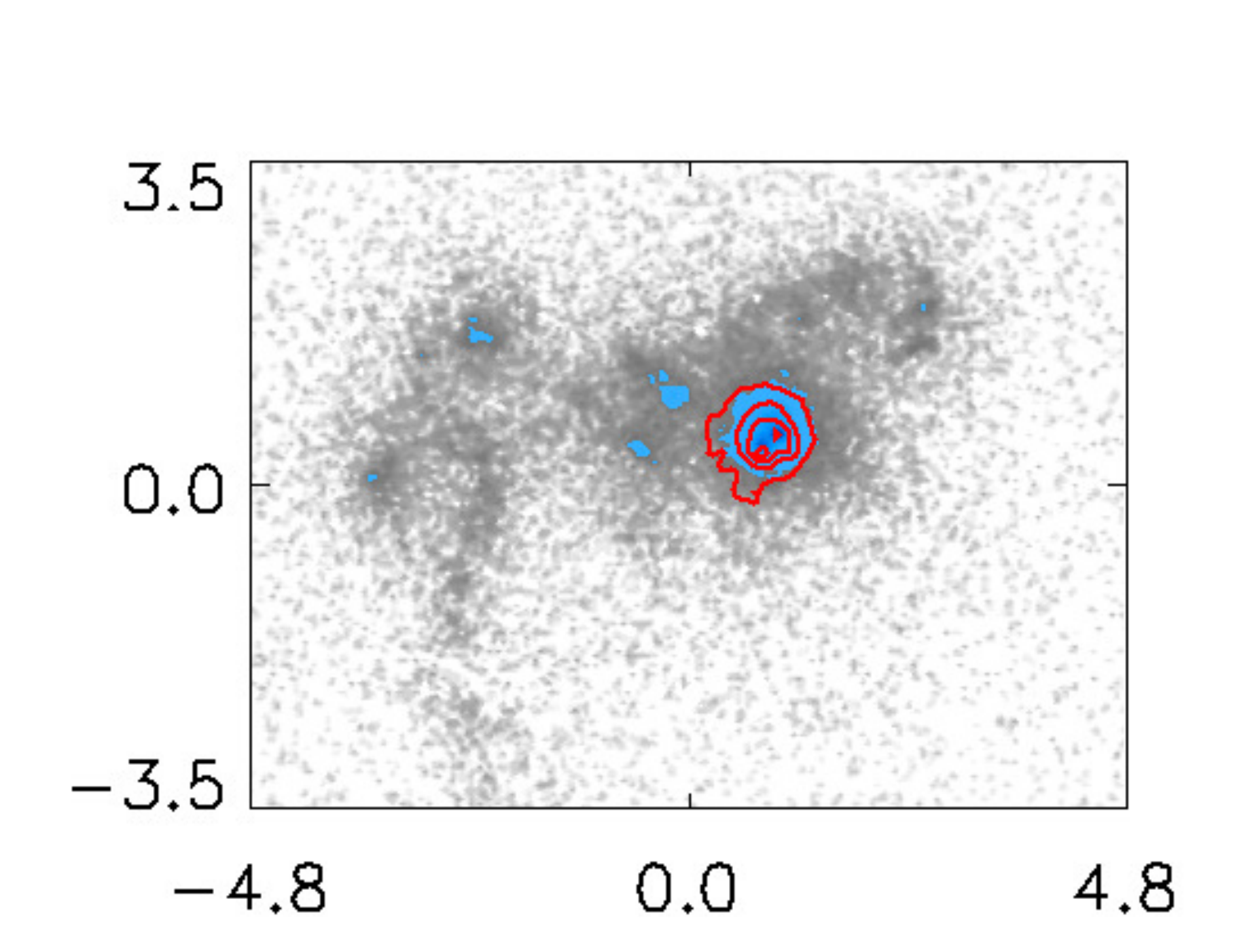} \hskip.05in
\includegraphics[width=.19\linewidth]{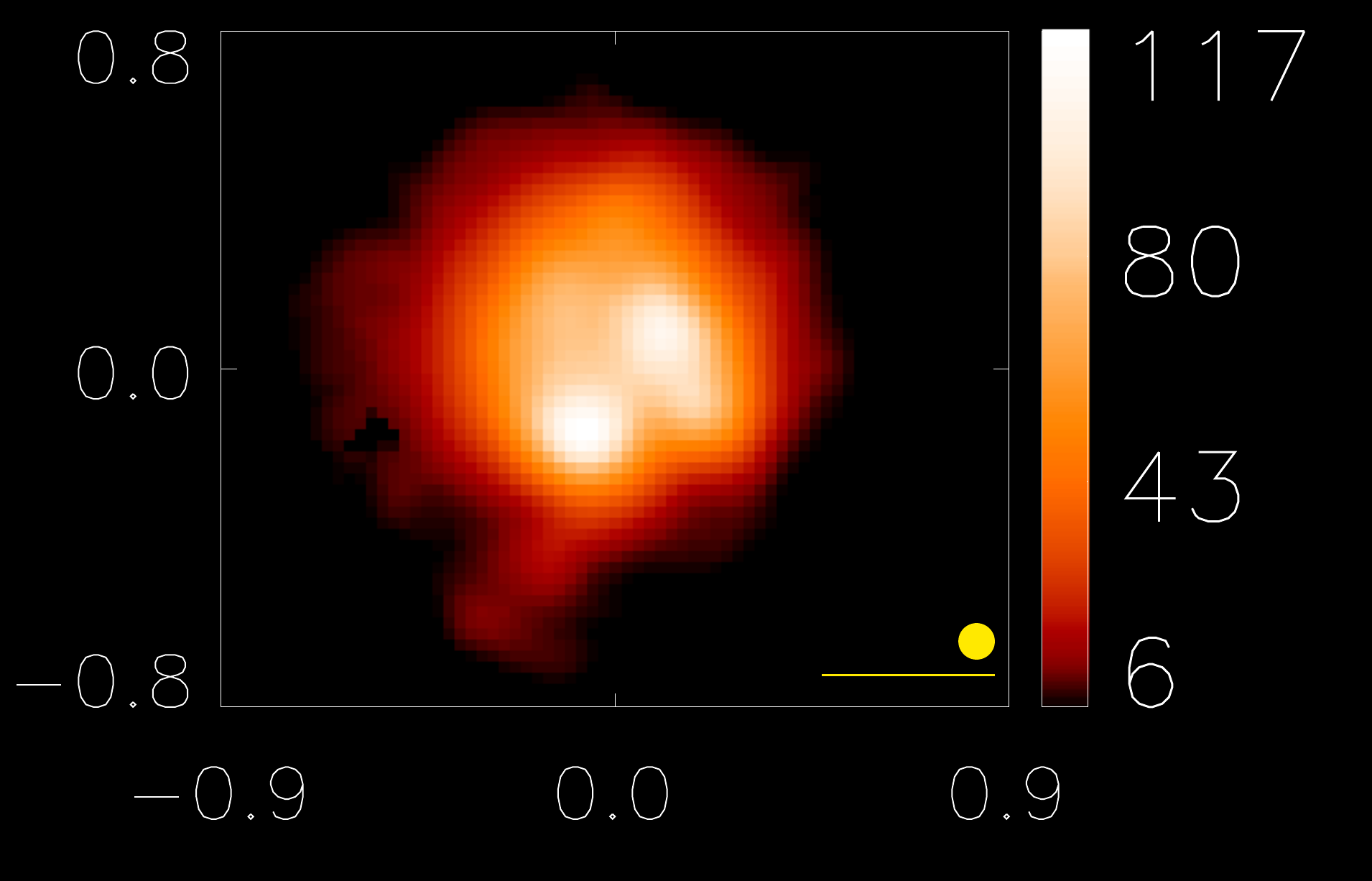} \hskip.05in
\includegraphics[width=.19\linewidth]{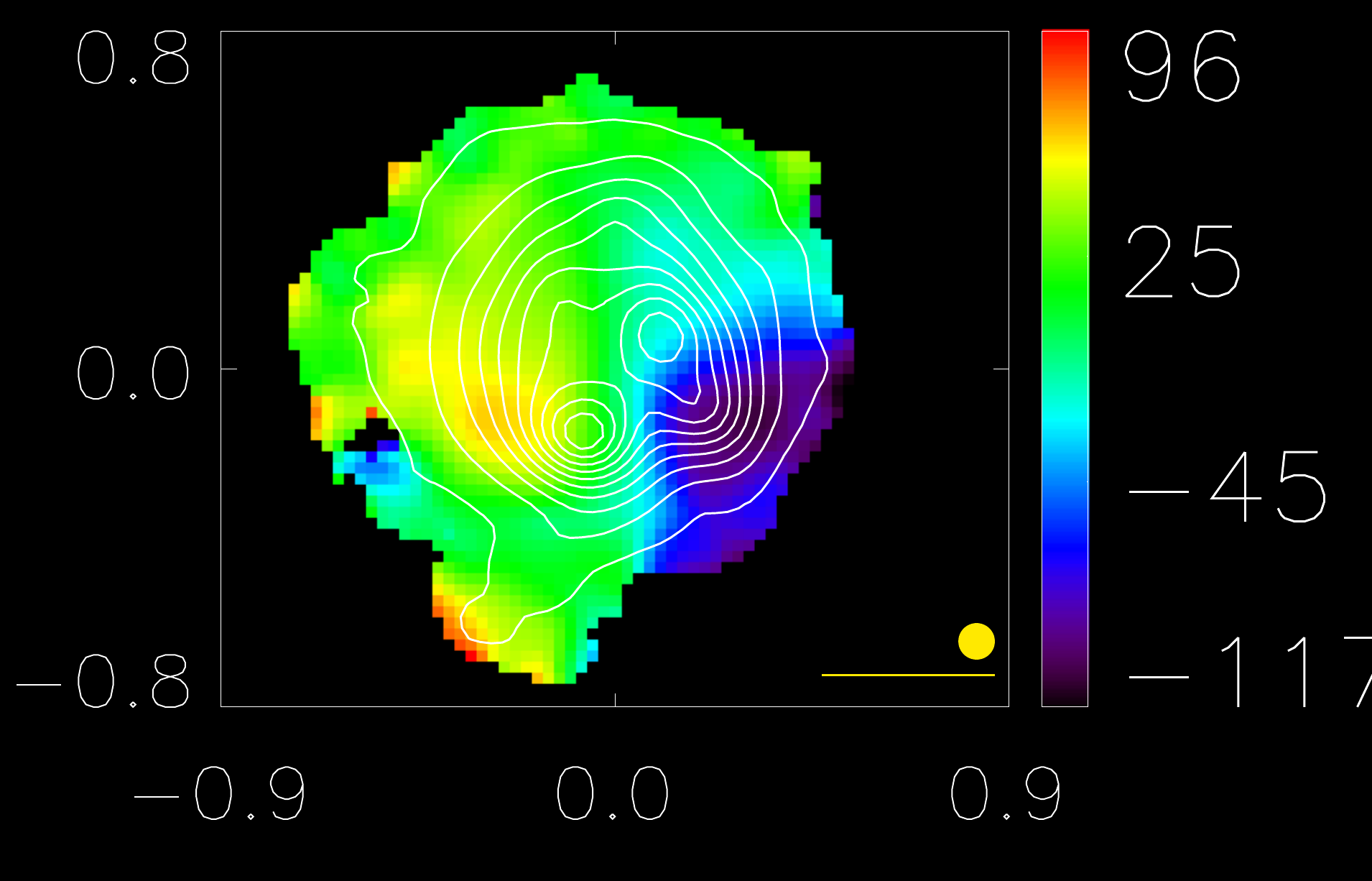} \hskip.05in
\includegraphics[width=.19\linewidth]{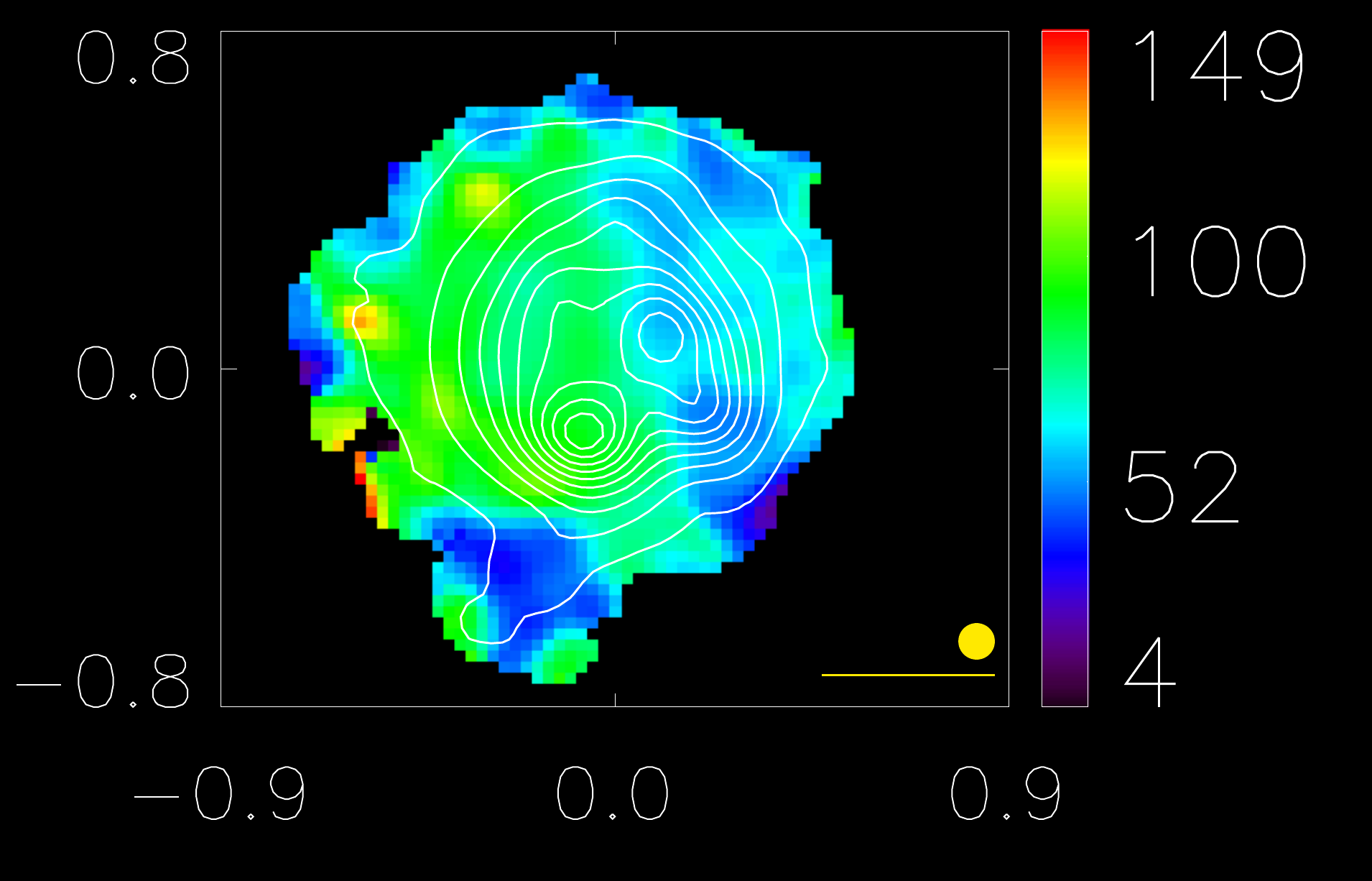} \hskip.05in
\vskip .1 in
\includegraphics[width=.19\linewidth]{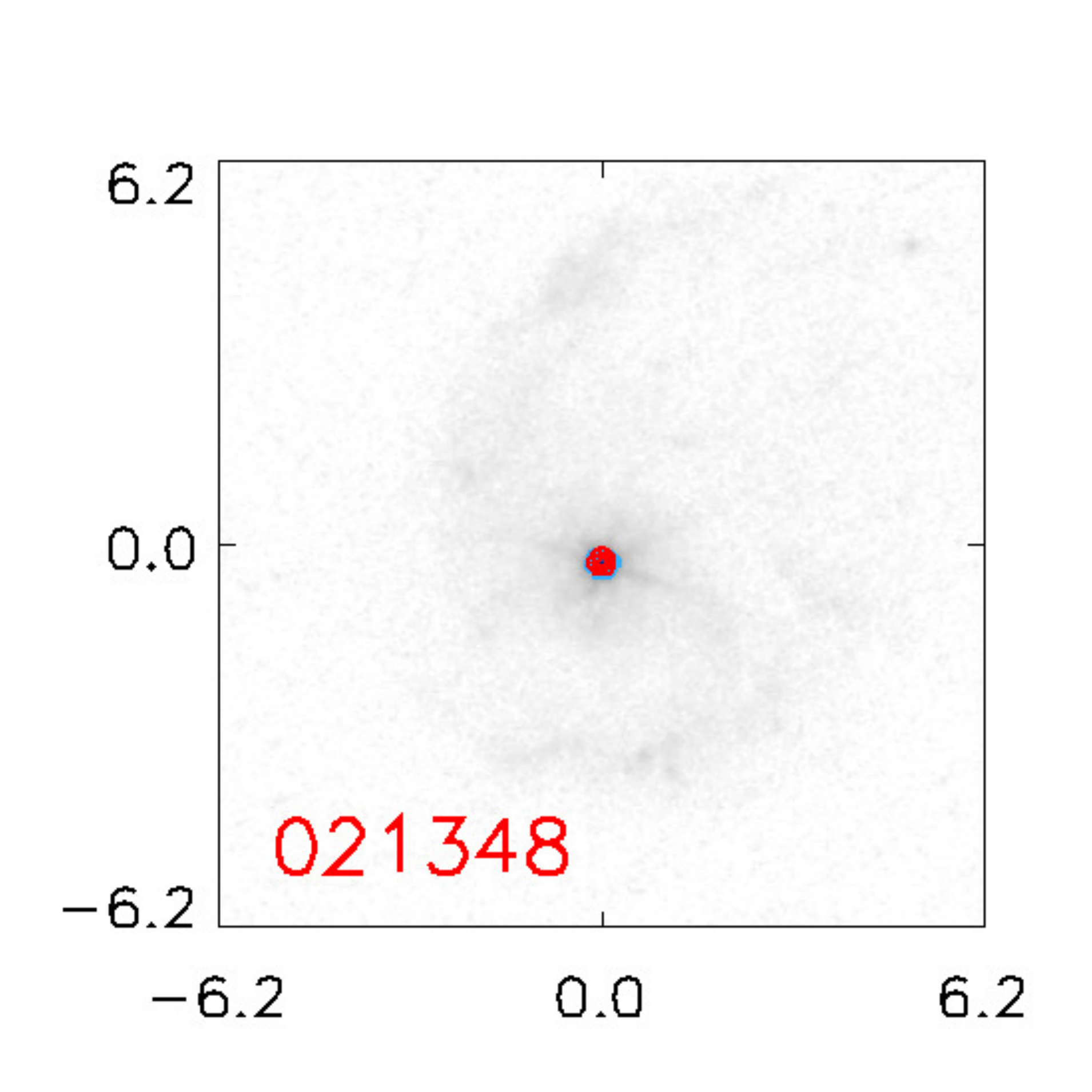} \hskip.05in
\includegraphics[width=.19\linewidth]{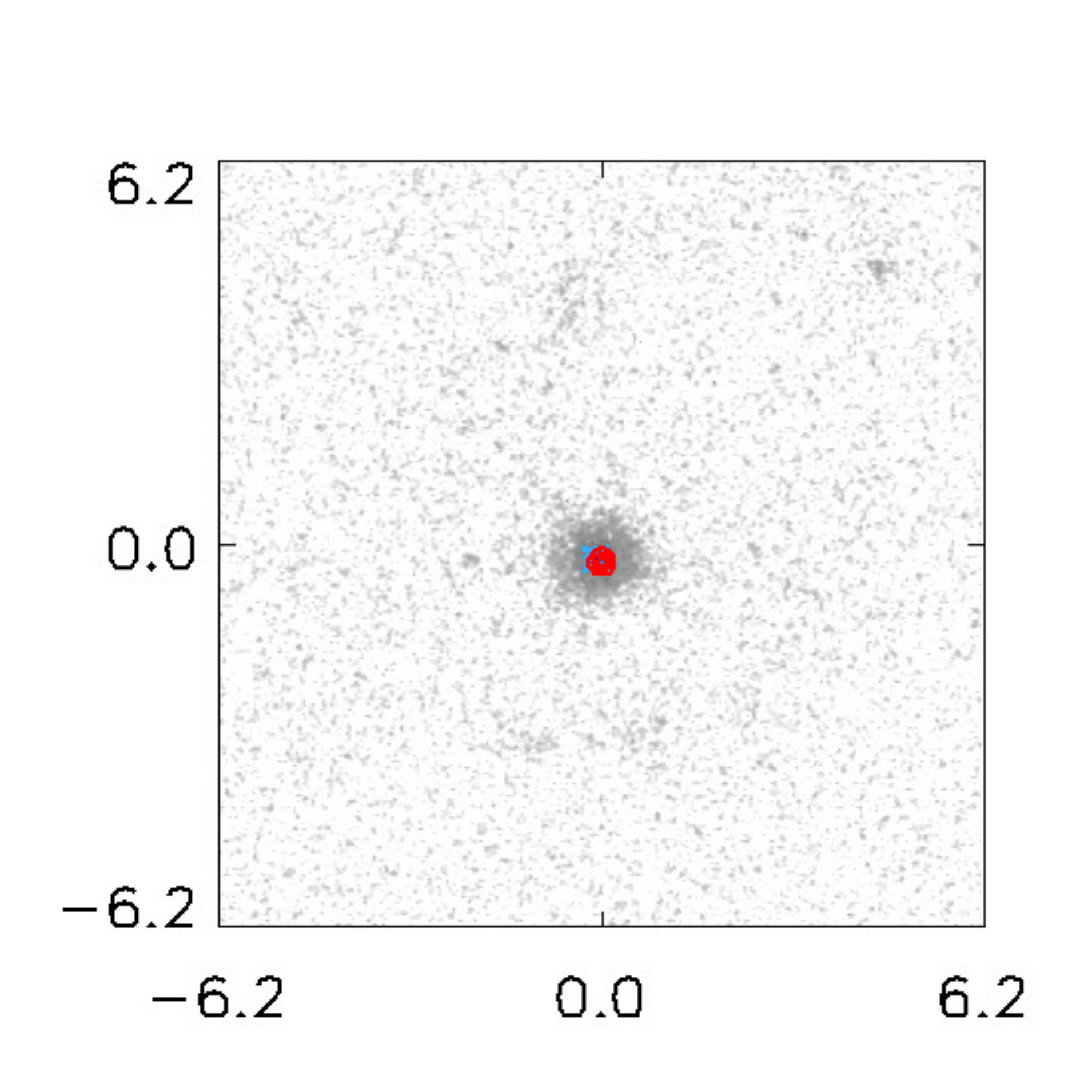} \hskip.05in
\includegraphics[width=.19\linewidth]{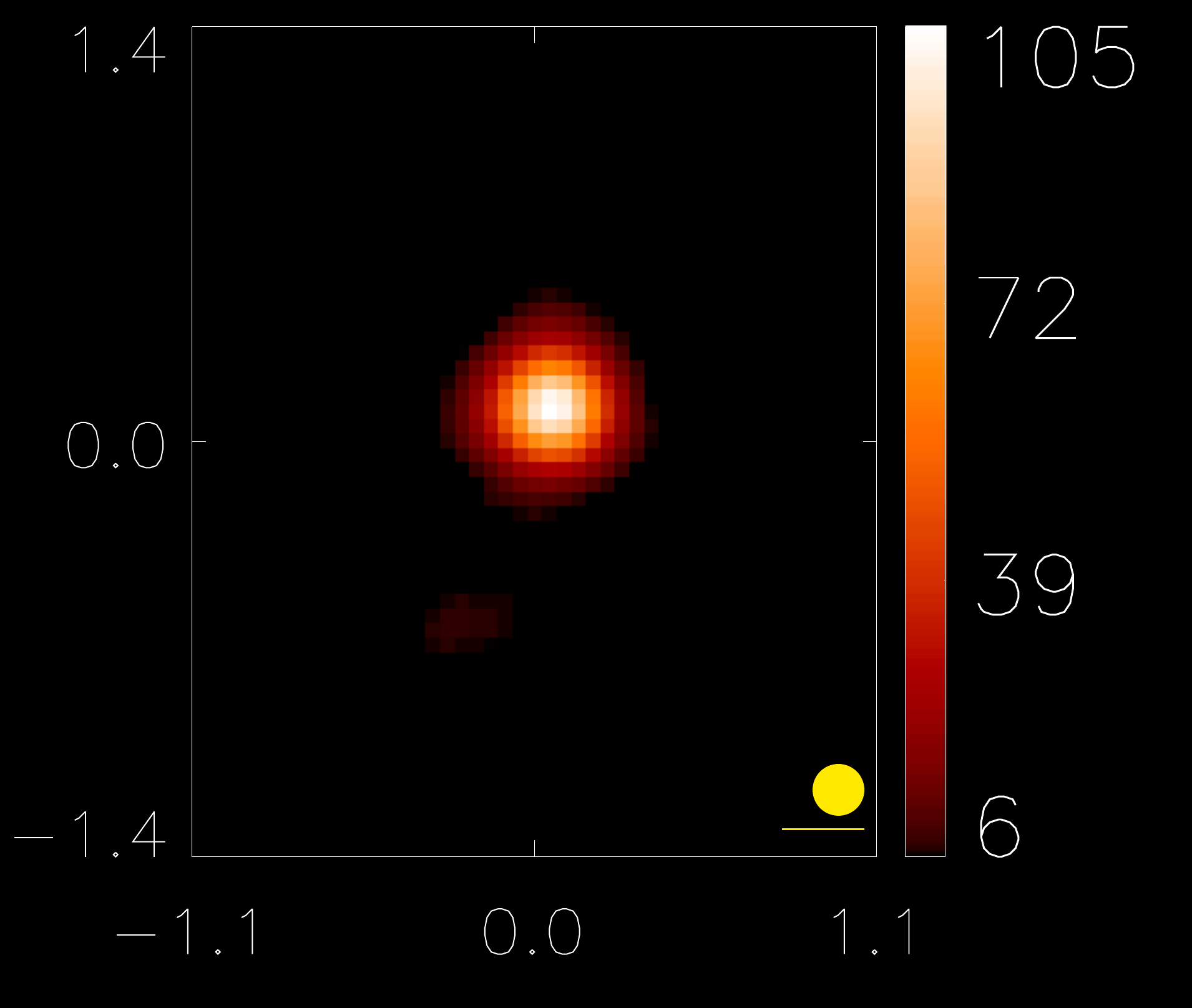} \hskip.05in
\includegraphics[width=.19\linewidth]{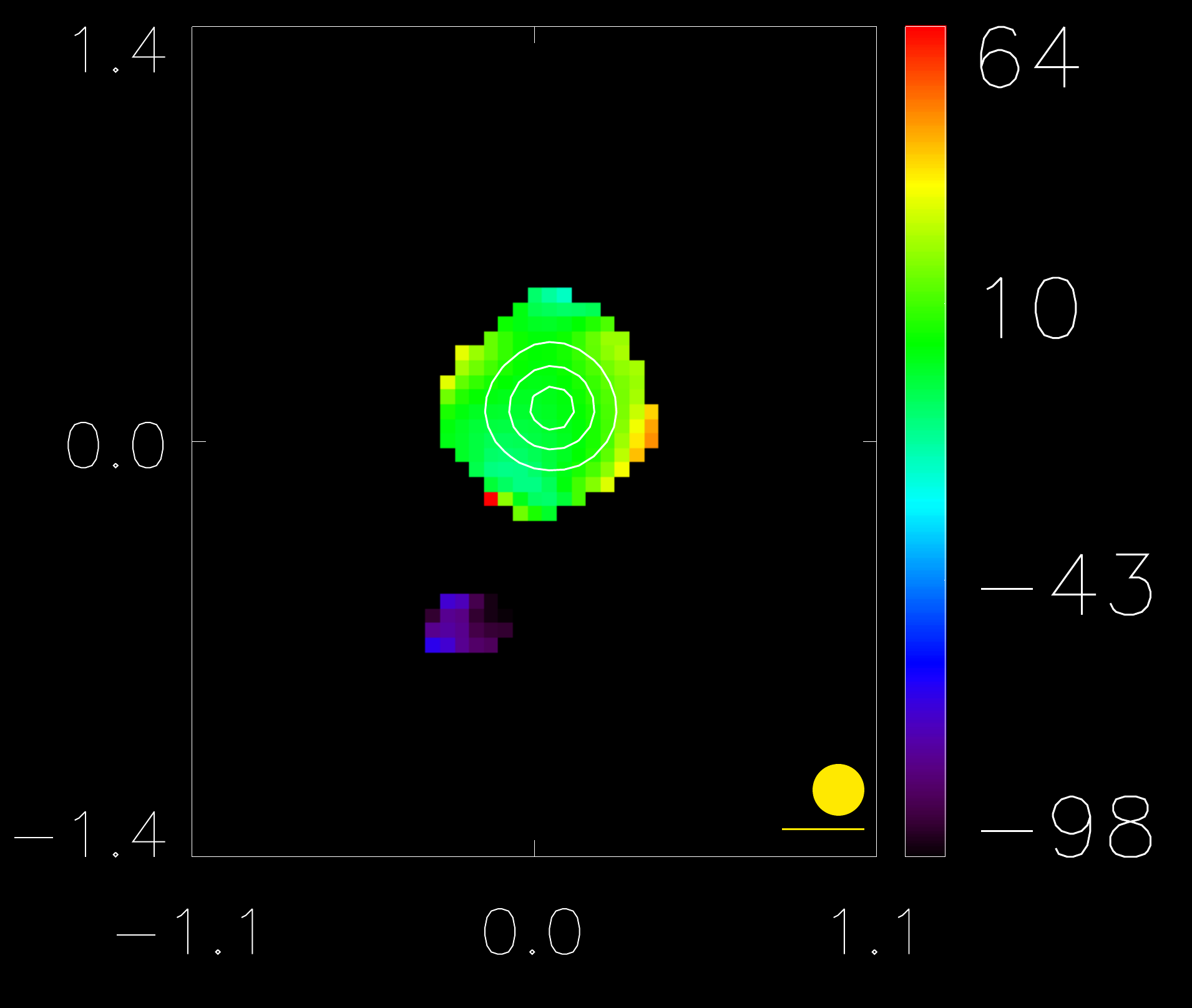} \hskip.05in
\includegraphics[width=.19\linewidth]{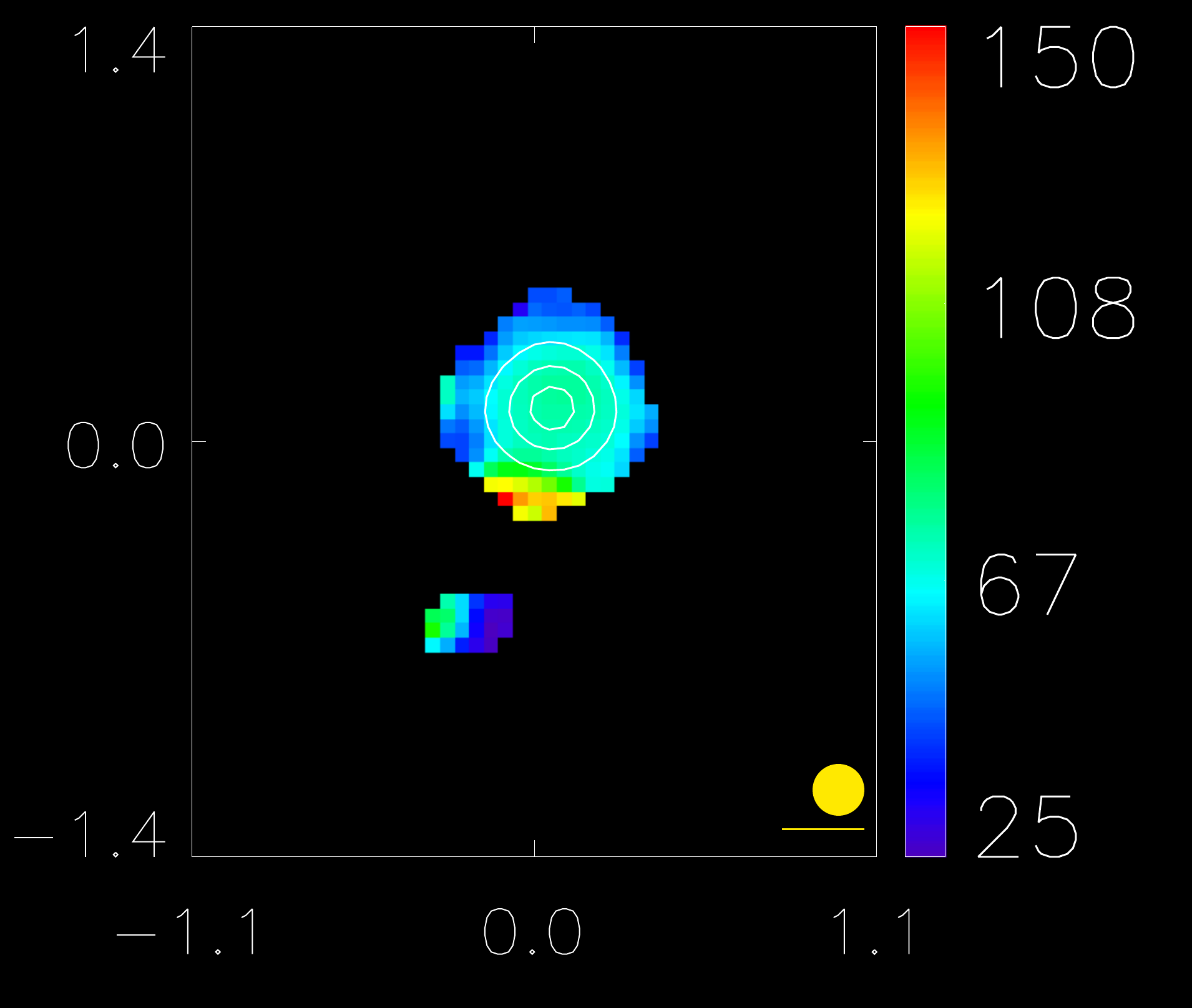} \hskip.05in
\vskip .1 in
\includegraphics[width=.19\linewidth]{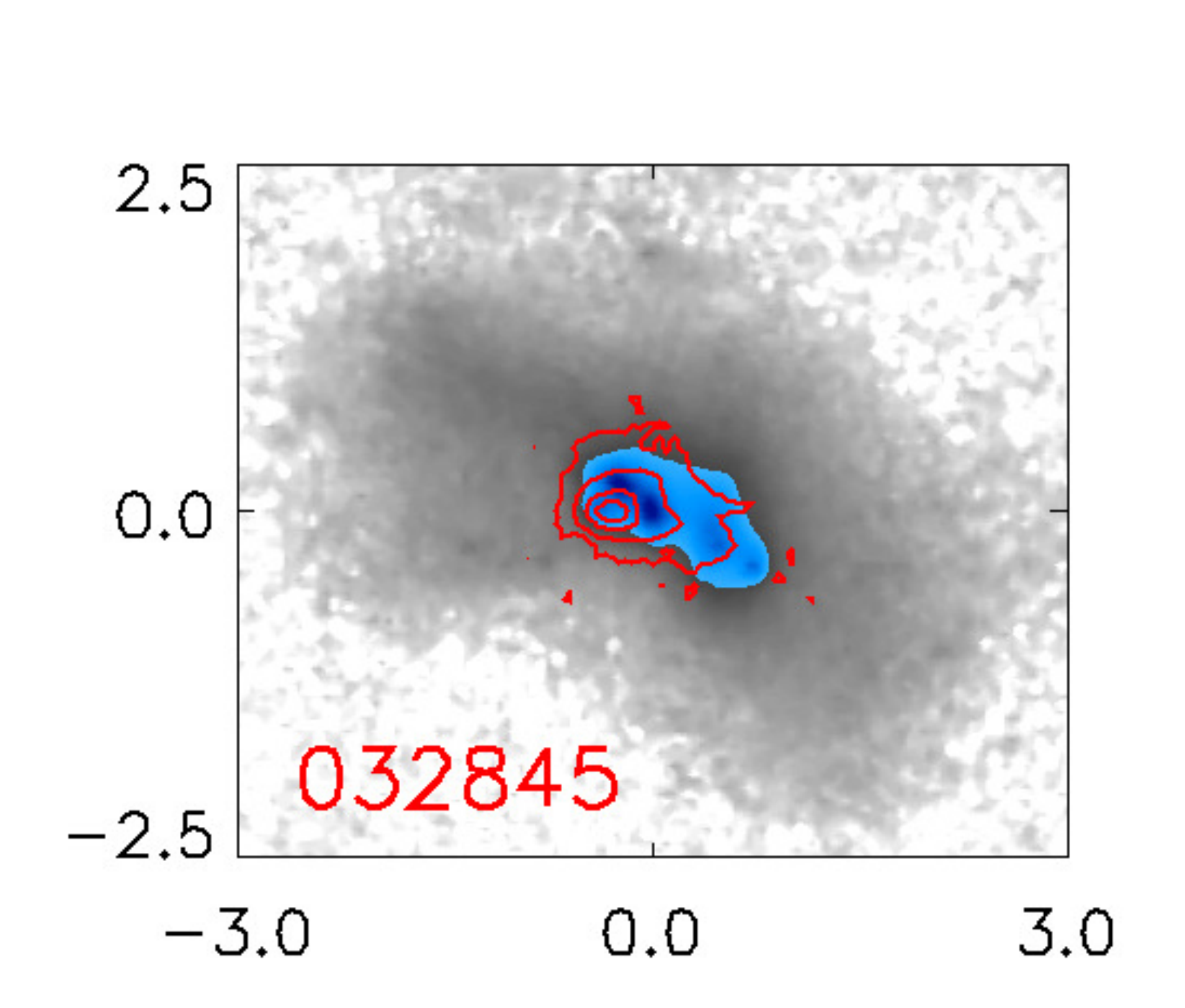} \hskip.05in
\includegraphics[width=.19\linewidth]{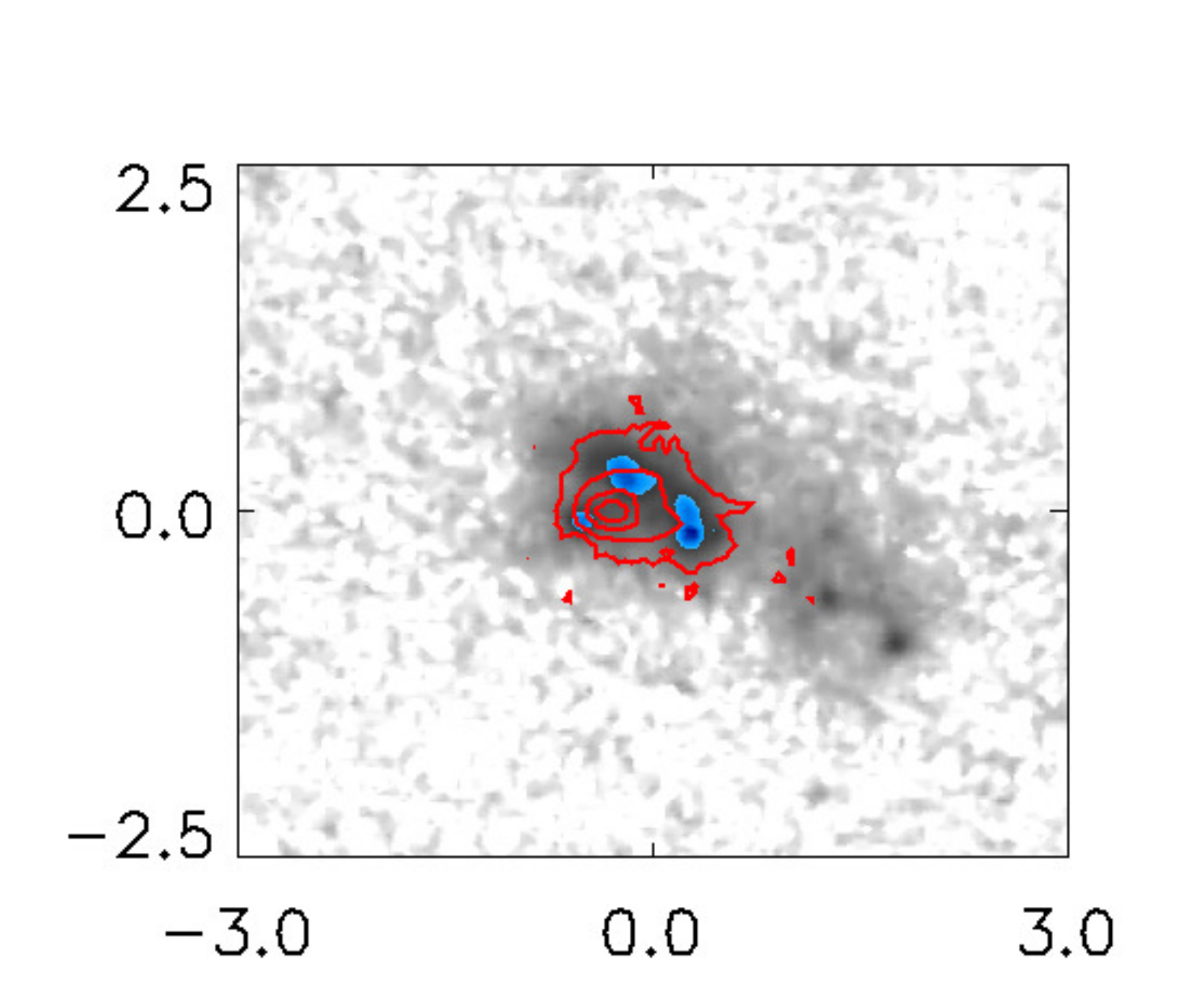} \hskip.05in
\includegraphics[width=.19\linewidth]{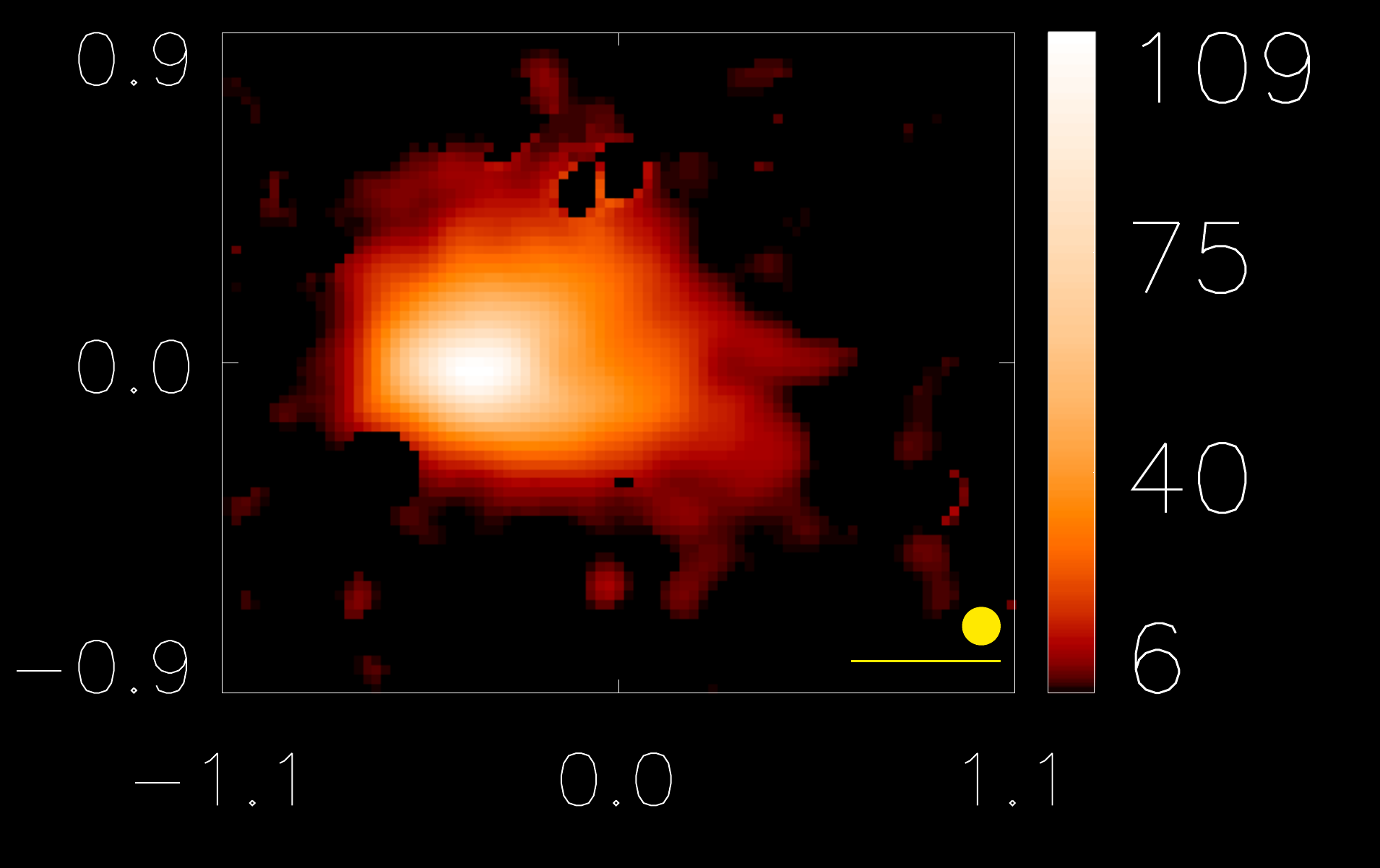} \hskip.05in
\includegraphics[width=.19\linewidth]{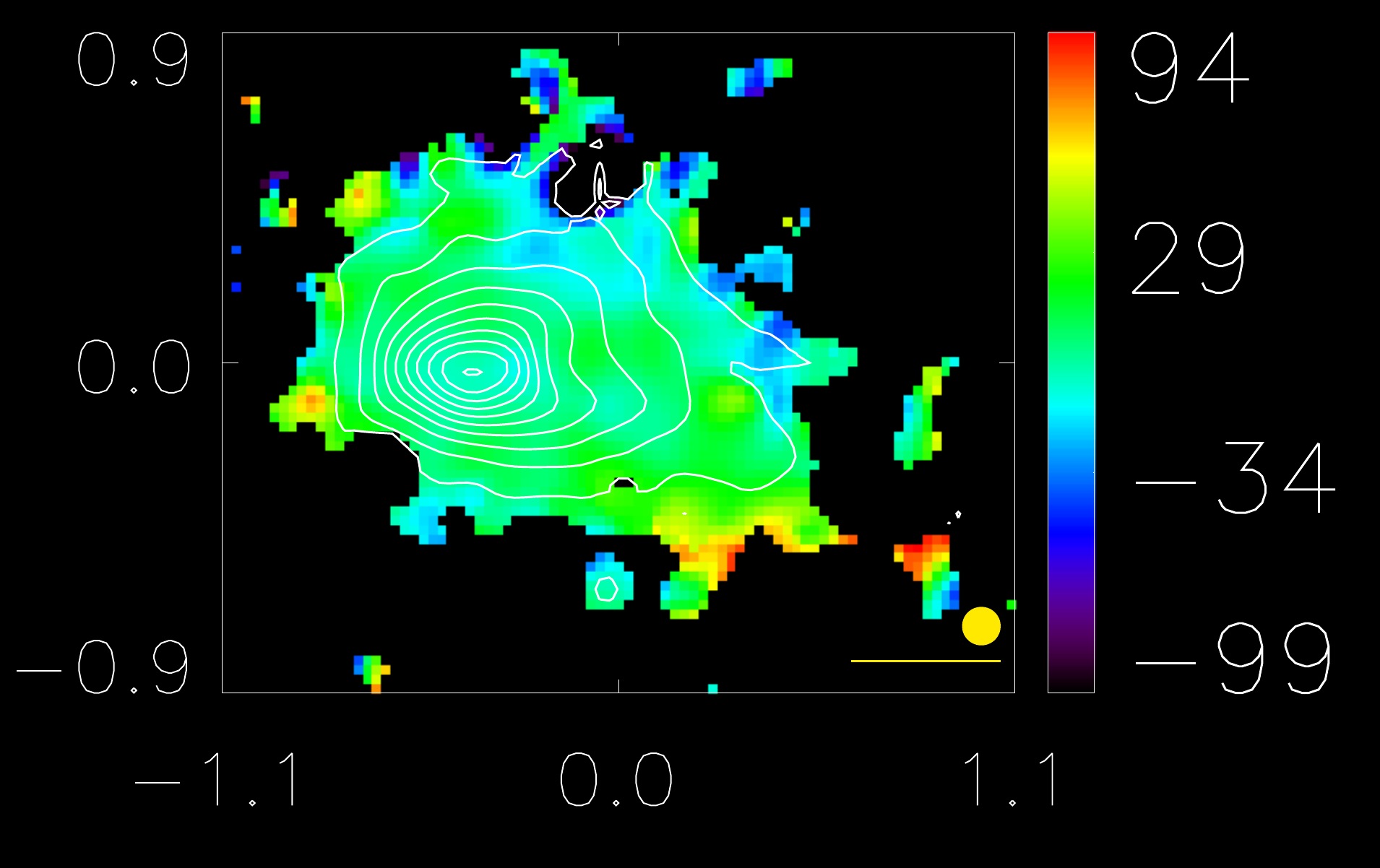} \hskip.05in
\includegraphics[width=.19\linewidth]{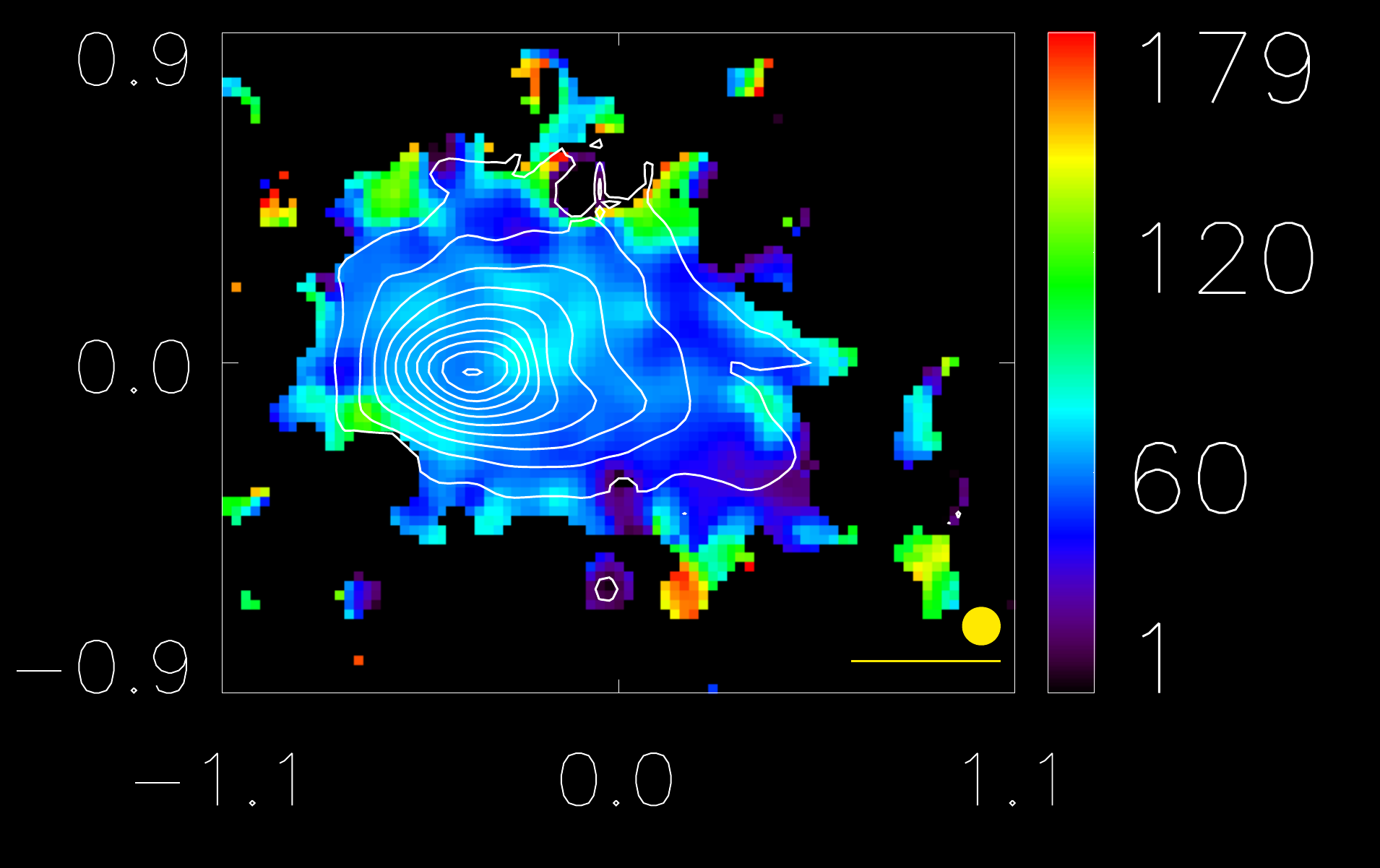} \hskip.05in
\vskip .1 in
\includegraphics[width=.19\linewidth]{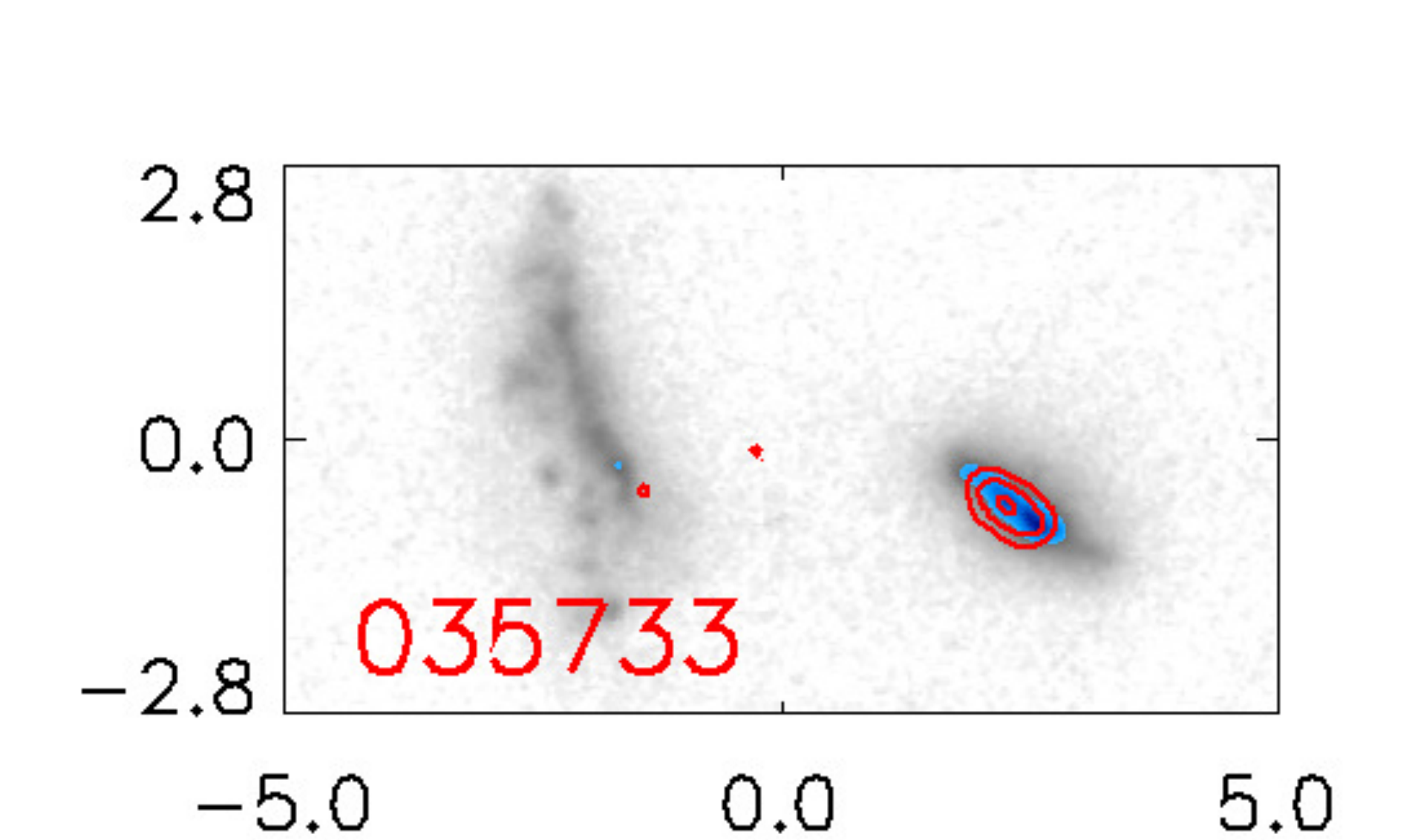} \hskip.05in
\includegraphics[width=.19\linewidth]{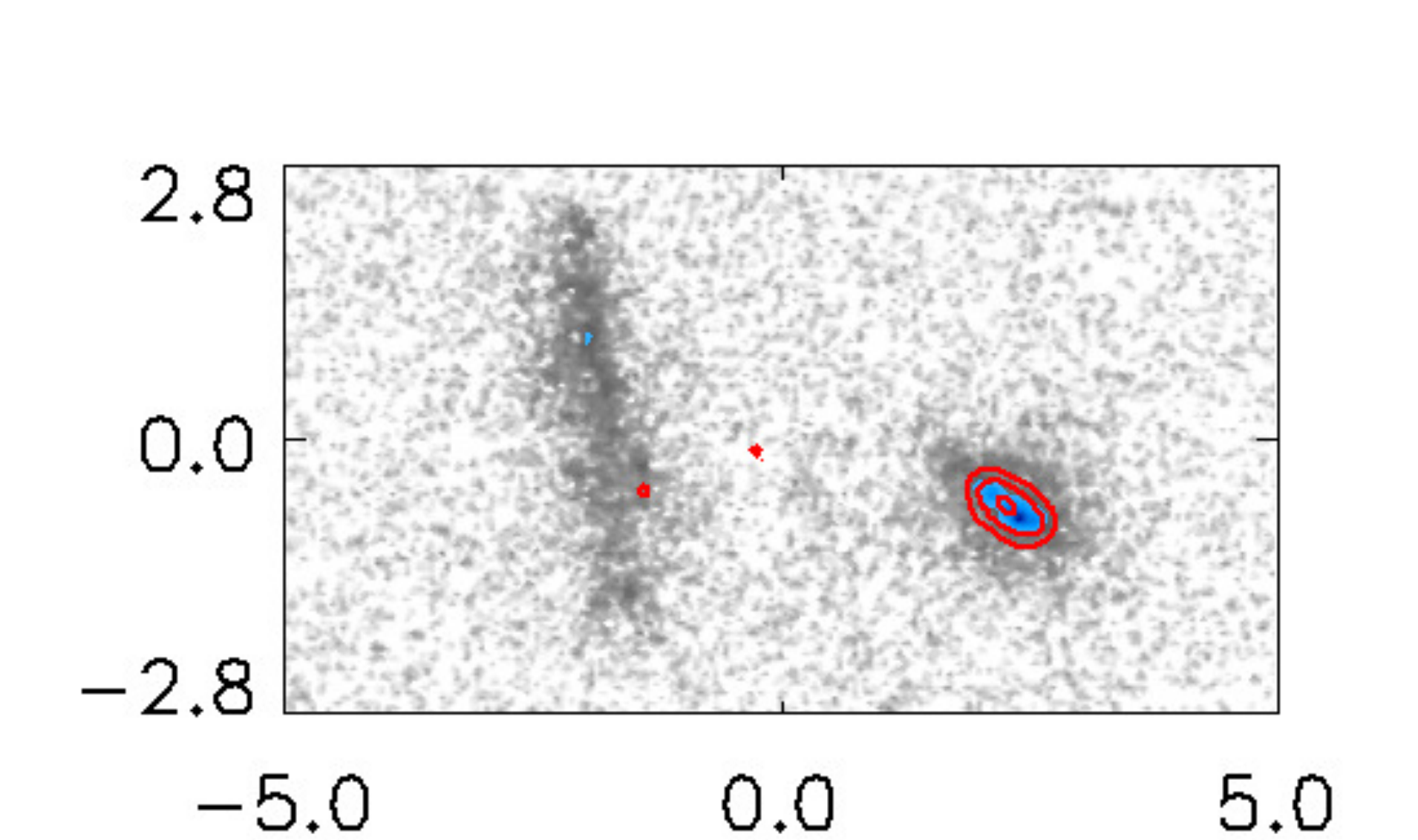} \hskip.05in
\includegraphics[width=.19\linewidth]{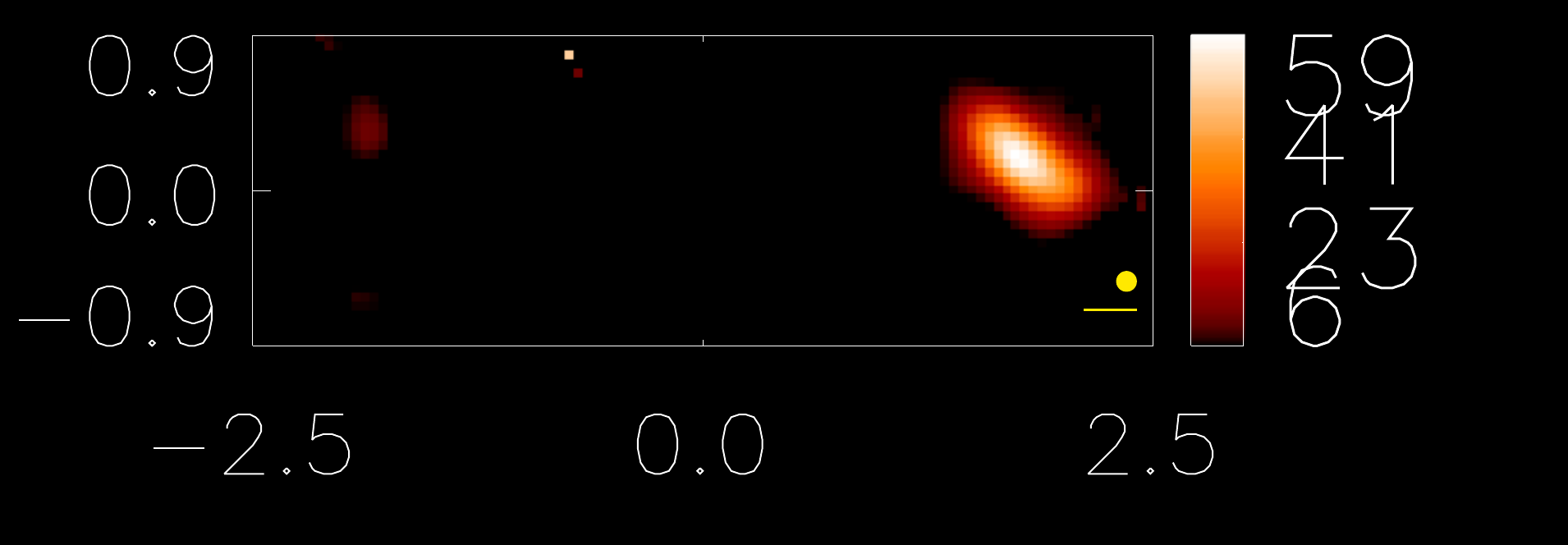} \hskip.05in
\includegraphics[width=.19\linewidth]{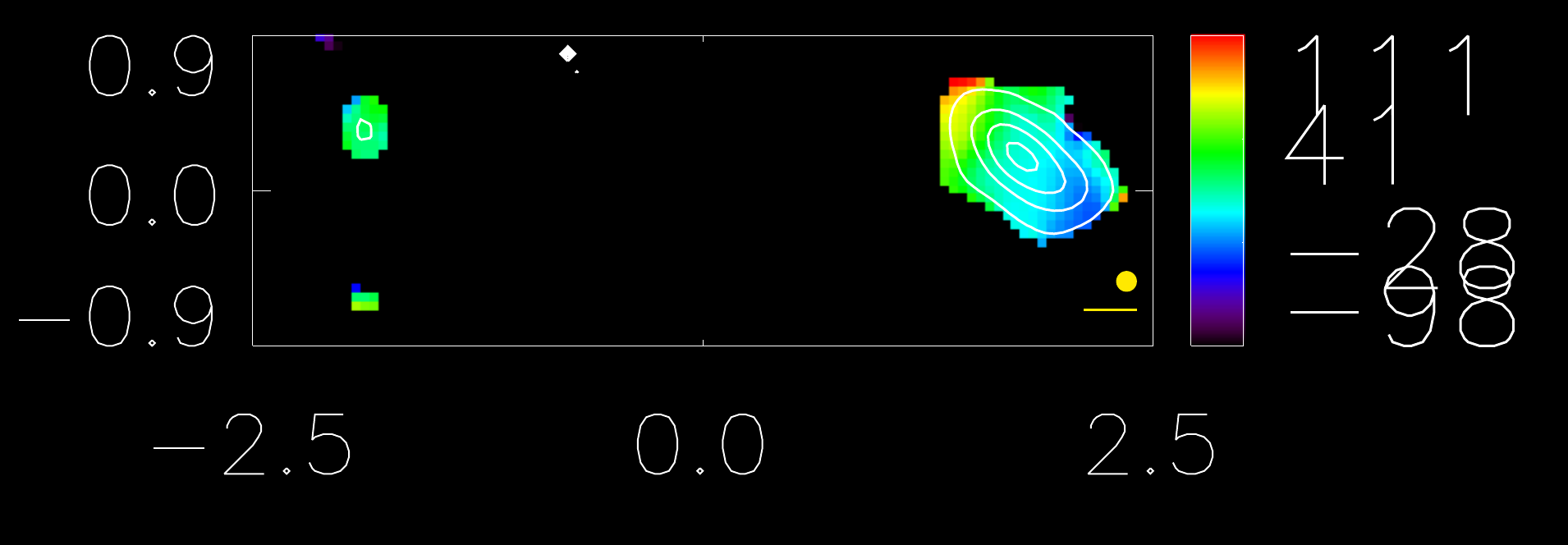} \hskip.05in
\includegraphics[width=.19\linewidth]{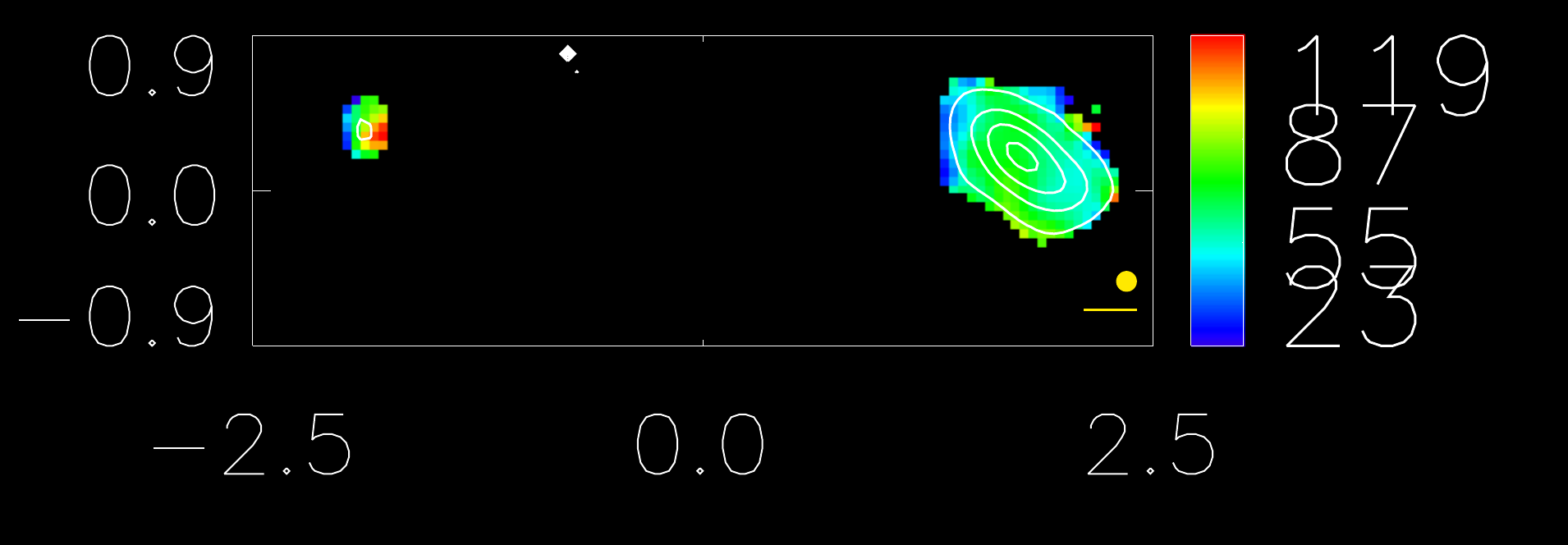} \hskip.05in
\vskip .1 in
\includegraphics[width=.19\linewidth]{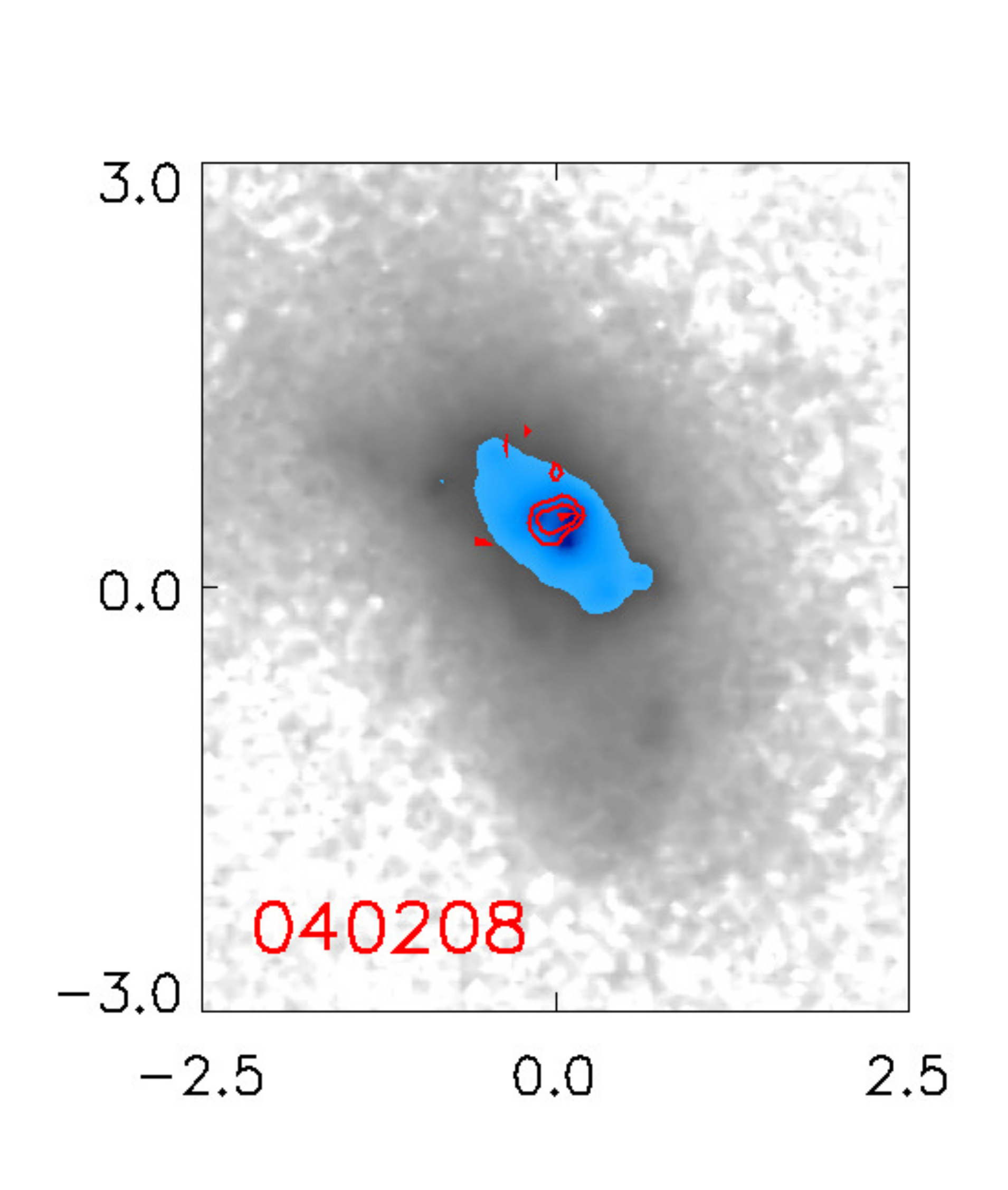} \hskip.05in
\includegraphics[width=.19\linewidth]{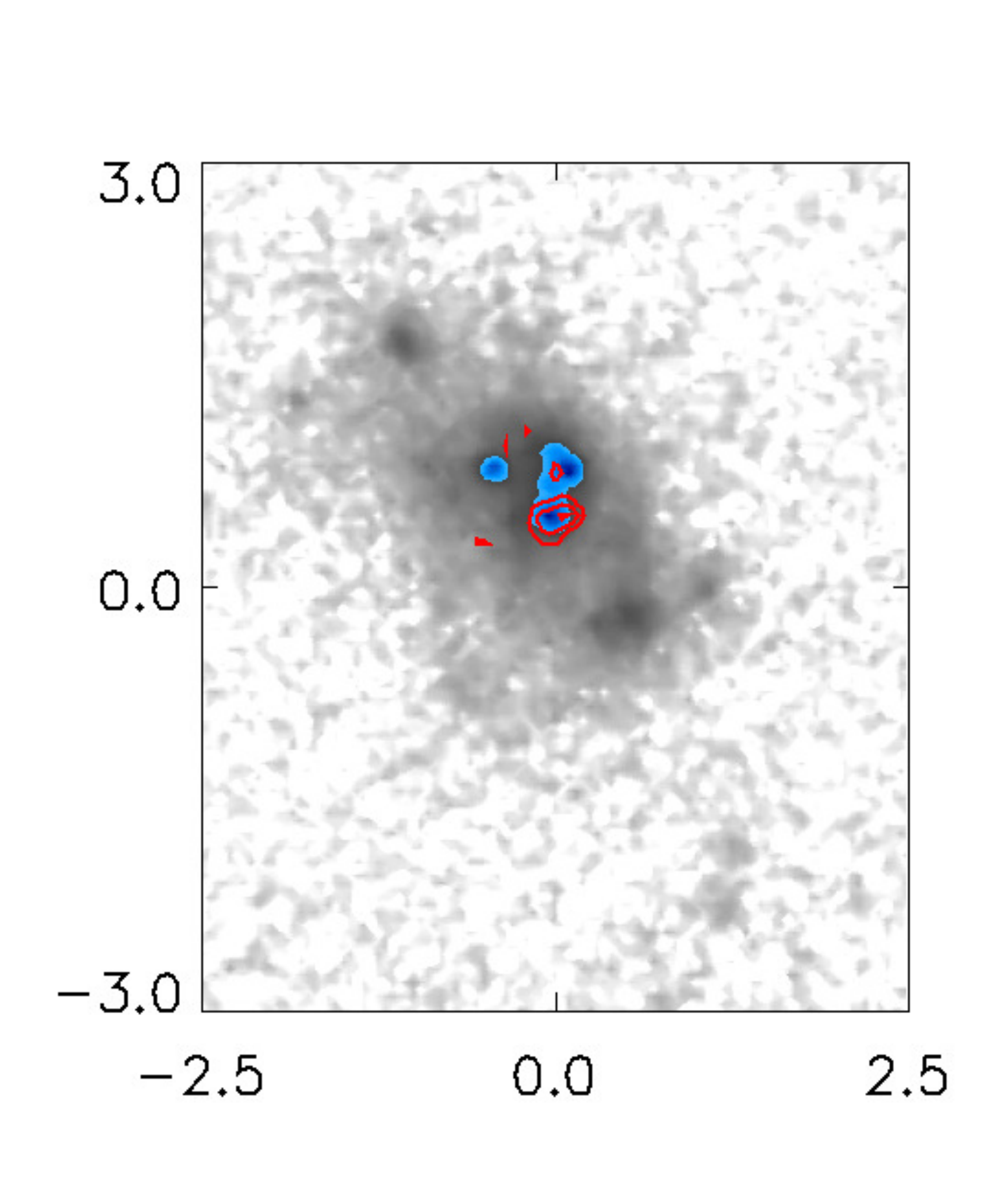} \hskip.05in
\includegraphics[width=.19\linewidth]{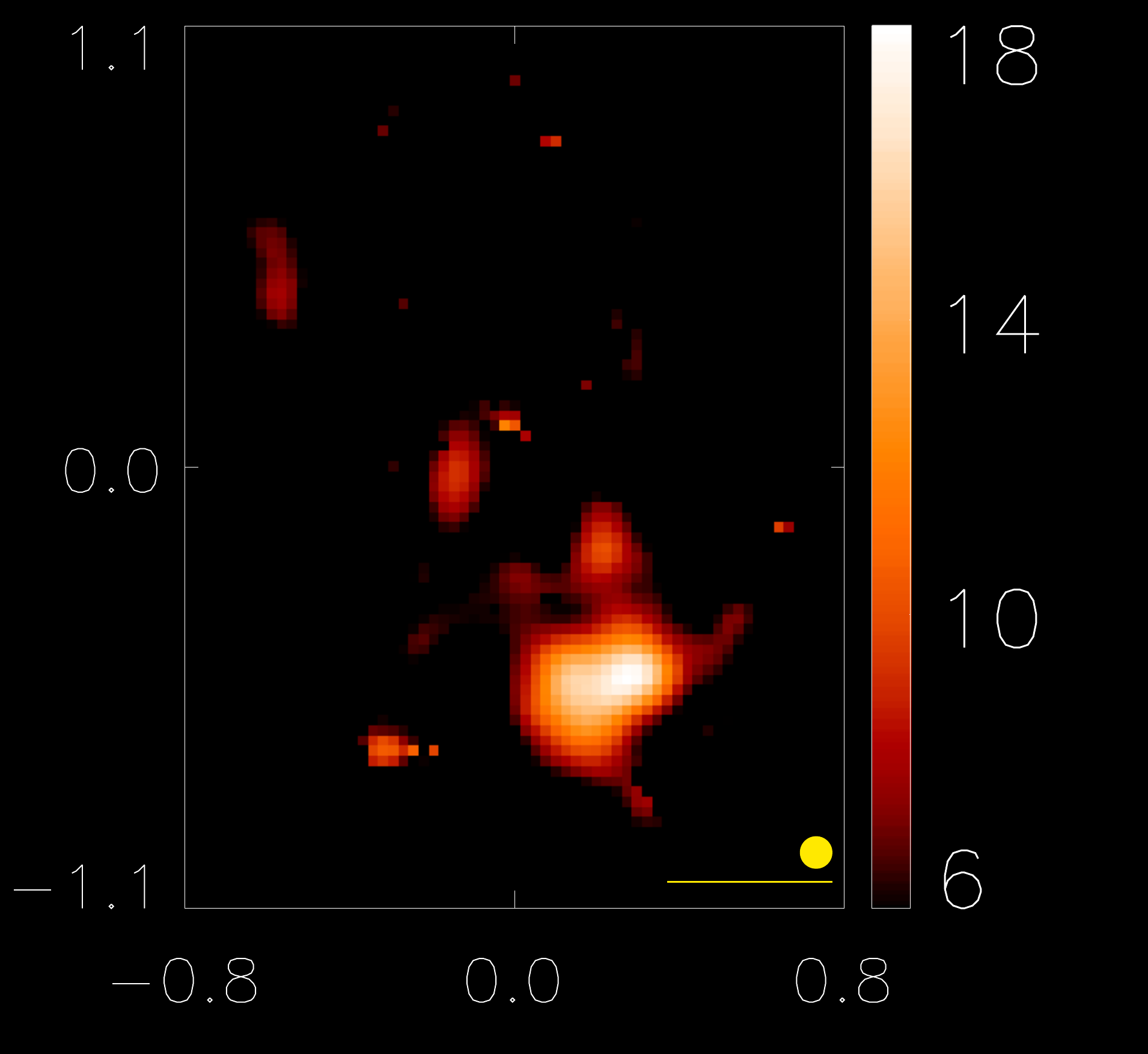} \hskip.05in
\includegraphics[width=.19\linewidth]{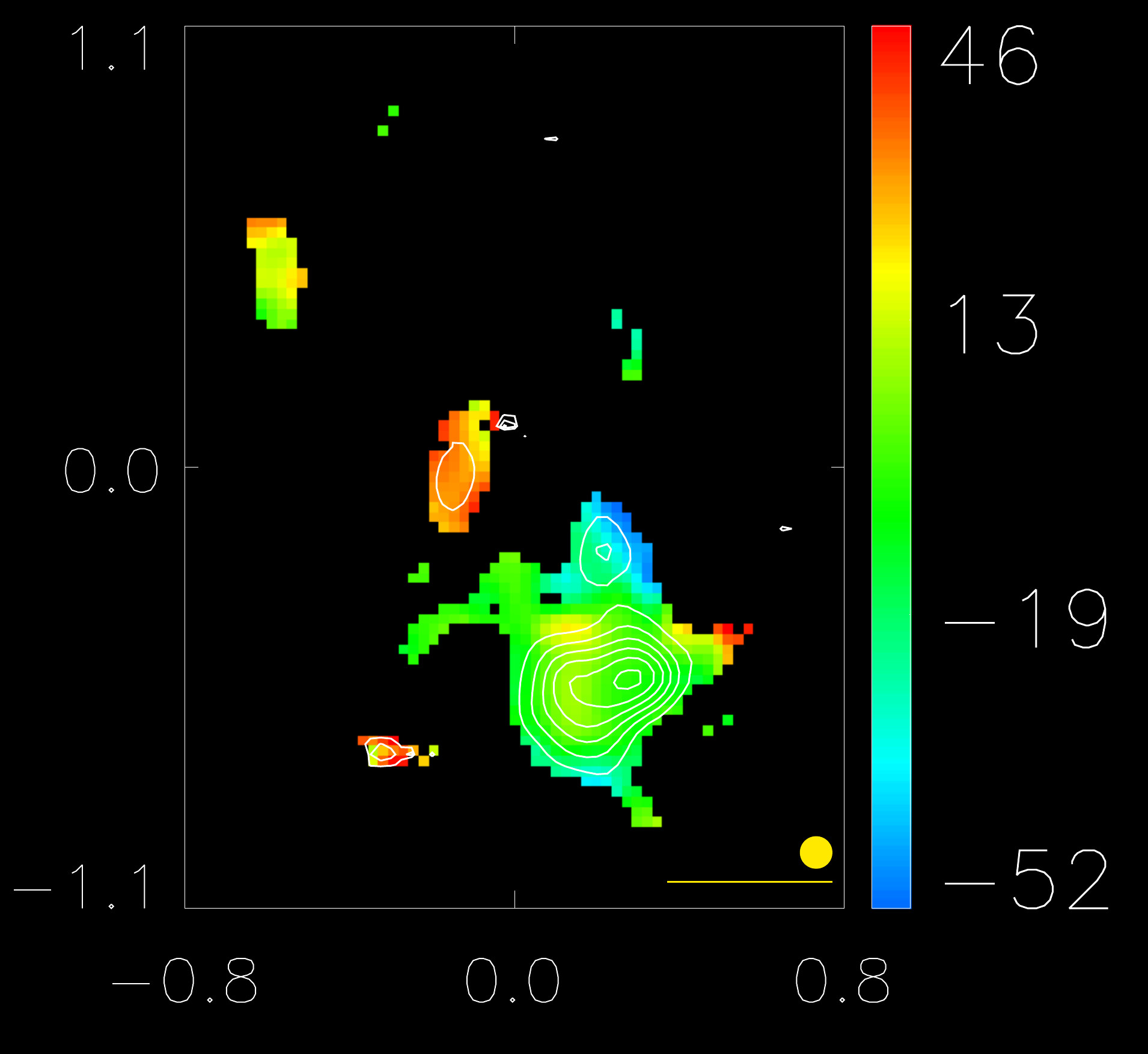} \hskip.05in
\includegraphics[width=.19\linewidth]{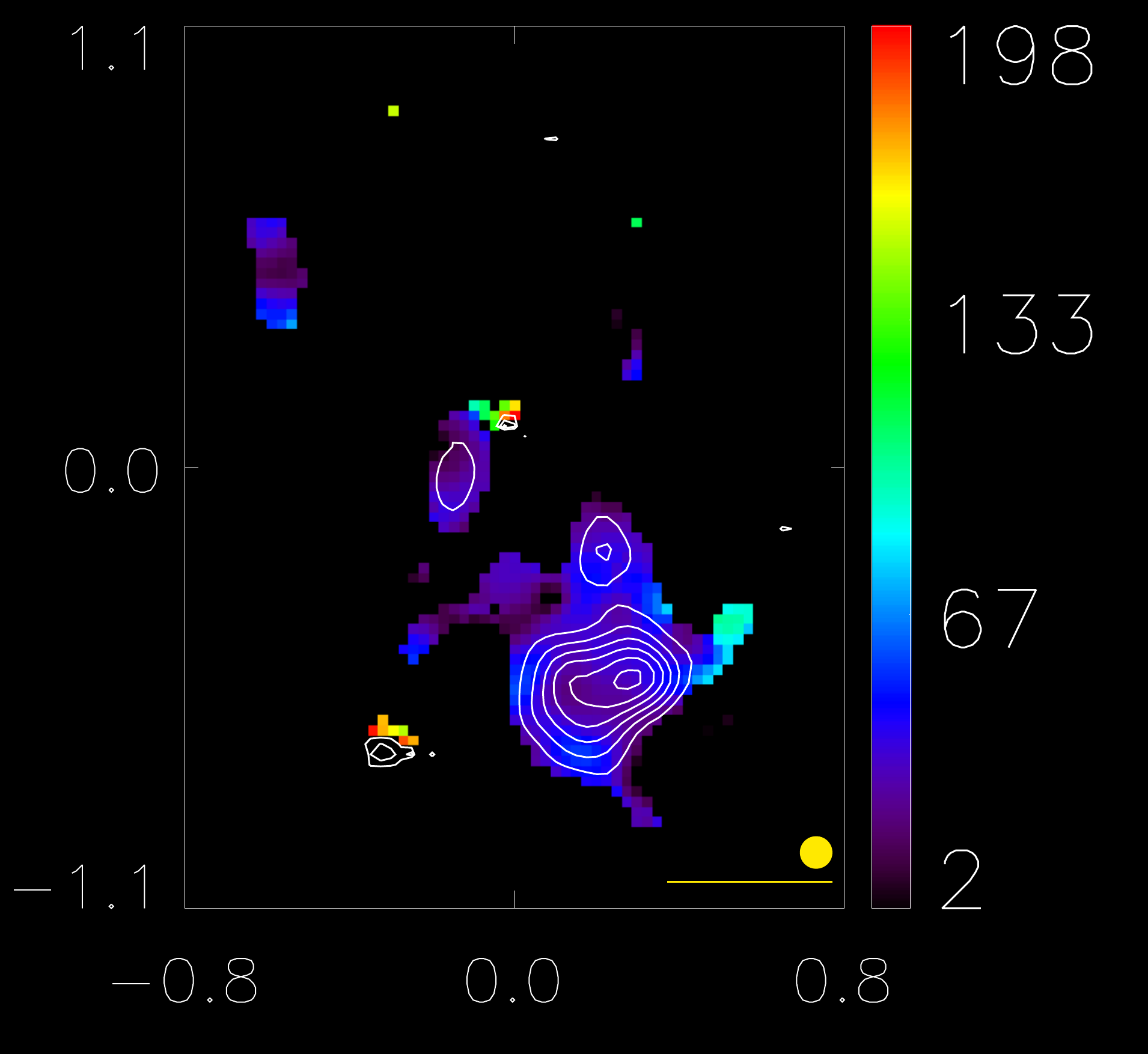} \hskip.05in

\caption{
We show here the velocity moment maps for all galaxies observed for this work. The two leftmost figures show  the HST rest-frame optical ({\rm left}) and UV ({\rm right}) morphologies, with logarithmic (black) and linear (blue) stretches. The Pa-$\alpha$ S/N levels are overlaid in red. There is no UV image available for 101211. The following images show, from left to right, the signal-to-noise ratios, line-of-sight velocity in km s$^{-1}$ and line-of-sight velocity dispersion, also in km s$^{-1}$. For the latter two we overplot S/N contours in white. The axes show the angular scale in arcsec; orientation is the same in every panel, with north pointing up and east to the left. We indicate in each panel the FWHM of a point source as a proxy for spatial resolution and the physical scale corresponding to 1kpc at the redshift of each galaxy. \label{fig:vdmaps}
}
\end{figure*}

\addtocounter{figure}{-1}
\begin{figure*}[ht]
\includegraphics[width=.19\linewidth]{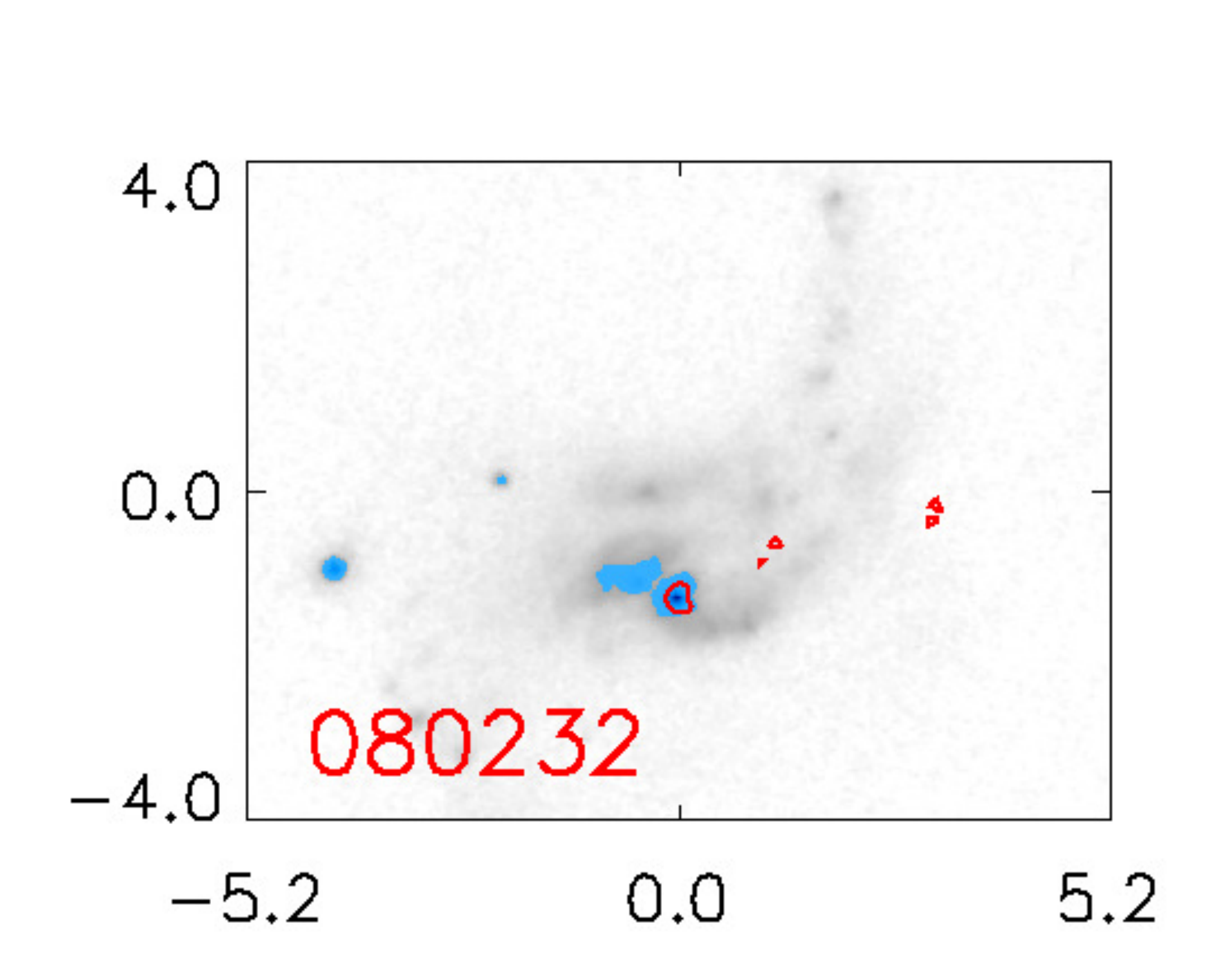} \hskip.05in
\includegraphics[width=.19\linewidth]{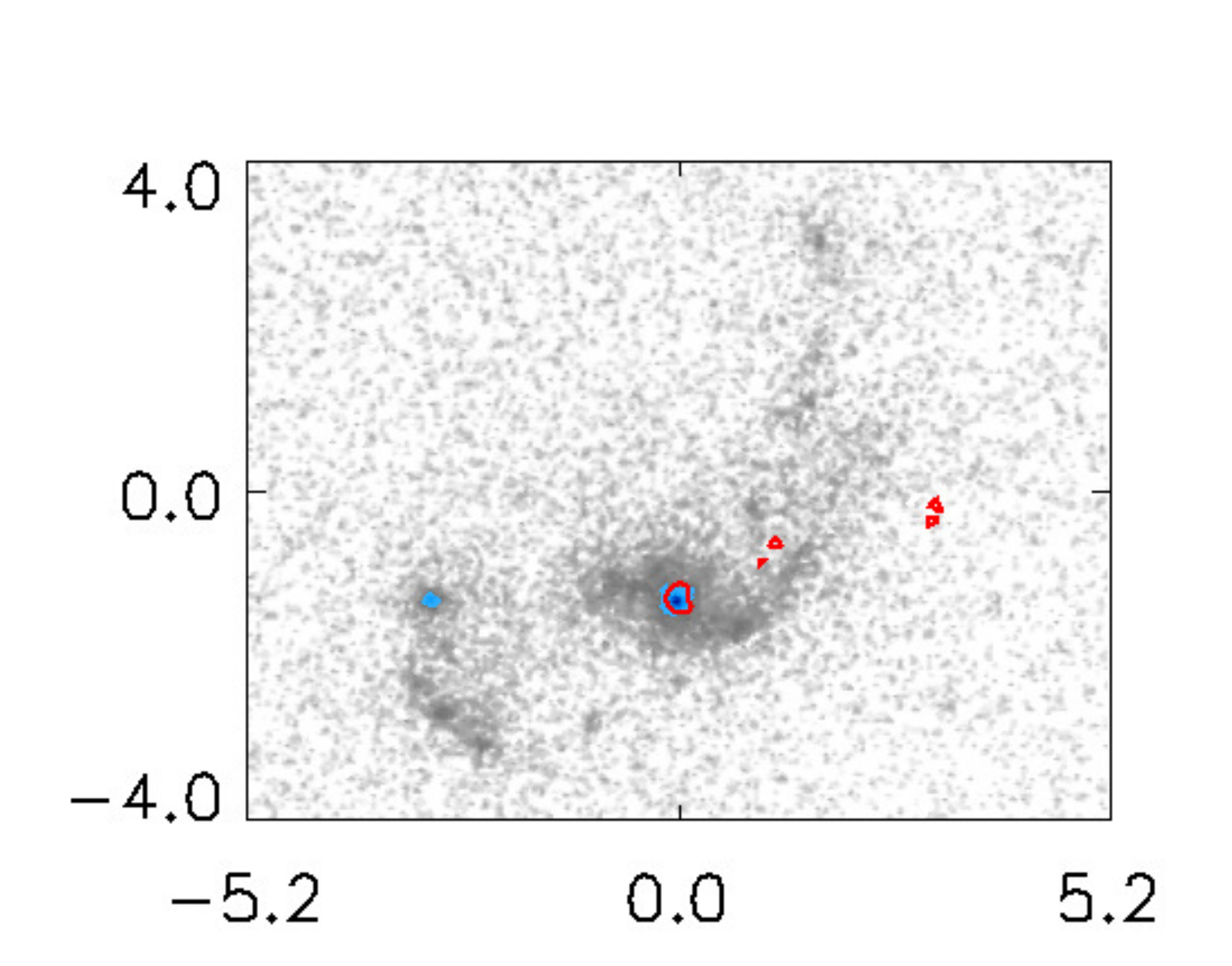} \hskip.05in
\includegraphics[width=.19\linewidth]{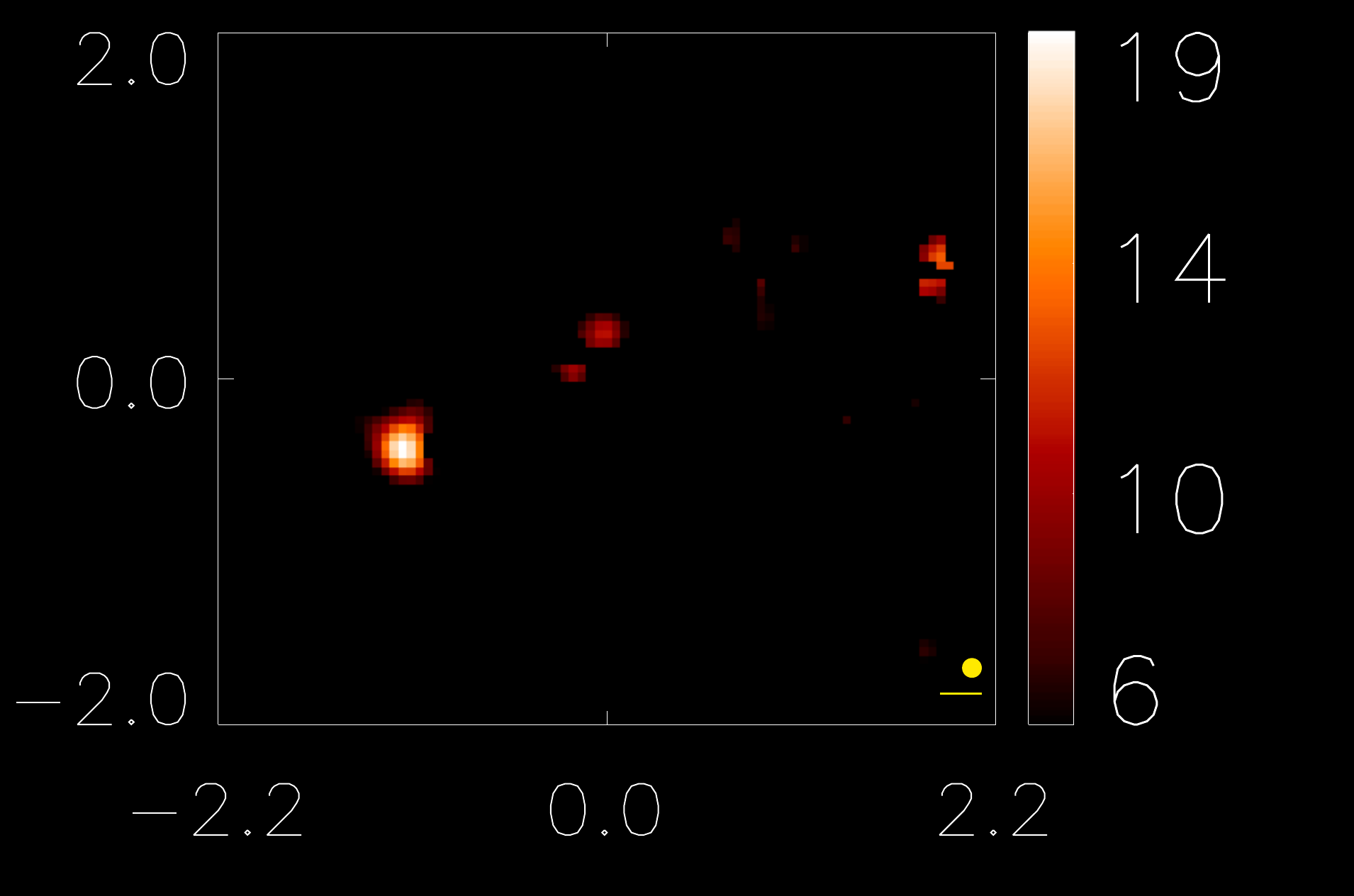} \hskip.05in
\includegraphics[width=.19\linewidth]{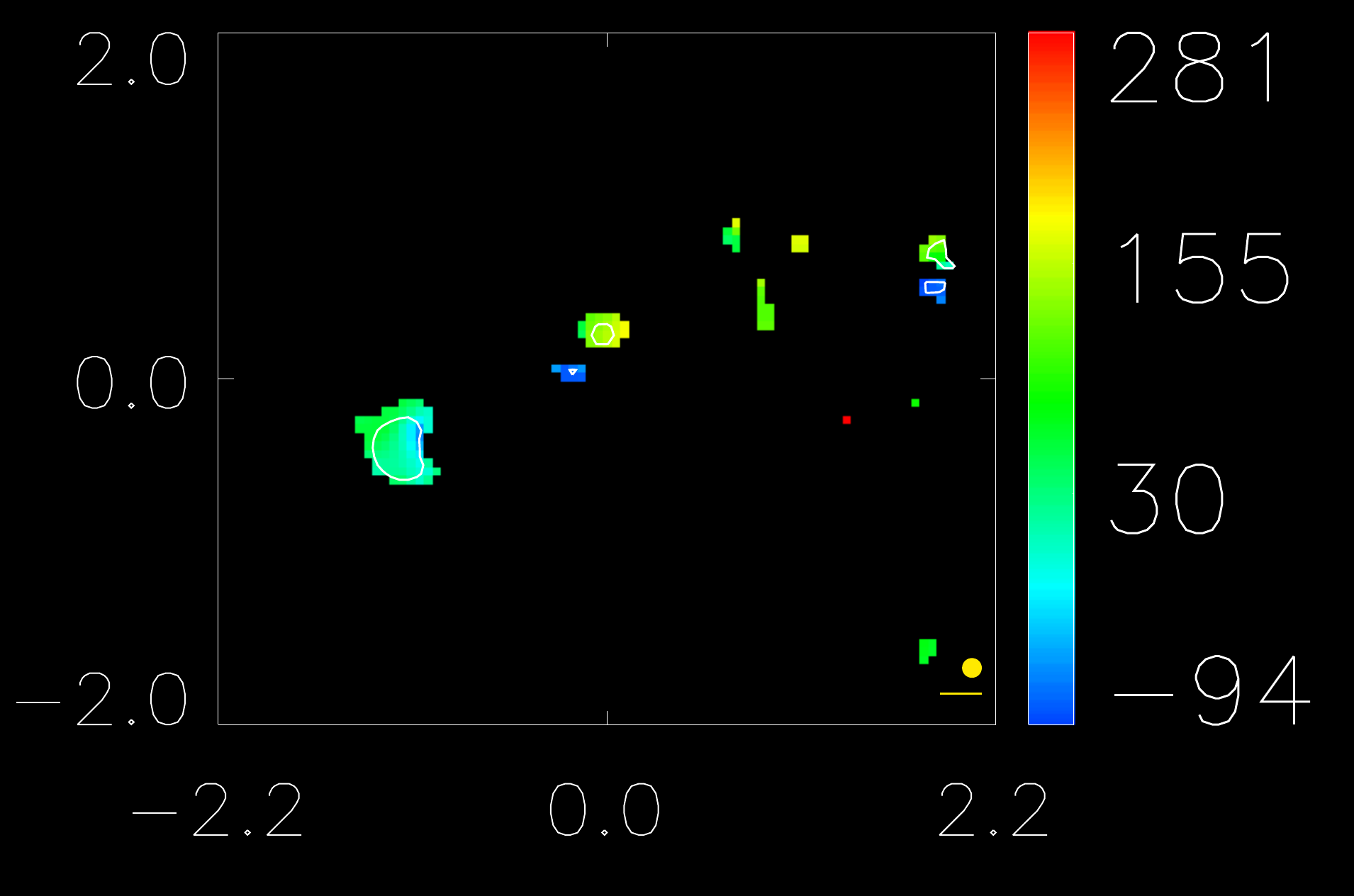} \hskip.05in
\includegraphics[width=.19\linewidth]{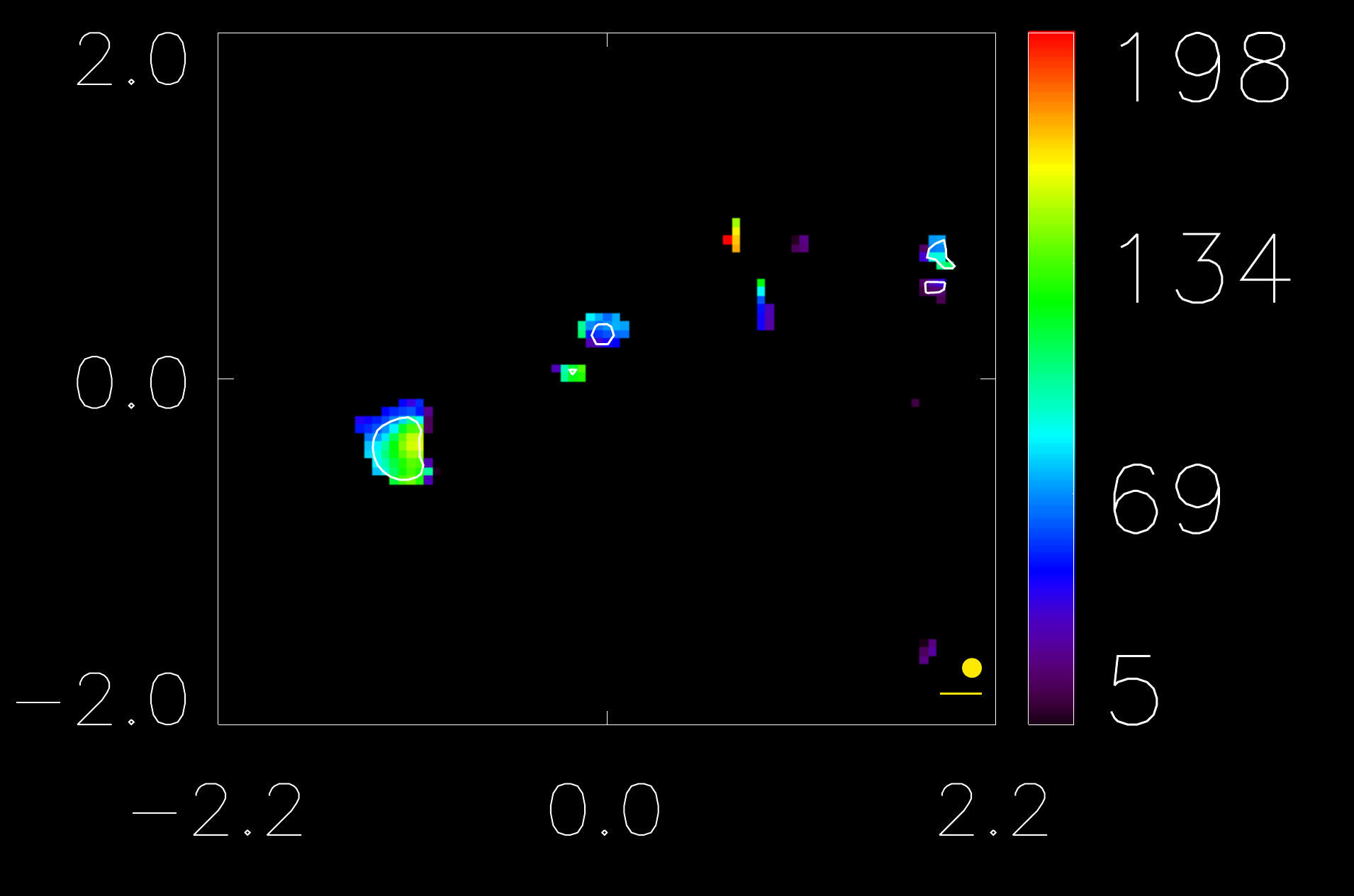} \hskip.05in
\vskip .1 in
\includegraphics[width=.19\linewidth]{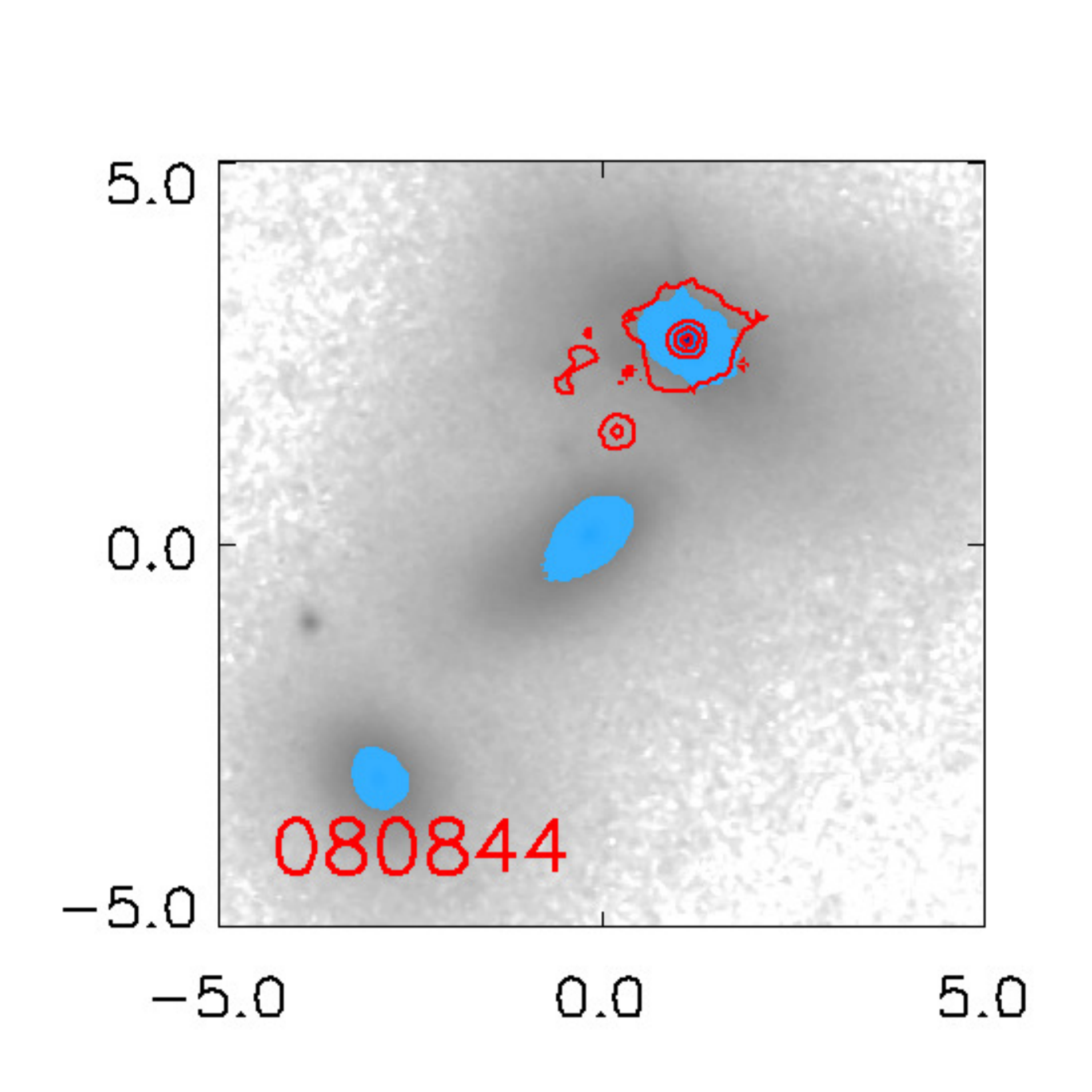} \hskip.05in
\includegraphics[width=.19\linewidth]{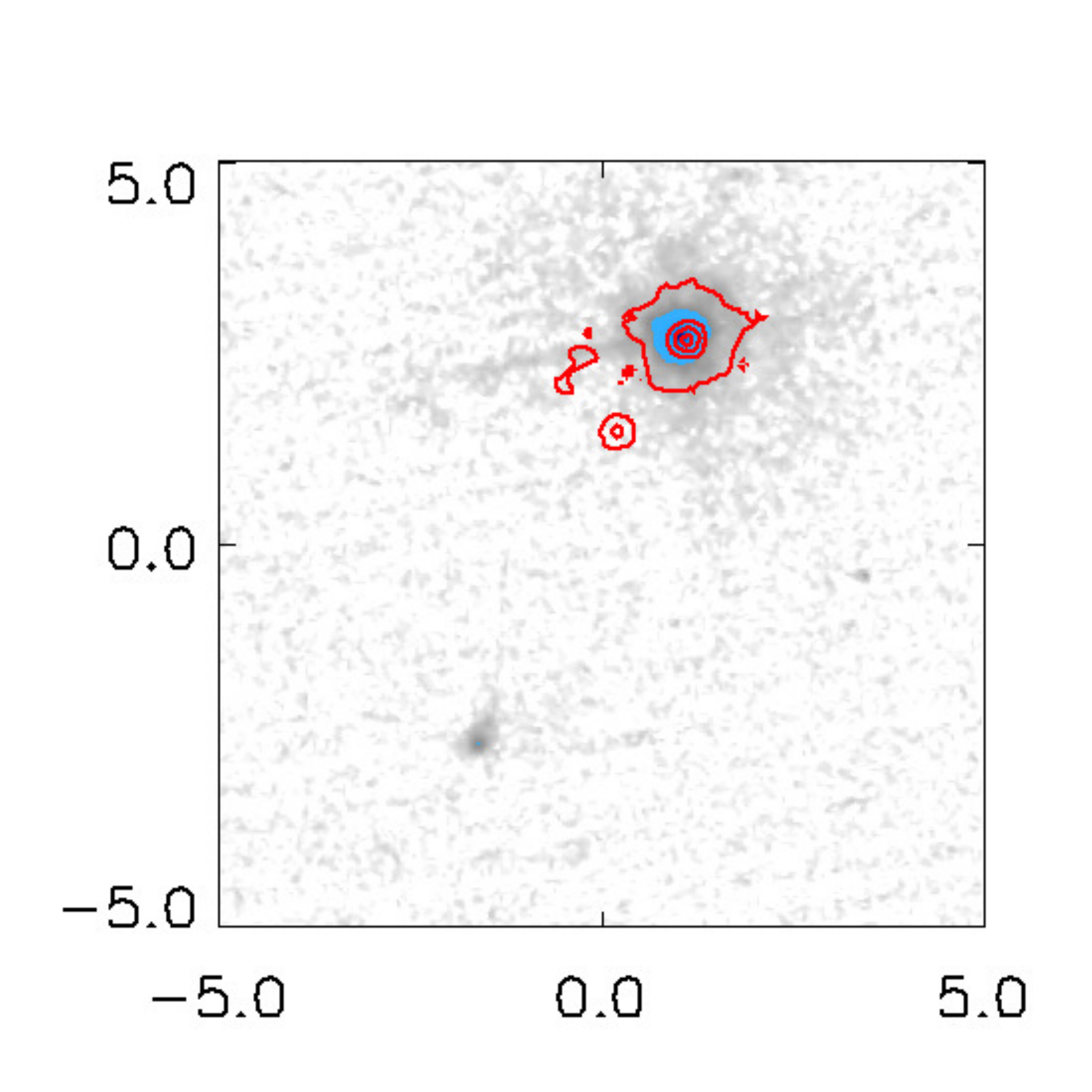} \hskip.05in
\includegraphics[width=.19\linewidth]{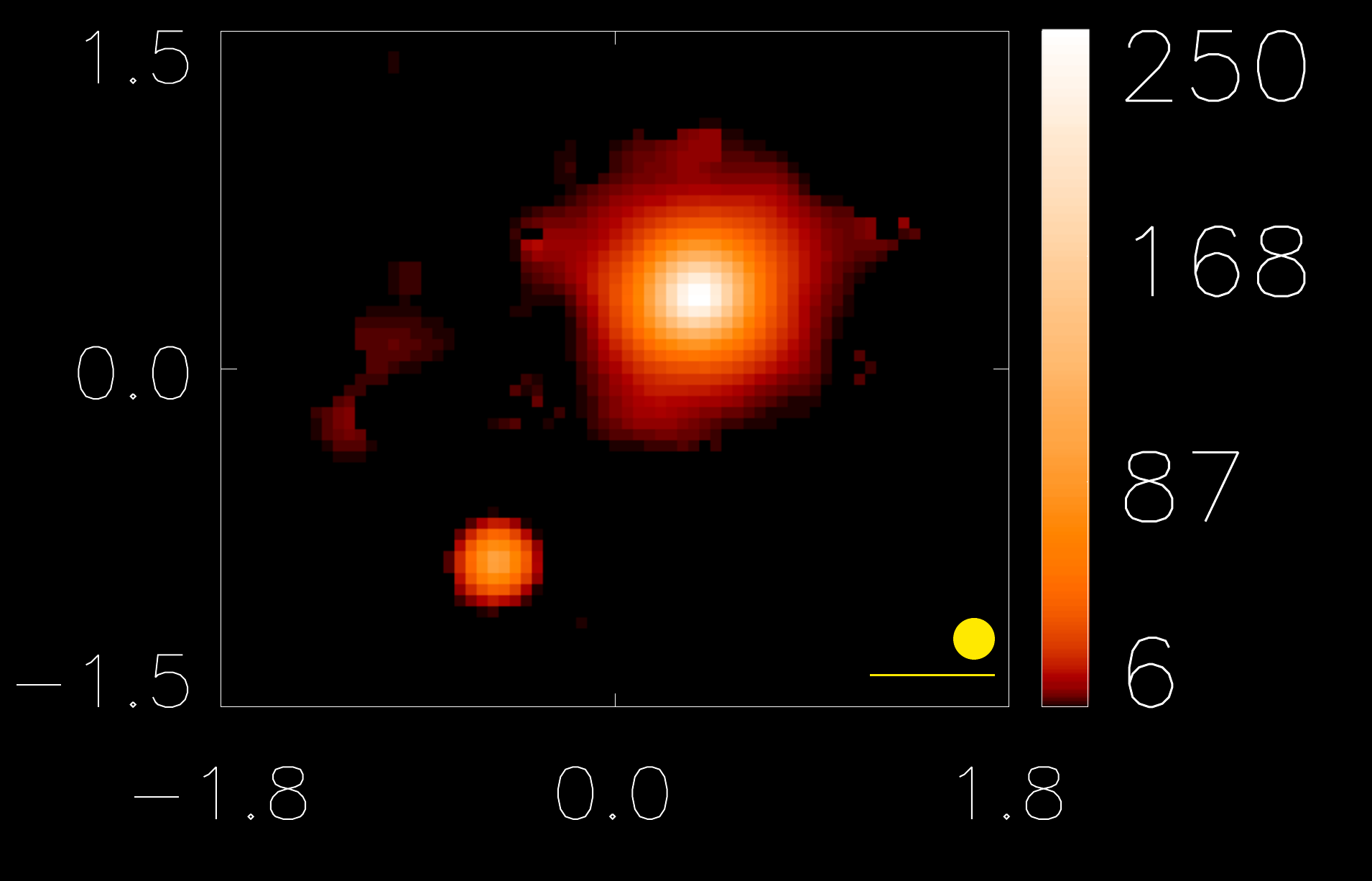} \hskip.05in
\includegraphics[width=.19\linewidth]{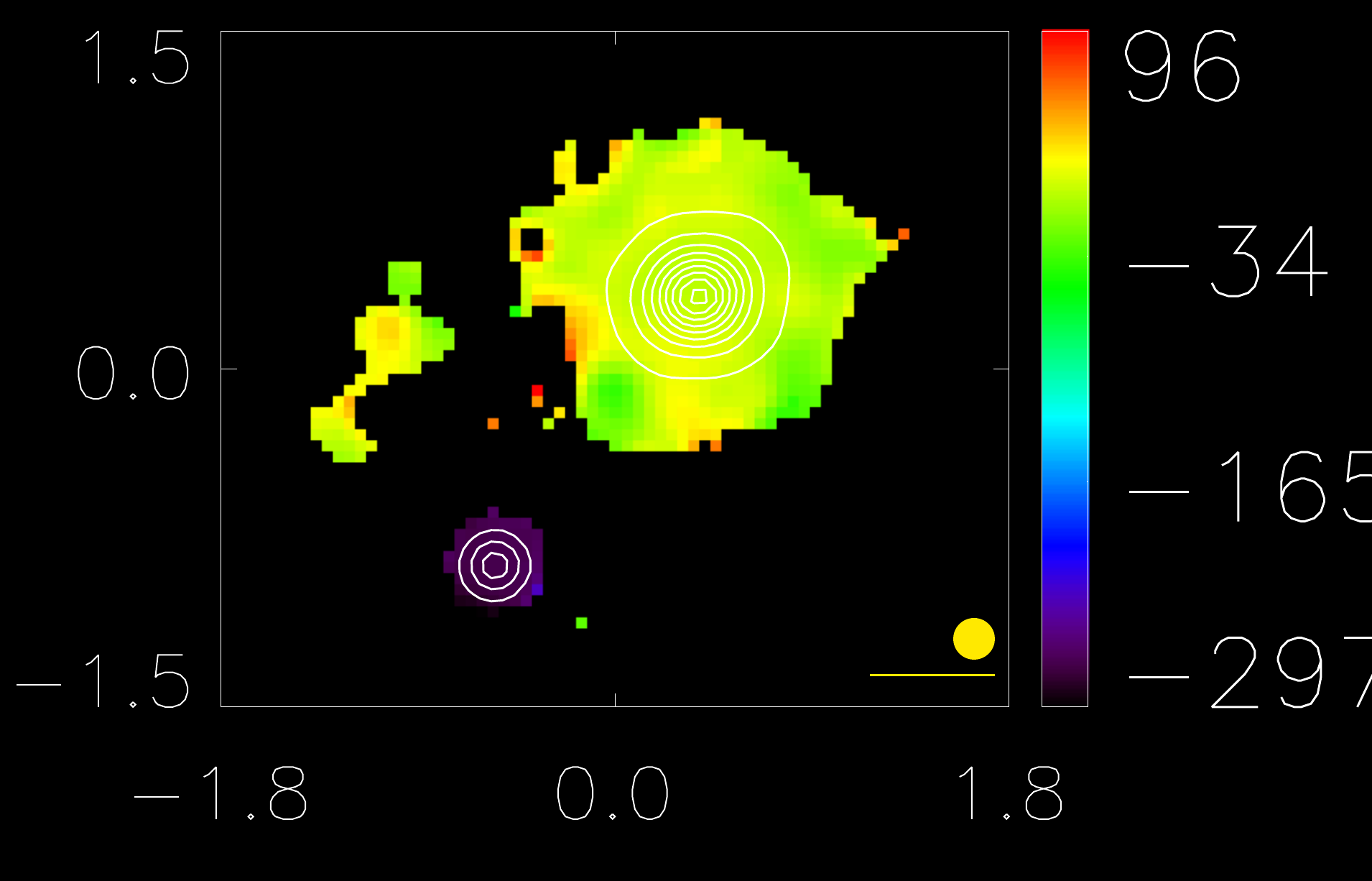} \hskip.05in
\includegraphics[width=.19\linewidth]{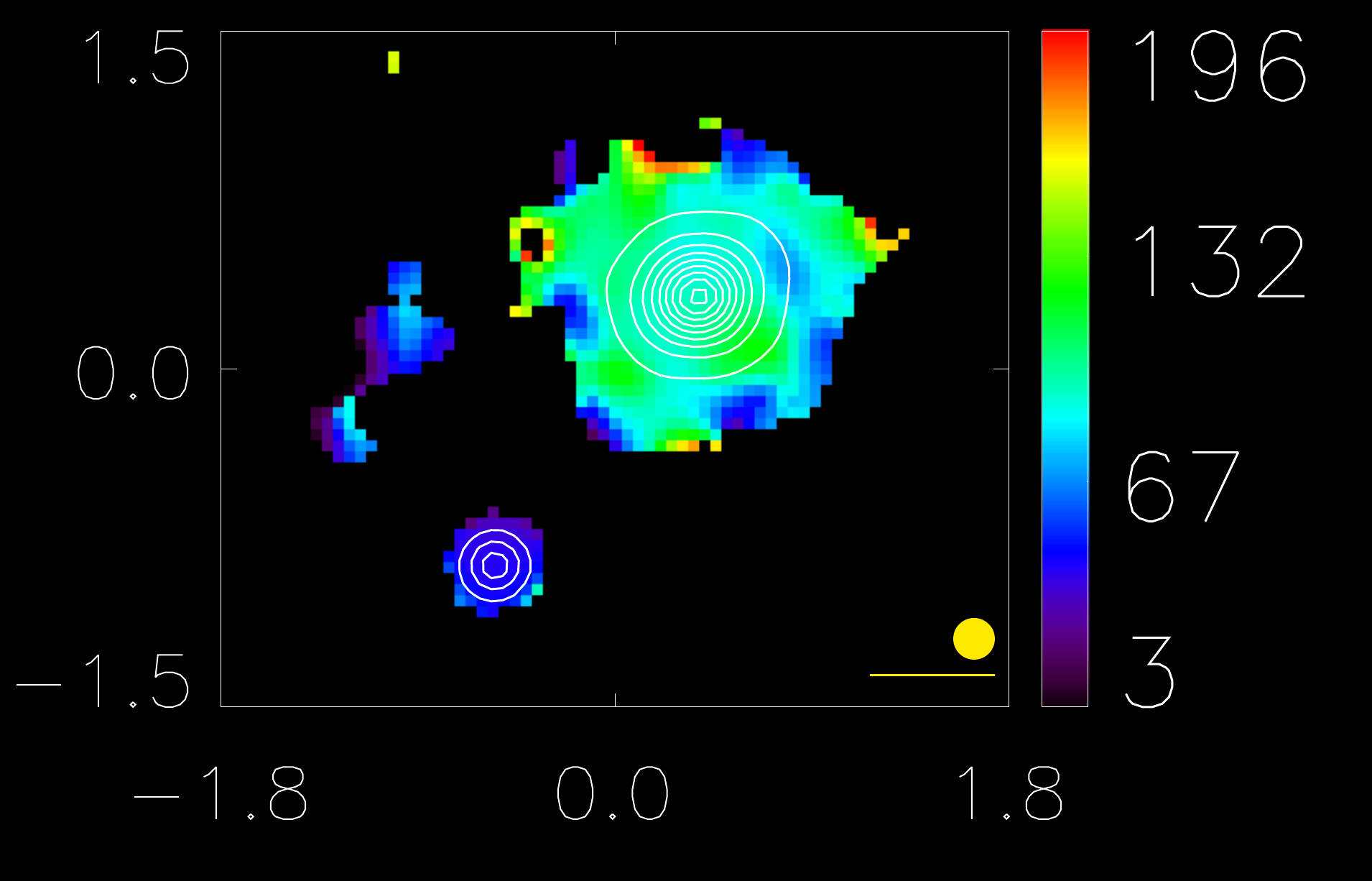} \hskip.05in
\vskip .1 in
\includegraphics[width=.19\linewidth]{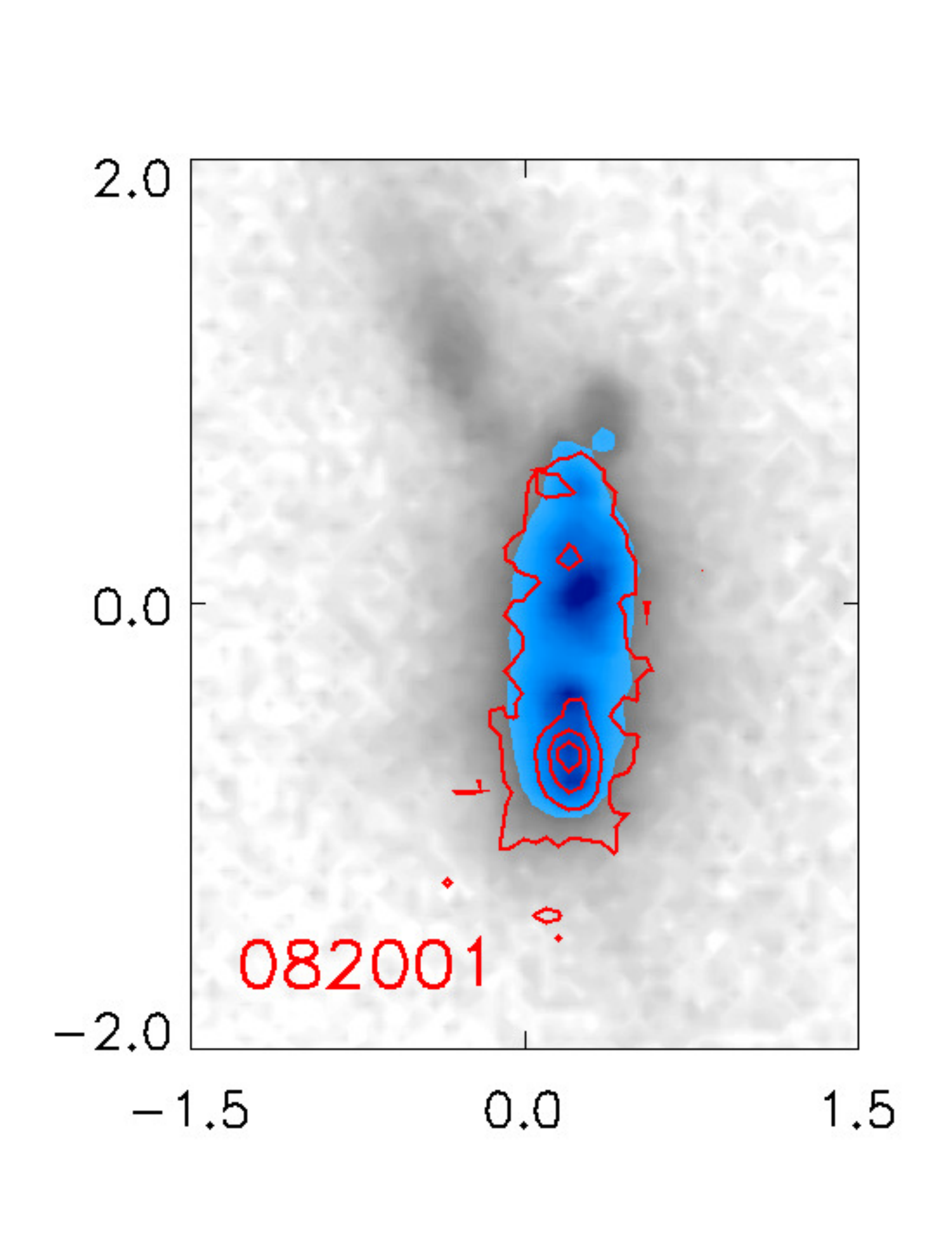} \hskip.05in
\includegraphics[width=.19\linewidth]{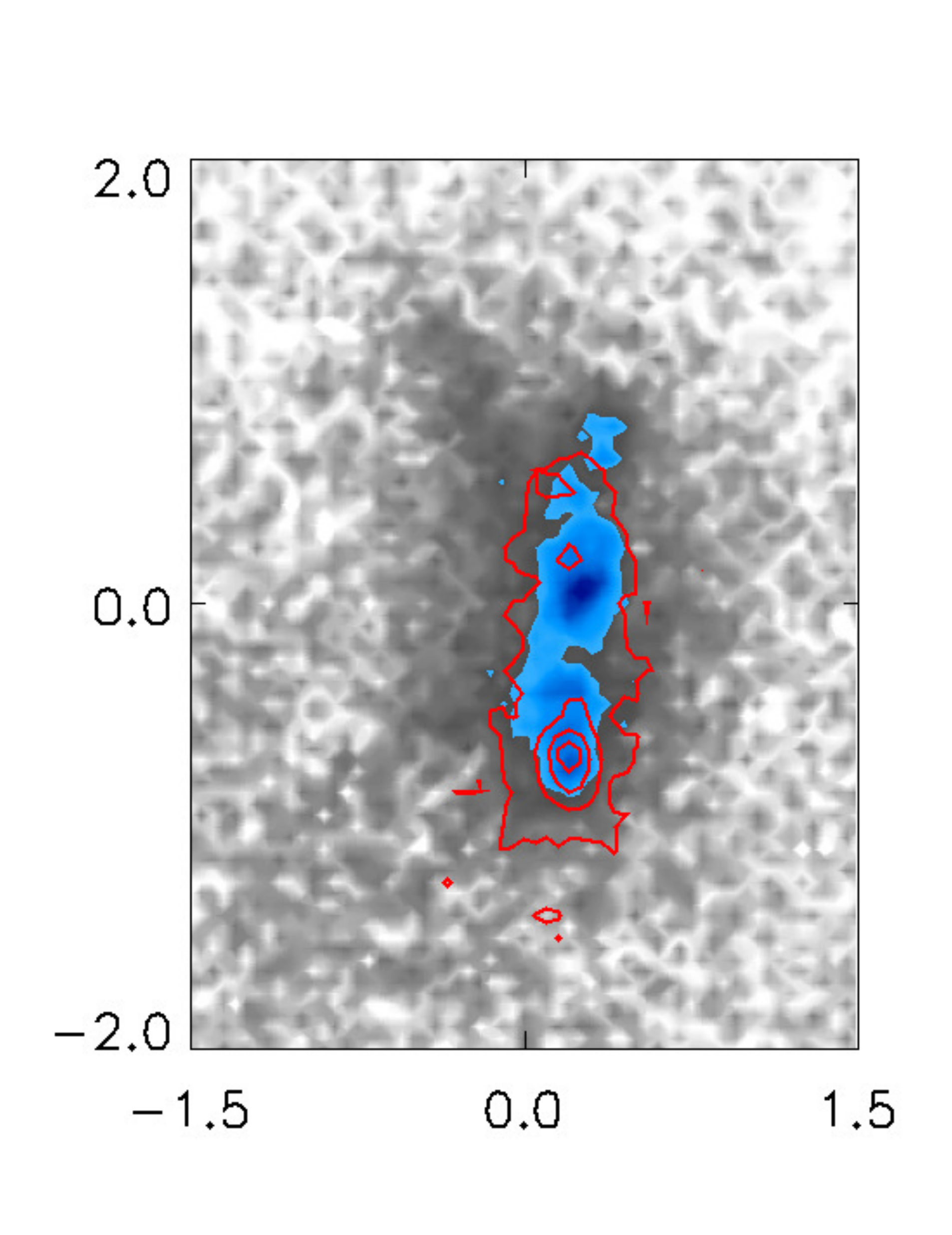} \hskip.05in
\includegraphics[width=.19\linewidth]{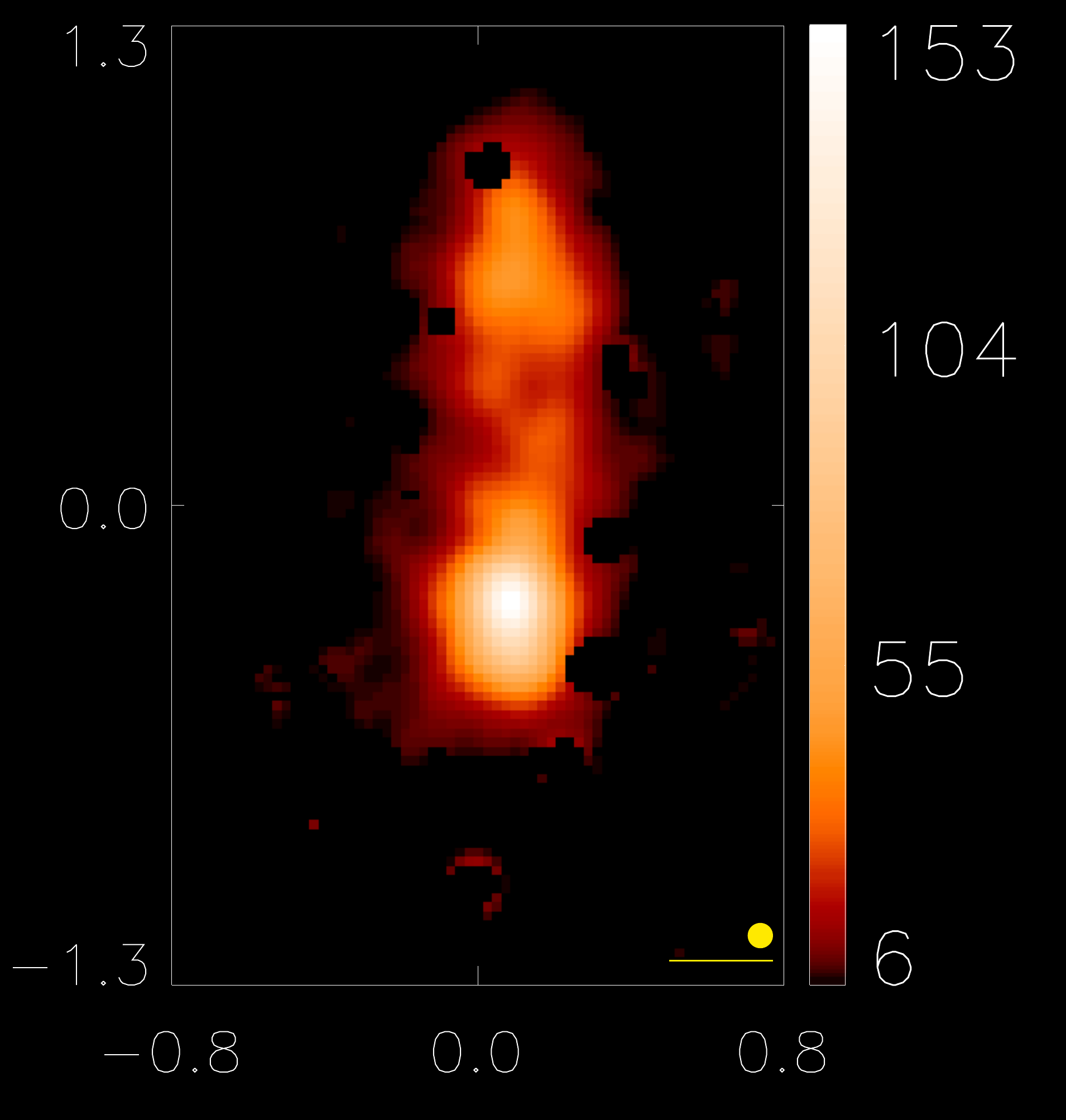} \hskip.05in
\includegraphics[width=.19\linewidth]{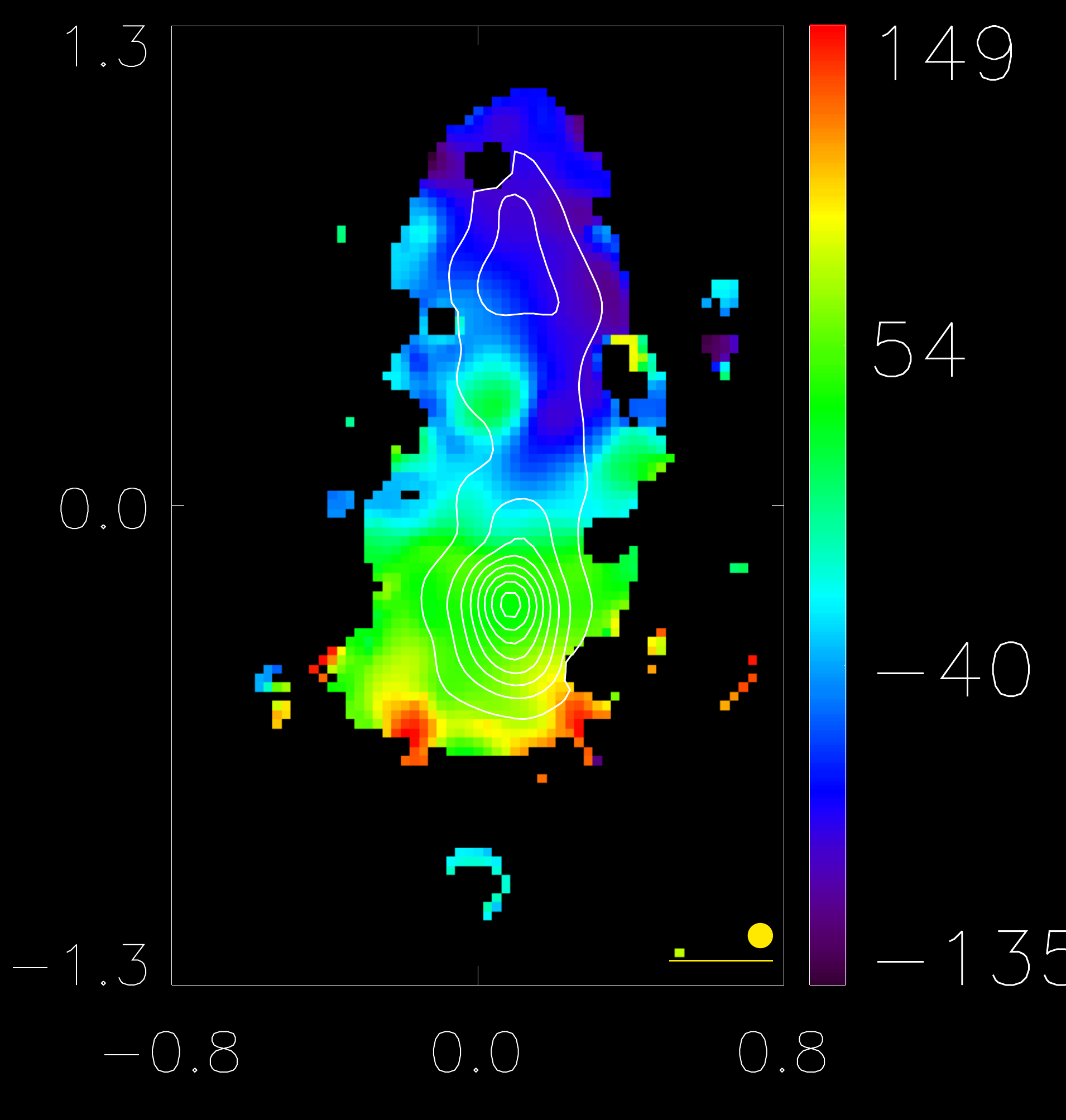} \hskip.05in
\includegraphics[width=.19\linewidth]{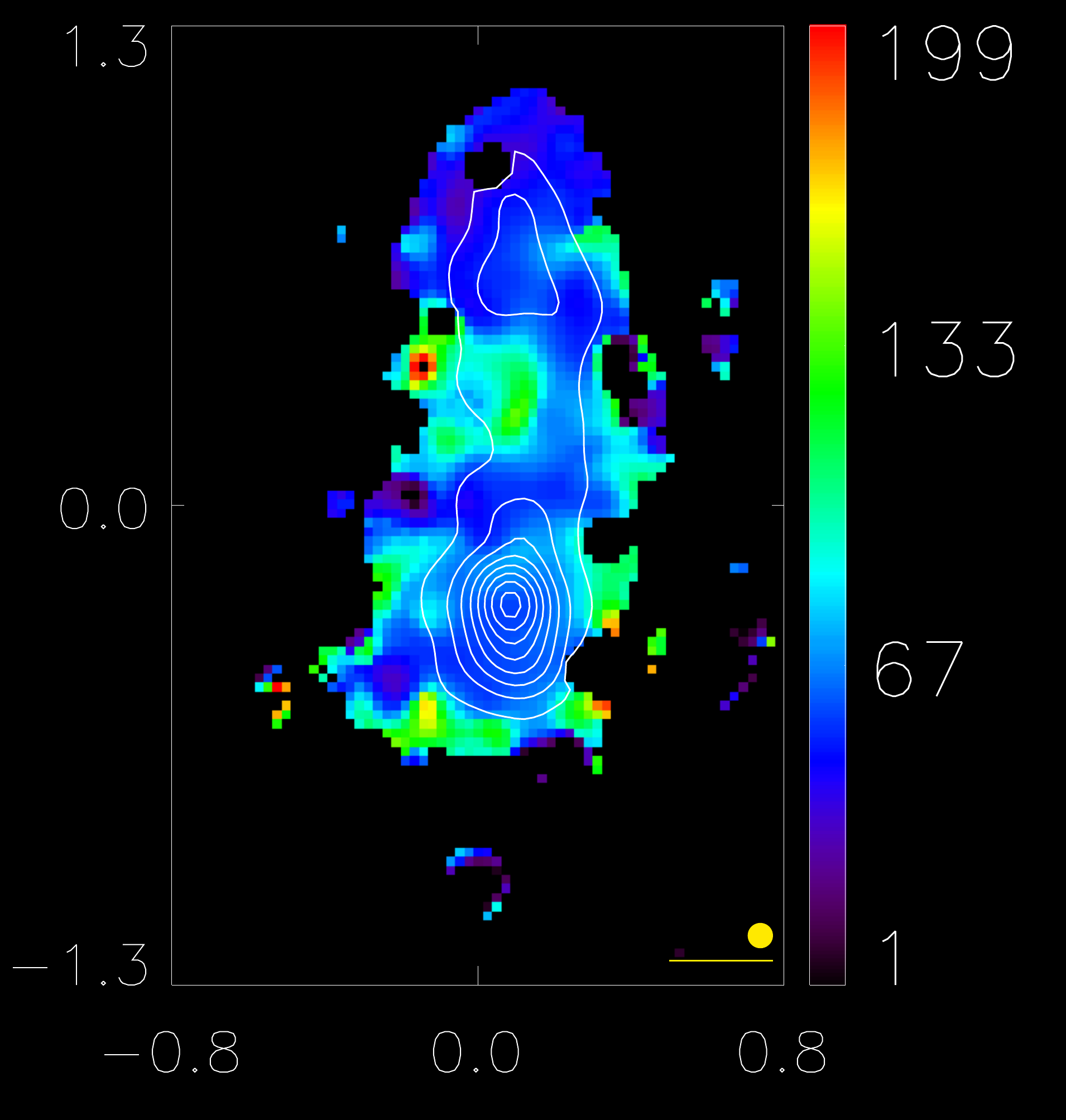} \hskip.05in
\vskip .1 in
\includegraphics[width=.19\linewidth]{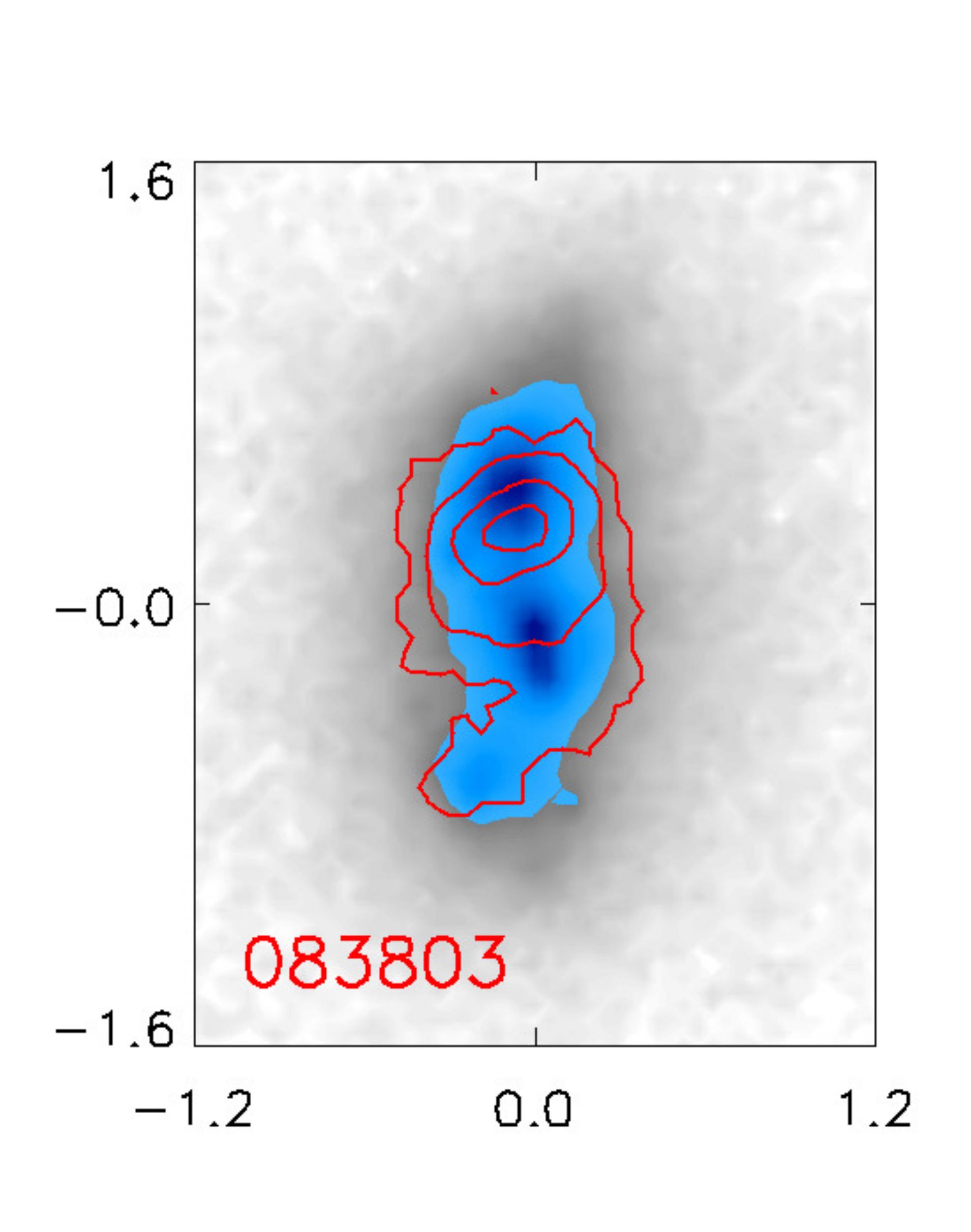} \hskip.05in
\includegraphics[width=.19\linewidth]{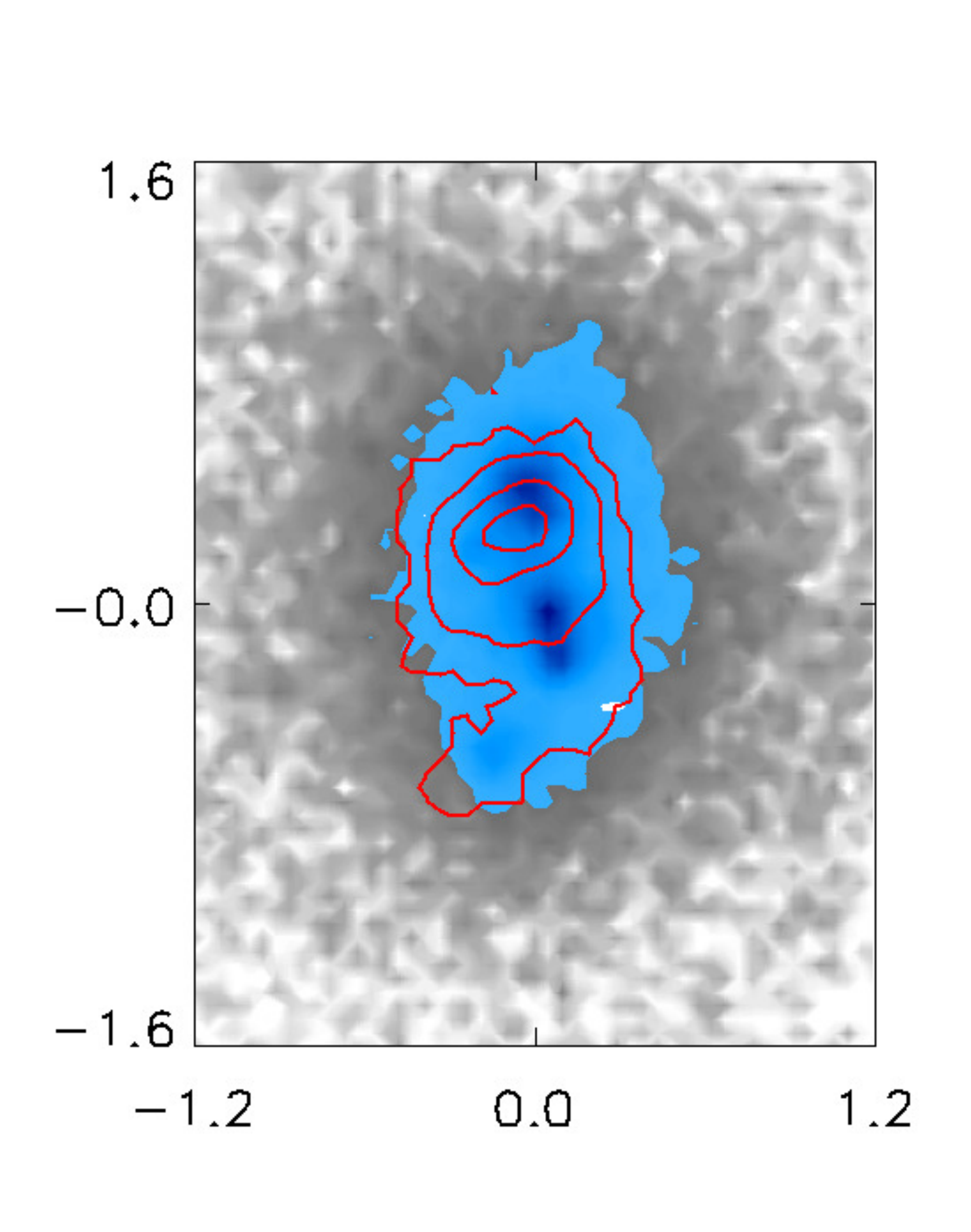} \hskip.05in
\includegraphics[width=.19\linewidth]{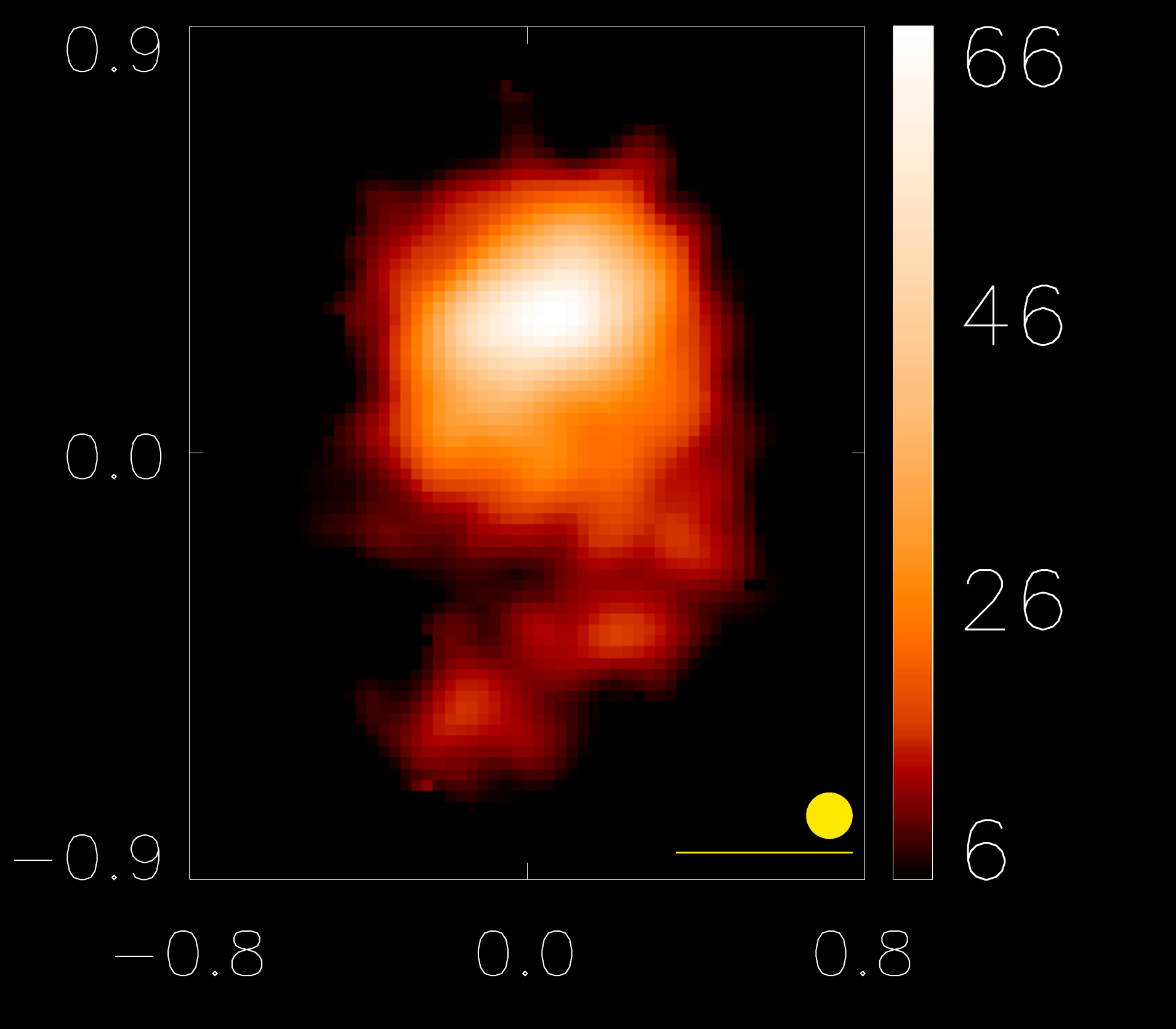} \hskip.05in
\includegraphics[width=.19\linewidth]{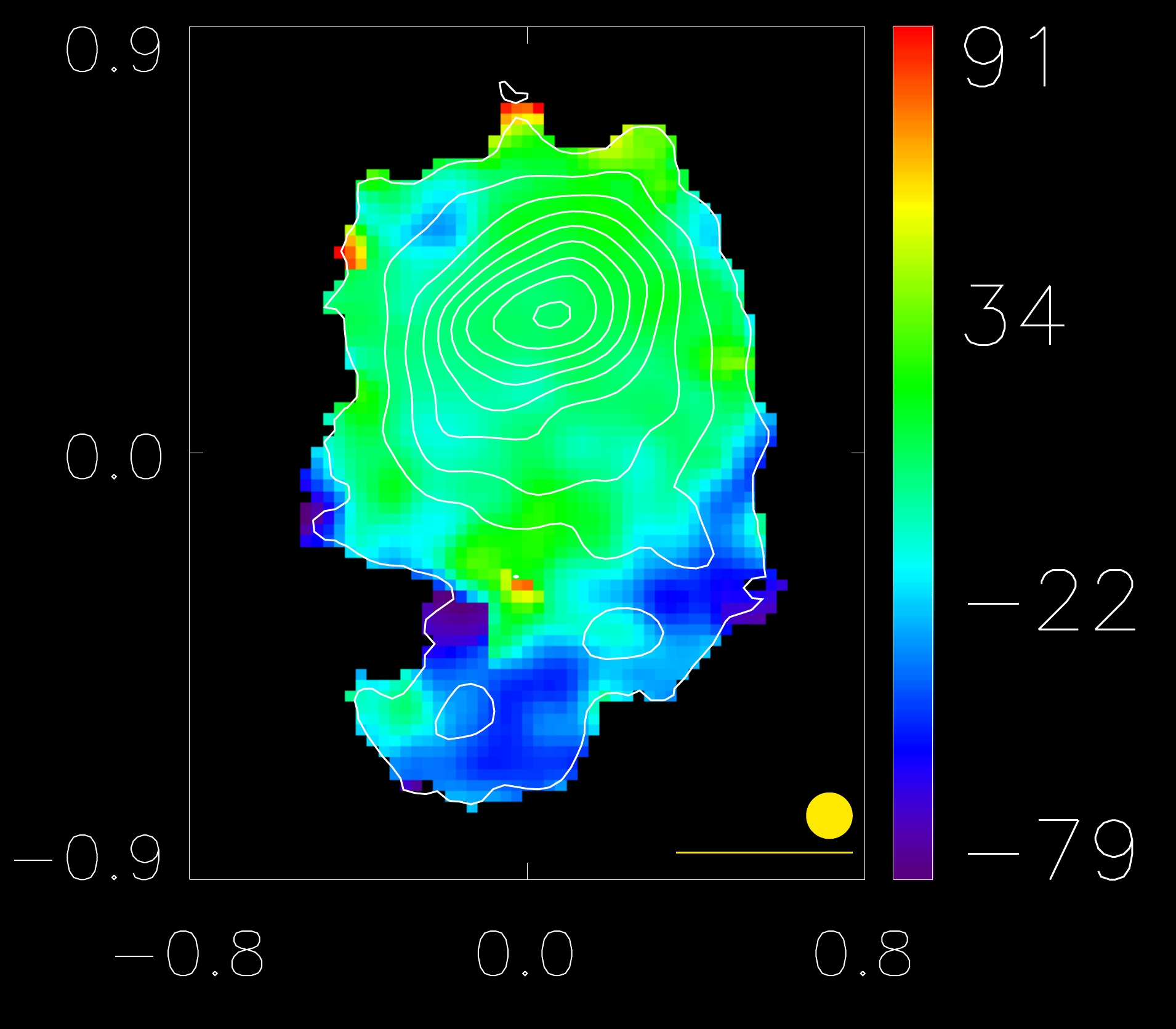} \hskip.05in
\includegraphics[width=.19\linewidth]{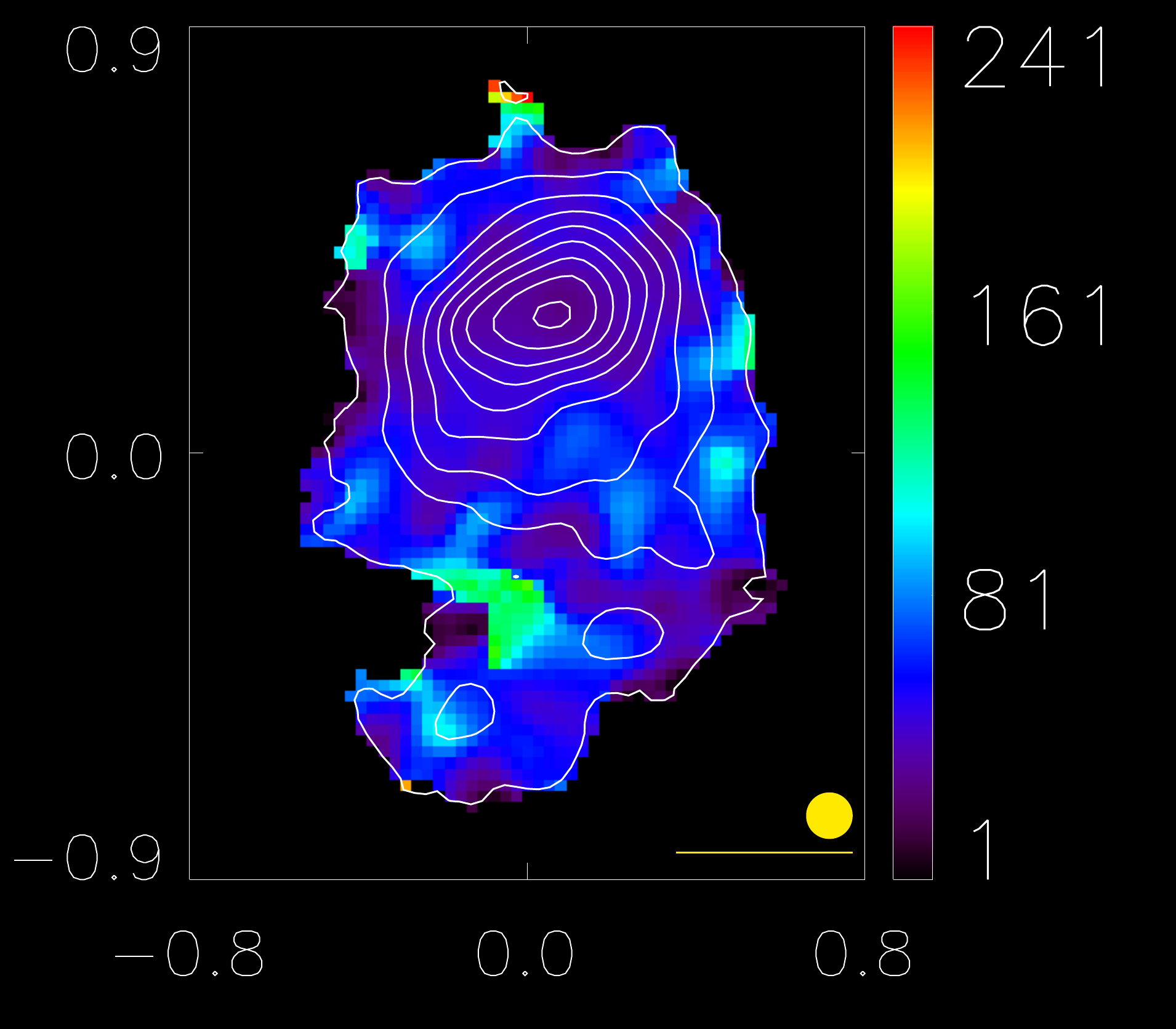} \hskip.05in

\caption{
continued.\label{fig:vdmaps2}
}
\end{figure*}

\addtocounter{figure}{-1}
\begin{figure*}[ht]

\includegraphics[width=.19\linewidth]{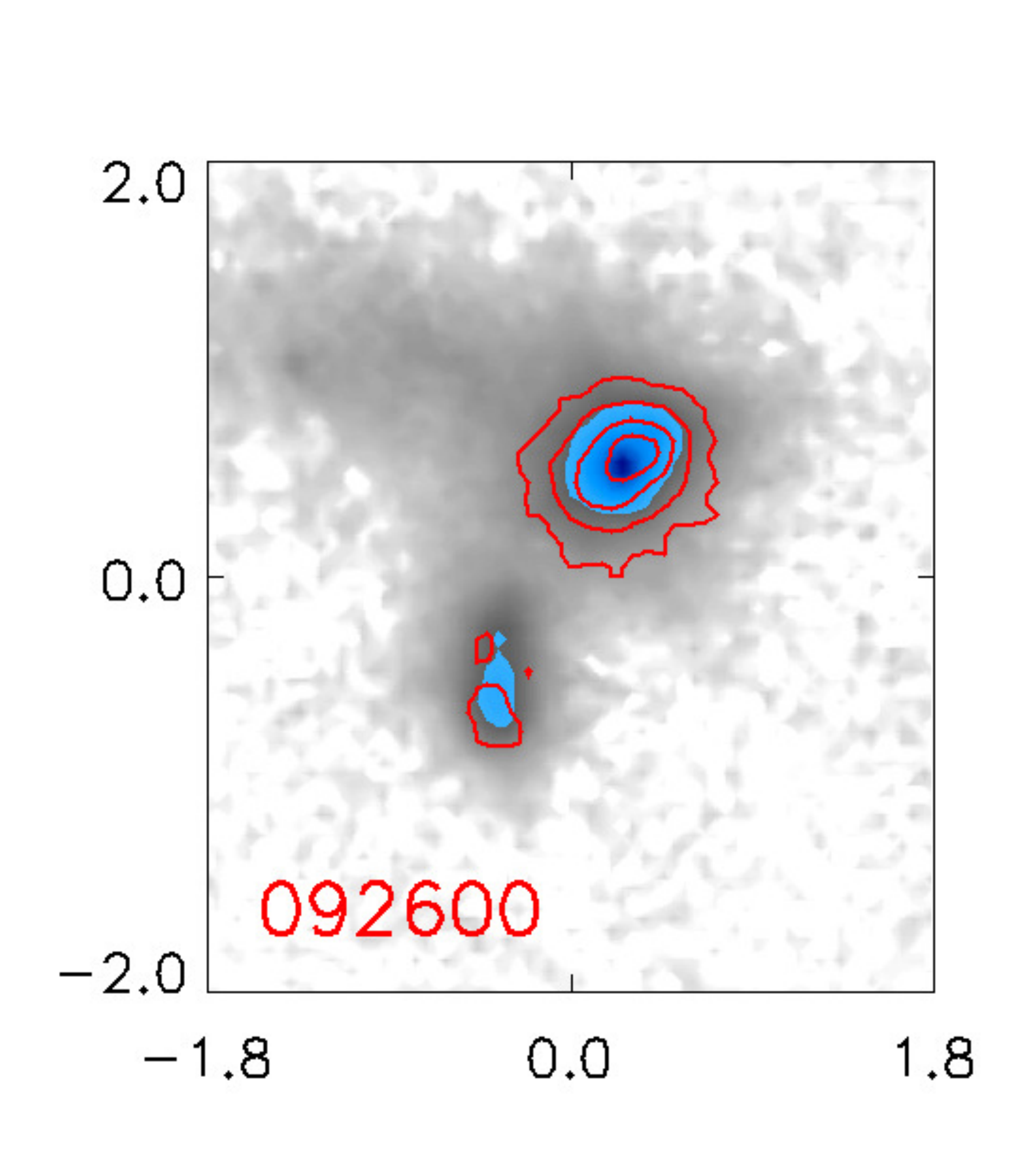} \hskip.05in
\includegraphics[width=.19\linewidth]{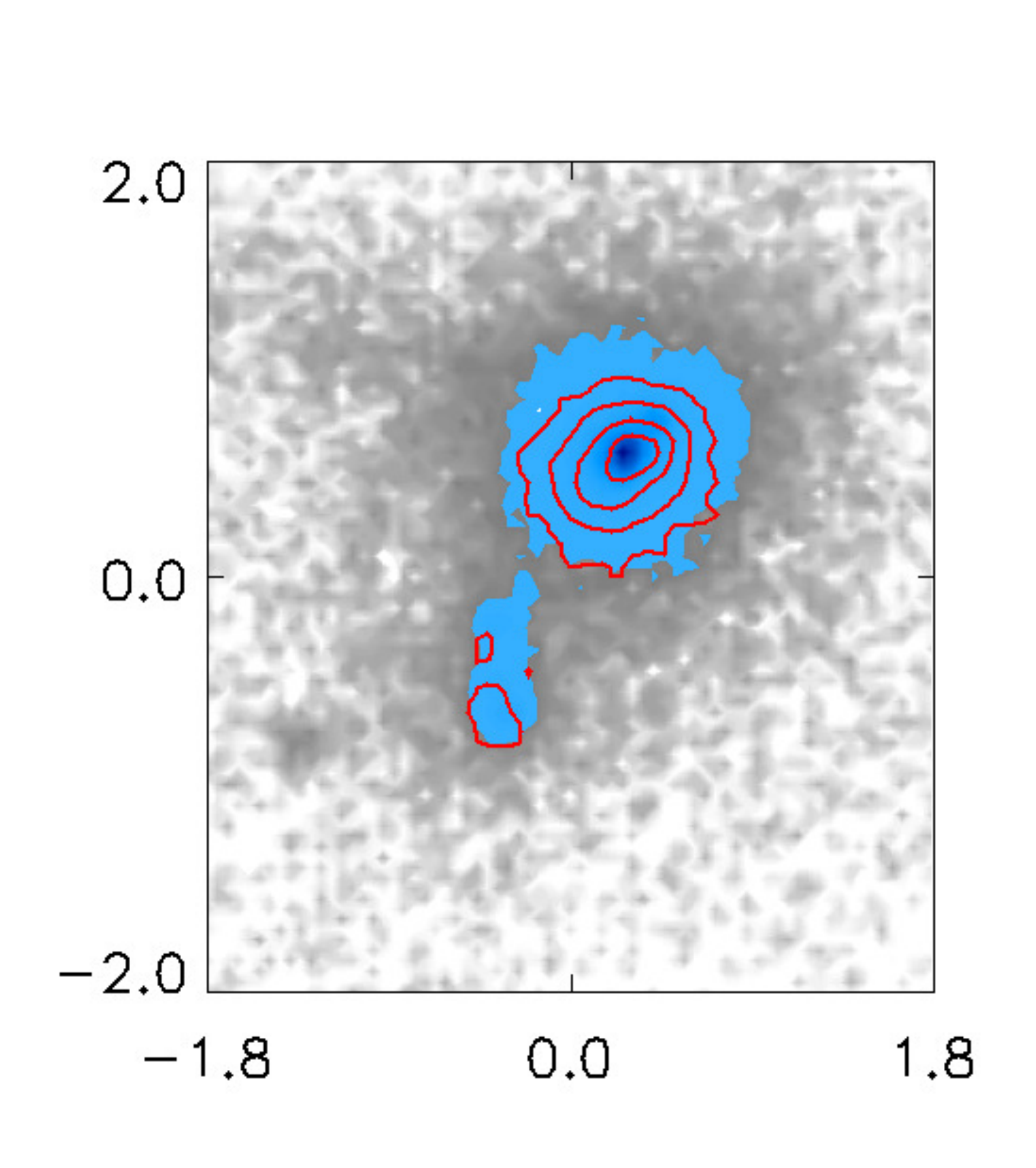} \hskip.05in
\includegraphics[width=.19\linewidth]{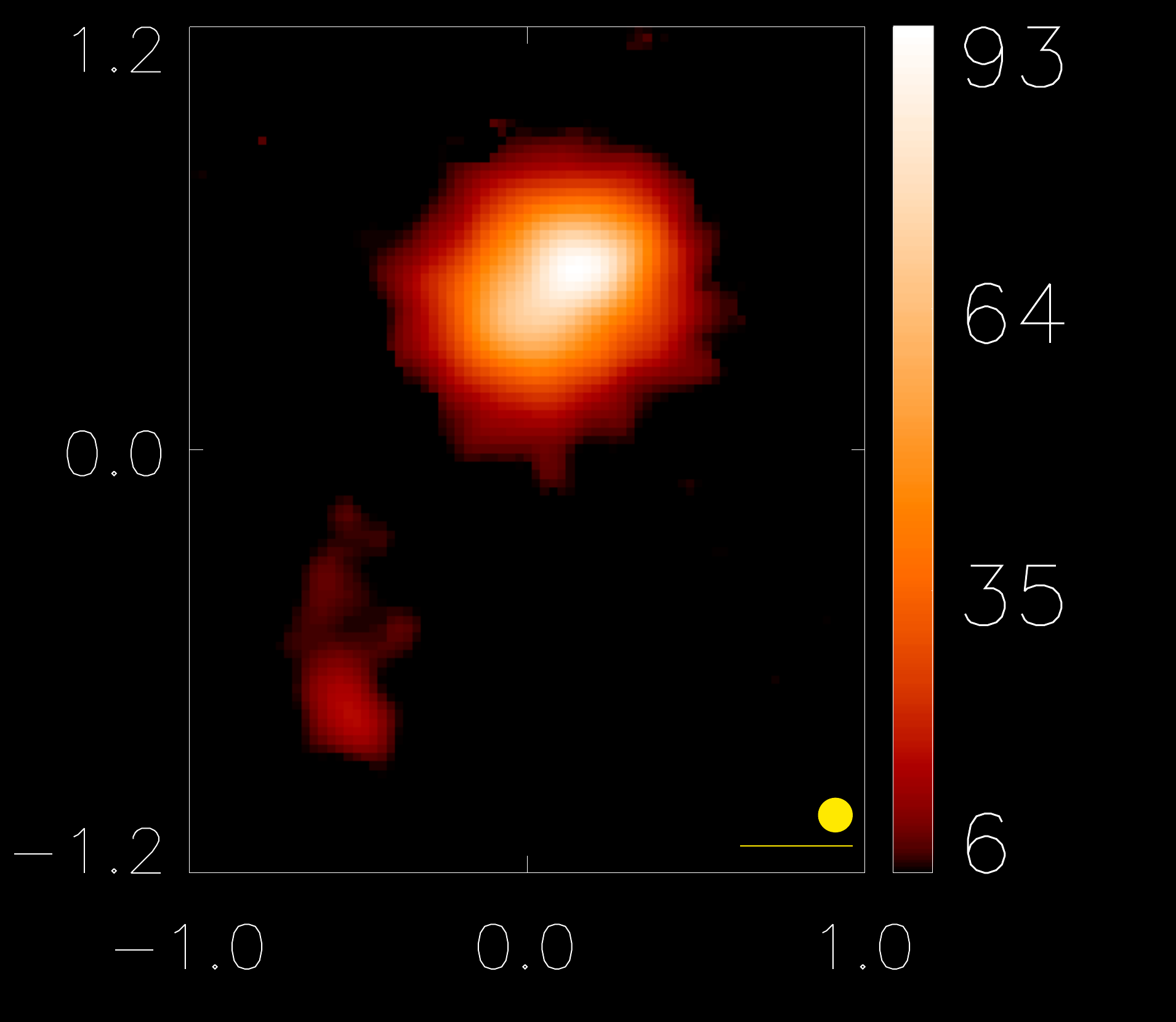} \hskip.05in
\includegraphics[width=.19\linewidth]{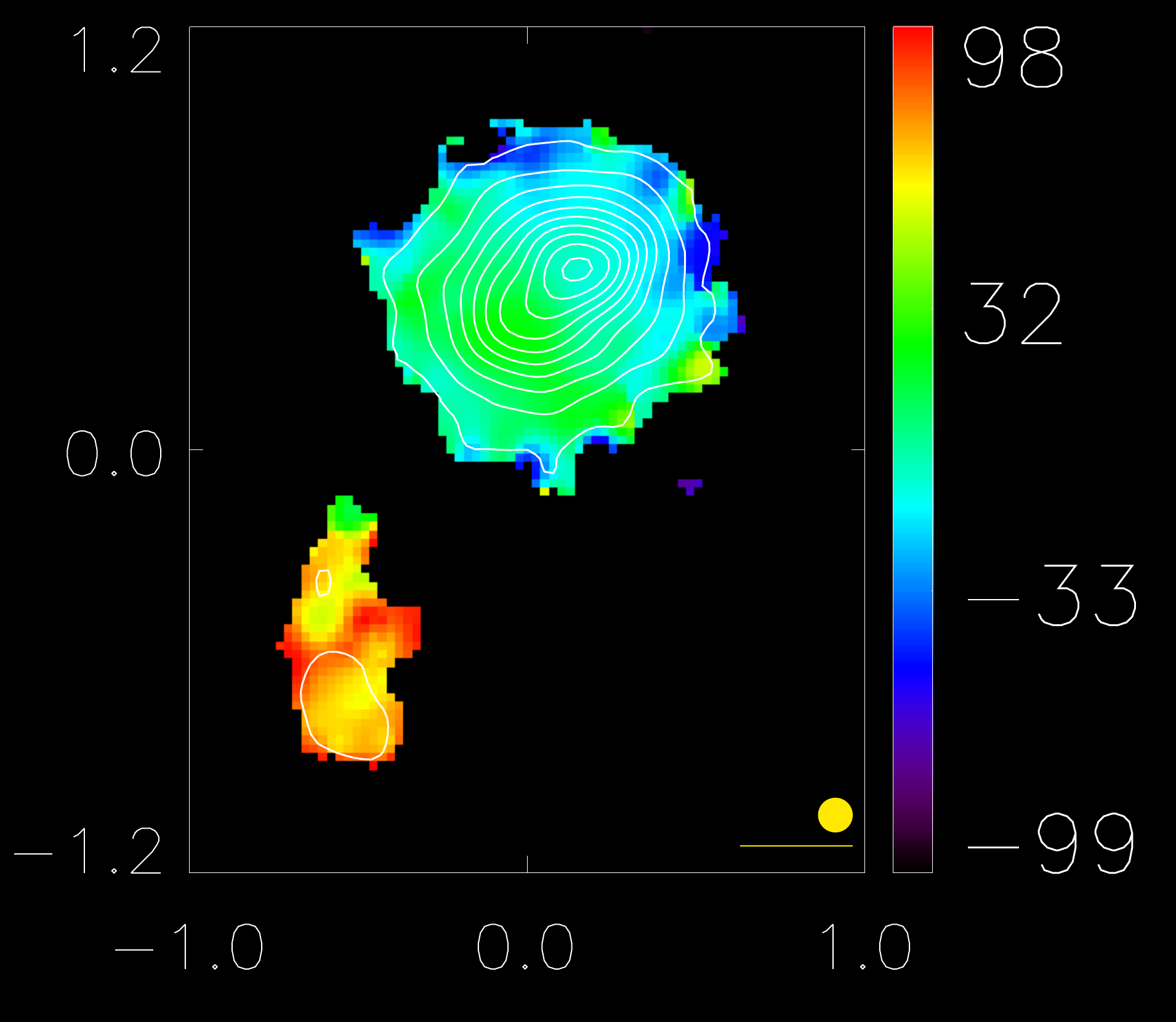} \hskip.05in
\includegraphics[width=.19\linewidth]{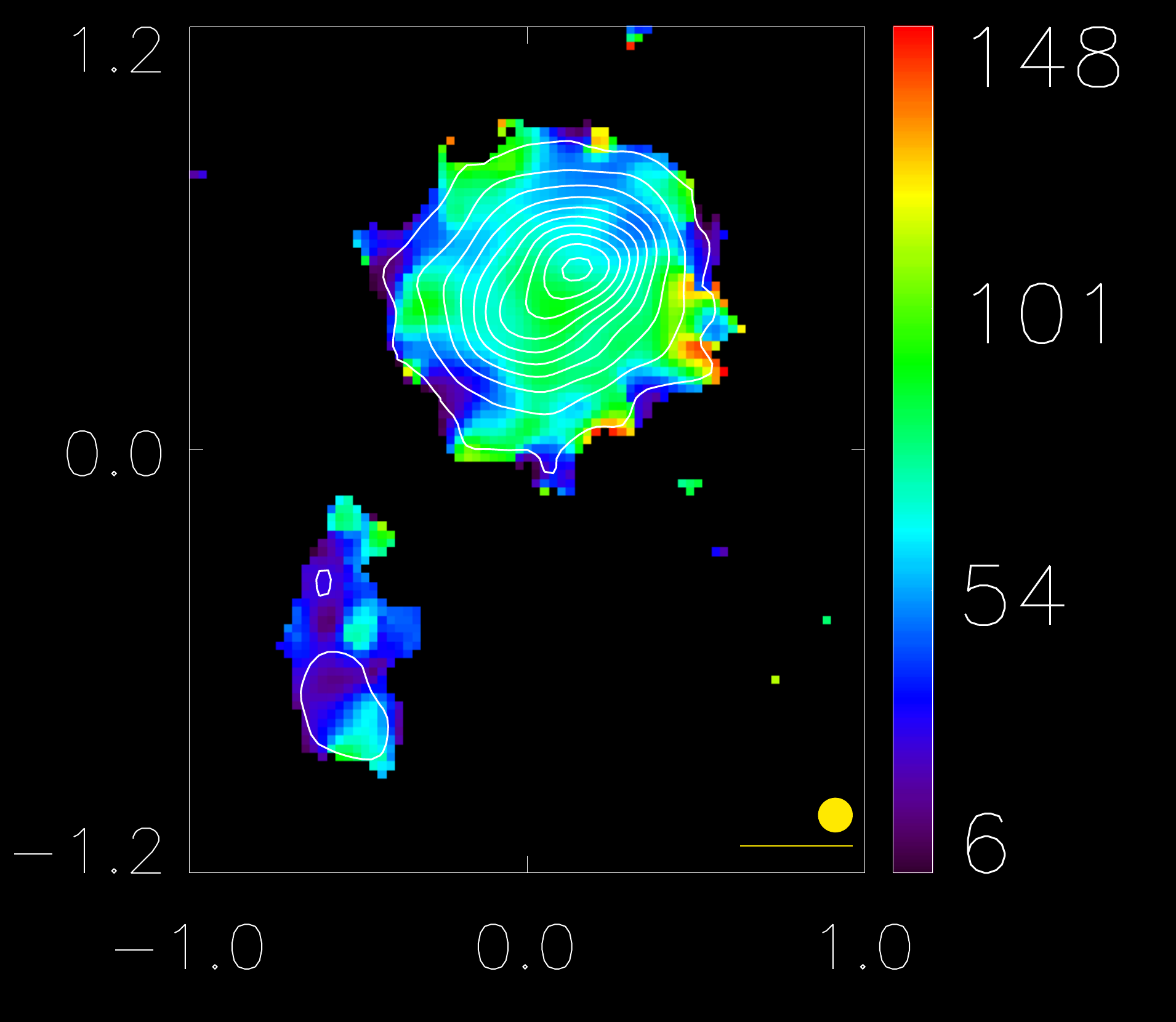} \hskip.05in
\vskip .1 in
\includegraphics[width=.19\linewidth]{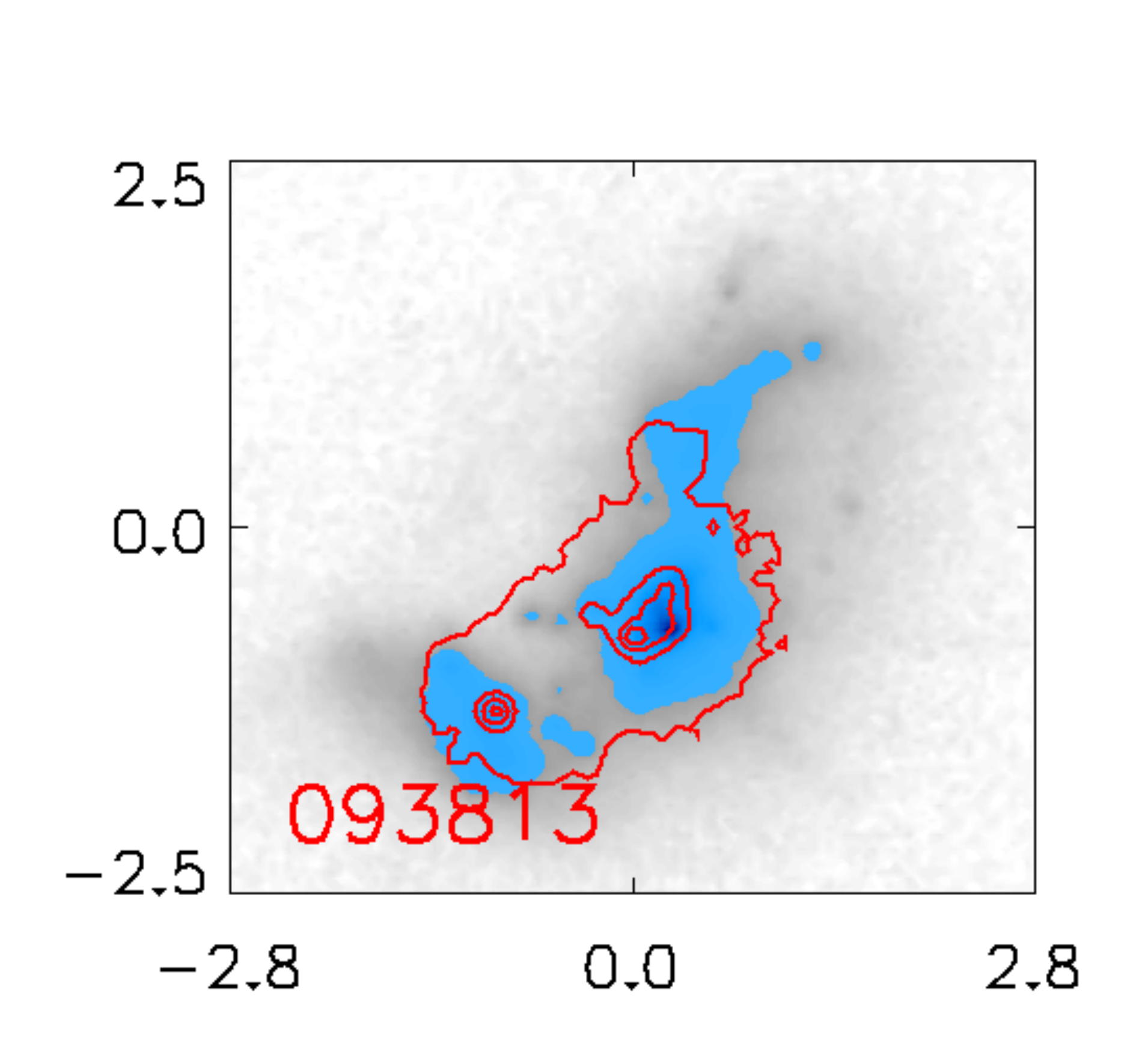} \hskip.05in
\includegraphics[width=.19\linewidth]{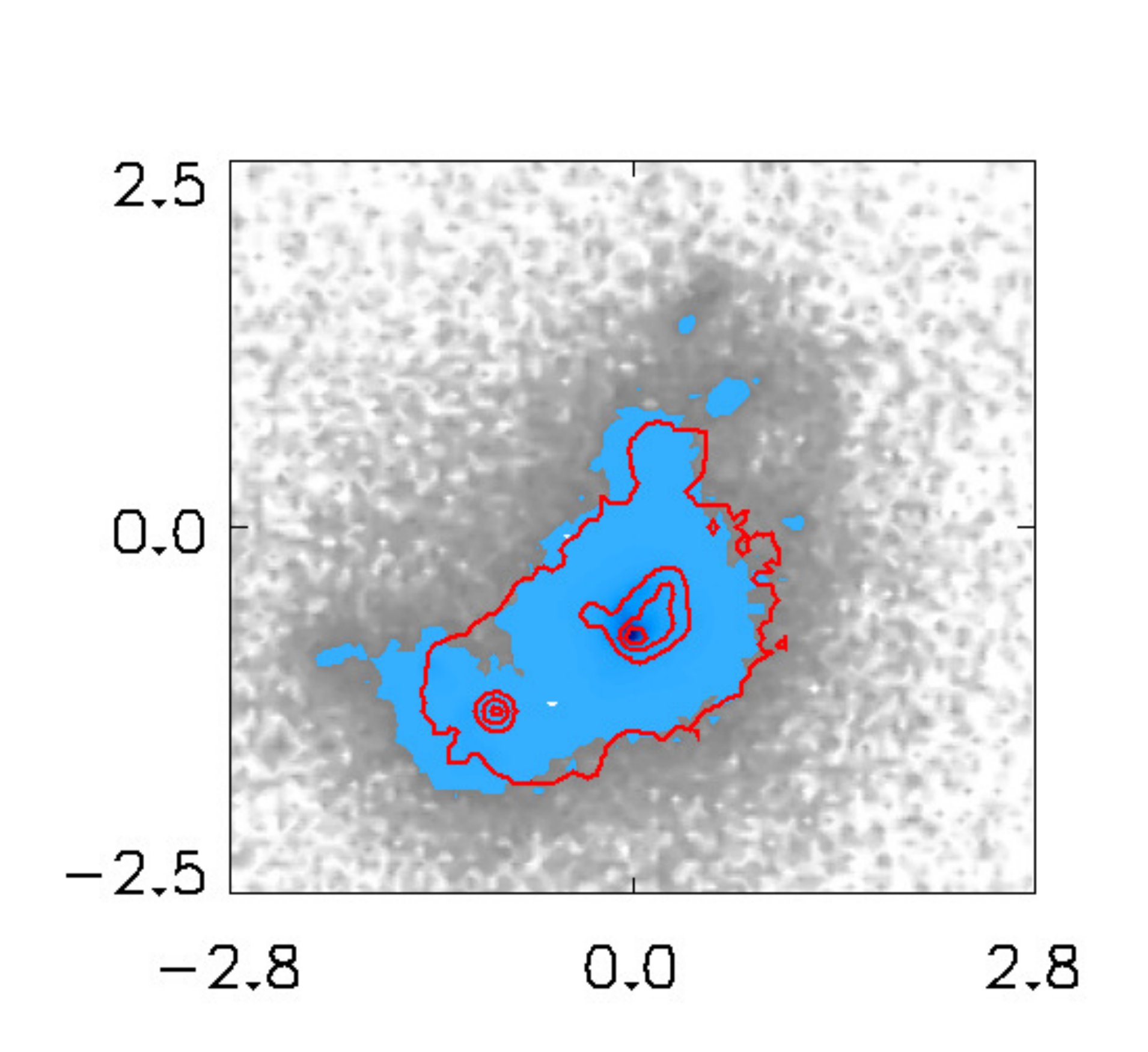} \hskip.05in
\includegraphics[width=.19\linewidth]{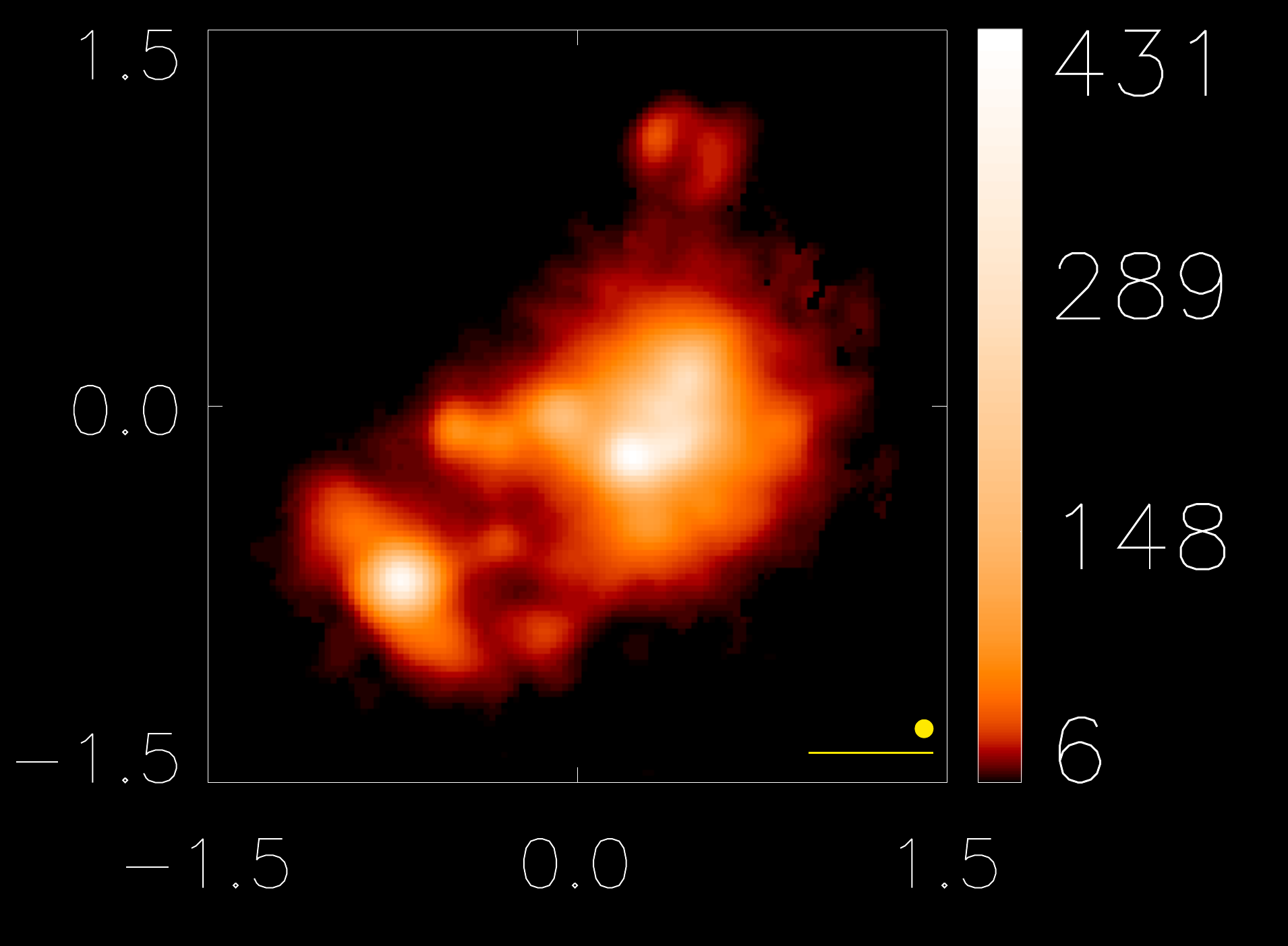} \hskip.05in
\includegraphics[width=.19\linewidth]{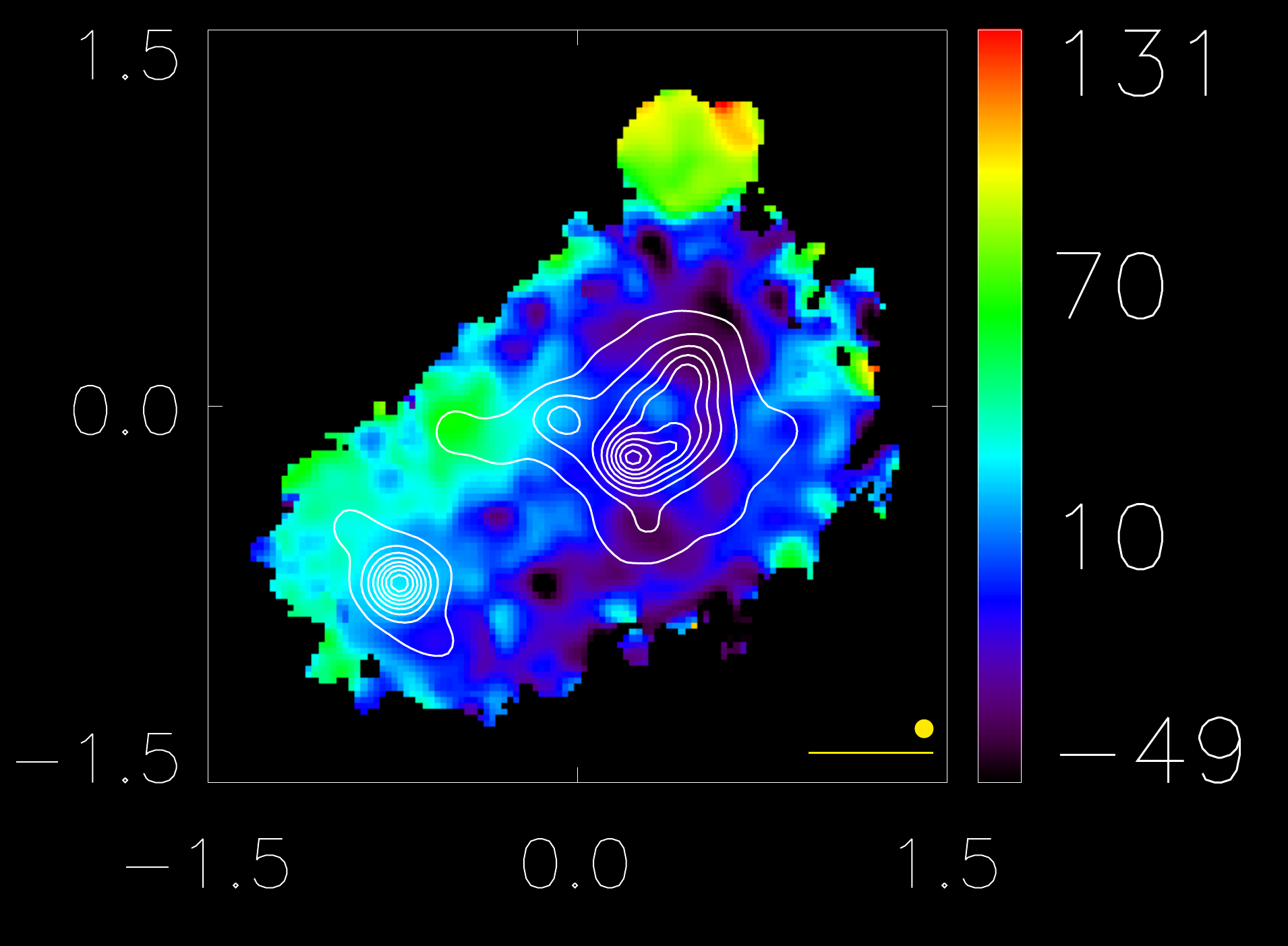} \hskip.05in
\includegraphics[width=.19\linewidth]{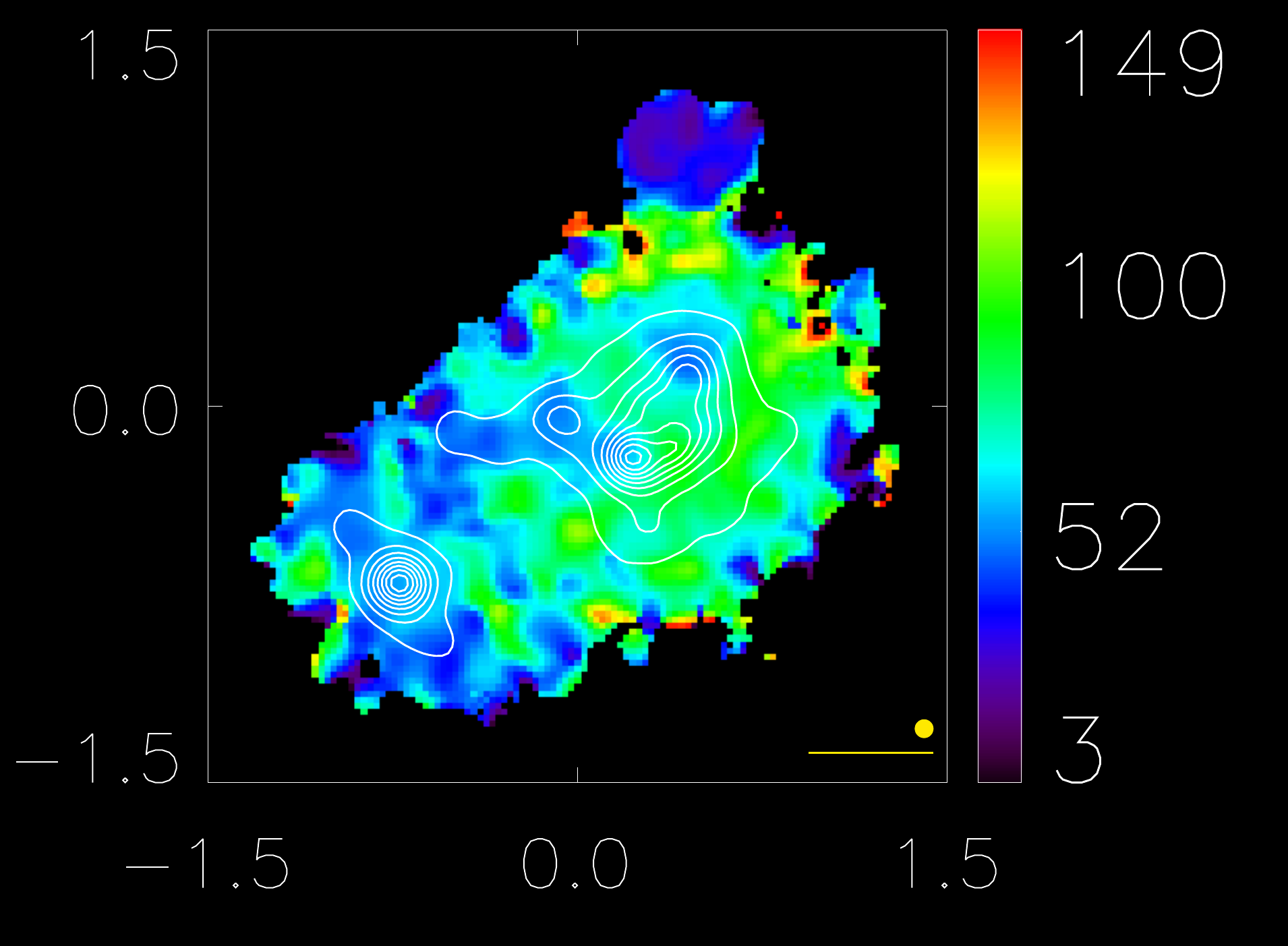} \hskip.05in
\vskip .1 in
\includegraphics[width=.19\linewidth]{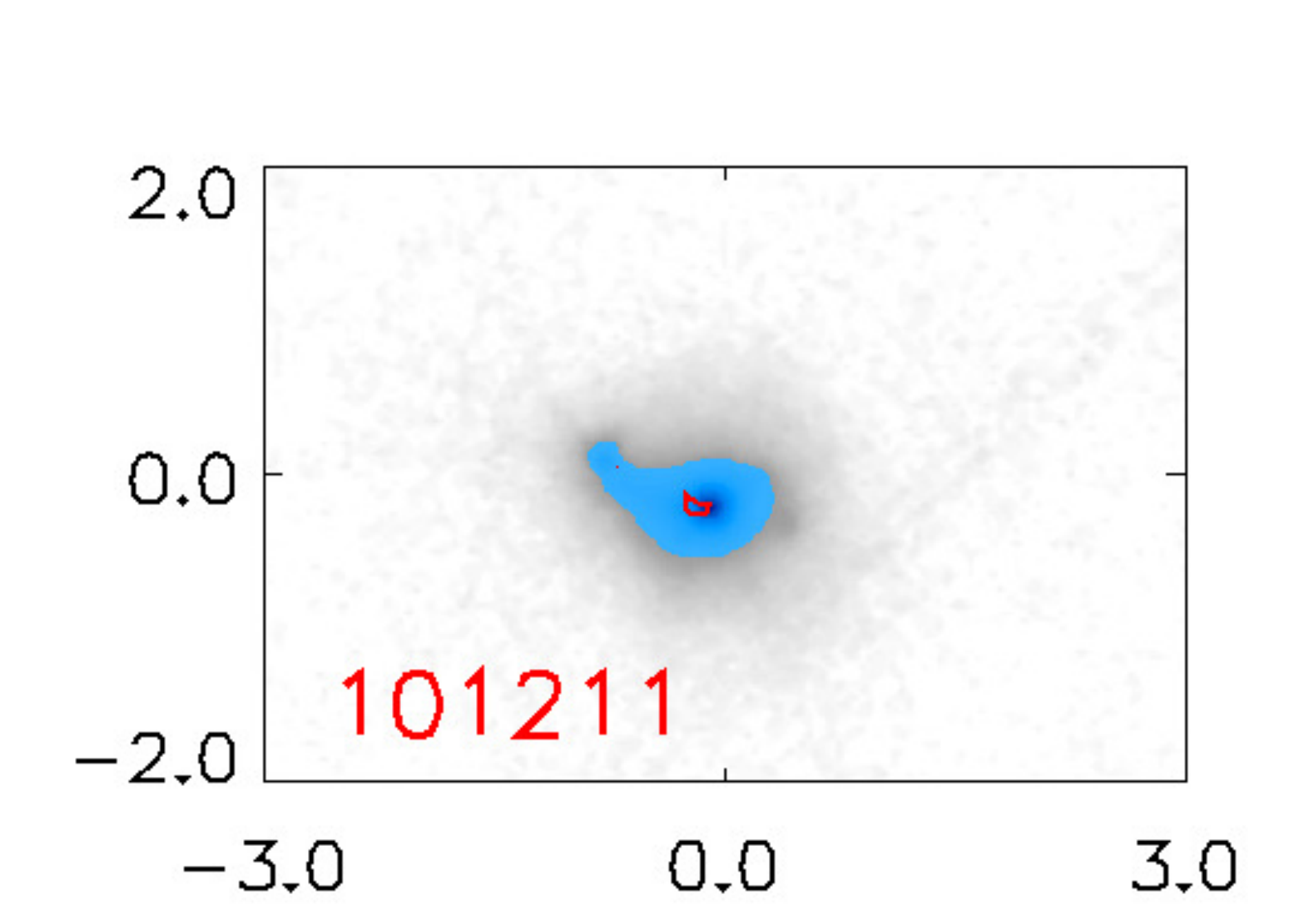} \hskip.05in
\hskip .19\linewidth \hskip.05in
\includegraphics[width=.19\linewidth]{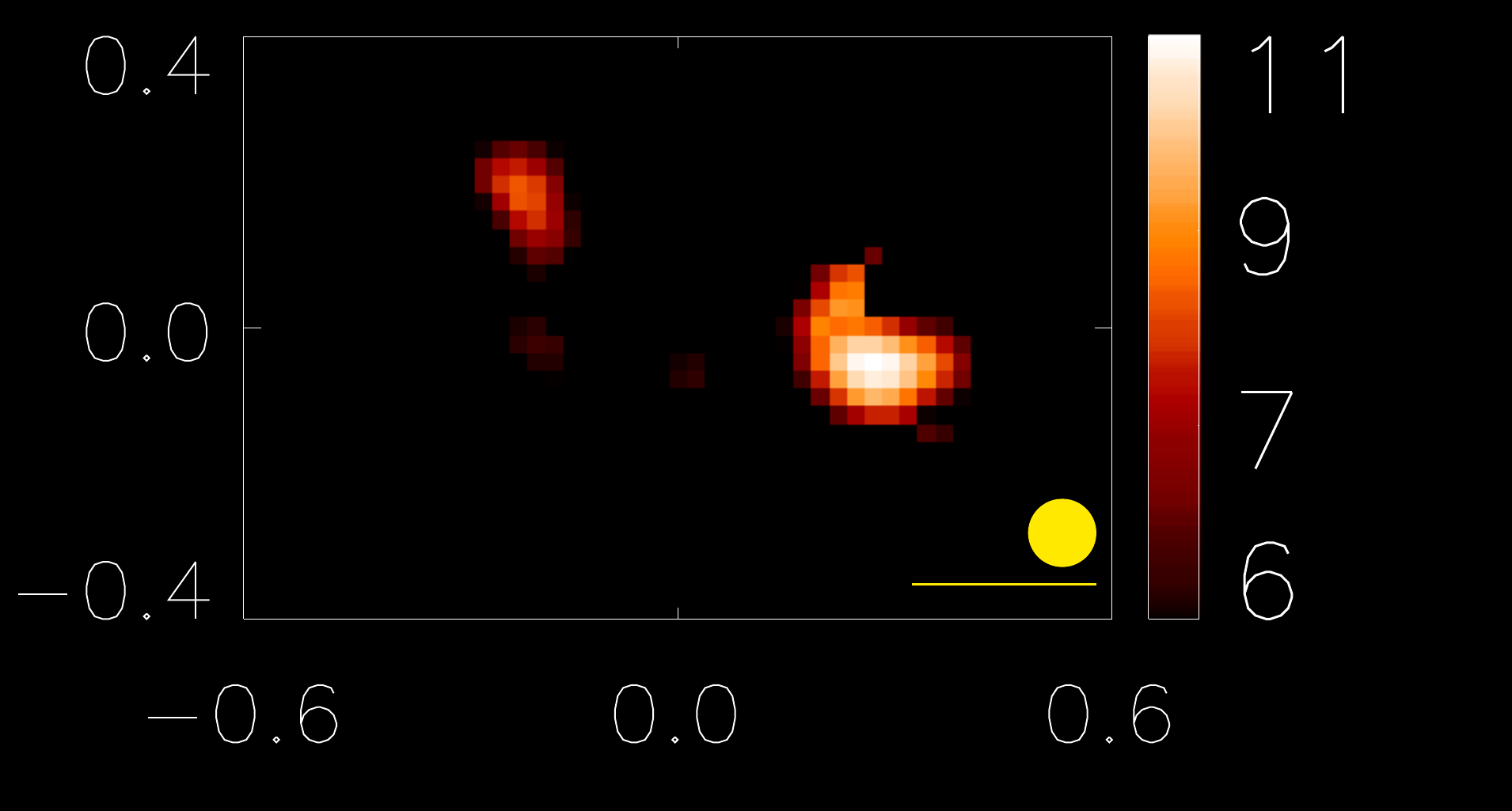} \hskip.05in
\includegraphics[width=.19\linewidth]{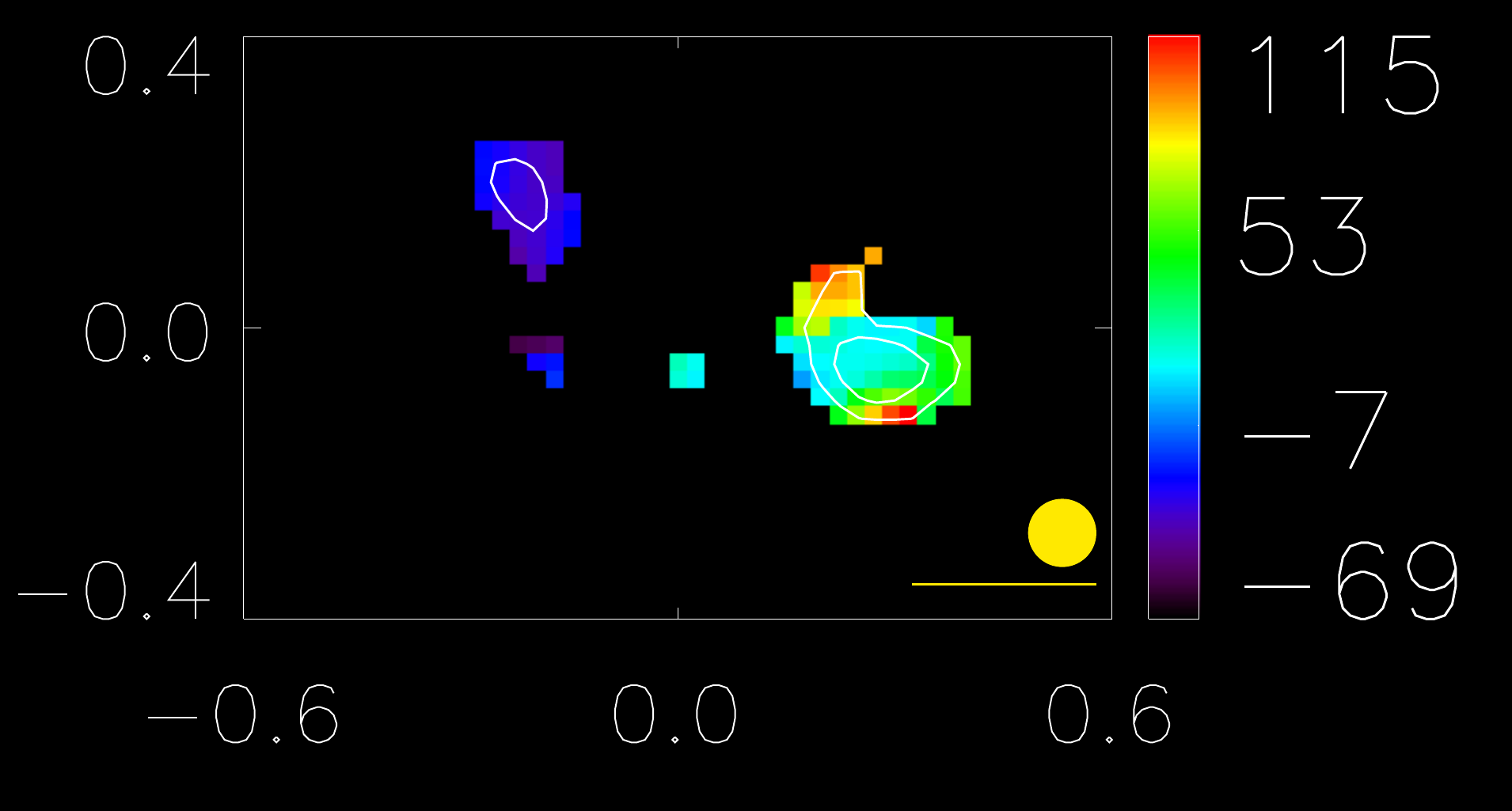} \hskip.05in
\includegraphics[width=.19\linewidth]{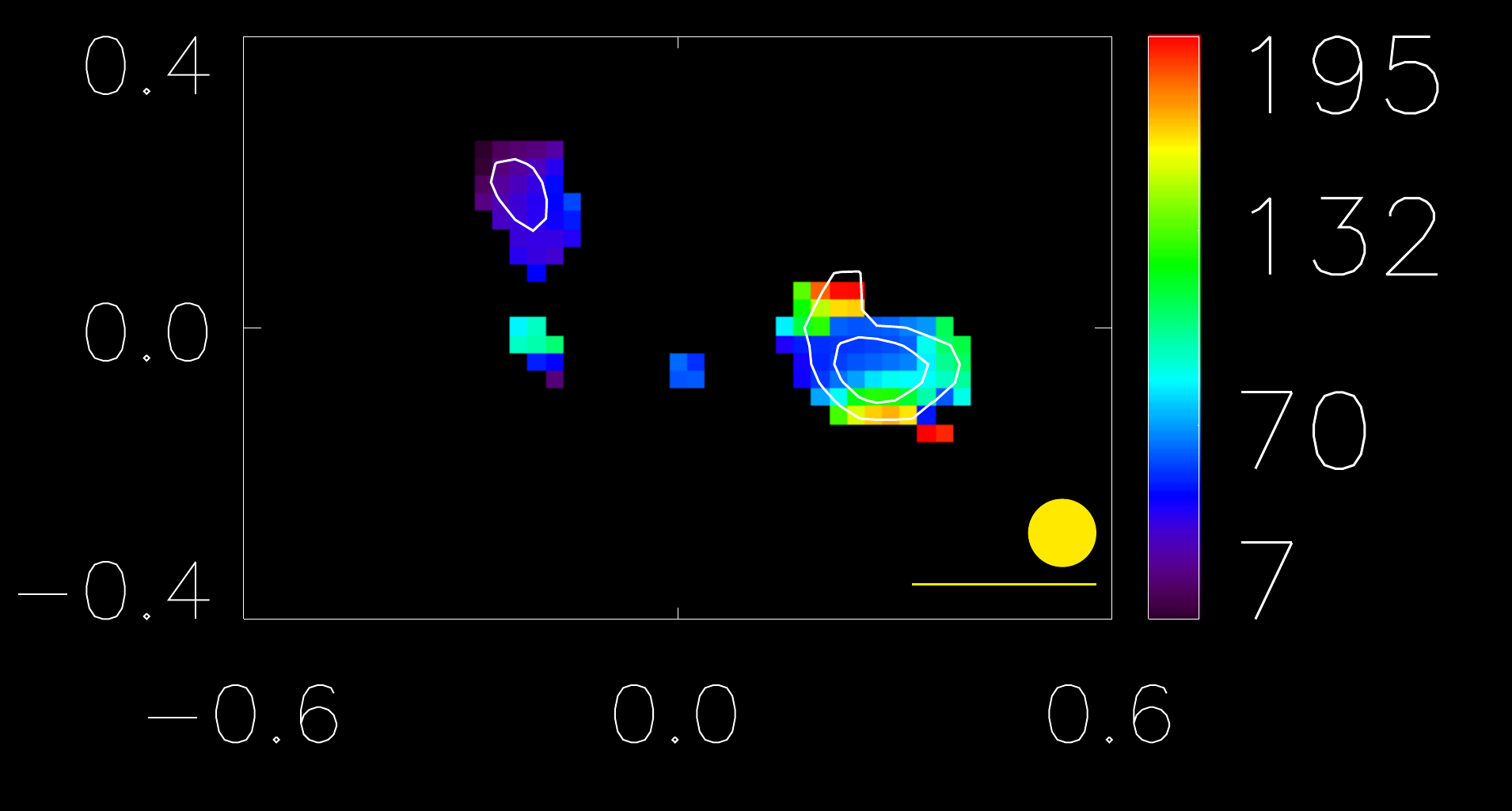} \hskip.05in
\vskip .1 in
\includegraphics[width=.19\linewidth]{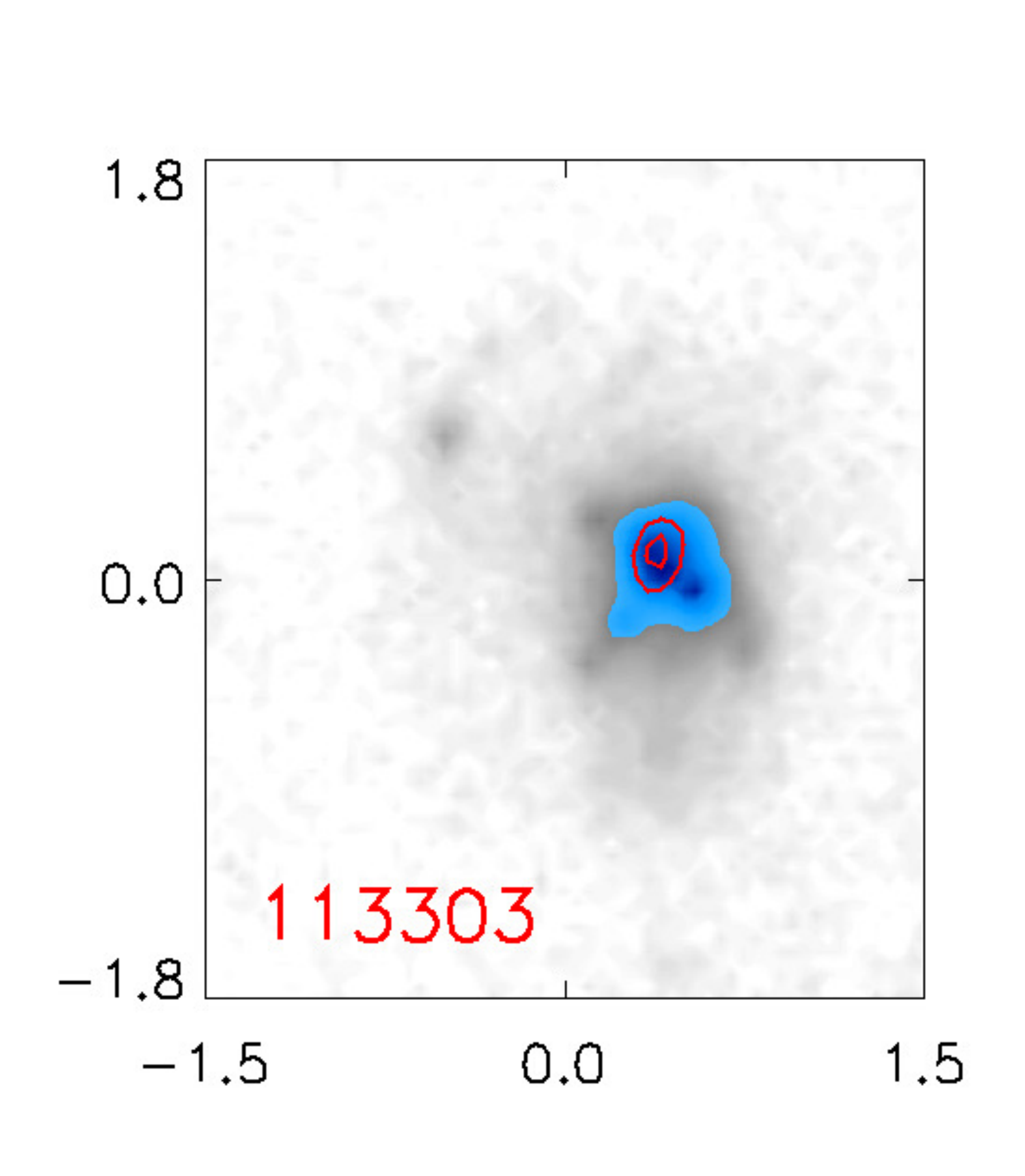} \hskip.05in
\includegraphics[width=.19\linewidth]{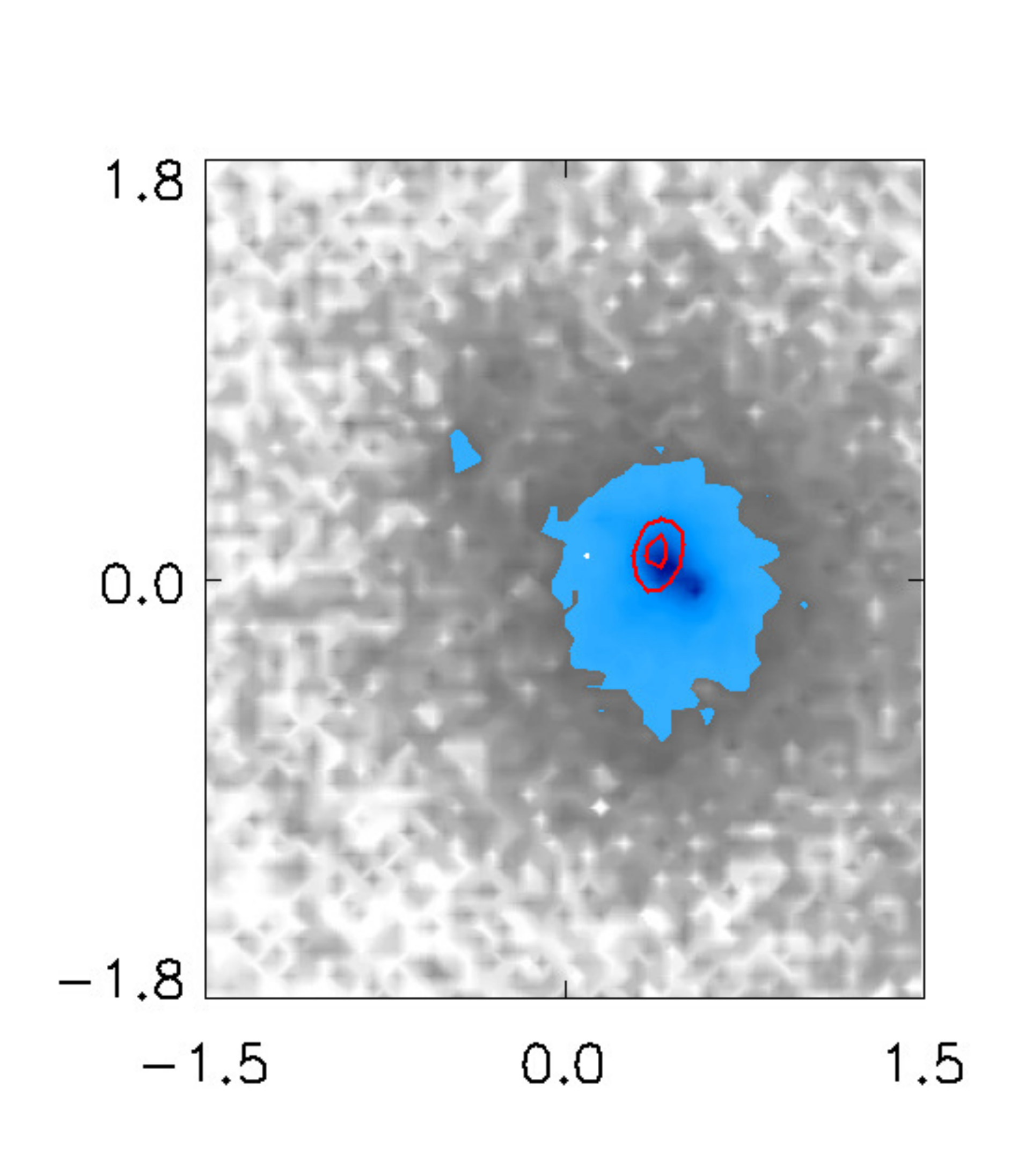} \hskip.05in
\includegraphics[width=.19\linewidth]{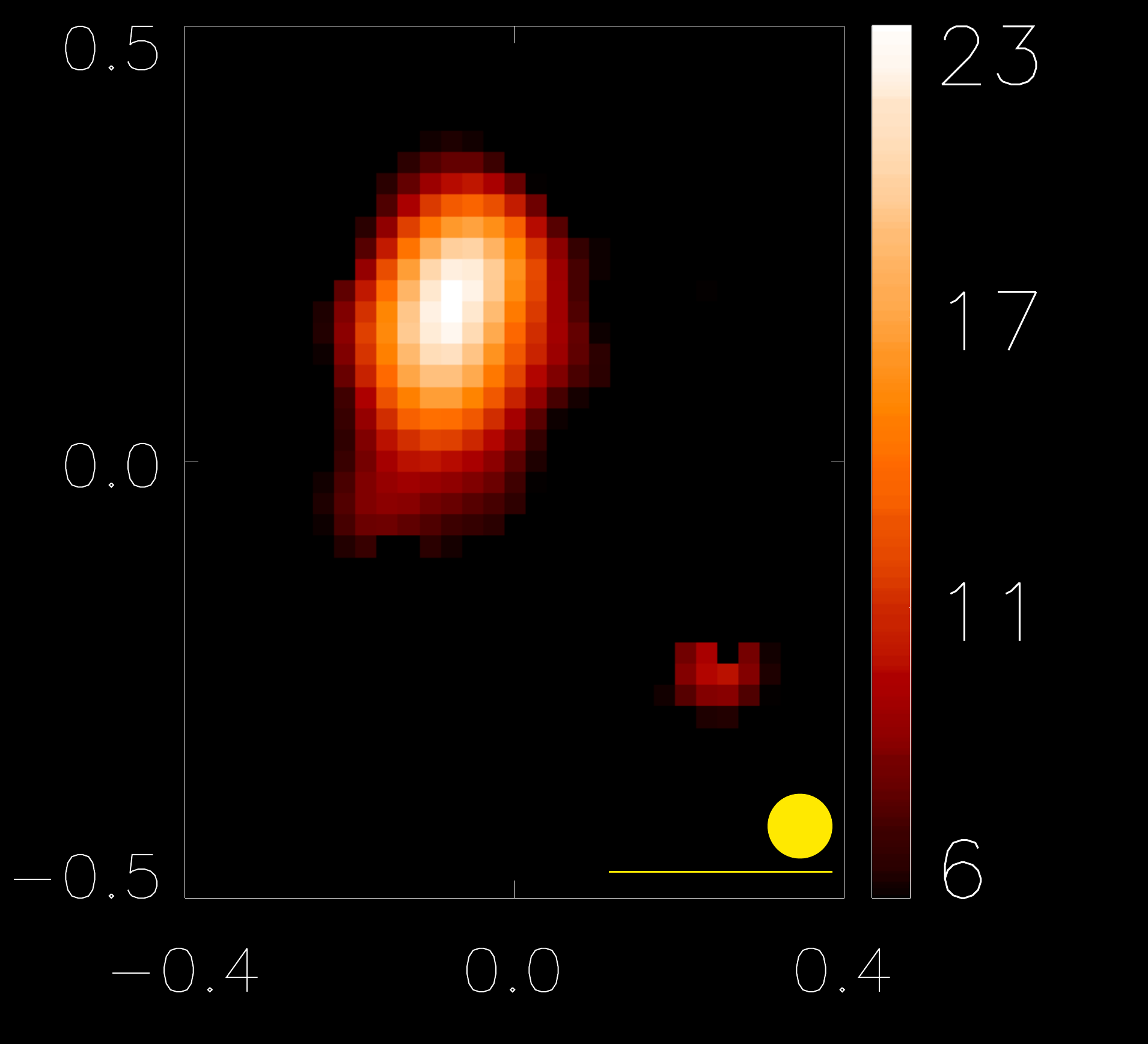} \hskip.05in
\includegraphics[width=.19\linewidth]{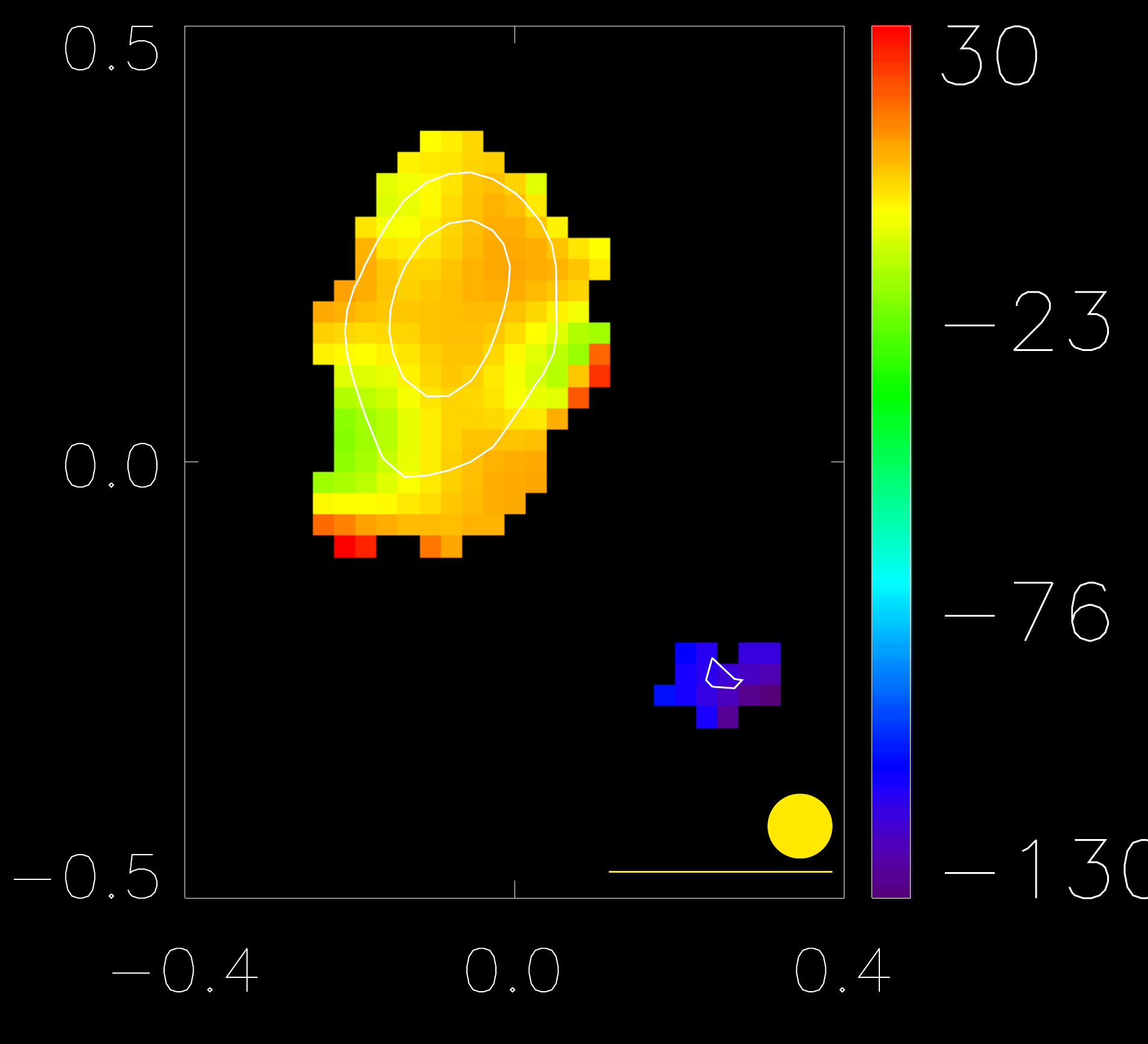} \hskip.05in
\includegraphics[width=.19\linewidth]{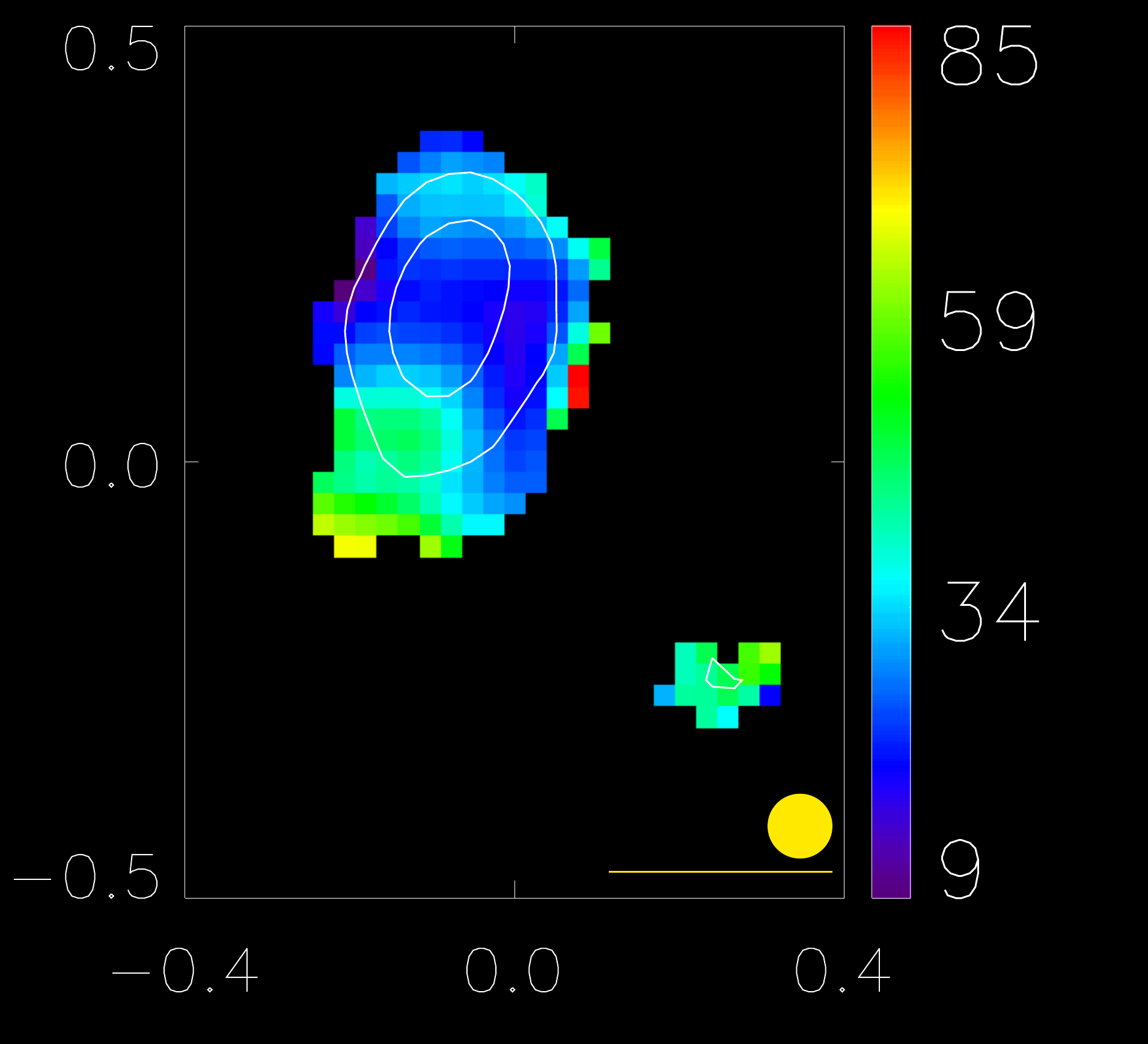} \hskip.05in
\vskip .1 in
\includegraphics[width=.19\linewidth]{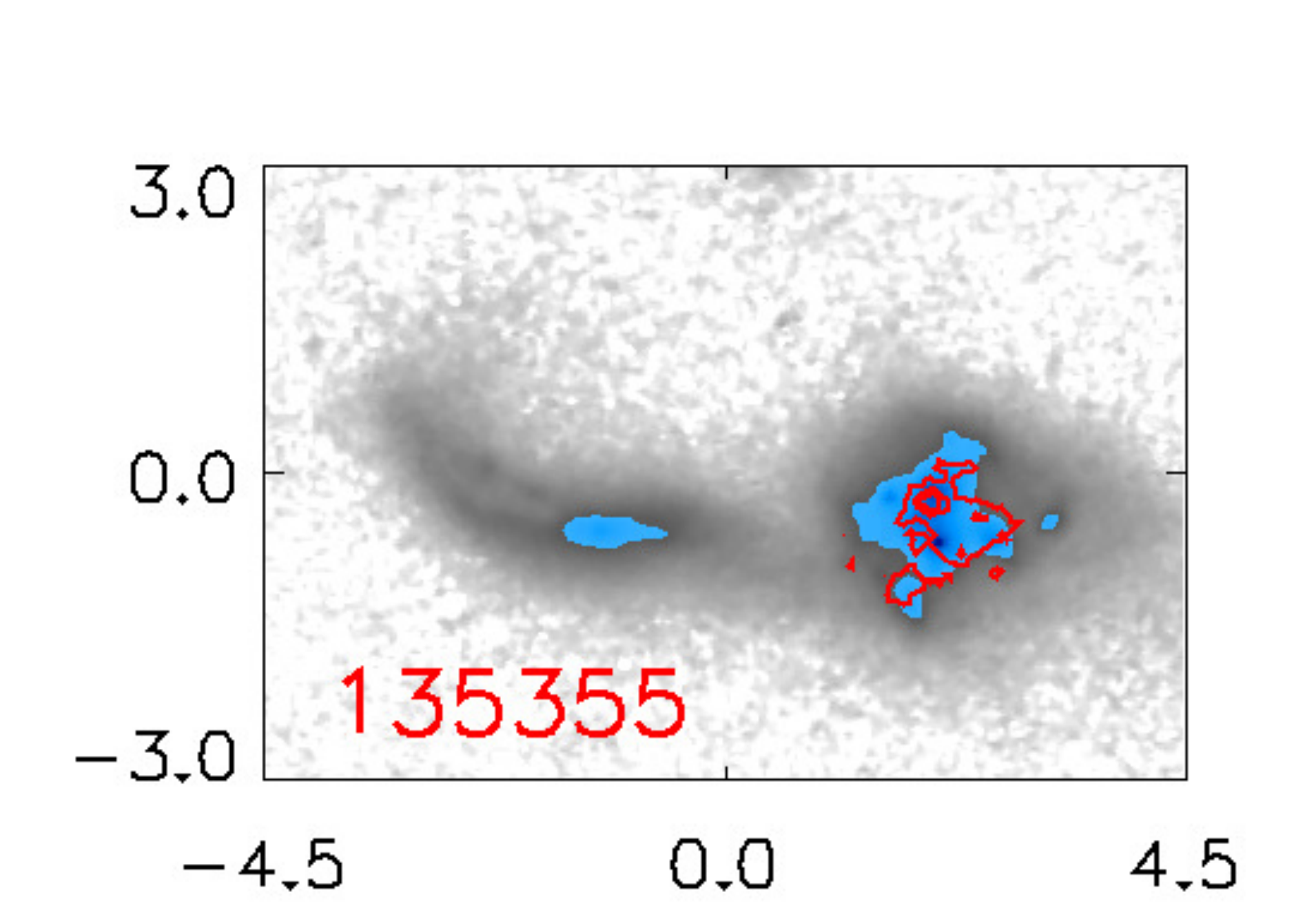} \hskip.05in
\includegraphics[width=.19\linewidth]{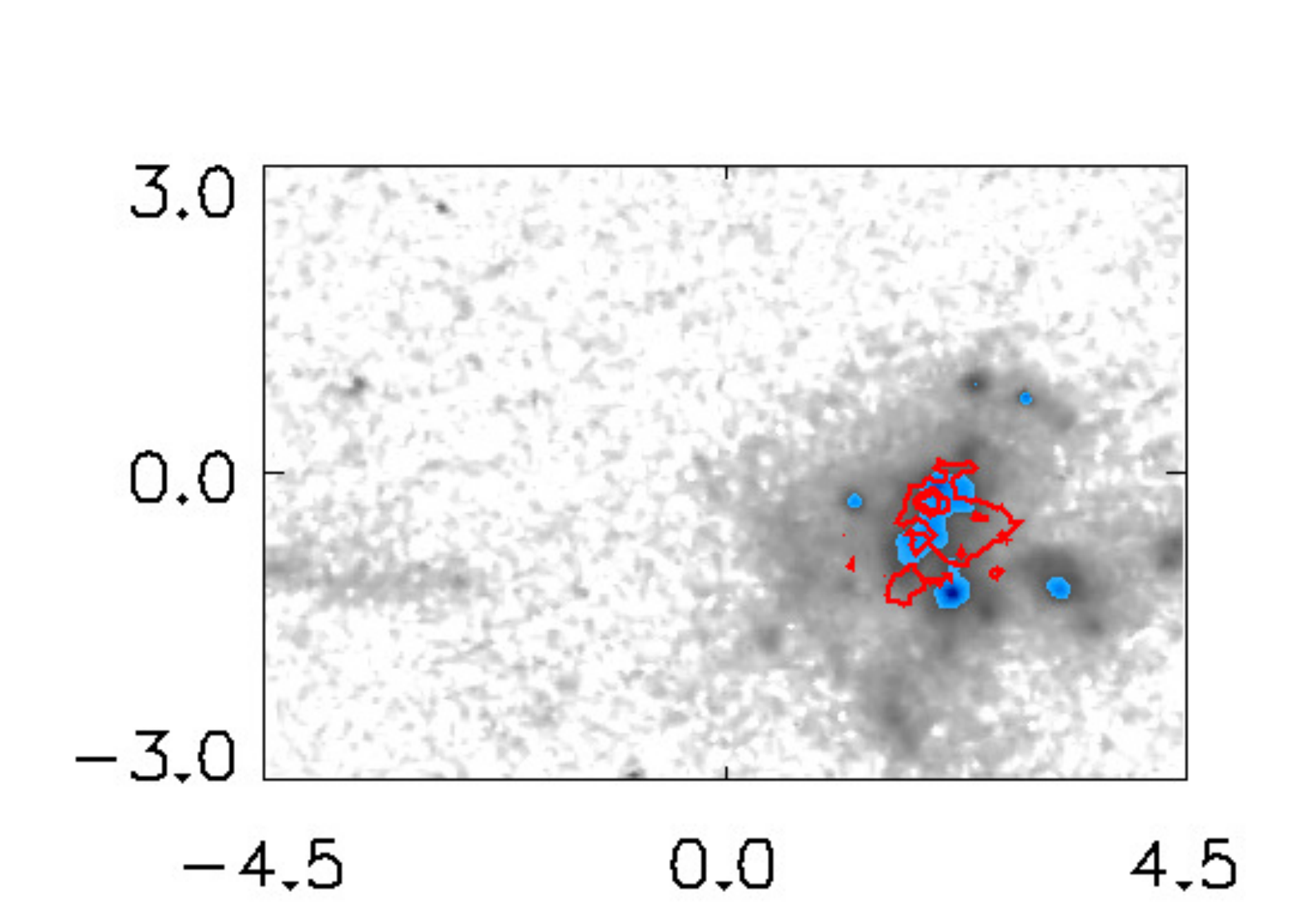} \hskip.05in
\includegraphics[width=.19\linewidth]{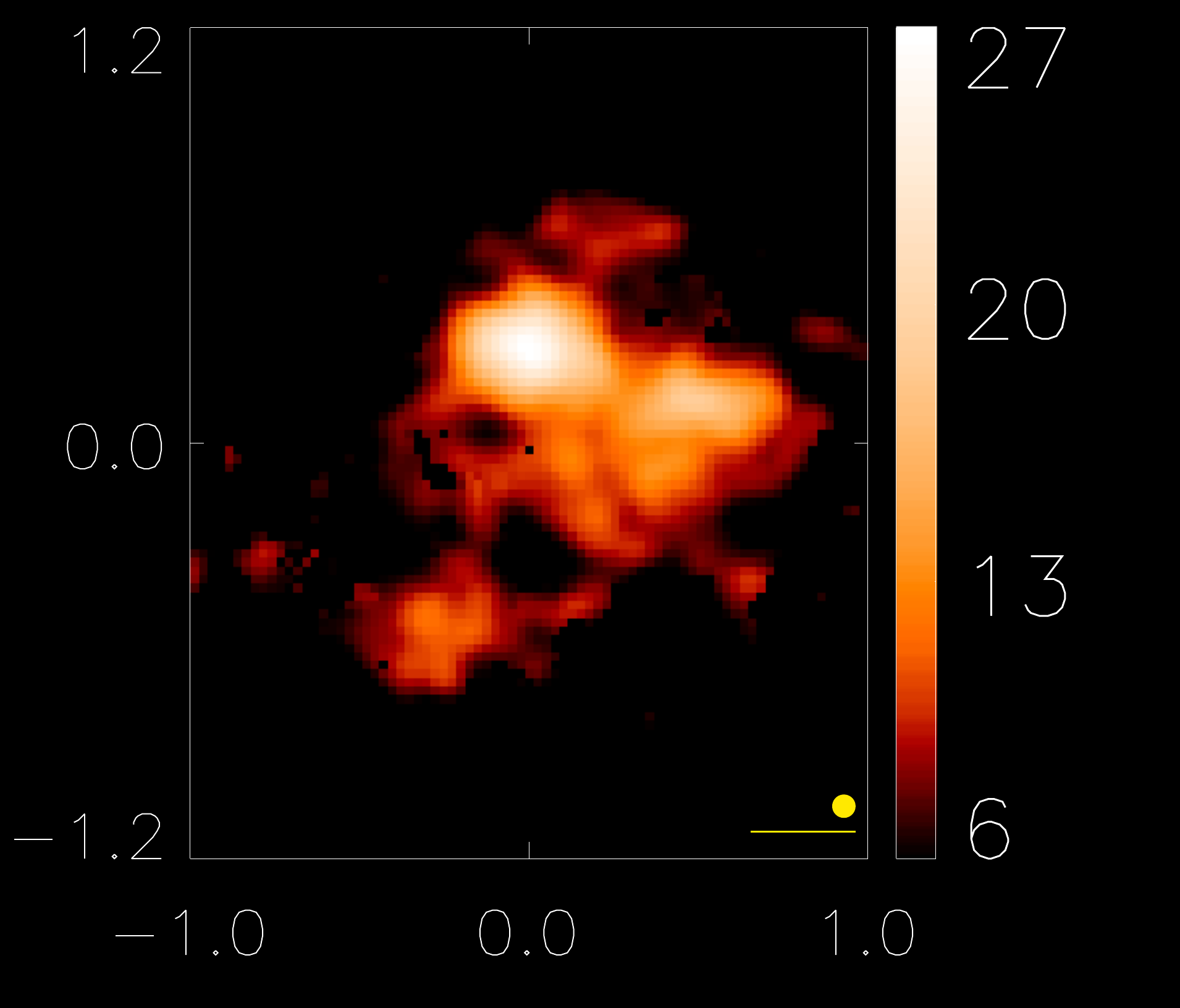} \hskip.05in
\includegraphics[width=.19\linewidth]{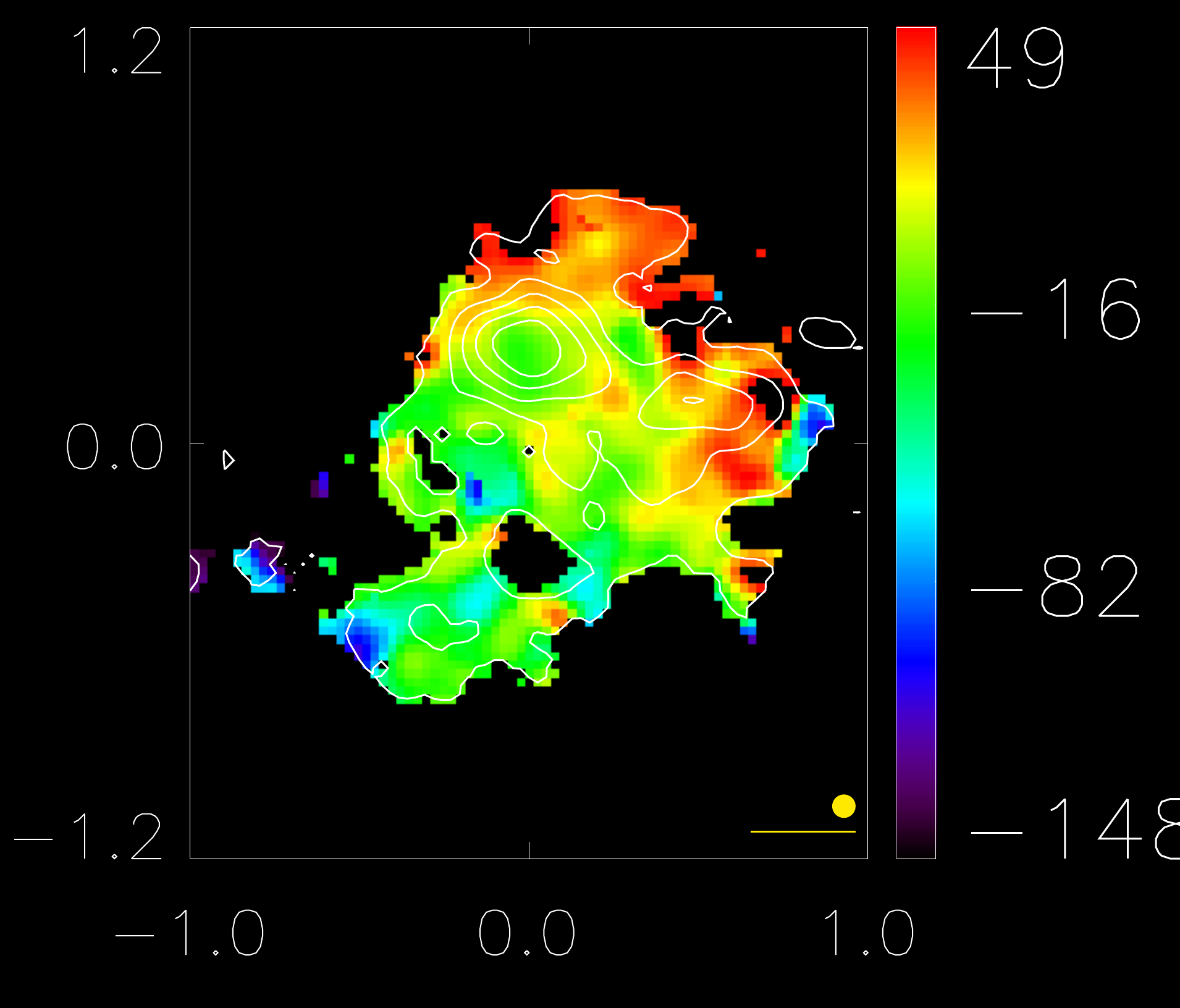} \hskip.05in
\includegraphics[width=.19\linewidth]{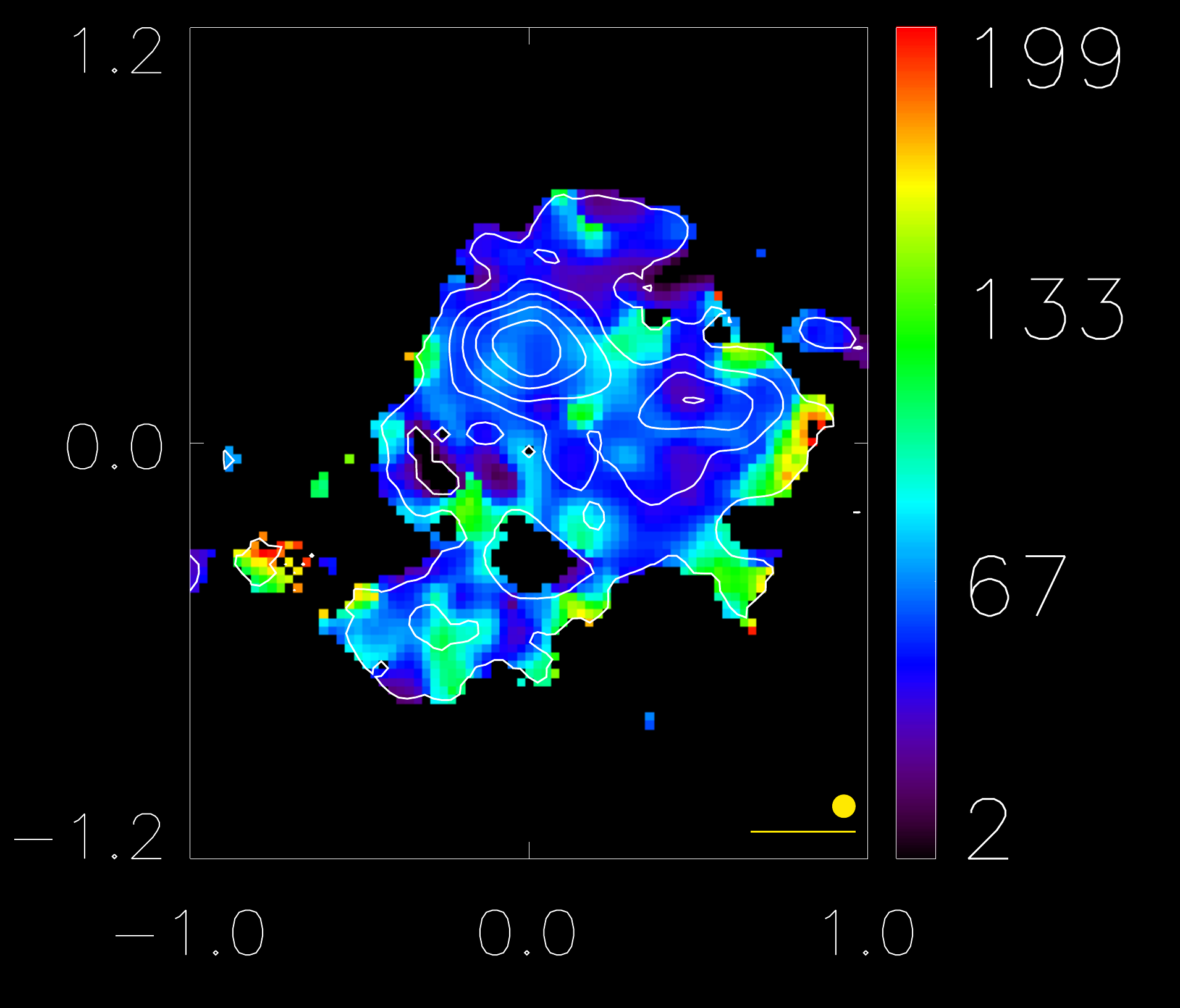} \hskip.05in
\vskip .1 in
\includegraphics[width=.19\linewidth]{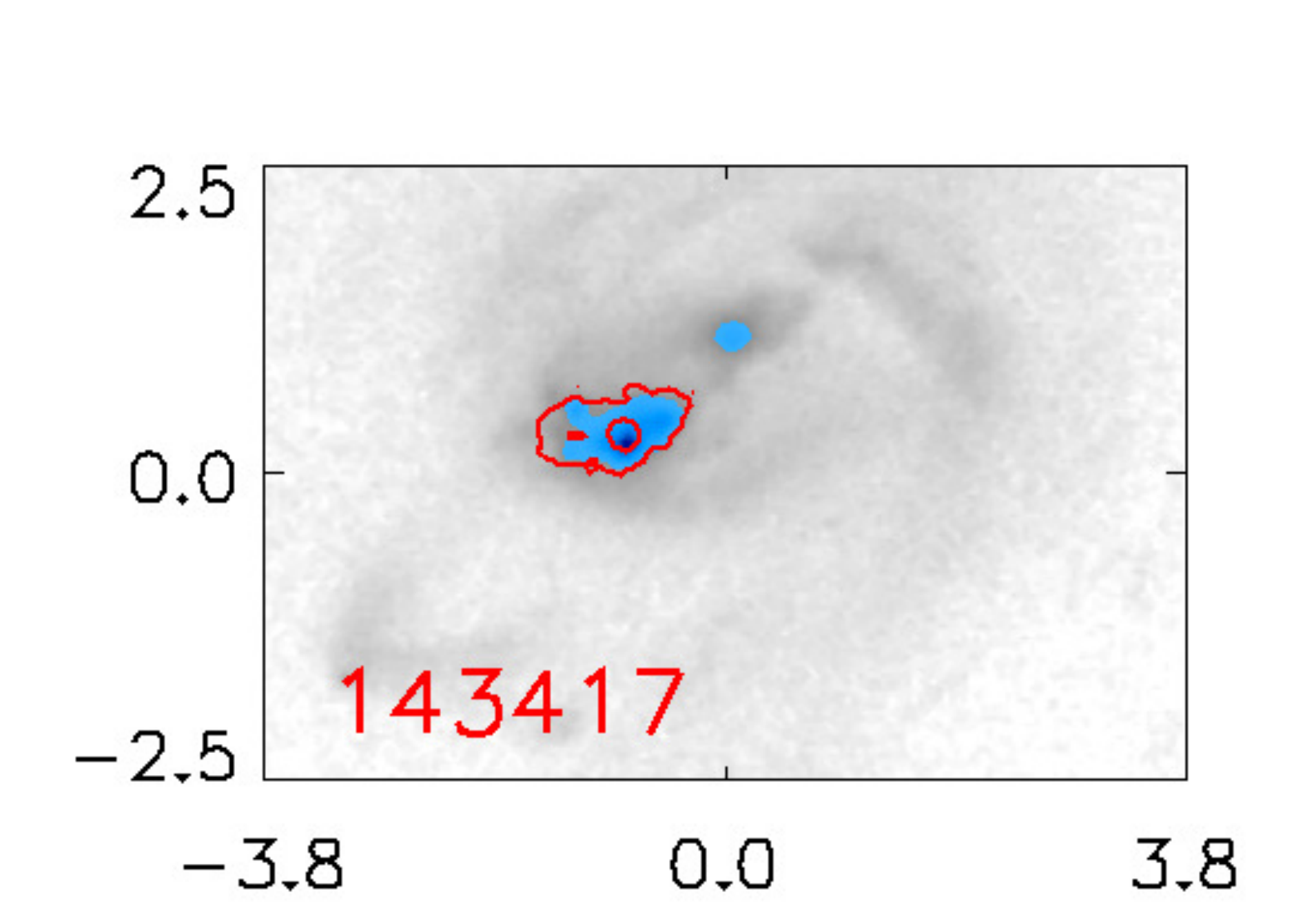} \hskip.05in
\includegraphics[width=.19\linewidth]{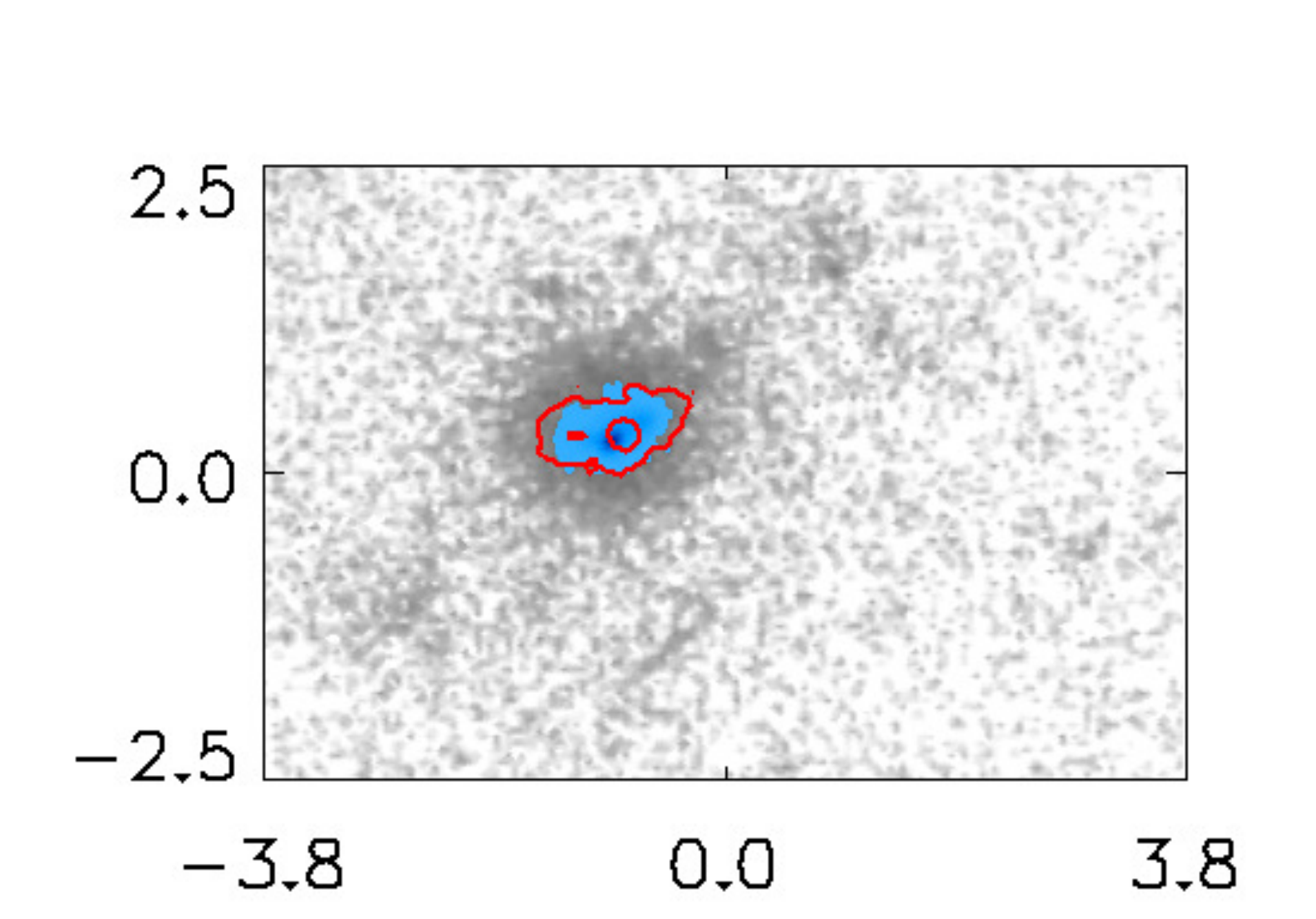} \hskip.05in
\includegraphics[width=.19\linewidth]{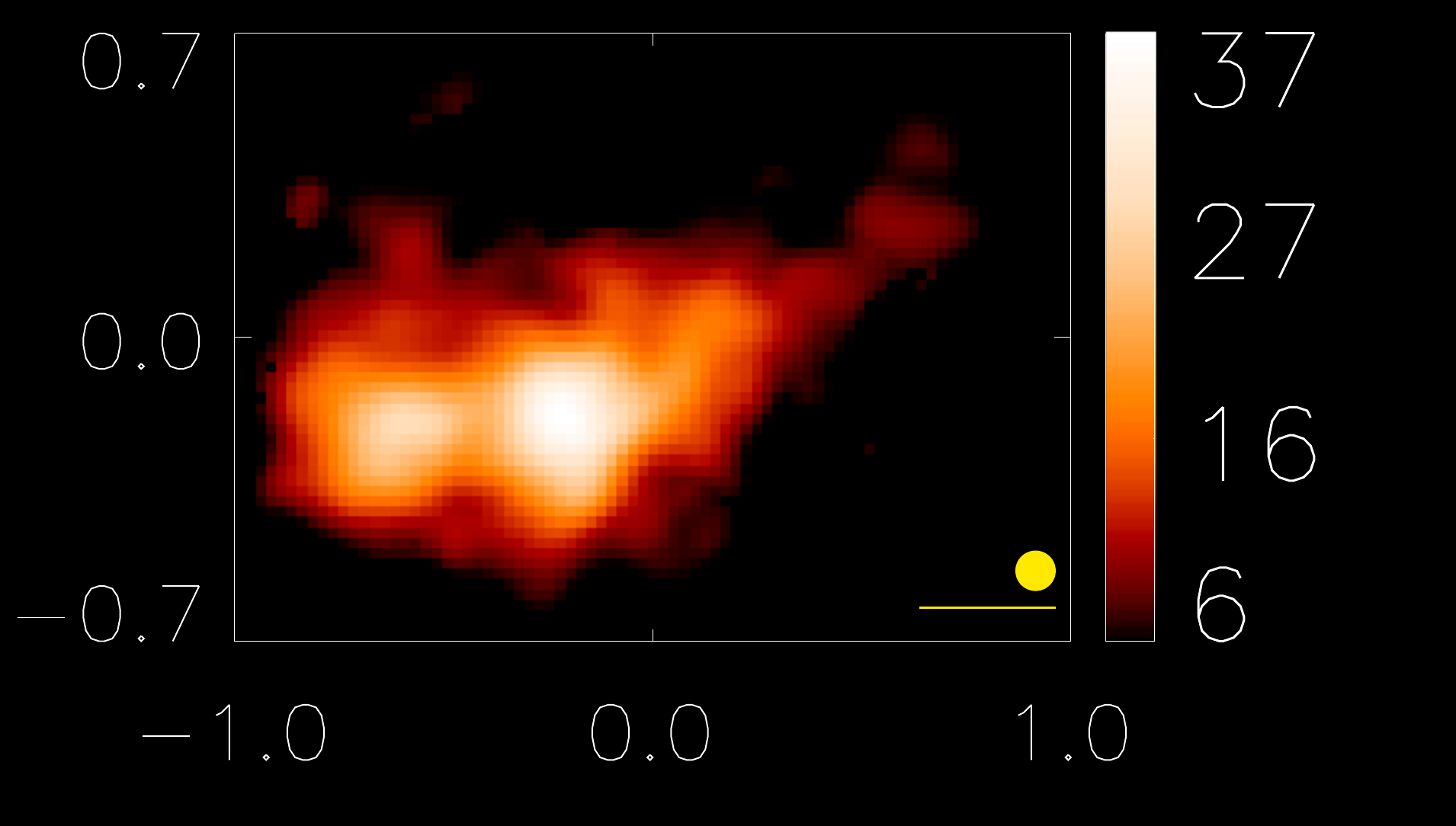} \hskip.05in
\includegraphics[width=.19\linewidth]{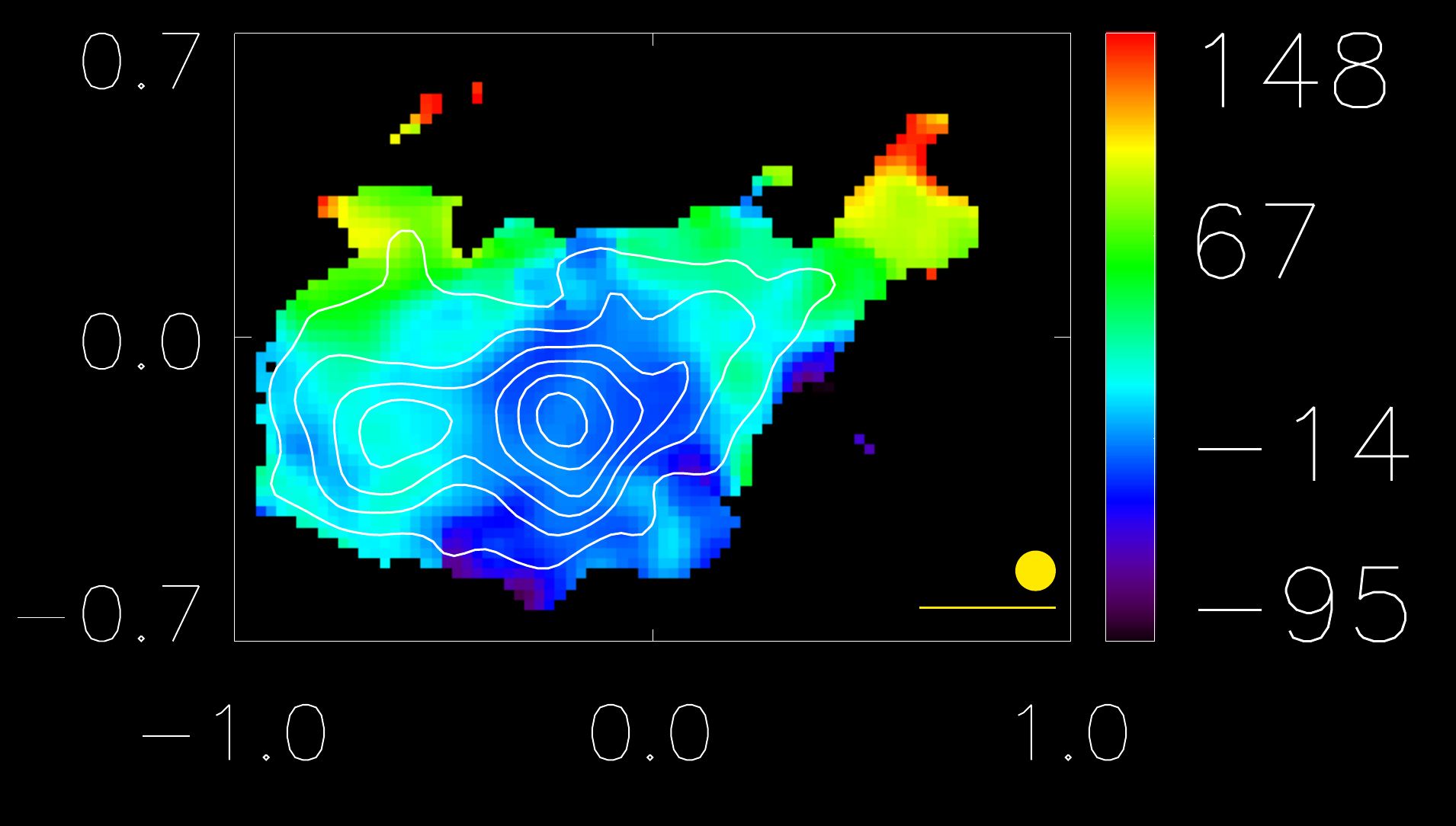} \hskip.05in
\includegraphics[width=.19\linewidth]{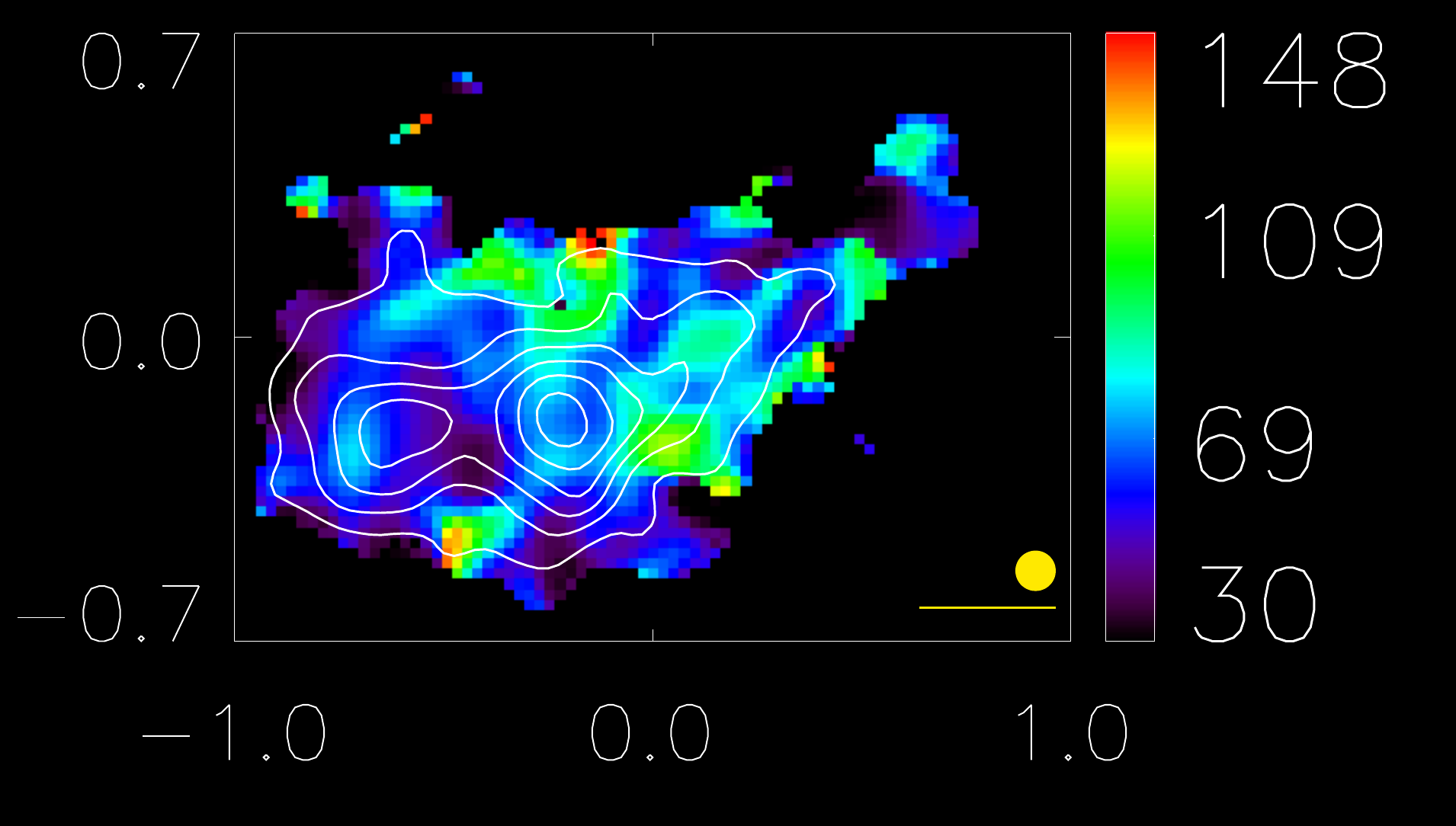} \hskip.05in

\caption{
continued.\label{fig:vdmaps4}
}
\end{figure*}

\addtocounter{figure}{-1}
\begin{figure*}[ht]

\includegraphics[width=.19\linewidth]{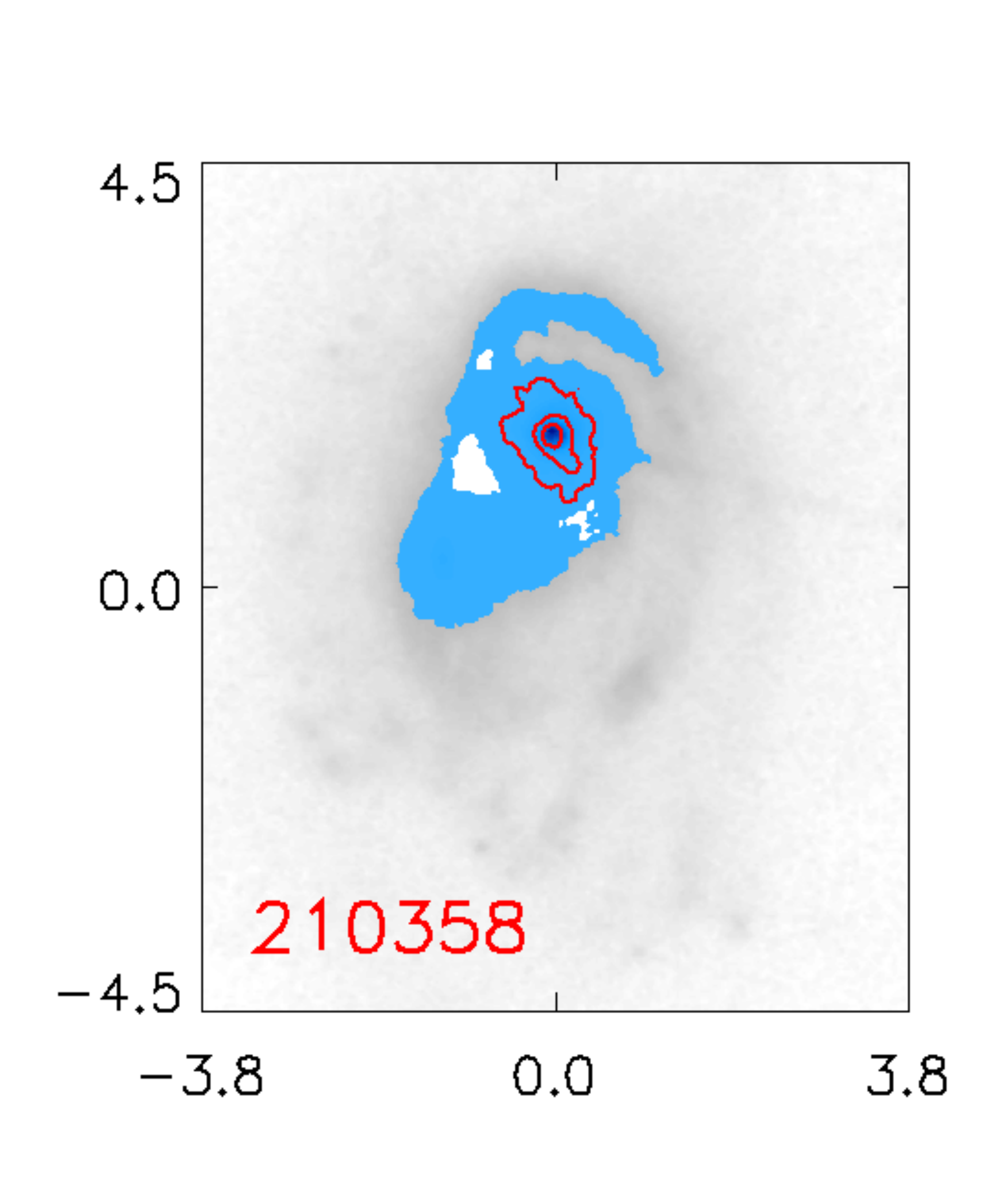} \hskip.05in
\includegraphics[width=.19\linewidth]{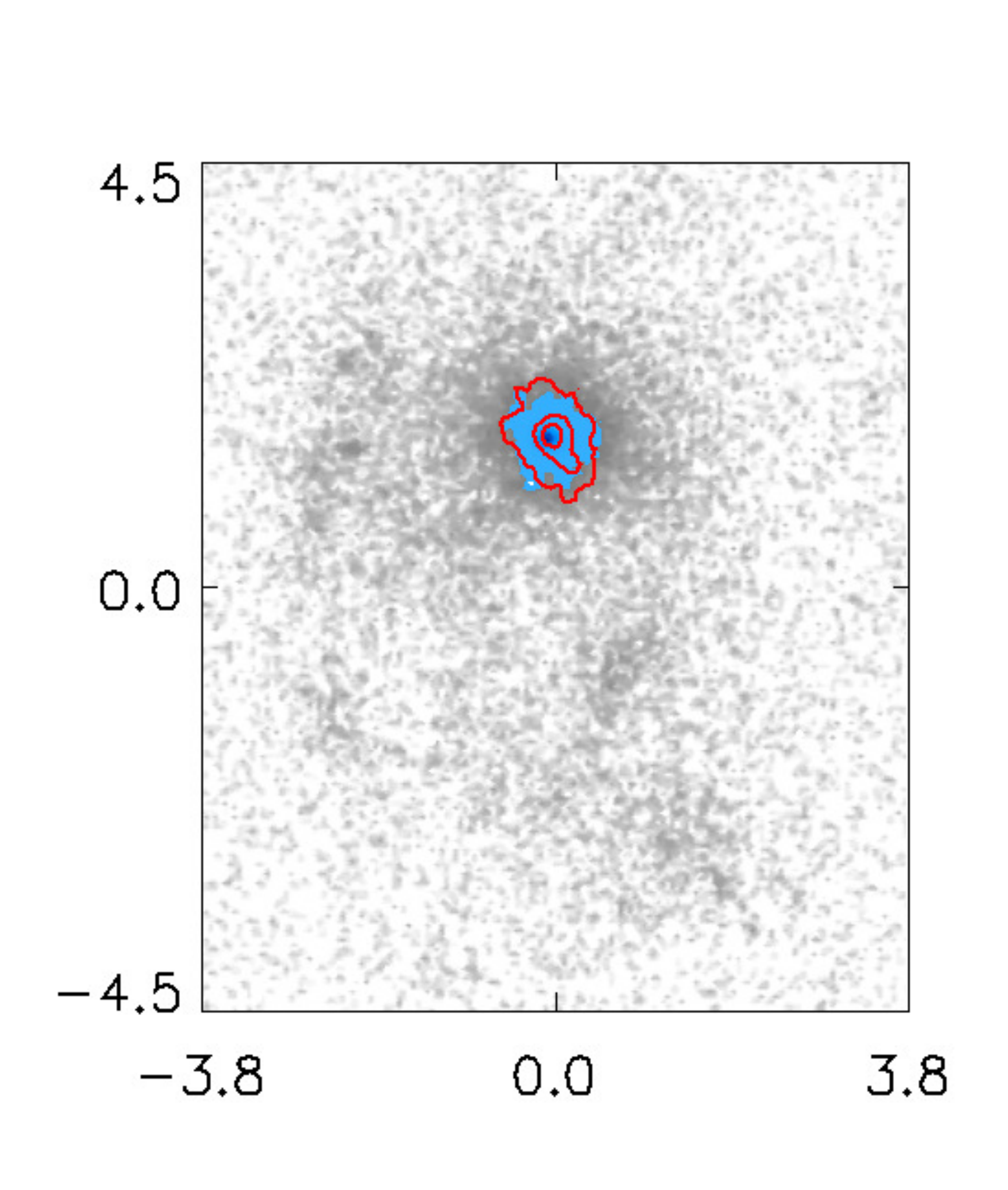} \hskip.05in
\includegraphics[width=.19\linewidth]{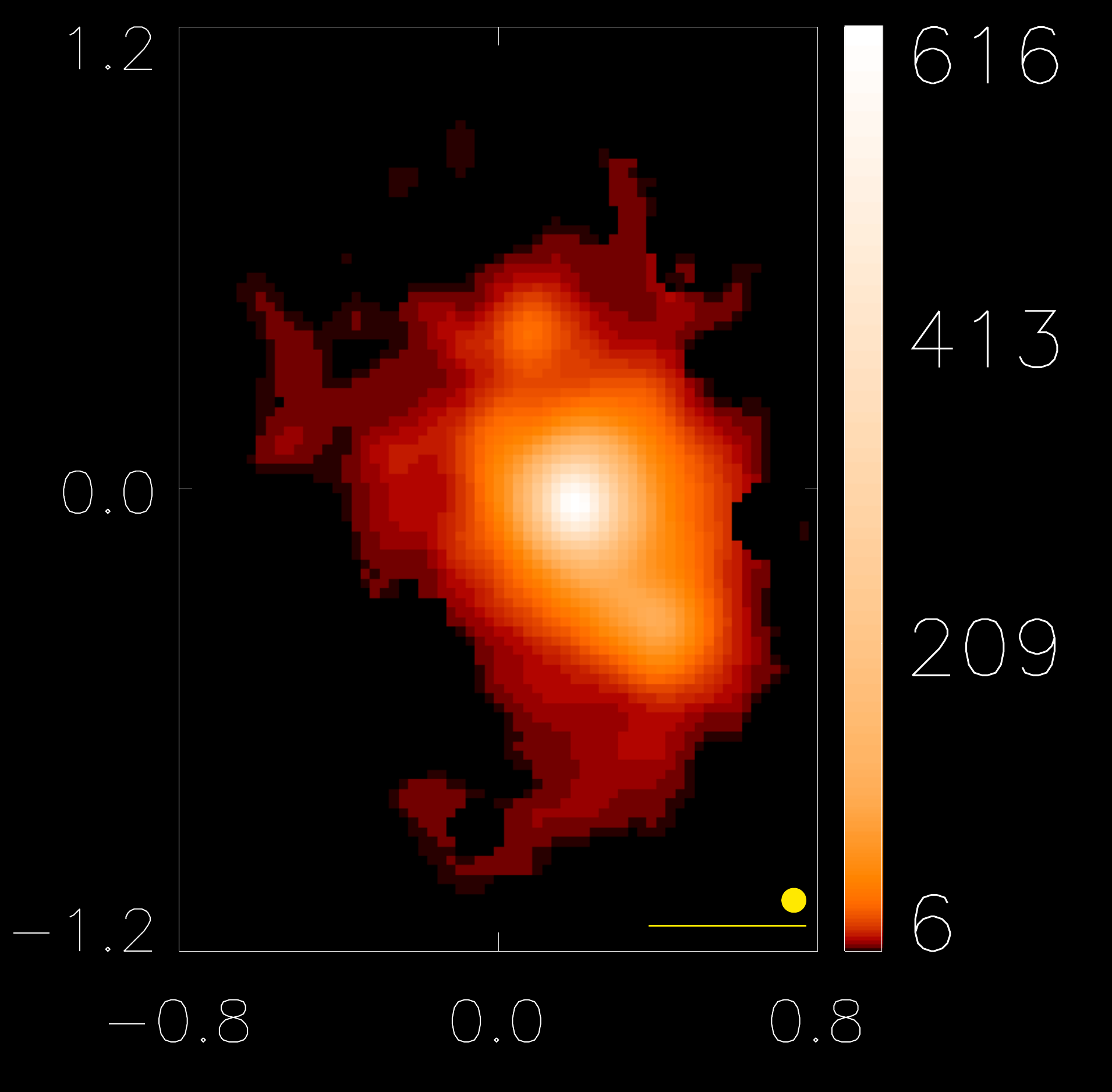} \hskip.05in
\includegraphics[width=.19\linewidth]{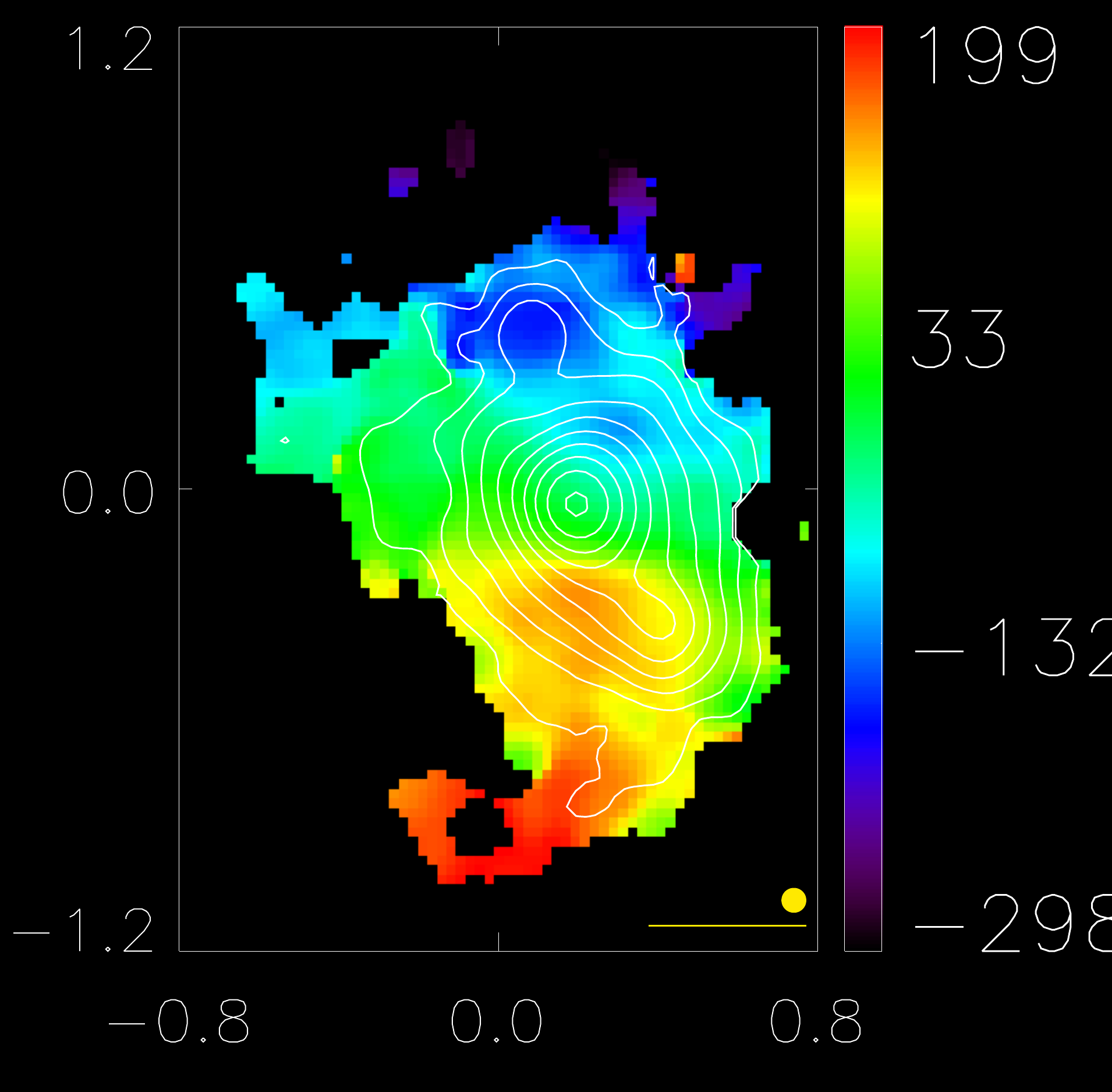} \hskip.05in
\includegraphics[width=.19\linewidth]{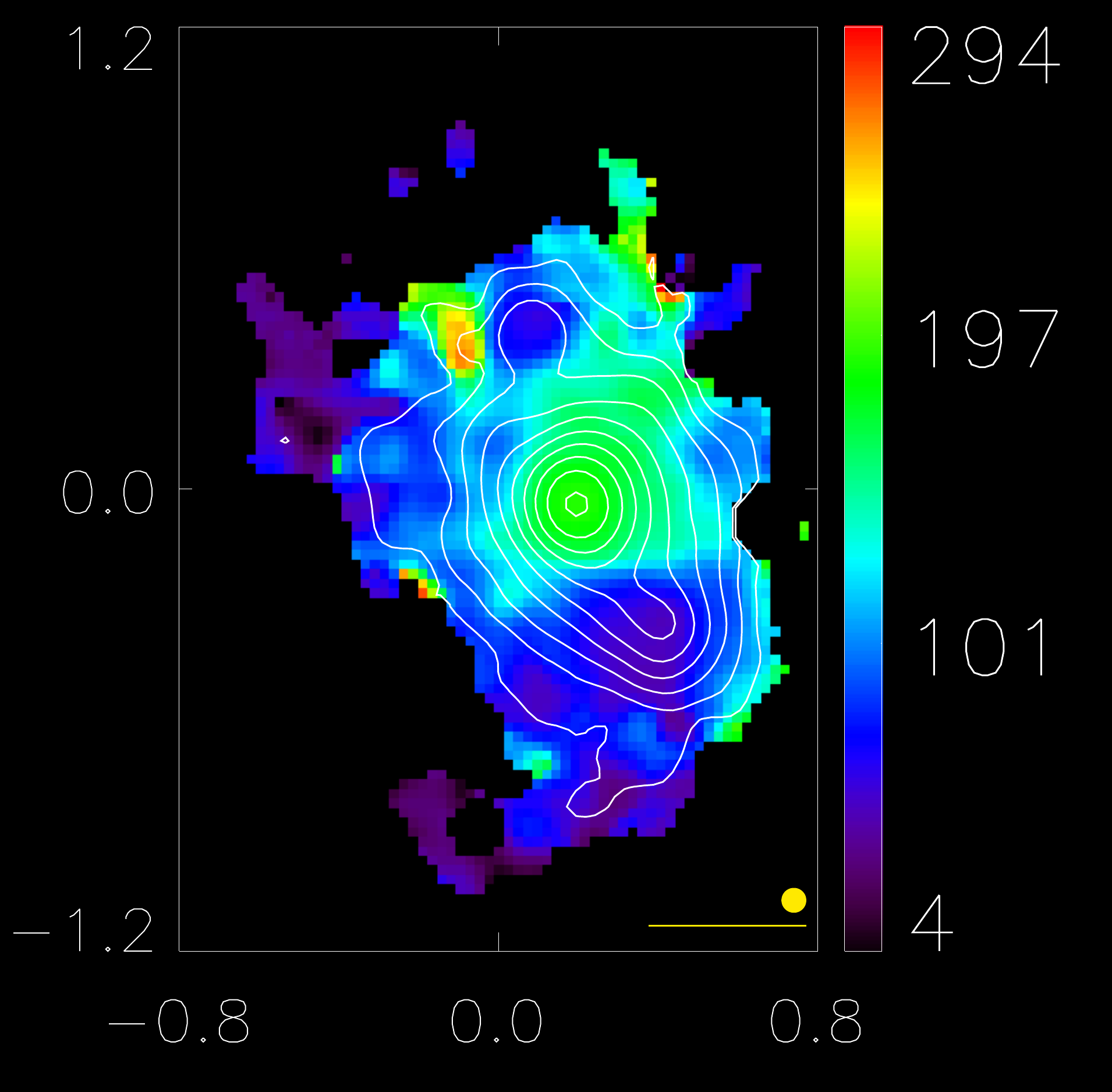} \hskip.05in
\vskip .1 in
\includegraphics[width=.19\linewidth]{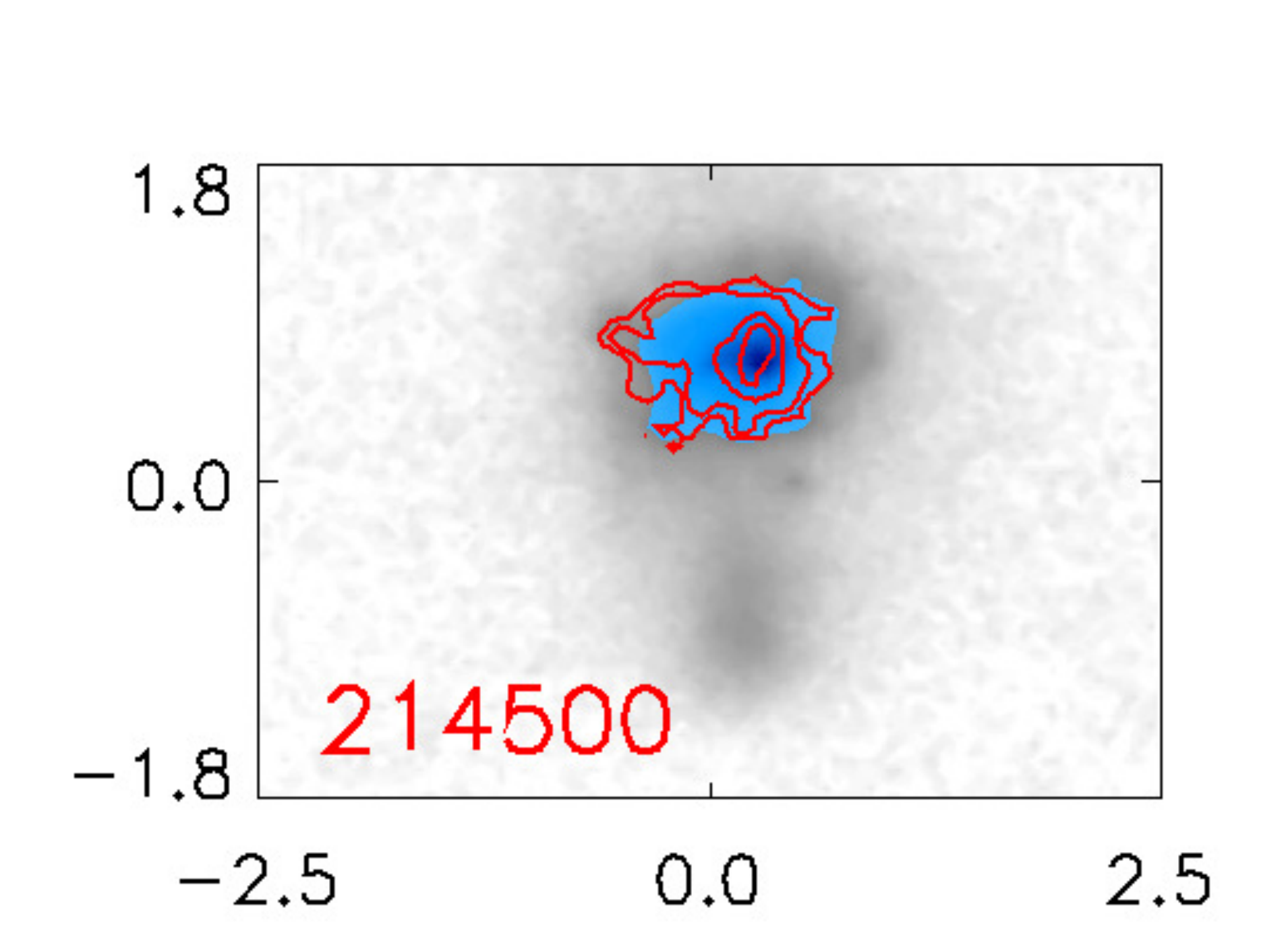} \hskip.05in
\includegraphics[width=.19\linewidth]{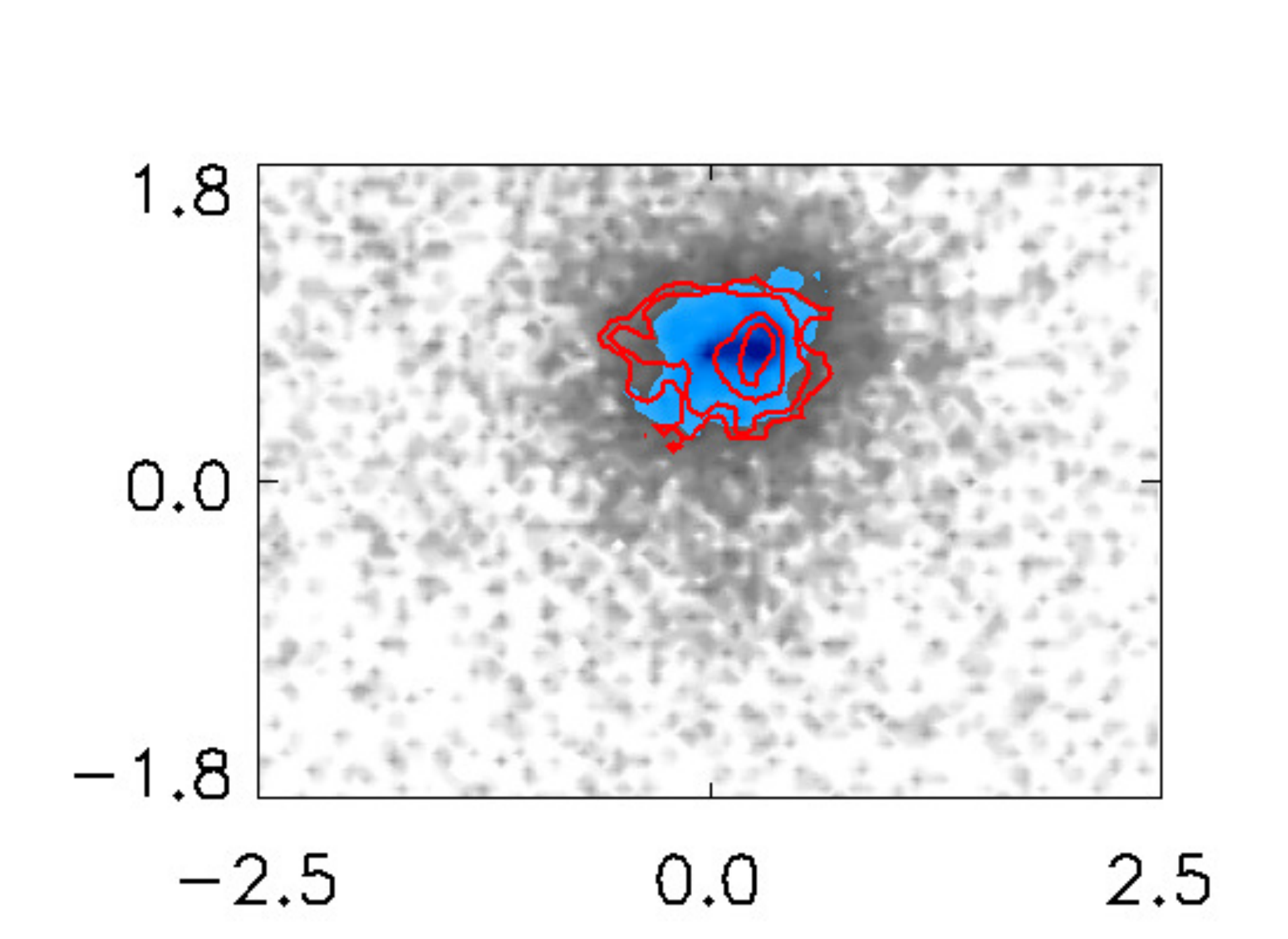} \hskip.05in
\includegraphics[width=.19\linewidth]{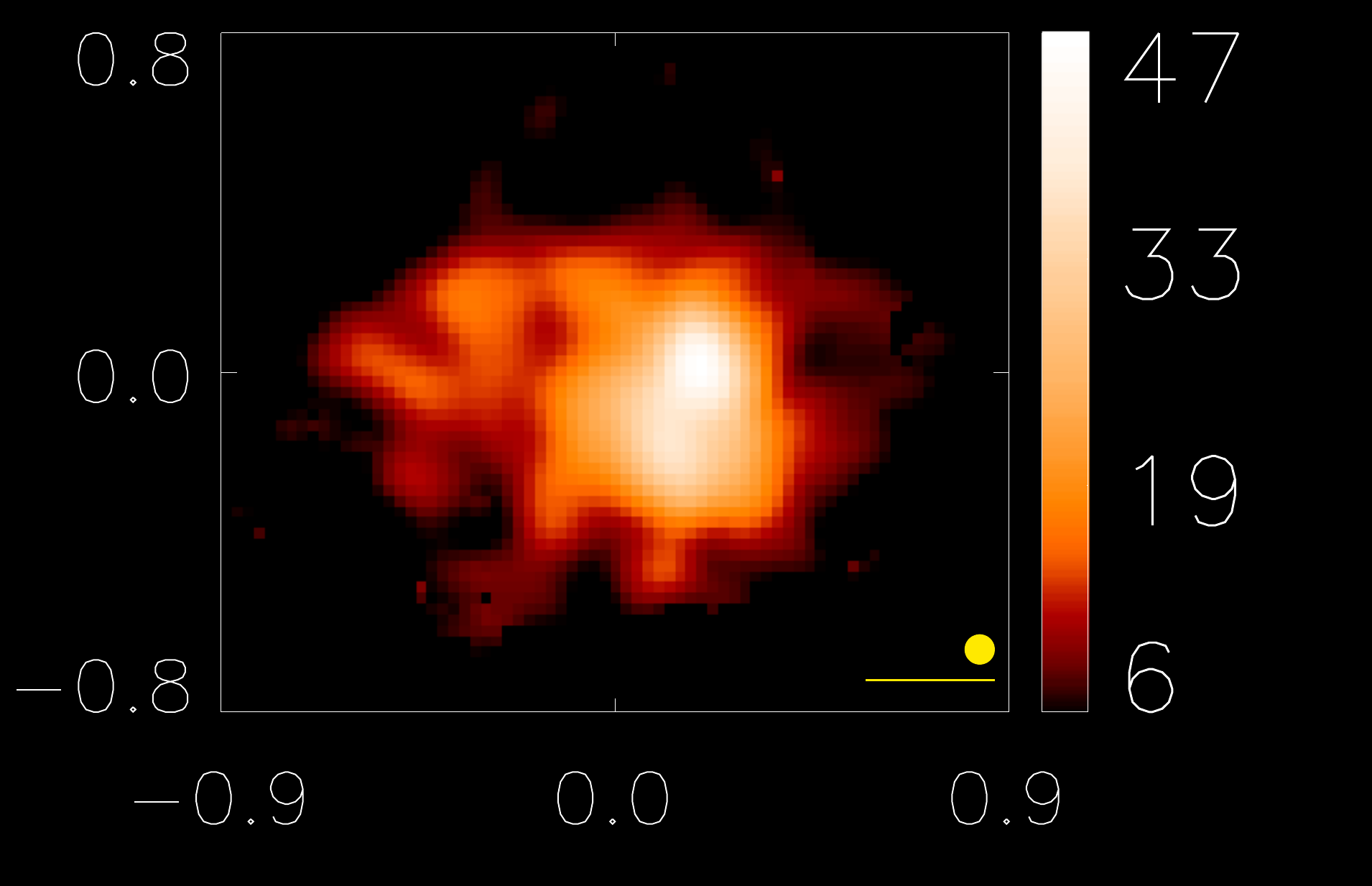} \hskip.05in
\includegraphics[width=.19\linewidth]{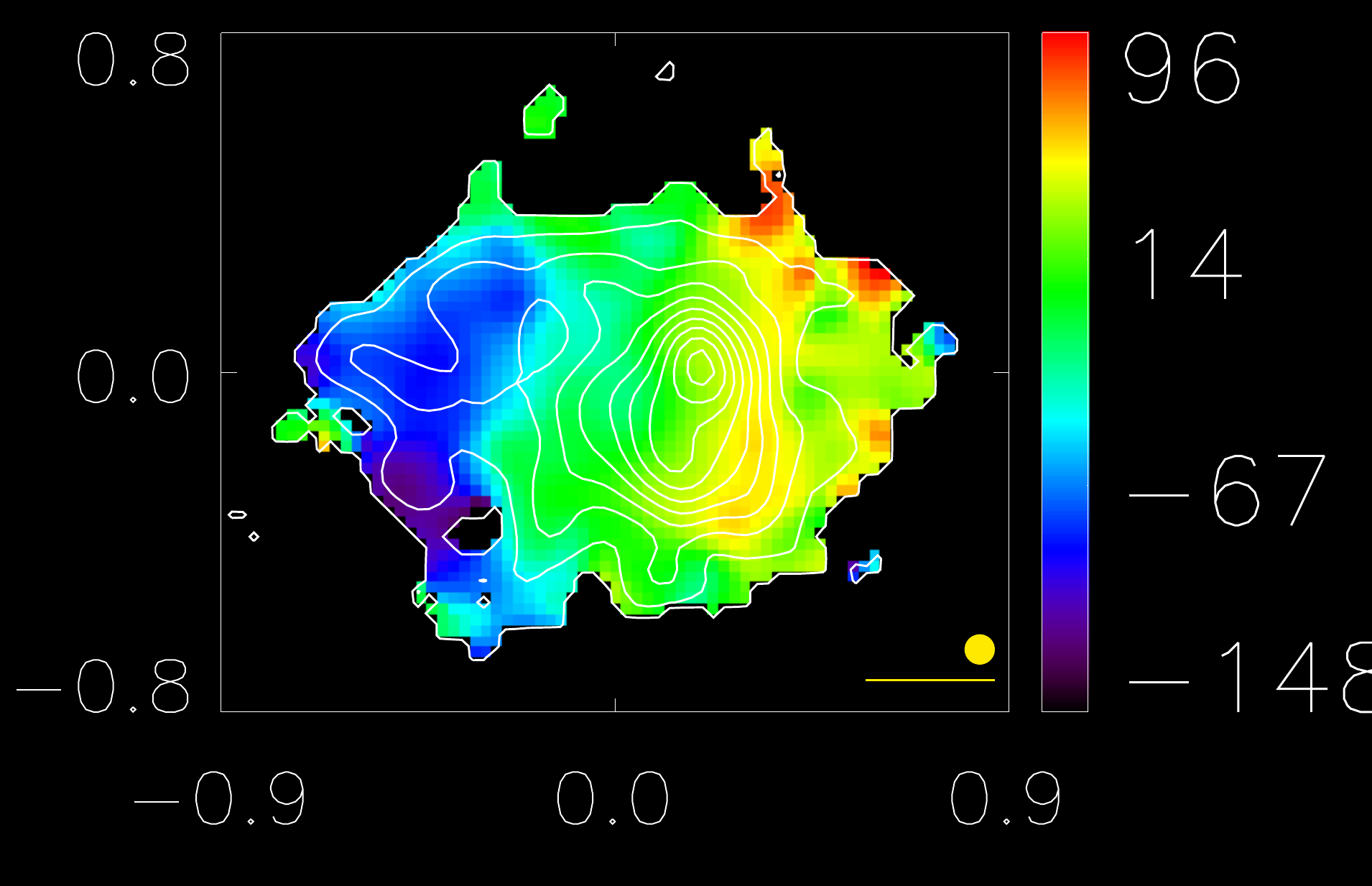} \hskip.05in
\includegraphics[width=.19\linewidth]{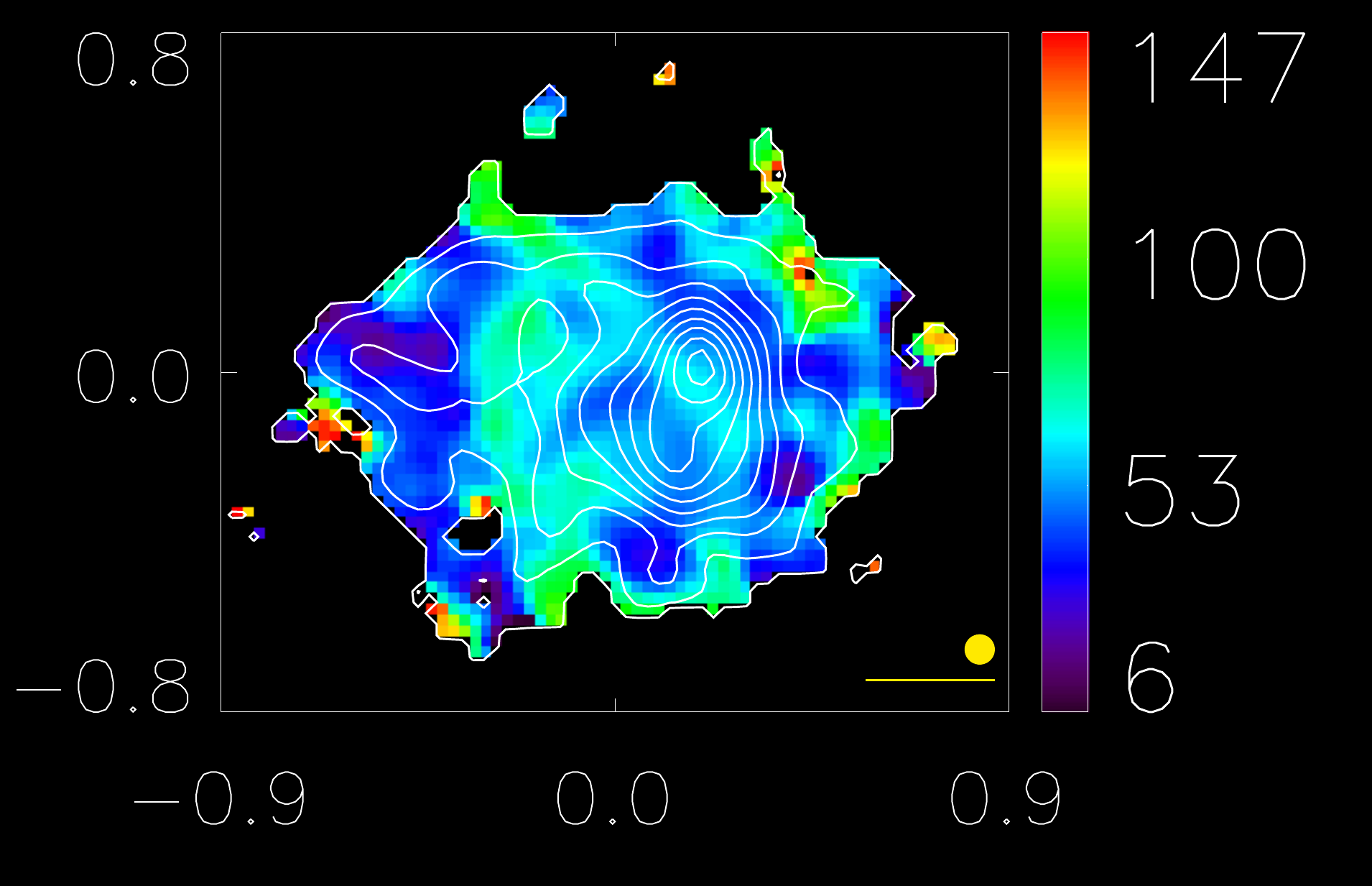} \hskip.05in
\vskip .1 in
\includegraphics[width=.19\linewidth]{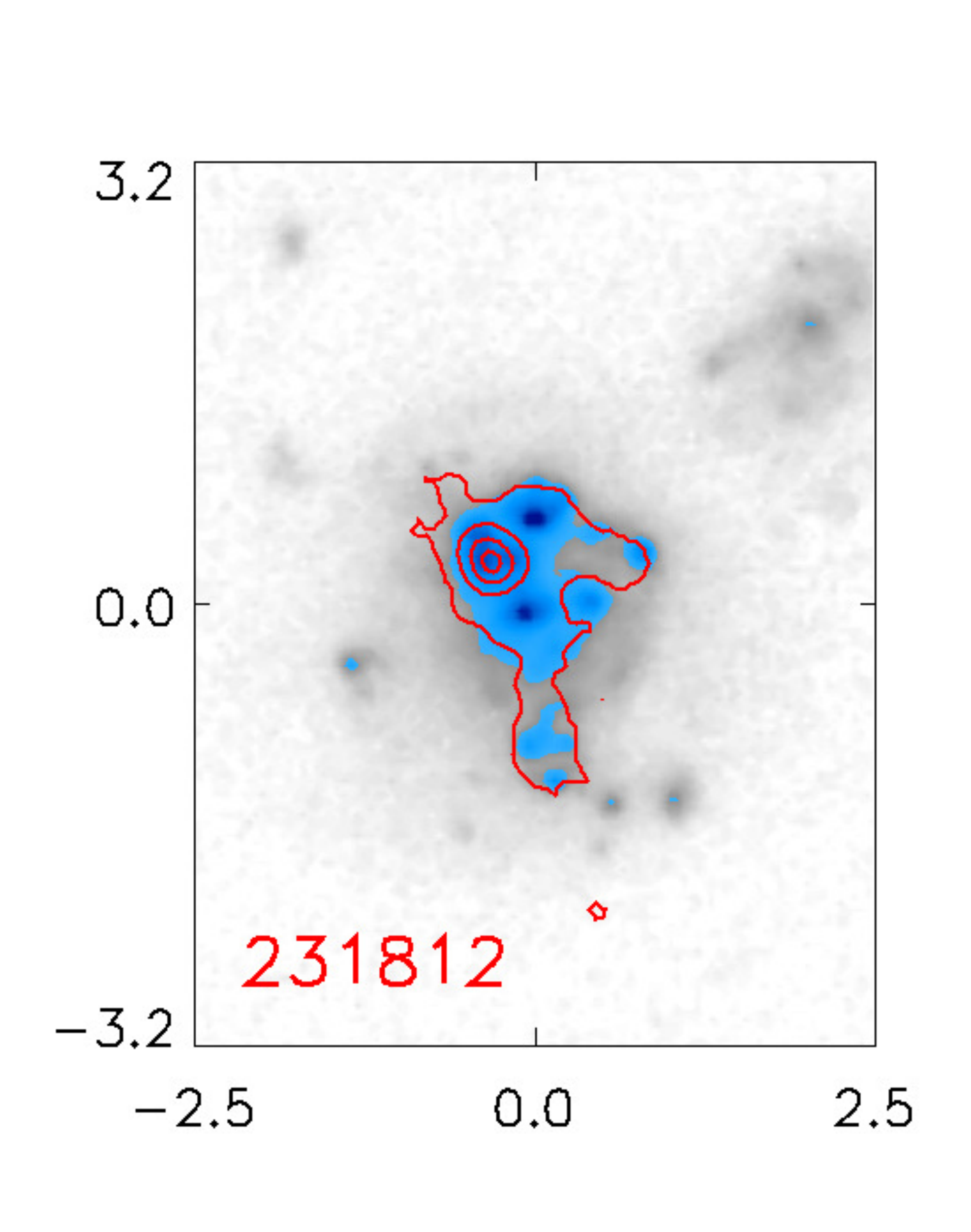} \hskip.05in
\includegraphics[width=.19\linewidth]{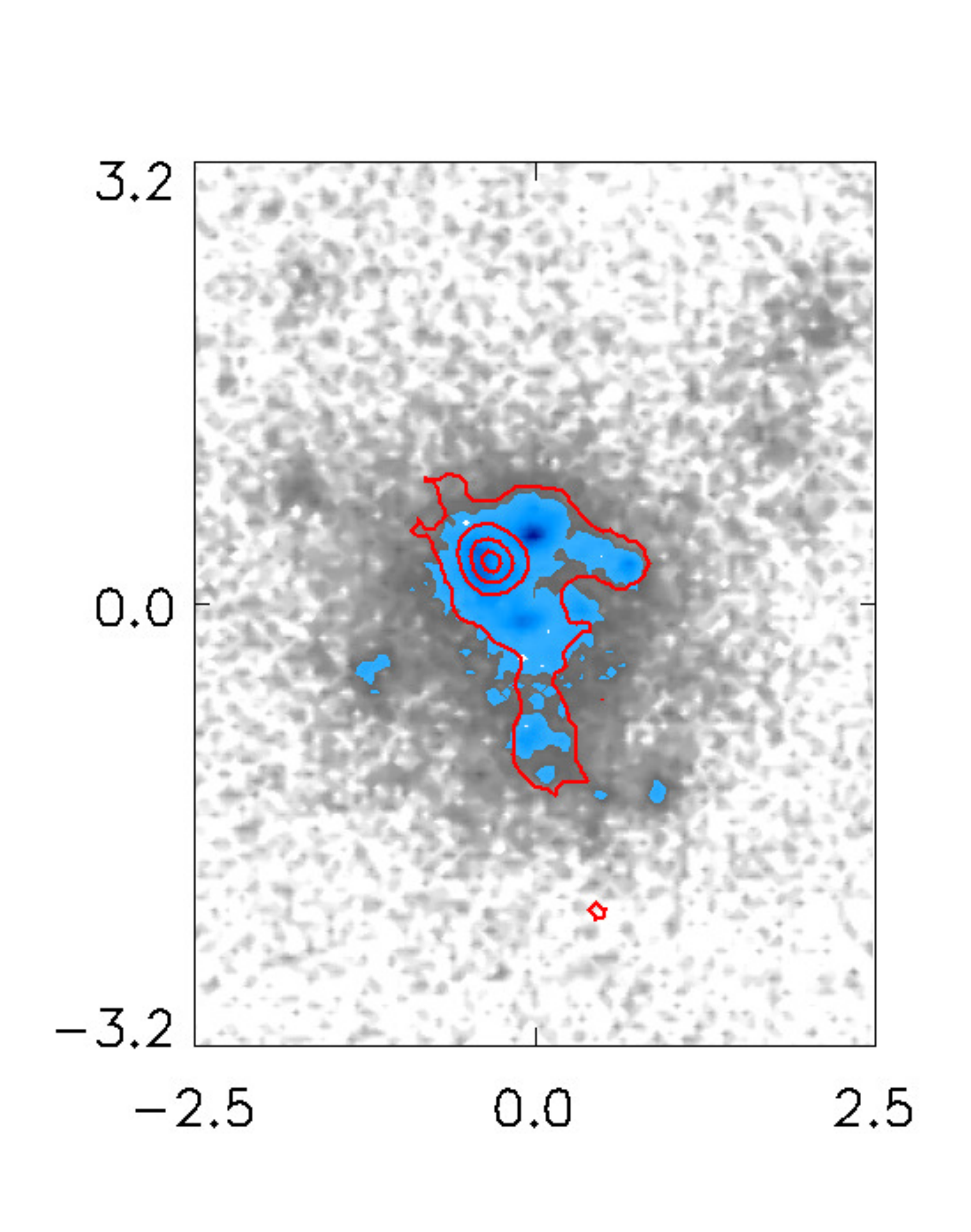} \hskip.05in
\includegraphics[width=.19\linewidth]{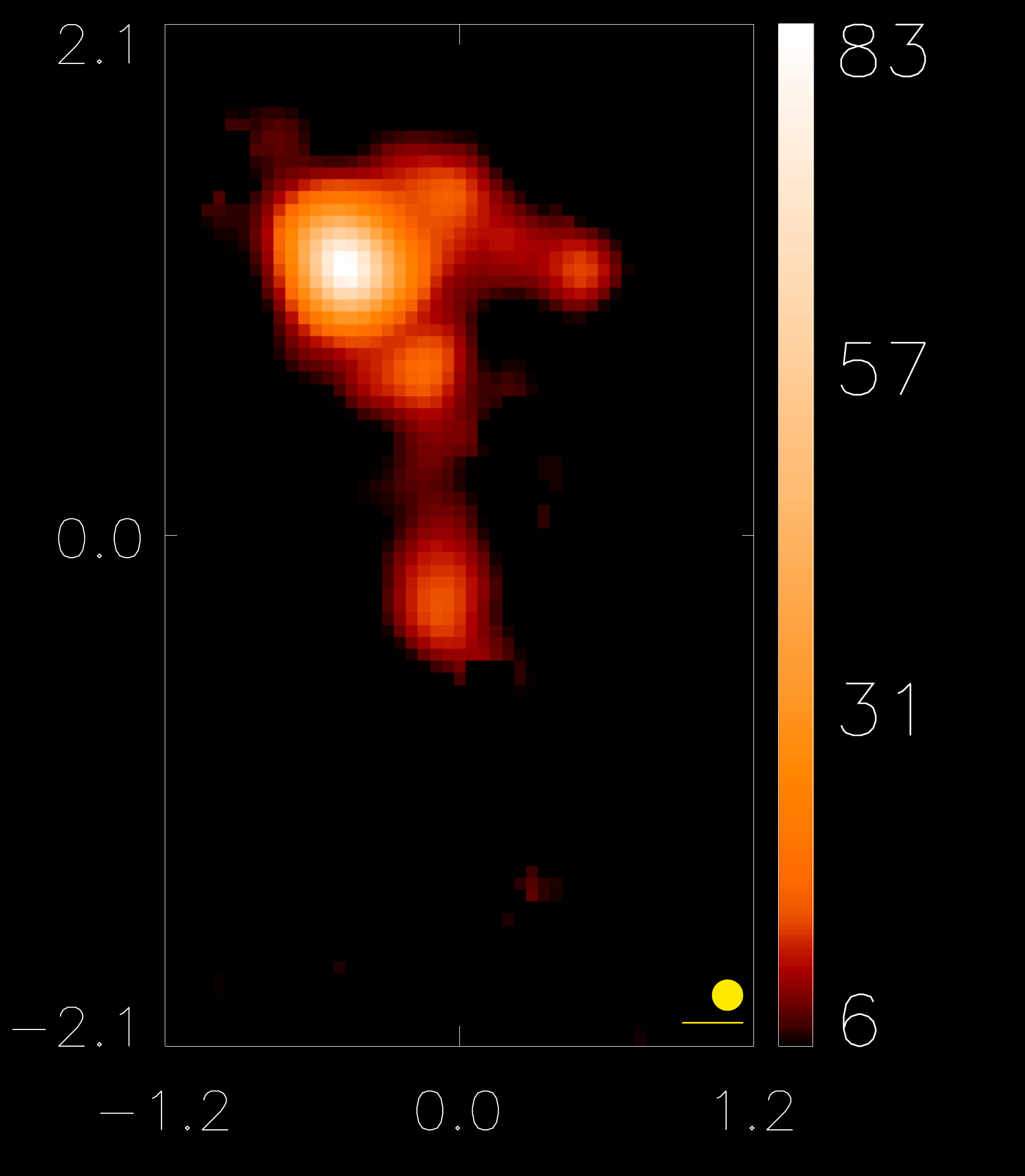} \hskip.05in
\includegraphics[width=.19\linewidth]{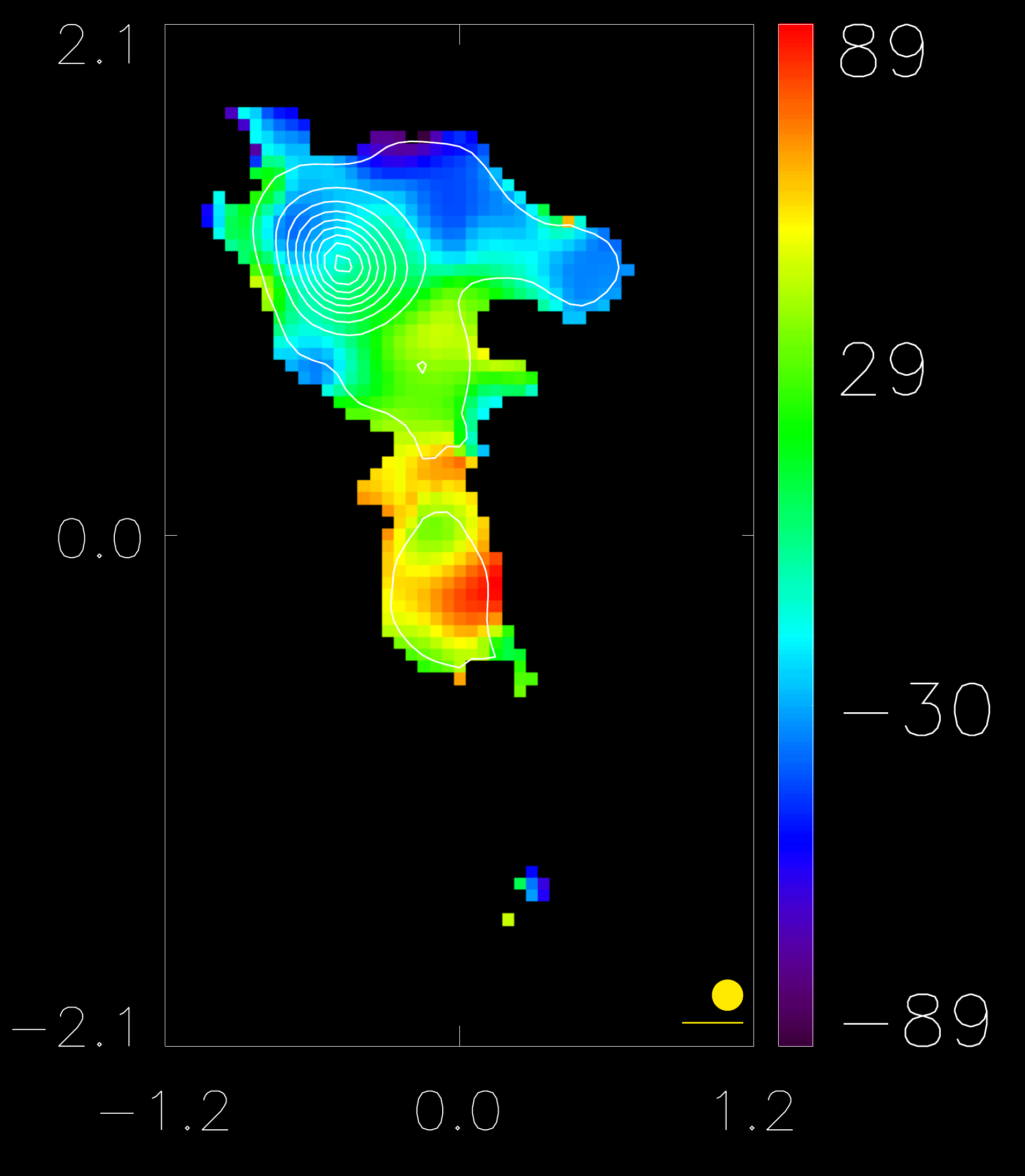} \hskip.05in
\includegraphics[width=.19\linewidth]{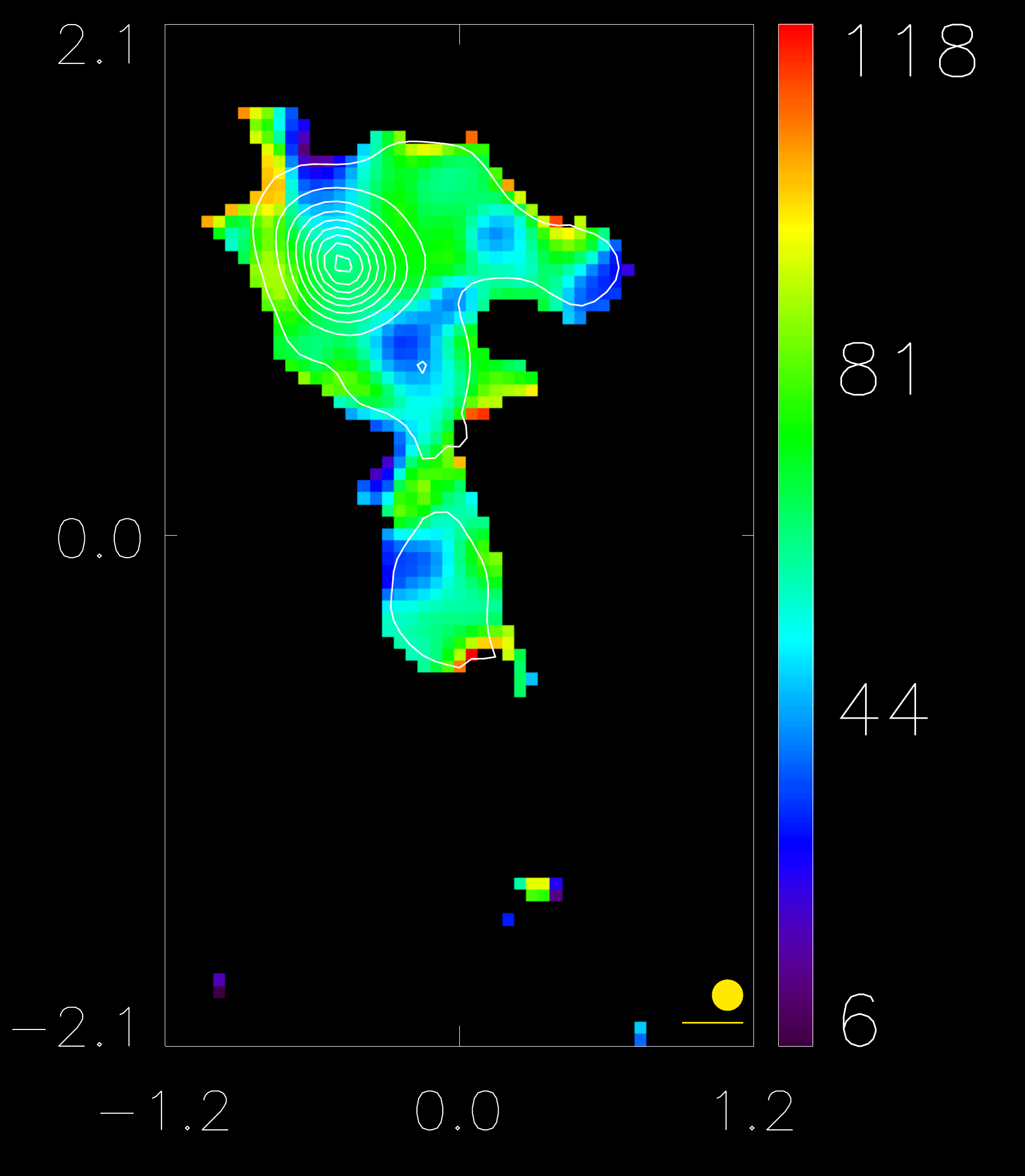} \hskip.05in

\caption{
continued.\label{fig:vdmaps5}
}
\end{figure*}

\subsection{Simulation to High Redshift}\label{section:hiz}

As briefly discussed in section \ref{section:intro}, LBAs have been defined on the basis of UV luminosity and surface brightness thresholds as appropriate for high-redshift Lyman break galaxies. Previous studies have supported the analogy, finding both apparent and physical properties consistent with those of their high-z counterparts. In this section we investigate the parallel in terms of gas kinematics of LBAs compared to LBGs.

In order to allow for a direct comparison between kinematics of LBA- and LBG-type systems, we have artificially redshifted all our galaxies to $z\sim 2.2$\footnote[2]{This precise redshift was chosen to avoid major OH emission lines.}, and reobserved them with the simulated IFU prescriptions described by \citet{Law2006}. At this redshift, these galaxies would be observed in H$\alpha$. ÊWe scale our observed Pa-$\alpha$ flux maps to the total H$\alpha$ fluxes determined by SDSS. One should notice that the code used to artificially redshift our sample represents the exact same instrument, observational setup and reduction software as in \citet{Law2009}, and has been shown to appropriately reproduce actual OSIRIS observations. This ensures the robustness of comparisons between the LBA sample and that of LBGs presented in \citet{Law2009}

\begin{figure*}[ht]

\includegraphics[height=1.55in]{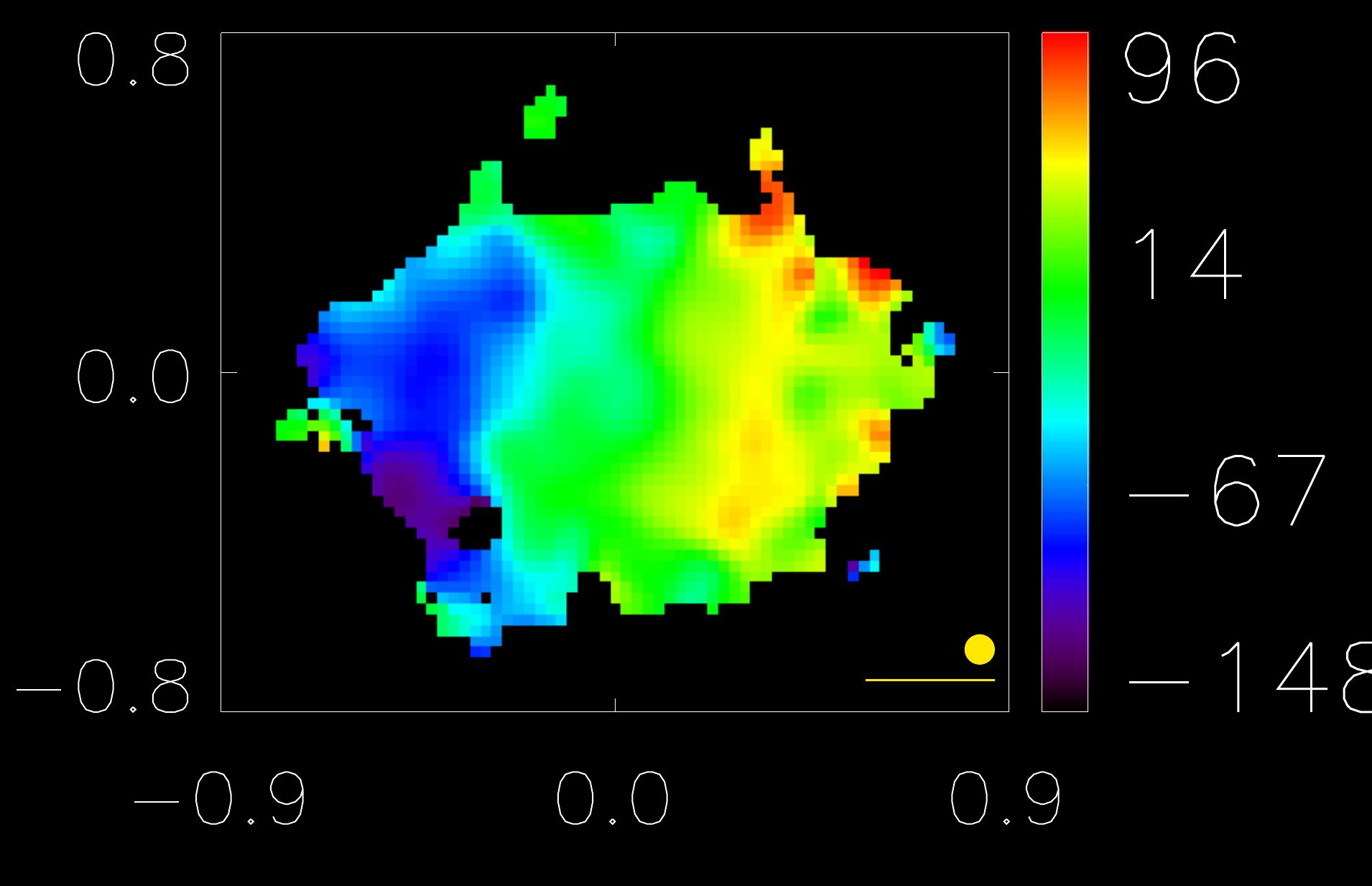} \hskip.1in
\includegraphics[height=1.55in]{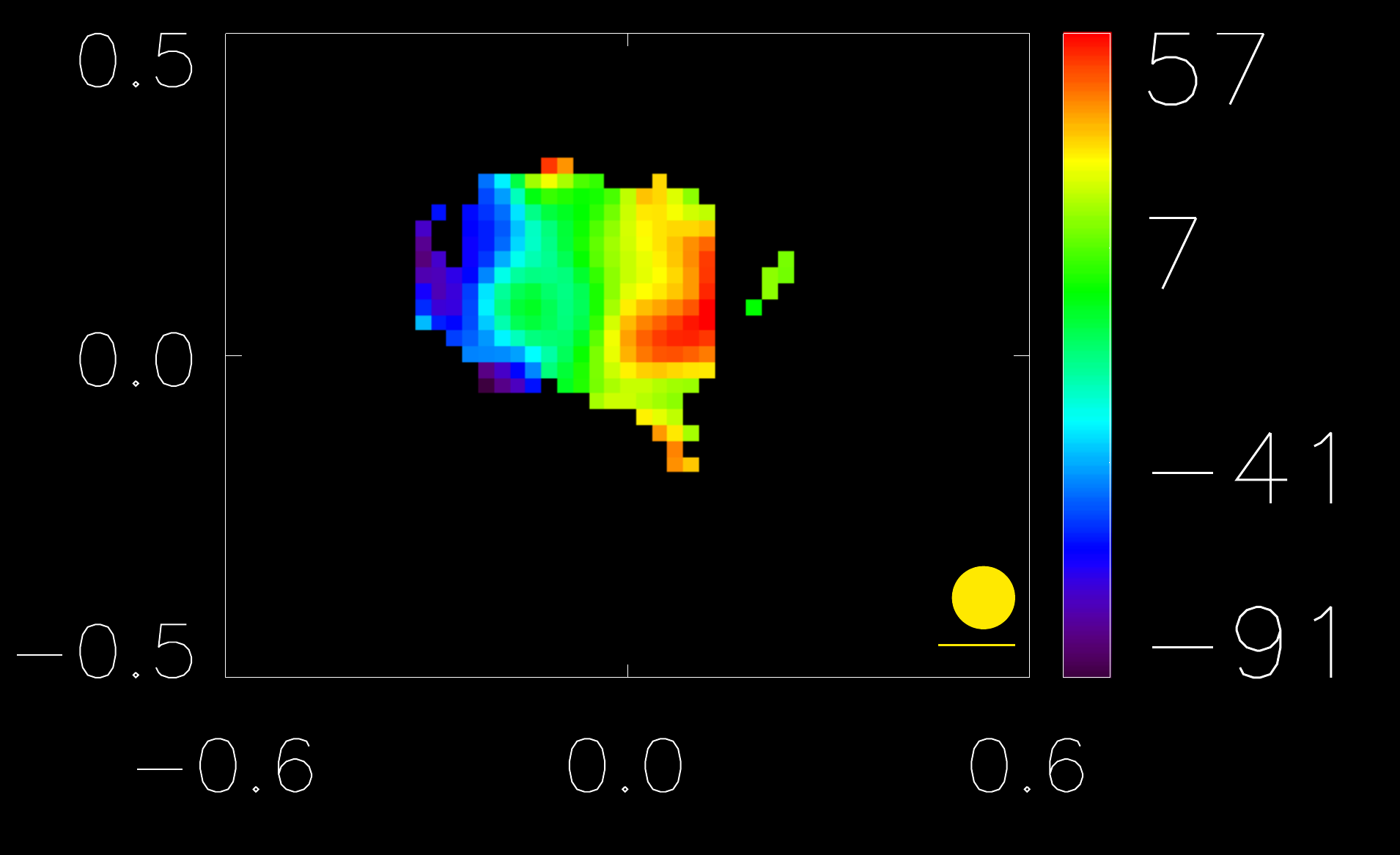} \hskip.1in
\includegraphics[height=1.55in]{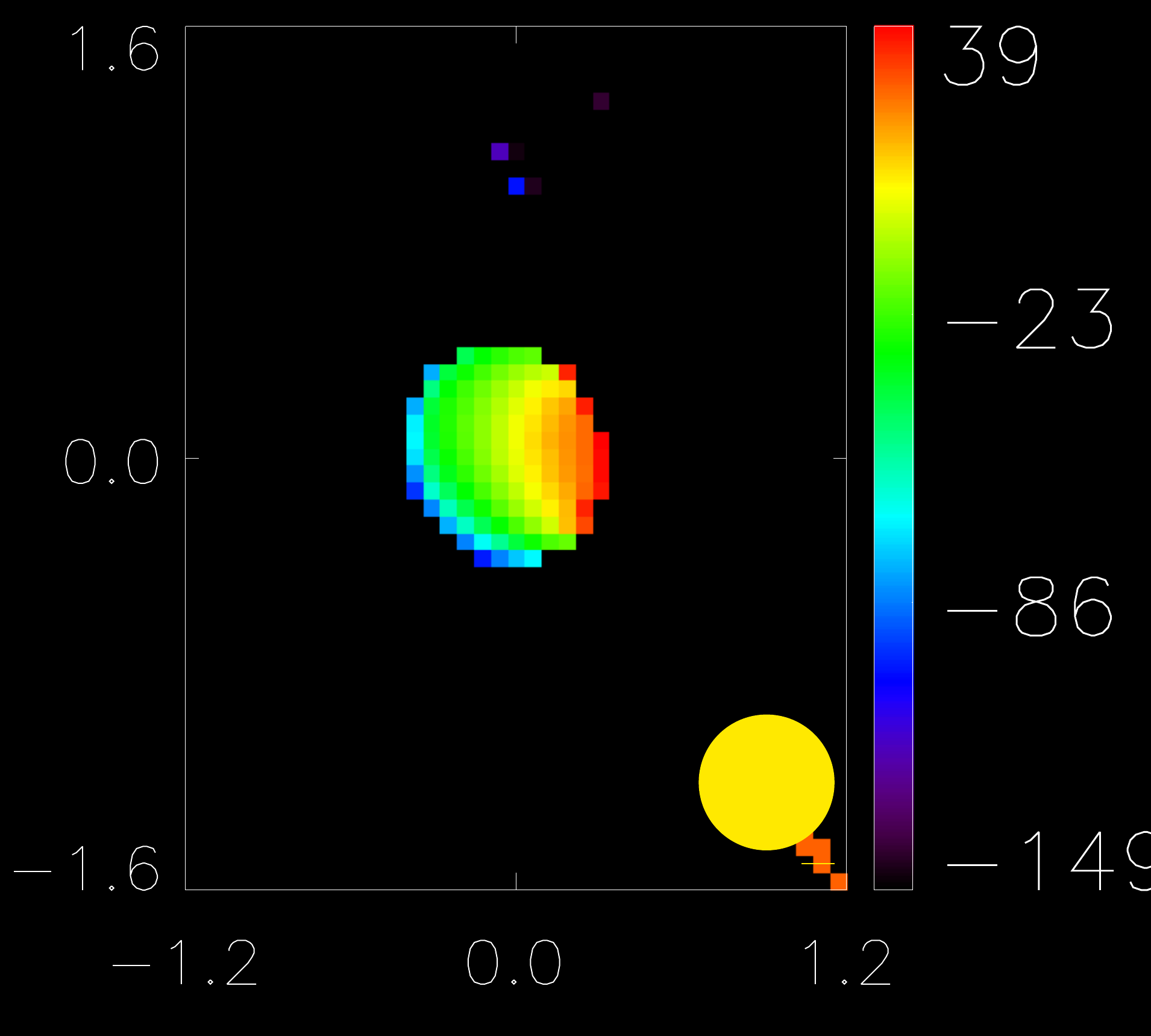}

\caption{Velocity maps of 214500 at its intrinsic redshift ({\it left}) and artificially redshifted to $z\sim 2.2$, as observed by OSIRIS ({\it center}) and SINFONI ({\it right}). Legends are the same as in Figure (\ref{fig:vdmaps}). As expected, spatial resolution is lower, and low surface brightness features are harder to distinguish.} \label{fig:maps_hiz}

\end{figure*}

We have also artificially redshifted our data and simulated observations with the SINFONI instrument, in non-AO mode. In this case, the optimal hydrogen line-emission surface brightness detection limits in \citet{ForsterSchreiber2009} is comparable to our sample: on one hand the instrument is more sensitive, H$\alpha$ is brighter and there is no loss due to the adaptive optics system; on the other hand, cosmological surface brightness dimming would make sources up to 200 times fainter per solid angle unit. Therefore, we simply degrade our spatial resolution with a 0.5" gaussian kernel, rebinning our datacubes to the nominal 0.125" pixel scale of SINFONI, while simultaneously reducing the total angular size of the galaxy as determined by the ratio of angular diameter distances at $z=2.2$ and their actual redshift. Examples for the resulting velocity maps can be seen in Fig. \ref{fig:maps_hiz}.

As discussed in \citet{Overzier2010}, where a similar technique was used for HST images, much detail is not observed due to loss of spatial resolution and/or surface brightness dimming. As in the case of HST observations, the loss in spatial resolution causes different star-forming regions to be confused into one larger clump. This might lead to misinterpreting multiple clumps with velocity differences as one larger, smoother rotating disk, with implications for inferences about its formation mechanism (see sections \ref{section:results}, \ref{section:discuss}). This is particularly true for the simulated SINFONI data, in which case many LBAs are not even spatially resolved.

These simulations will be used below when comparing kinematical measurements of LBAs and actual high redshift galaxies observed. These comparisons, along with implications for the analogy between LBAs and starbursts at high redshift, will be discussed in detail in section \ref{section:results}.

\section{Analysis of Individual Objects}\label{section:individual}

In the following sections we briefly describe each object in more detail. Two of these objects (021348 and 080232) are not resolved even with adaptive optics. They present dominant central objects (DCOs) as discussed in \citet{Overzier2009}. A third object (101211) is too faint, and no extended structure is detected. We exclude these three objects from the kinematic analysis in subsequent sections.

\subsection{005527}\label{section:0055}

This is the only object observed in broadband mode. Velocity dispersion is rather uniform across the whole galaxy, at about 100 km s$^{-1}$. The optical morphology is evidently much more extended than the Pa-$\alpha$ emitting region, which might indicate an underlying, more extended, older stellar structure.

\subsection{015028}\label{section:0150}

This is an object showing two clearly distinct star forming regions. There is also a clear velocity shear in the east-west direction, which is not aligned with the axis connecting the two bright clumps. Velocity dispersion is higher in the eastern half of the galaxy. In addition, there is some additional emission to the south, at higher velocity than the rest of the galaxy; it is unclear whether this represents a spiral arm or a tidal tail from an ongoing interaction.

\subsection{021348}\label{section:0213}

This is the faintest object observed, and we have only been able to detect an unresolved point source in the center of the galaxy, in addition to a low S/N region ($S/N < 10$) to the south. It is the first of five objects observed with OSIRIS that were classified as having a DCO, according to \citet{Overzier2009}. Since we cannot make any inferences about the resolved kinematic structure of the galaxy, we have excluded it from any further analysis.

\subsection{032845}\label{section:0328}

032845 is a bright object, and a significant amount of structure is detected. However, velocity shear is remarkably small, and velocity dispersion is, again, relatively homogeneous across the galaxy. The HST optical image shows an antenna-like structure, with distinct nuclei, in what appears to be a merger.

\subsection{035733}\label{section:0357}

We have been able to detect not only the brightest component, but the faint companion to the east, where line emission is evidently weaker. A comparison with the HST image shows a much more extended structure than what is seen here. The western region, however, is clearly defined, and shows a definite velocity shear across its major axis, resembling a rotating disk, but still with line-of-sight velocity dispersion values of approximately 70 km s$^{-1}$, close to the value of the velocity shear across the major axis. The companion to the east is at the same systemic velocity as the main component.

\subsection{040208}\label{section:0402}

This is one of the faintest galaxies we have observed (SFR$= 2.5 M_\odot$ yr$^{-1}$), therefore the signal-to-noise ratio is considerably smaller. There are a number of star-forming regions northeast of the main component, and the velocity offset between them is rather small. 

\subsection{080232}\label{section:0802}

This is another DCO, like 021348. Again, we detect very little emission besides a bright point source in the center of the galaxy. This object is also excluded from further analysis.

\subsection{080844}\label{section:0808}

This is another DCO, but in this case we were able to detect emission from the companion to the southeast. There is little velocity structure within the main component, but the companion is offset more than 200 km s$^{-1}$ from the point source.

\subsection{082001}\label{section:0820}

082001 is one of the most elongated objects in our sample, which leads to the assumption that it might be disk-like structure seen edge-on. The velocity structure seems to confirm this hypothesis, with a strong shear across the major axis. We are able to detect multiple components, indicating there are distinct star-forming regions within this disk.

\subsection{083803}\label{section:0838}

This object shows a main emission region larger than a kpc across, with little velocity structure. In addition, we were able to detect emission from a fainter structure to the south, with a velocity offset from the main component of $\sim 50$ km s$^{-1}$. This structure is also seen in the HST image.

%
%
\subsection{092600}\label{section:0926}

This is another example of an LBA with a companion structure, also evident in the HST image. The companion presents a $\sim$50 km s$^{-1}$ shift with respect to the main structure. Also evident is a velocity shear across the main region itself, albeit small -- $\sim$50 km s$^{-1}$ -- especially when compared to the velocity dispersion of approximately $\sim$100 km s$^{-1}$ found in the galaxy. This is the least massive of our objects (log $M_*/M\odot = 9.1$) and has also been described in \citet{Basu-Zych2009}.

\subsection{093813}\label{section:0938}

This is one of the galaxies with strongest line emission in our sample (the Pa-$\alpha$ line is detected at S/N $\gtrsim$ 400 in some regions), and therefore we are able to detect the substructure with great amount of detail. Multiple components are observed, with velocity offsets greater than 100 km s$^{-1}$ between them. Showing signs of a recent or ongoing strong merger event in the HST optical data, the velocity dispersion seems higher where the merging galaxies appear to meet, to the west, where Pa-$\alpha$ emission is strongest.

\subsection{101211}\label{section:1012}

The emission is weak, and little structure is detected beside a faint companion to the northeast. Due to lesser data quality in comparison with other galaxies in our sample, we do not use this object for our subsequent analysis.

\subsection{113303}\label{section:1133}

This galaxy shows a remarkable lack of velocity structure within the main component, with a shear of a few tens of km s$^{-1}$, comparable to the instrument resolution itself. However, we were able to detect some faint emission from a component to the southwest, offset from the main region at approximately 100 km s$^{ -1}$.

\subsection{135355}\label{section:1353}

135355 is composed of a large number of small star-forming regions, each measuring a few hundred pc across. These components show a gradual velocity shear at a 45 degree angle, indicative of a global velocity structure across the entire galaxy. In addition, there is an elongated component to the east, visible only in the optical HST data and which is likely a merging companion.

\subsection{143417}\label{section:1434}

This object presents two clearly distinct regions of star formation, along the east-west axis. The regions are at distinct velocities with respect to each other. In addition, we detect fainter emitting regions to the north and northwest, at very different velocities from the two brightest regions. These two regions are part of much more elongated structures, as can be seen in the HST image, which shows strong signs of an ongoing interaction. This has also been discussed in \citet{Basu-Zych2009}.

\subsection{210358}\label{section:2103}

This is the most massive object we have observed, and one with very unique features. It is one of the DCO objects as described in \citet{Overzier2009}, and we confirm the existence of a bright, unresolved region in the center of the galaxy. This region has high Pa-$\alpha$ surface brightness, with values above 10$^{-13}$ erg s$^{-1}$ cm$^{-2}$ arcsec$^{-2}$. This galaxy presents the strongest velocity shear across its major axis, $v_{\rm shear}\sim 250$ km s$^{-1}$. This is the third object presented in \citet{Basu-Zych2009}.

\subsection{214500}\label{section:2145}

This galaxy presents high velocity shear across its major axis, uncommonly so for its low stellar mass (see section \ref{section:dynamics}). However, its structure is not smooth, and there are undetected stellar components to the south, seen in the HST image. Likewise, the velocity dispersion map is not as well structured as other disk-like galaxies. This may indicate a recent merger event.

\subsection{231812}\label{section:2318}

This is one of the largest galaxies in our sample, and therefore was observed with the 100 mas spaxel scale to maximize its field of view. It shows a bright component with fainter structure to the south and west. The star-forming region to the south has a velocity offset of $\sim 75$ km s$^{-1}$ from the brightest part of the galaxy.

\section{Results}\label{section:results}

In this section we discuss some of the analytic results obtained from our observations, describing the methodology used to calculate each of the quantities presented.

\subsection{Sizes}\label{section:sizes}

Previous studies of LBAs and high-redshift starbursts infer kpc-scale sizes for the star forming regions, from rest-frame UV continuum as observed with HST \citep[and references therein]{Overzier2010} and the emission line regions as observed with IFU instruments \citep{Law2009,ForsterSchreiber2009}. Here we present our calculations for sizes of LBAs as seen with OSIRIS, comparing these figures with results from the above-mentioned studies.

\begin{figure}[ht]

\includegraphics[width=\linewidth]{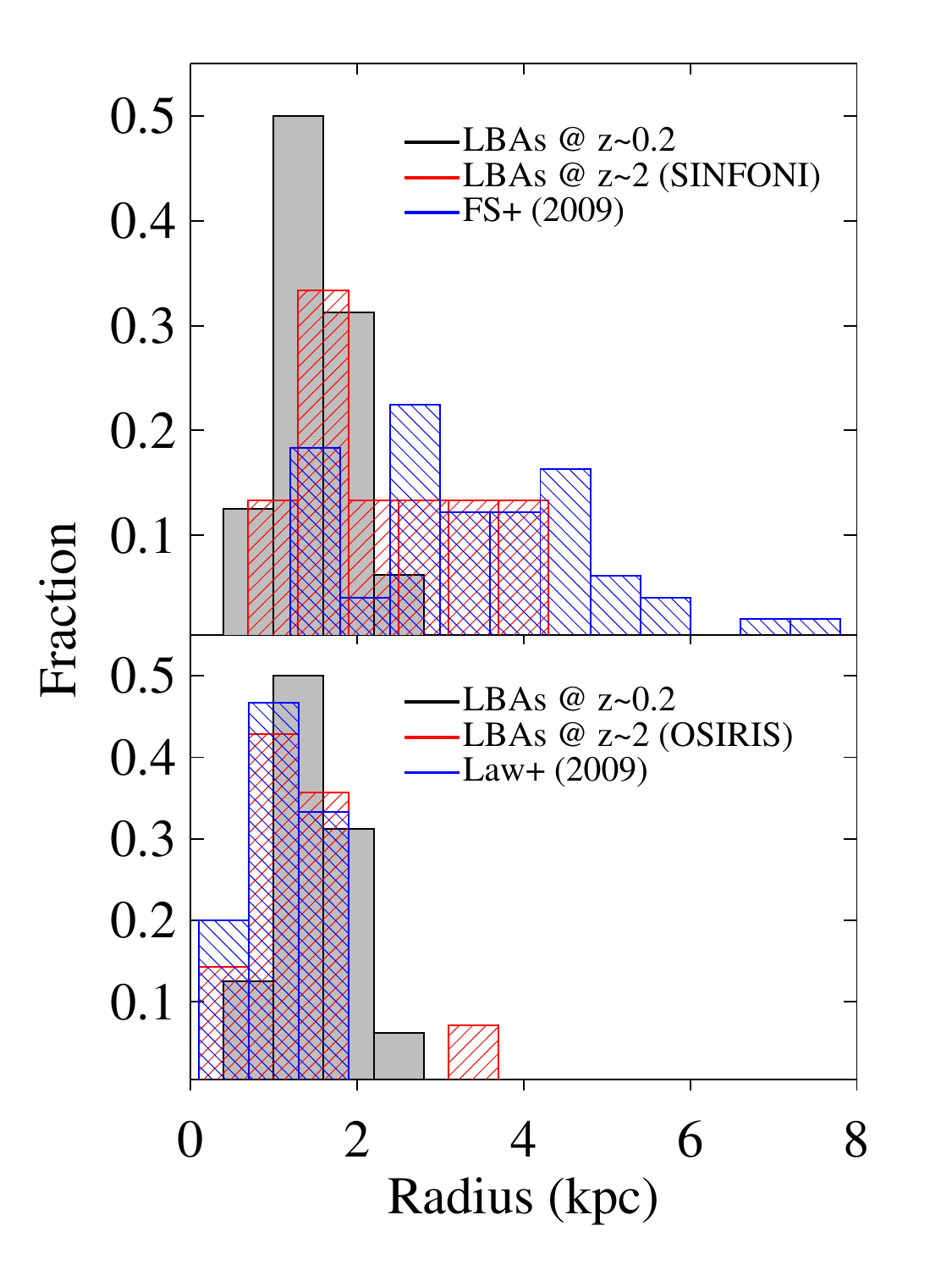}

\caption{Distribution of sizes for LBAs. Black filled histograms represent intrinsic values, while red hashed histograms represent simulated values at high redshift as would be detected with SINFONI ({\it top}) and OSIRIS ({\it bottom}). Blue histograms indicate size distributions of actual high-redshift starburst galaxies as measured with corresponding instruments by \citet{ForsterSchreiber2009} and \citet{Law2009}.} \label{fig:lba_radius}

\end{figure}

We replicate the method described in \citet{Law2007}, to allow for a direct comparison with results for LBGs utilizing the same instrument. This comprises counting the number of spaxels $N$ above a certain $S/N$ threshold to represent the size of the star-forming region. We use the same threshold, namely $S/N > 6$. To determine a radius, we assume galaxies are approximately circular, and therefore calculate a radius in spaxels as $r=\left(N/\pi\right)^{1/2}$. This number is corrected for the PSF size in each case (see section \ref{section:maps}) and later converted to a physical size at the corresponding spaxel scale and redshift of each object. We only use contiguous spaxels connected to the brightest region of the galaxy in order to exclude companions. Finally, we repeat the process for our simulated high-redshift observations of LBAs. Errors are typically 0.1 kpc (low-z), 0.4 kpc (OSIRIS) and 2.0 kpc (SINFONI), given by half the PSF size in each case. The results are presented in Table \ref{table:kinematics} and shown in Figure \ref{fig:lba_radius} along with actual measurements for high-redshift galaxies. Most LBAs present sizes between 1 and 2 kpc, consistent with findings from \citet{Overzier2010}.

From Figure \ref{fig:lba_radius} we notice this method yields smaller sizes at high-redshift in AO-assisted observations; this is caused be surface brightness dimming of objects, which prevents detection of emission at the outer radii of galaxies. Sizes for our OSIRIS simulated observations are remarkably similar to those found in \citet{Law2009} for LBGs, which have been calculated in an identical manner; a two-sided Kolmogorov-Smirnov test yields a 97\% probability of both samples being drawn from the same parent population.

Non-AO observations, however, produce different results. Galaxies are apparently larger, in many cases due to blending of different components, combined with somewhat improved sensitivity at the outer radii, aside from obvious loss of resolution. Still, simulated LBAs look smaller than galaxies in the SINS survey \citep{ForsterSchreiber2009} - many, in fact, show sizes smaller than the inferred uncertainty, which means they are essentially unresolved. When comparing HST sizes of LBAs with those found for BzK galaxies, \citet{Overzier2010} conclude both samples have similar rest-frame UV sizes, but the latter has larger rest-frame optical sizes; therefore it is not unreasonable to assume galaxies in the SINS survey might be intrinsically larger than both LBAs and LBGs.

\subsection{Kinematics and Dynamics of Star Forming Galaxies}\label{section:dynamics}

The ionized gas in LBAs exhibit very high velocity dispersions\footnote[3]{The global velocity dispersion $\sigma$ measured for the each galaxy is an average of each spaxel, weighted by flux. This allows for a more accurate measurement than simply measuring the velocity dispersion of the whole cube, since it does not incorporate the intrinsic velocity shear within the galaxy.}, with median $\sim 67$ km s$^{-1}$ and some galaxies reaching values above 100 km s$^{-1}$. This is much higher than those observed in ordinary local star forming galaxies (typical gas velocity dispersions of 5--15 km $s^{-1}$, e.g., Dib et al. 2006) but analogous to the increased velocity dispersions observed in local (ultra-)luminous infrared galaxies \citep[e.g.][]{Arribas2008, Monreal-Ibero2010}. These values are also in good agreement with high-redshift star-forming galaxies, as observed both in single-slit spectroscopy \citep[e.g.][]{Pettini2001, Erb2006} and the aforementioned integral field studies \citep{Law2009, ForsterSchreiber2009}.

We also measure the velocity shear within each galaxy. Since we cannot always precisely define an axis of rotation, we simply determine the difference between the maximum and minimum velocities observed within the main body of the galaxy (excluding companions in order to probe for intrinsic rotation of one star-forming region). We determine $v_{\rm max}$ and $v_{\rm min}$ as the median of the 5-percentile at each end of the velocity distribution, so that outliers and artifacts are excluded. The velocity shear is then simply defined as $v_{\rm shear}=\frac{1}{2}(v_{\rm max}-v_{\rm min})$. The values vary between a few tens of km s$^{-1}$ and over 200 km s$^{-1}$. These measurements are presented in Table \ref{table:kinematics}. In many cases, the velocity shear is not caused by actual rotation of the whole galaxy, since there is not a significant velocity gradient observed across the entire object.

\begin{deluxetable*}{lcccccccccccccc}
\tablewidth\linewidth
\tabletypesize{\scriptsize}
\tablecaption{Kinematic data for LBAs} 
\tablehead{
  \colhead{Name}  & \colhead{r$_{{\rm Pa-}\alpha}$} &  \colhead{$v_{\rm shear}$} &  \colhead{$\sigma$} &  \colhead{$v/\sigma$} & \colhead{r$_{{\rm Pa-}\alpha,{\rm hiz}}$} &  \colhead{$v_{\rm shear,hiz}$} & \colhead{$\sigma_{\rm hiz}$} & \colhead{r$_{{\rm Pa-}\alpha,{\rm hiz}}$} & \colhead{$v_{\rm shear,hiz}$} & \colhead{$\sigma_{\rm hiz}$} & \colhead{log $M_{dyn}$} & \colhead{$K_{\rm asym}$} & \colhead{$K_{\rm asym,hiz}$} & \colhead{$K_{\rm asym,hiz}$}\\
  & & & & & (OSIRIS) & (OSIRIS) & (OSIRIS) & (SINF) & (SINF) & (SINF) & (M$_\odot$) & & (OSIRIS) & (SINF)\label{table:kinematics}
  }
\startdata
$005527$ & 1.2 & 42 & 104 & 0.41 & 1.0 & 35 & 89 & 2.5 & 18 & 122 & 10.2 & 0.77 & 0.46 & 0.93\\
$015028$ & 1.5 & 78 & 74 & 1.05 & 1.0 & 57 & 73 & 1.7 & 31 & 82 & 10.0 & 0.21 & 0.19 & 0.09\\
$032845$ & 1.5 & 73 & 68 & 1.08 & 0.6 & 13 & 46 & 3.9 & 101 & 78 & 9.9 & 1.60 & 0.63 & 0.59\\
$035733$ & 1.4 & 50 & 66 & 0.76 & 1.9 & 28 & 47 & 1.5 & 19 & 62 & 9.9 & 0.27 & 0.26 & 0.25\\
$040208$ & 0.6 & 53 & 50 & 1.06 & N/A & N/A & N/A & 0.7 & 23 & 35 & 9.3 & 0.89 & N/A & 0.44\\
$080844$ & 1.2 & 27 & 92 & 0.30 & $<1.5$ & 16 & 95 & 3.1 & 14 & 117 & 10.1 & 2.16 & 0.46 & 0.41\\
$082001$ & 1.9 & 119 & 67 & 1.78 & 1.6 & 85 & 65 & 3.1 & 67 & 91 & 10.0 & 0.17 & 0.11 & 0.08\\
$083803$ & 1.4 & 41 & 49 & 0.83 & 0.8 & 28 & 29 & 0.9 & 13 & 45 & 9.6 & 1.38 & 0.50 & 0.53\\
$092600$ & 1.4 & 36 & 71 & 0.51 & 1.3 & 23 & 54 & 1.5 & 25 & 69 & 9.9 & 0.61 & 0.94 & 0.19\\
$093813$ & 2.1 & 63 & 67 & 0.94 & 1.4 & 38 & 63 & 4.2 & 30 & 85 & 10.0 & 0.55 & 0.25 & 1.85\\
$113303$ & 0.7 & 14 & 30 & 0.45 & 0.1 & 30 & 41 & $<2.0$ & 66 & 66 & 8.9 & 0.66 & 0.60 & 0.09\\
$135355$ & 1.6 & 77 & 67 & 1.15 & 1.1 & 46 & 51 & 1.9& 50 & 82 & 9.9 & 0.70 & 0.43 & 0.19\\
$143417$ & 1.7 & 73 & 67 & 1.09 & 0.8 & 41 & 65 & 1.9 & 37 & 67 & 10.0 & 1.19 & 0.21 & 0.29\\
$210358$ & 1.6 & 183 & 136 & 1.35 & 1.0 & 72 & 161 & 3.0 & 109 & 210 & 10.5 & 0.17 & 0.12 & 0.16\\
$214500$ & 1.8 & 81 & 55 & 1.47 & 1.3 & 56 & 58 & 2.4 & 64 & 96 & 9.8 & 0.18 & 0.20 & 0.19\\
$231812$ & 2.8 & 70 & 63 & 1.11 & 3.6 & 55 & 55 & 2.9 & 47 & 65 & 10.1 & 0.28 & 0.33 & 0.10
\enddata
\end{deluxetable*}

\begin{figure}[ht]

\includegraphics[width=\linewidth]{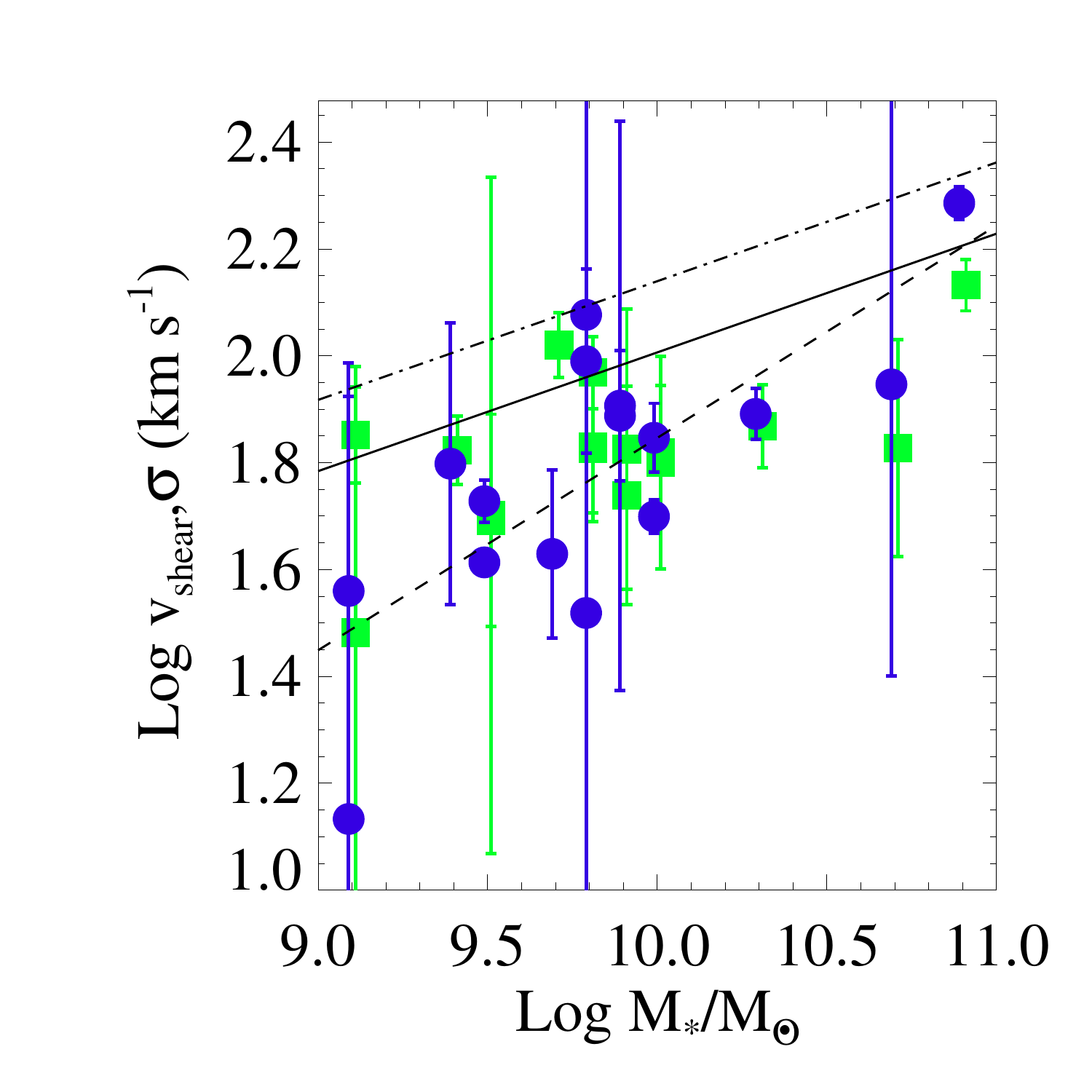}

\caption{Velocity shear $v_{\rm shear}$ ({\em blue circles}) and velocity dispersion $\sigma$ ({\em green squares}) as a function of stellar mass. The plot shows clearly how more massive galaxies show a stronger velocity shear than less massive ones, particularly the ones above $\sim 10^{10}$ M$_\odot$. The same trend, albeit weaker, exists for velocity dispersion $\sigma$. Dashed line shows a power-law fit to our data, while the solid line is the Tully-Fisher relation at $z \sim 0$ according to \citet{Bell2001}. The dotted-dashed line shows the Tully-Fisher relation at $z \sim 2$ according to \citet{Cresci2009}.}\label{fig:mstar_vshear}

\end{figure}

There is a strong trend of velocity shear with stellar mass: more massive objects tend to show greater velocity differences between distinct regions of ionized gas. This can be seen in detail in Figure (\ref{fig:mstar_vshear}). Velocity dispersion $\sigma$, also correlates with stellar mass, albeit with a shallower slope. For comparison, we also show in Figure \ref{fig:mstar_vshear} the local Tully-Fisher relation derived in \citet{Bell2001}, corrected for an average inclination factor of $\left< \sin i \right> = 0.79$ \citep[see Appendix in][]{Law2009}. Although an inference for such a relation for LBAs is not reasonable, since these objects are not necessarily rotating disks, this serves as a comparison with velocity shear in local spirals. These values are slightly smaller for a given stellar mass, especially at lower masses (up to a factor of 2). Also shown is the derived relation for star-forming galaxies at $z\sim 2.2$ from \citet{Cresci2009}, which shows higher $v_{\rm circ}$ values than spirals in the present day; however, in the former, the galaxies studied are more massive ($ M_* \gtrsim 2-3\times10^{10} M_\odot$), and were pre-selected to look like rotating disks.

\begin{figure*}[ht]

\includegraphics[width=\linewidth]{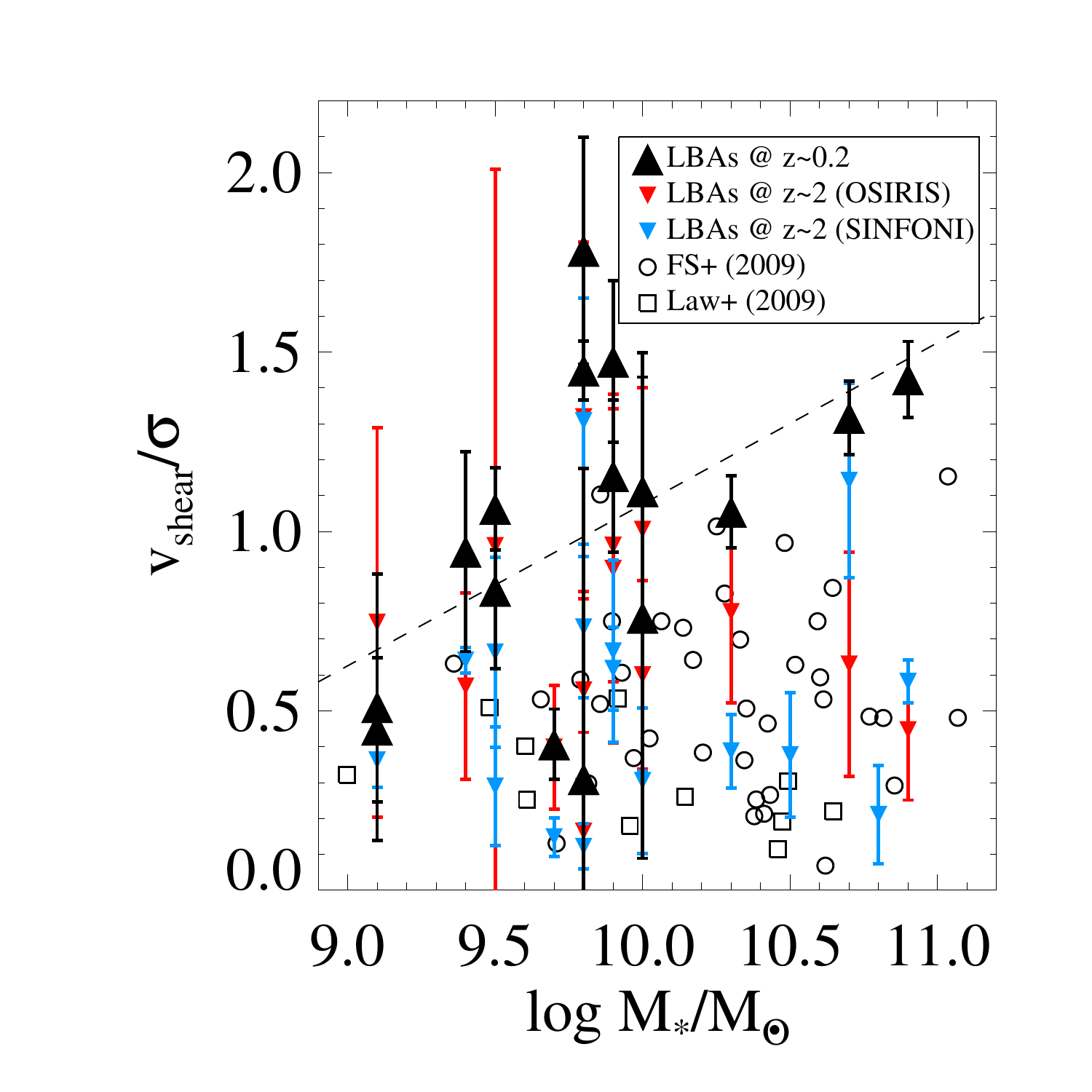}

\caption{Ratios between velocity shear and velocity dispersion $v_{\rm shear}/\sigma$ as a function of stellar mass ({\em black triangles}). The dashed line shows a fit to the $v_{\rm shear}/\sigma$ data at their intrinsic redshift. We see a moderate trend, indicating more massive galaxies have a stronger rotational dynamical component than less massive ones. Also shown as downward triangles are $v_{\rm shear}/\sigma$ values for galaxies artificially redshifted to $z \sim 2$ (see text in section \ref{section:hiz}). In this case, red triangles represent the OSIRIS simulated data, and blue triangles represent SINFONI non-AO simulations. Values from actual high-redshift observations are presented as hollow symbols, representing data from \citet{ForsterSchreiber2009}(open circles) and \citet{Law2009}(open squares). The LBAs form an upper envelope with respect to the observational data at high redshift, but span a very similar range of parameter space when
simulated at $z=2$.}\label{fig:mstar_vsigma}

\end{figure*}

Due to the difference in slopes, the ratio between velocity shear and velocity dispersion ($v/\sigma$) is also a function of stellar mass (black triangles in Figure \ref{fig:mstar_vsigma}). A Spearman's $\rho$ correlation test shows a $\sim$6\% null-hypothesis probability of $M_*$ and $v/\sigma$ not being correlated. This indicates that more massive LBAs have a stronger component of rotational support against gravitational collapse, as opposed to less massive ones, which are more dispersion dominated.

When artificially redshifted, the $v/\sigma$ ratio decreases, from a combination of two effects: on one hand, surface brightness dimming causes the high-velocity values at the outskirts of the galaxy to be undetected - this is particularly true for the artificial OSIRIS high-z data (shown as red downward triangles in Figure \ref{fig:mstar_vsigma}). On the other hand, loss of spatial resolution, especially for non-AO observations performed with instruments such as SINFONI (blue downward triangles in Figure \ref{fig:mstar_vsigma}), causes blending of features and inner velocity values to dominate, due to higher signal-to-noise. The net result is lower $v_{\rm shear}$ values. Although our observed $v/\sigma$ values are higher than high-redshift ones (open circles and squares in Figure \ref{fig:mstar_vsigma}), when artificially redshifted these galaxies look very similar to high-z star-forming galaxies, with 72\% chance of being drawn from the same parent population according to a standard Kolmogorov-Smirnov test. We present all relevant values in Table \ref{table:kinematics}, along with measurements at their real redshift. We caution the reader, however, to the fact that the observed ratios at low redshift are still much smaller than found in local spiral galaxies, which have $v/\sigma$ values of 10-20.

The main kinematic difference when comparing LBAs and local spirals comes from gas velocity dispersions, indicating that LBAs have a dynamically thick structure, disk or otherwise. We find it unlikely that the dynamics in all of the LBAs is actually dominated by rotation, given low overall $v/\sigma$ values. Instead, the trend with stellar mass might simply indicate a colder, less random dynamical structure in the process of forming a disk from the dynamically hot gas in more massive galaxies.

Another quantity one can infer from gas kinematics is the dynamical mass of the galaxy, assuming the velocity dispersion in the nebular gas is dominated by random motions within a gravitational field. In that case,
\begin{equation}
M_{dyn} = \frac{C \sigma^2 r}{G},
\label{eq:mdyn}
\end{equation}
where $G$ is the gravitational constant, $r$ represents the size of the galaxy and C is a proportionality constant related to the geometry of the galaxy; for a disk, $C=3.4$, whereas for a uniform sphere, $C=5$ \citep{Erb2006}. The geometry is not always well determined in LBAs, but we assume dispersion-dominated dynamics with $C=5$ to allow for a direct comparison with LBGs as discussed in \citet{Law2009}. We use Pa-$\alpha$ radius determined in section \ref{section:sizes}. The results are presented in Table \ref{table:kinematics}).

\begin{figure}[ht]

\includegraphics[width=\linewidth]{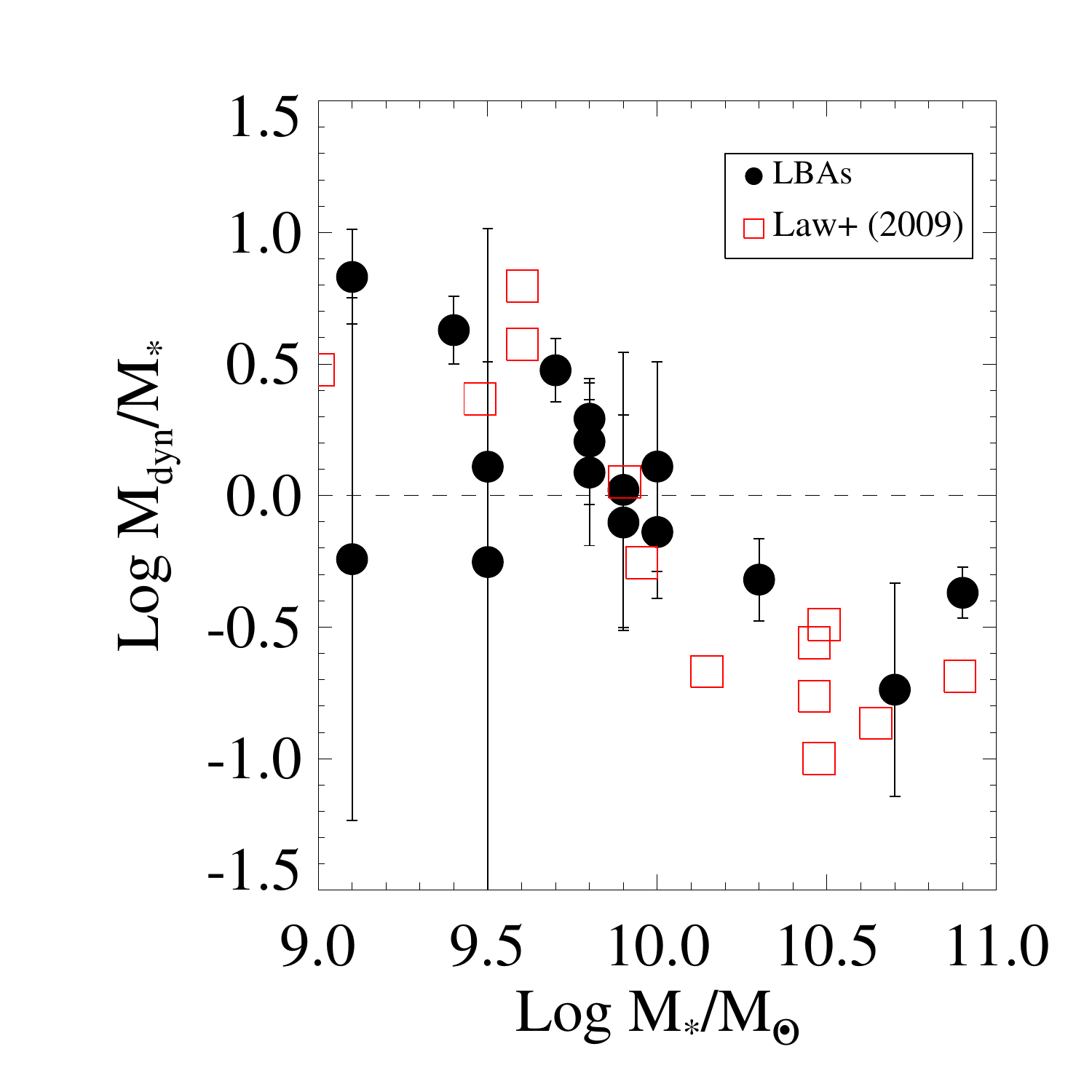}

\caption{Ratio between dynamical and stellar masses as a function of stellar mass. Black filled circles indicate the LBAs in this paper, while the hollow red squares are the high-z LBGs from \citet{Law2009}. The dashed line indicates $M_{dyn} = M_*$. More massive galaxies, both at low and high redshifts, present lower ratios, with $M_{dyn} \sim 1/10 M_*$ for galaxies with $\log M_* \sim 10^{11} M_\odot$.}\label{fig:mdyn}

\end{figure}

Dynamical masses are well correlated with stellar masses. $M_{dyn}$ agrees with stellar masses within a factor of two (0.3 dex) in 63\% of the galaxies in our sample, and they agree within a factor of three for 81\% of the objects. This means that for most LBAs, the high observed velocity dispersions can be explained simply by random motion of the gas given the observed masses. In Figure \ref{fig:mdyn}, we present the ratios between dynamical mass and stellar mass as a function of stellar mass. We notice that these ratios are larger for less massive galaxies, with the implied dynamical masses an order of magnitude smaller than the observed stellar mass for the most massive objects. This supports the hypothesis that the dynamical support offered by the rotating disk is more significant in the most massive star forming galaxies. It is also interesting to notice the same trend, with very similar dynamical and stellar mass values, for high-redshift star-forming galaxies, as indicated by red squares. This reinforces the analogy between LBAs and LBGs. 

\subsection{Kinemetry Measurements}\label{section:kinemetry}

Another way of assessing the presence of a rotational component within the dynamics of the gas in each galaxy is provided by the kinemetry method, as introduced by \citet{Krajnovic2006}. The method comprises a decomposition of the velocity moment maps into its Fourier components, that is, for a given ellipse:
\begin{eqnarray}
K(\psi) = A_0 + A_1 \sin(\psi) + B_1 \cos(\psi) + A_2 \sin(2\psi) + \nonumber \\
+ B_2 \cos(2\psi) + ...,\label{eq:fourier}
\end{eqnarray}
where $\psi$ is the azimuthal angle along which one measures a given velocity moment $K$ (in our case, velocity $v$ or velocity dispersion $\sigma$). Written in another way, 
\begin{equation}
K(r,\psi) = A_0(r) + \sum_{n=1}^{N} k_{n}(r)\cos\left[n\left(\psi-\phi_n(r)\right)\right], \label{eq:fourier2}
\end{equation}
where the expansion terms have been redefined as $k_n = \sqrt{A_n^2 + B_n^2}$ and  $\phi_n = \arctan\left(A_n/B_n\right)$. For a detailed discussion of the method, see \citet{Krajnovic2006} and \citet{Shapiro2008}.

For an ideal rotating disk, one would expect the velocity profile to be perfectly antisymmetric, that is, the $B_1$ term would dominate the Fourier expansion. Likewise, the velocity dispersion map is expected to be perfectly symmetric, and therefore all terms with the exception of $A_0$ would vanish.

\citet{Shapiro2008} have used this method to analyze the dynamics of high-redshift star-forming galaxies observed with the SINFONI instrument. In quantifying the asymmetry of the velocity moment maps, they have defined the quantities
\begin{equation}
v_{\rm asym} = \left<\frac{k_{\rm avg,v}}{B_{1,v}}\right>_r \label{eq:vasym}
\end{equation}
and
\begin{equation}
\sigma_{\rm asym} = \left<\frac{k_{\rm avg,\sigma}}{B_{1,v}}\right>_r. \label{eq:sigasym}
\end{equation}
By using local galaxies and numerical models as if observed at high redshift as templates for disk versus merger events, they have found the threshold of
\begin{equation}
K_{\rm asym}=\sqrt{v_{\rm asym}^2+\sigma_{\rm asym}^2}=0.5
\end{equation}
to distinguish between rotating disks and mergers. Galaxies previously identified by eye as rotating disks were correctly classified as disks by the kinemetry method, as were galaxies previously identified as mergers.



We have used the same IDL code as presented in \citet{Krajnovic2006}, that, at each semi-major axis, determines values for inclination and ellipticity of the curve that will minimize asymmetry. The ellipse center was determined as a flux-weighted average of the main body of the galaxy, again excluding companions not connected to the brightest star-forming region. The ellipses defined using the velocity map would then be used with the velocity dispersion map.

\begin{figure}[ht]

\includegraphics[width=\linewidth]{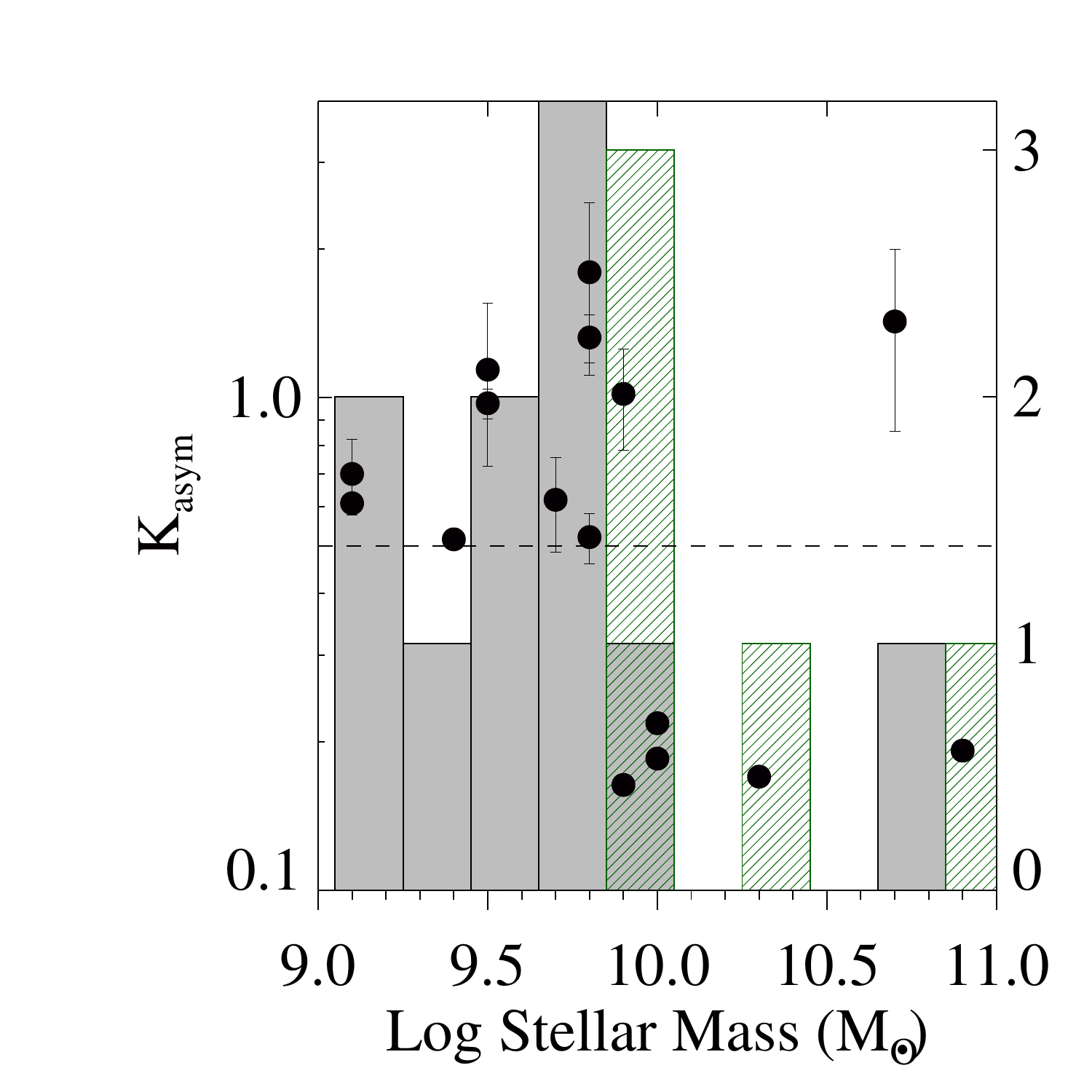}

\caption{Kinemetric asymmetry measurements as a function of galactic stellar mass. Left y-axis shows values of $K_{\rm asym}$, while right y-axis shows quantities for histograms. Gray histogram shows number of galaxies that would be classified as mergers in \citet{Shapiro2008}, while the green hatched histogram shows the number of galaxies that would be classified as disks. Galaxies with high $K_{\rm asym}$ are predominantly less massive, but the lowest value of stellar mass for a galaxy with $K_{\rm asym}<0.5$ is 9.9.}\label{fig:mstar_kinemetry}

\end{figure}

In Figure \ref{fig:mstar_kinemetry} we show average kinemetric asymmetry as a function of stellar mass. As before, we notice a trend in which the most symmetric objects tend to be those with high stellar mass. We use the same threshold of $K_{\rm asym}=0.5$ to distinguish between two categories of symmetry. The histogram in the plot shows that more asymmetric galaxies (gray bars) are predominantly less massive, with one single exception; the symmetric galaxies, on the other hand (green hatched bars), are typically more massive, the least massive object having log $M_*/M\odot = 9.9$. According to a standard Kolmogorov-Smirnov test, there is a 0.7\% probability that stellar masses from $K_{\rm asym}>0.5$ are drawn from the same parent population as the $K_{\rm asym}<0.5$ ones.

Evidently, the threshold of $K_{\rm asym}=0.5$ is a simplification; in fact, an inspection of Figure 7 in \citet{Shapiro2008} shows an overlap of disks and mergers in the region where $0.1\lesssim K_{\rm asym}\lesssim 1.0$, which might indicate instead a transition region between disk galaxies and mergers in the kinemetry plot. This region is where most star-forming galaxies at high-redshift lie, as is the case for the LBAs.

\begin{figure}[ht]

\includegraphics[width=\linewidth]{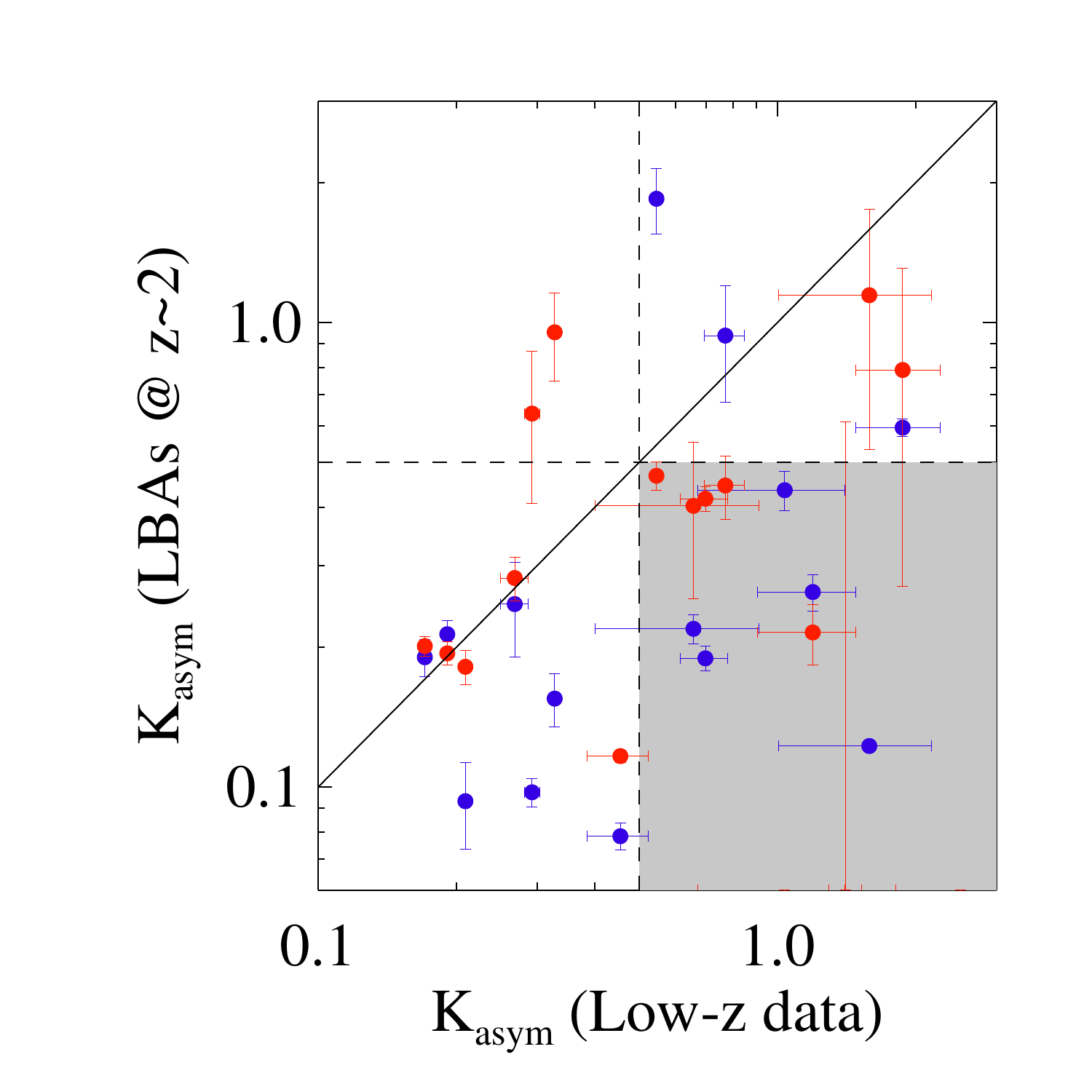}

\caption{Kinemetry measurements for our high-redshift simulations as a function of ''intrinsic'' values measured at low redshift. Dashed lines show the same threshold of $K_{\rm asym}=0.5$. Red points represent OSIRIS-AO simulations, while blue points represent the SINFONI non-AO simulations. The gray shaded area indicates the region of the plot where one finds LBAs having high-asymmetry values at low redshift but low values at $z\sim2.2$ (lower-right quadrant). In the classification scheme of \citet{Shapiro2008} these objects would likely be classified as rotationally-supported ``disks''.} \label{fig:kinemetry_loz_hiz}

\end{figure}

Finally we compare our results with the high redshift simulations from Section \ref{section:hiz}. In Figure \ref{fig:kinemetry_loz_hiz}, we show the measurements based on the simulations and compare them to the ``intrinsic" values measured in the original (i.e. low redshift) data. The dashed lines show the same threshold of $K_{\rm asym}=0.5$ used to distinguish between disks and mergers. In general, galaxies at high redshift present smaller values of $K_{\rm asym}$, i.e. they appear more symmetric than they actually are. One-third of the galaxies would be classified differently at high redshift (lower-right quadrant). The net effect is that the percentage of galaxies classified as mergers drop from $\sim$70\% to $\sim$38\%. This is a combined effect of signal loss at larger radii (where kinematics are less symmetric) and confusion and blending, smoothing out features that would otherwise show departures from a rotating disk.

%
%

%
%

\section{Discussion}\label{section:discuss}

In this section we briefly discuss some of the current models for galaxy formation, and how they may explain the observed properties of both LBAs and LBGs in terms of their morphologies and kinematics. In addition, we discuss some of the implications of the stellar mass dependence of observables discussed in the previous section.

\subsection{Ionized gas kinematics as a diagnostic for galaxy formation mechanisms}

In light of new techniques and integral-field instruments, recent studies of the kinematics of ionized gas at high redshifts have been used as diagnostics for galaxy formation models attempting to explain the distinctive properties observed in star-forming galaxies at $z \sim2-3$. In particular, the existence of a large number of rotating gas disks with high velocity dispersions at these redshifts has been pointed out to support the hypothesis of cold gas flows from the IGM directly feeding vigorous star formation at the center of sufficiently massive dark matter haloes \citep{Dekel2006, Dekel2009, Keres2009}. The high densities generated would then be Toomre unstable, leading to subsequent fragmentation into multiple regions and the observed clumpiness of star-forming galaxies and, in particular, the so-called ``clump-cluster galaxies'' at such redshifts \citep{Immeli2004, Elmegreen2005, Bournaud2008}. These galaxies are ultimately expected to coalesce, with individual clumps migrating inwards and creating a bulge at the center of the galaxy \citep{Noguchi1999, Immeli2004, Elmegreen2008, Elmegreen2009, Genzel2008}.

\begin{figure}[ht]

\includegraphics[width=\linewidth]{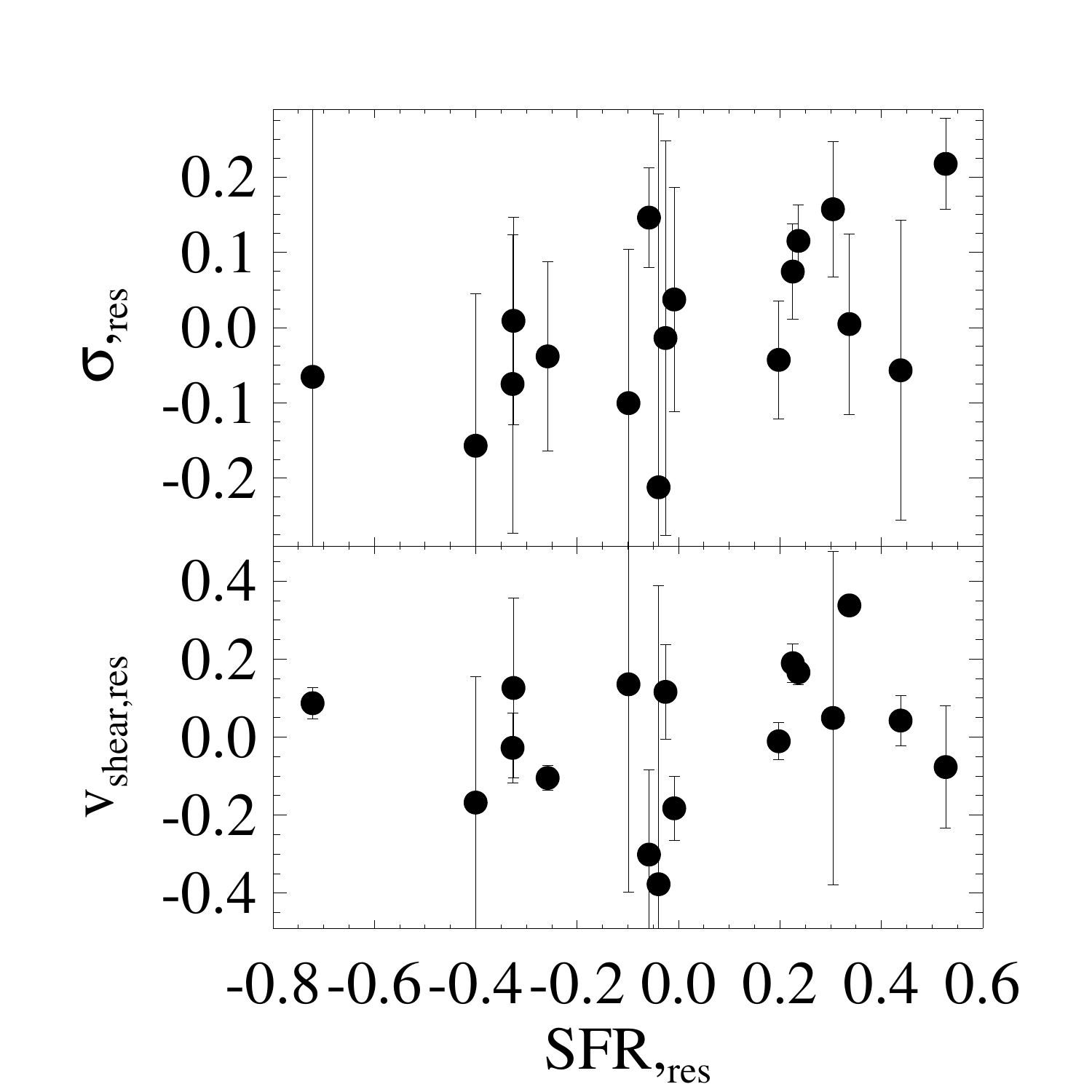}

\caption{Velocity shear and velocity dispersion as a function of star formation rates. Values are the residuals from a fit with respect to stellar mass ({\it see text}).}\label{fig:sfr_vsigma}

\end{figure}

It should be noted that the emission lines are produced by ionized gas close to star-forming regions, and might therefore not be ideal tracers for the dynamics of the galaxy as a whole. In fact, comparison between our HST rest-frame optical and Pa-$\alpha$ images show we are tracing regions that contain approximately a third of the total stellar mass in the galaxy (section \ref{section:hst}). Furthermore, the gas in these regions is subject to a number of local feedback effects from stellar winds and turbulence, and thus may not always represent motion of the bulk of the dynamical mass within. \citet{Lehnert2009} argue that the high velocity dispersion values could not be sustained simply by cosmological gas accretion; instead, self-gravity drives the early stages of galaxy evolution until dense clumps collapse, at which point star formation is self-regulated by mechanical output of massive stars. Although the inferred dynamical masses from velocity dispersion can be explained for the most part as a result of random motions within the potential well of the galaxy (section \ref{section:dynamics}), there is a second-order effect in LBAs that supports the influence of star formation-driven turbulence; in Figure \ref{fig:sfr_vsigma} we show $v_{\rm shear}$ and $\sigma$ as a function of star formation rates. Since all variables correlate with stellar mass, we present the residuals of a power-law fit for all of them with respect to $M_*$, in order to exclude any induced correlations. $v_{\rm shear}$ is independent of star formation rates (52\% null-hyphotesis probability according to Spearman's $\rho$ test), but more star-forming galaxies show stronger velocity dispersion (2\% null-hypothesis probability), supporting the idea that star formation has an effect on $\sigma$ values, generating turbulence from the energy output in processes related to star formation.

Another relevant point in this discussion is that rotating kinematics do not exclude the possibility of a merger-triggered starburst. In a hydrodynamical simulation of a gas-rich major merger, \citet{Robertson2006} show that for wet mergers rotating disks may form approximately $\sim$100 Myr after the final coalescence. At this point, a large, rotating gaseous disk is formed, with kinemetric asymmetry indices that would in principle rule out the merger scenario \citep{Robertson2008}. The signature for interactions might be as subtle as small displacements of the $\sigma$ map peak from the center of the galaxy \citep{Flores2006}. \citet{Puech2010} have used similar arguments through IFU studies of clumpy, intermediate redshift ($z\sim 0.6$) galaxies to claim that interactions might be responsible for driving star formation at all redshifts. In these cases, high gas fractions are a fundamental ingredient in the formation of the disk. Indeed, a number of recent CO observations indicate that star-forming galaxies at high redshifts have a much larger gas fraction than standard spirals in the local universe \citep{Tacconi2010, Daddi2010}.

\begin{figure*}[ht]
\hskip .4 in
\includegraphics[height=.28\linewidth]{uvlg_2103_velmap.pdf} \hskip .12in
\includegraphics[height=.28\linewidth]{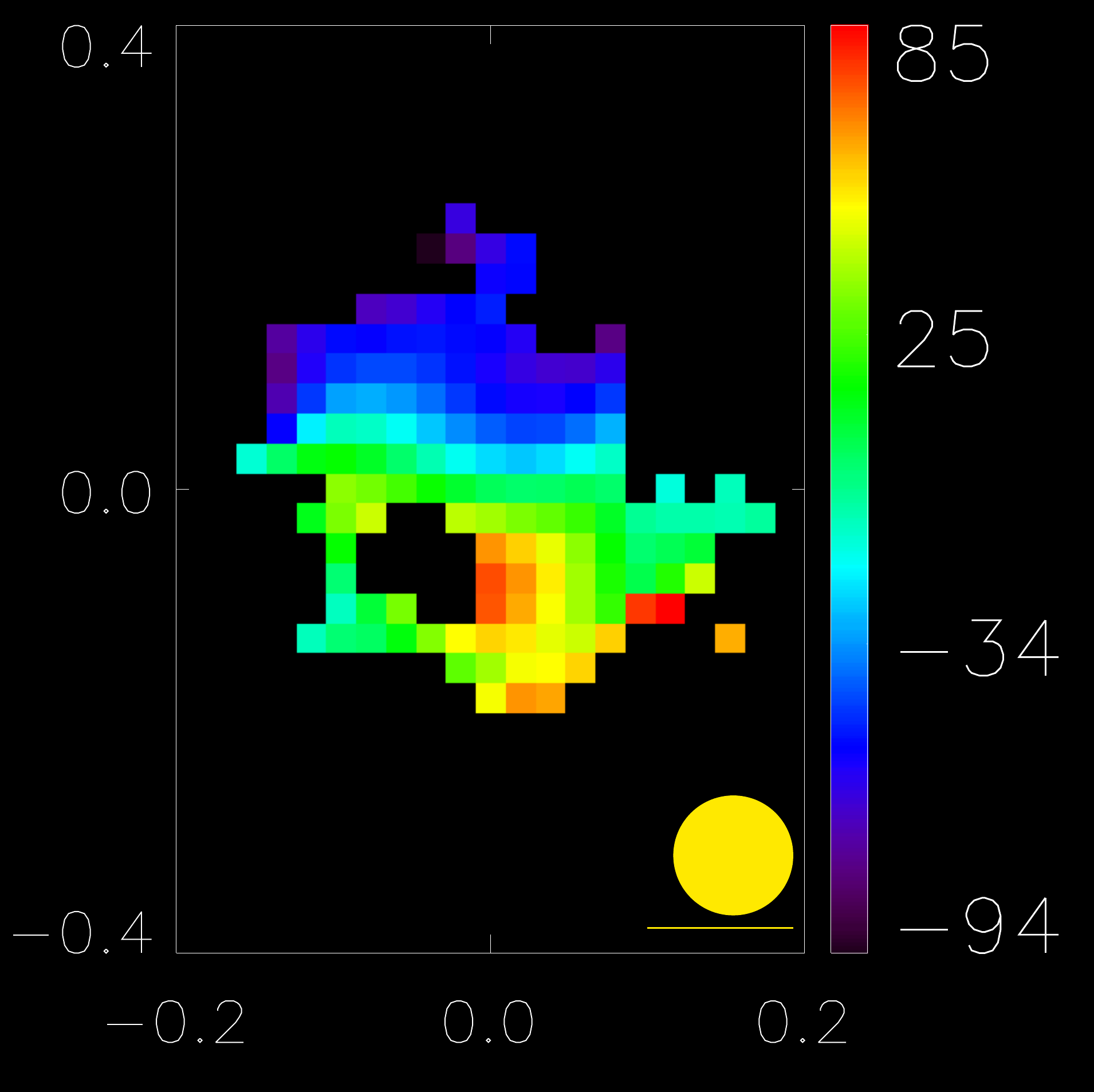} \hskip .12in
\includegraphics[height=.28\linewidth]{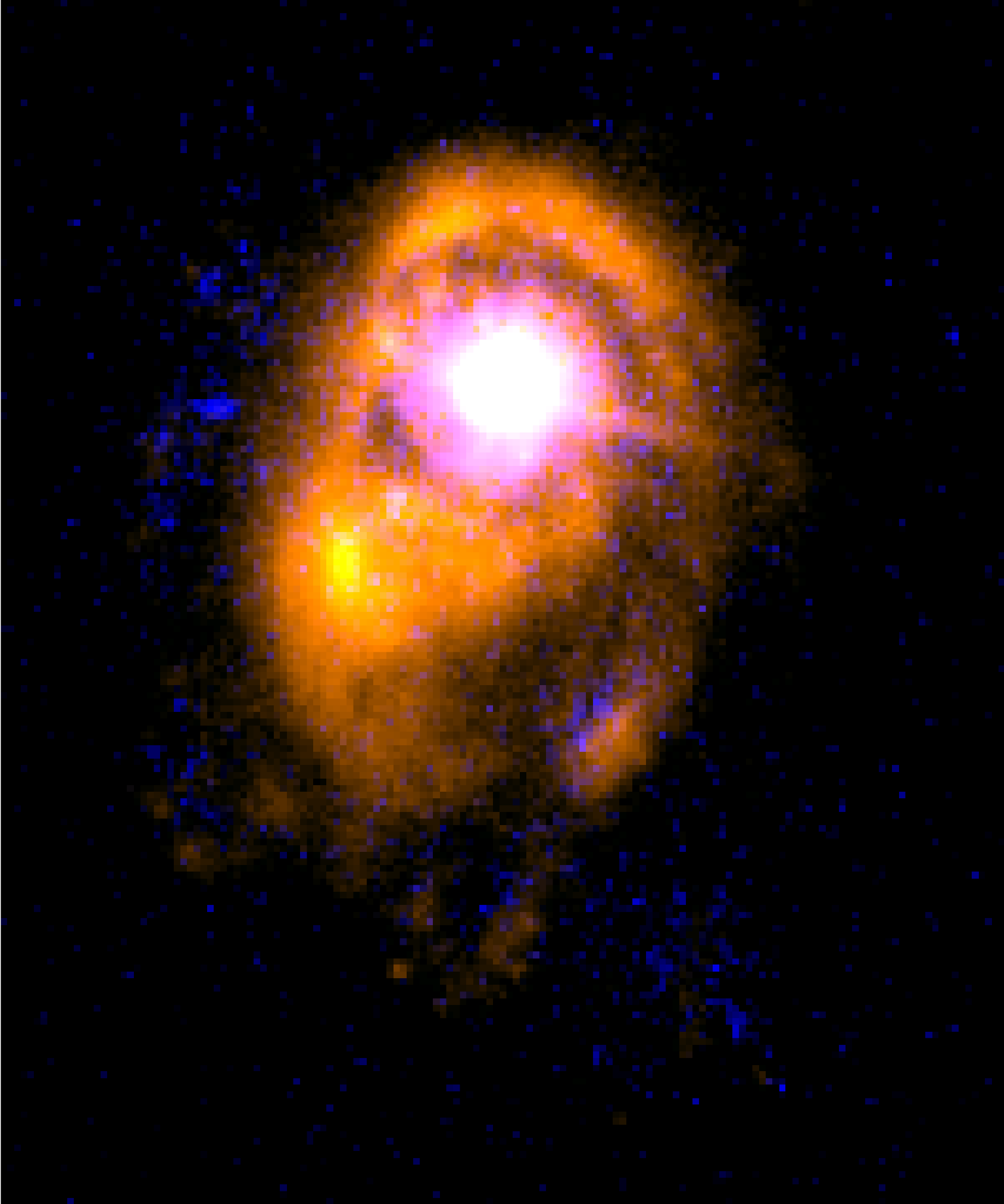}
\vskip .01 in
\hskip .4 in
\includegraphics[height=.22\linewidth]{uvlg_1353_velmap.pdf} \hskip.03in
\includegraphics[height=.22\linewidth]{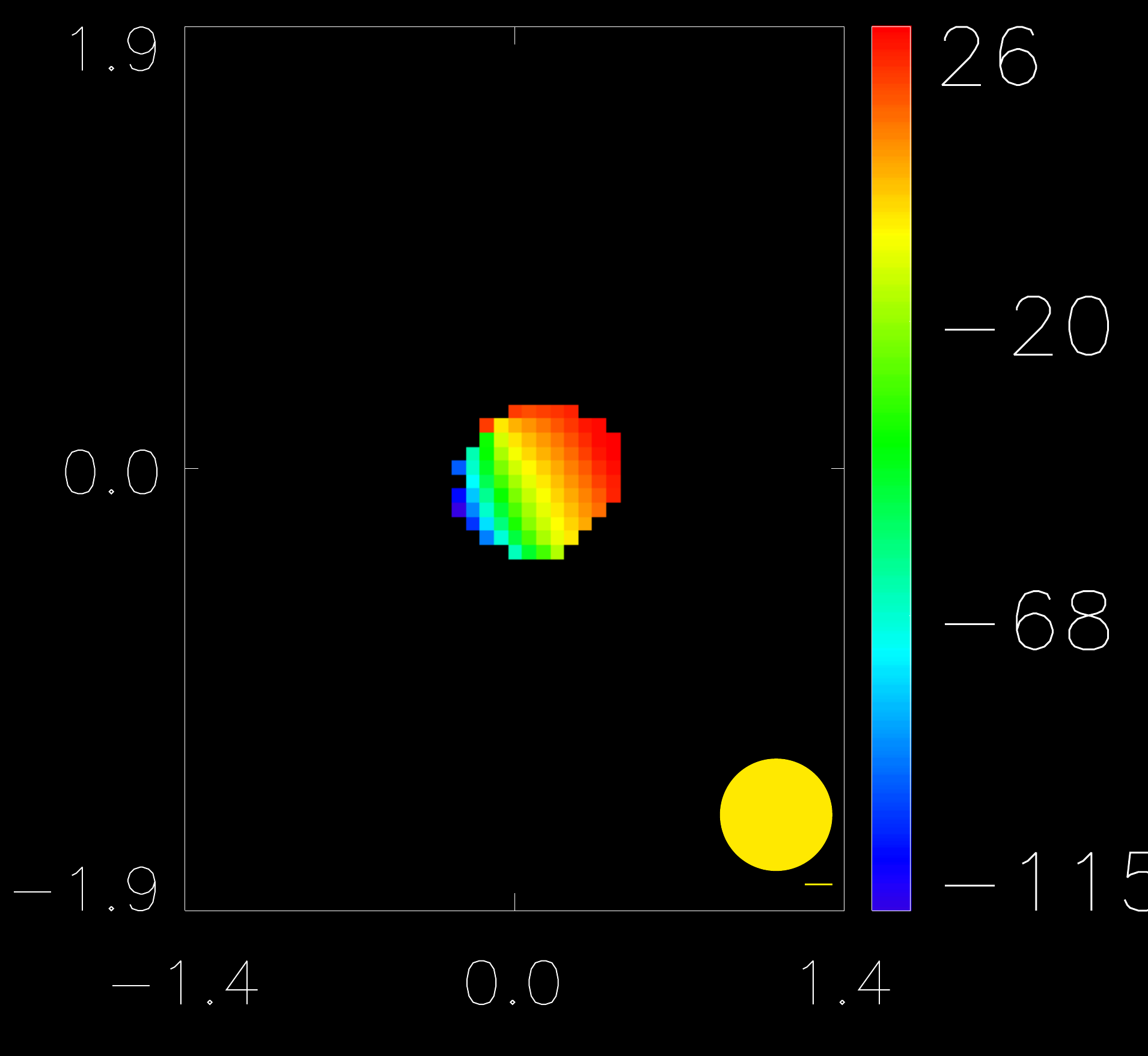} \hskip.03in
\includegraphics[height=.22\linewidth]{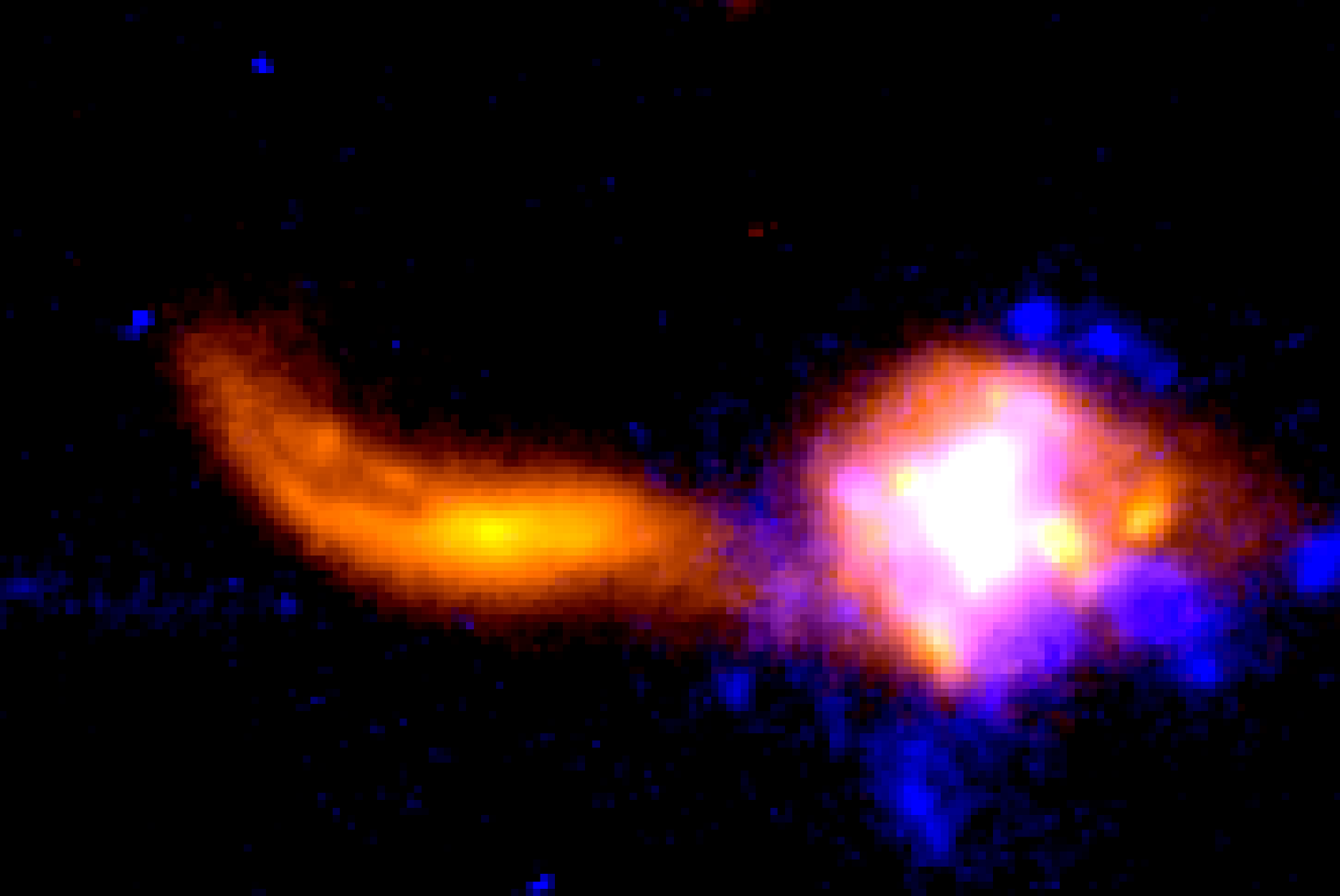}
\vskip .1in

\caption{
Velocity maps at low ({\rm left}) and high ({\rm center}) redshifts for 210358 and 135355. On the right we show the optical morphologies of each object as seen by HST, combining optical(orange) and ultraviolet(blue) data. High-z simulated map for 210358 is for OSIRIS data, while for 135355 this is the simulated SINFONI data. In the top case we see a galaxy for which a disk is apparent even at low redshift, while in the second case we notice the effect of loss of spatial resolution. Both these galaxies are classified as mergers through quantitative morphological analysis of the optical images. \label{fig:merger}
}
\end{figure*}

It is revealing to compare kinematics of some LBAs in our sample with their optical morphologies as observed with HST \citep{Overzier2009, Overzier2010}. In Figure \ref{fig:merger} we reproduce the velocity map of 210358 and 135355 and compare them with their optical image. Although they appear as rotating disks in Pa-$\alpha$, their optical morphologies indicate recent major merger events, with quantitative classification supporting that view. When simulated at higher redshift, both the morphology and the velocity structure appear much more regular (see Fig. 1 in \citet{Overzier2010} and Figs. \ref{fig:maps_hiz} and \ref{fig:merger} in this paper).

Also, on larger scales LBAs seem to support the idea of mergers as triggers for the high star formation rates observed in these galaxies, as these galaxies tend to pair with other galaxies more strongly than a random sample does \citep{Basu-Zych2009a}. Some studies at high-redshift have concluded that the galaxy number density is not high enough to account for mergers in all observed starbursts \citep{Conroy2008, Genel2008}, while some morphological studies indicate the merger fraction is high, up to 50\%, with $M_* > 10^{10} M_\odot$ galaxies undergoing $\sim$ 4 major mergers at $z>1$ \citep{Conselice2003, Conselice2006, Conselice2009}. Furthermore, \citet{Lotz2008} argue that the merger fraction might be even higher, since starbursts may outlast morphological asymmetries. On the other hand, the aforementioned morphological studies are all based on rest-frame UV images; these may differ dramatically from the rest-frame optical morphologies, which better traces the mass distribution of the stellar population -- the LBAs themselves presenting such contrast \citep{Overzier2010}. Whether or not the same mechanism is triggering star formation at either redshift is still unknown. We do not discard the possibility of an increasing fraction of galaxies at low redshift being created by major merger events, and that mergers of varying mass ratios may be taking place at either epoch.

\label{section:mergers}

%





\subsection{The dependence of rotational properties on stellar mass}\label{section:mass}

It has been shown that stellar mass in star-forming galaxies at high-redshift correlates with a number of physical properties, such as metallicity, star formation rates and age \citep{Erb2006, Erb2006a, Magdis2010}. Some results from kinematic studies of the H-$\alpha$ emission at high redshift also indicate this dependence, with more massive objects being more extended and presenting higher $v/\sigma$ ratios \citep{Law2009, ForsterSchreiber2009}.

In this work, we have shown that massive galaxies are more likely to present disk-like features, as evidenced by higher $v/\sigma$ ratios and higher levels of symmetric kinematics, while gas kinematics in less massive objects is dominated by random motions, as indicated by higher dynamical-to-stellar mass ratios as inferred from velocity dispersion measurements. This distinction is particularly important when taking into account the stellar mass function of LBGs at $z \sim 2-3$. \citet{Reddy2009} have found that the stellar mass function is particularly steep at these redshifts, which means an elevated contribution from less massive galaxies. That in turn would suggest more random dynamics for the majority of star-forming galaxies in the early universe, which are responsible for a significant fraction of the stars observed today -- $\sim$45\% of the present-day stellar mass have formed in galaxies with $L_{bol} < 10^{12} L_\odot$ \citep{Reddy2009}.

The dependence on mass is predicted even in more traditional star formation models. From a large $N$-body/gasdynamical simulation, \citet{Sales2009} have shown that the angular momentum in a $z\sim 2$ galaxy depends on halo mass. This dependence extends to stellar mass in the galaxy, albeit with varying amounts of scatter according to feedback efficiency (L. Sales, private communication). Whether that can be also a result of wet mergers remains unknown.

It should be noted that the stellar mass presented is the global value for the whole galaxy, while Pa-$\alpha$ traces a region containing a fraction of the stellar mass. It would be interesting to determine whether these relations still hold for the stellar mass contained within that small region, but for accurate measurements we need high-resolution near-infrared imaging, in order to trace stellar mass distribution at sub-kpc scales. Alternatively, longer exposures or more sensitive instruments capable of tracing {\rm stellar} dynamics instead of nebular gas could probe the kinematic properties at low surface brightness regions. Evidently, this is more difficult at high redshift, where cosmological dimming decreases surface brightness values by a factor of up to 200.

It is also unclear whether the trend with stellar mass represents an evolutionary effect or simply distinct formation scenarios. It is tempting to assume these galaxies keep forming stars for a period of time, increasing stellar mass while at the same time settling onto a rotating disk. However, this would mean that LBAs would necessarily keep elevated star formation rates for a period over 1 Gyr. Dynamical times of objects containing multiple star-forming regions, however, are too short (on the order of few tens of Myr), and the galaxy would coalesce much more rapidly. Therefore, a continuous inflow of gas or a sequence of minor mergers feeding star formation in these galaxies would be necessary to keep the observed star formation rates. Alternatively, it is possible that more massive galaxies have experienced more violent star formation episodes in the past, after which the dynamical structure has cooled down. An in-depth comparison with hydrodynamical simulations, with careful examination of star formation histories in LBAs, is required to examine each hypothesis in detail.

%
%

\section{Summary and Conclusions}\label{section:conc}

We have performed adaptive-optics assisted observations of 19 Lyman Break Analogs (LBAs) with the OSIRIS spectrograph at the Keck telescope. By studying spatially resolved Pa-$\alpha$ emission in these objects, we are able to draw the following conclusions:

(1) All galaxies show high velocity dispersions, indicating gas dynamics with a strong random component. Most galaxies show velocity shears of the same order of magnitude as velocity dispersions along the line of sight. This is consistent with our general picture of LBAs as dynamically young, starburst-dominated galaxies that are frequently undergoing mergers as shown by our HST data;

(2) The kinematics in LBAs are remarkably similar to high-redshift LBGs, which have also been the target of IFU studies. This is demonstrated by artificially redshifting the LBA sample to $z \sim 2$ and comparing simulated observations of these galaxies to observations of real $z \sim 2$ LBGs that have been presented by other groups \citep[see also][]{Basu-Zych2009}. All quantitative indicators of gas properties agree with those found for LBGs (e.g. Figures \ref{fig:lba_radius} and \ref{fig:mstar_vsigma}). This indicates that our identification of LBAs as being good local analogs of LBGs based on other, previously determined properties (e.g., SFR, mass, dust, size, metallicity and morphology) can be extended to include their emission line properties as well;

(3) As opposed to IFU observations of high-redshift star-forming galaxies, the proximity of LBAs allows for a more detailed picture of galactic dynamics. In particular, we have high physical resolution and are less subject to surface brightness dimming. We show that this bias at high can lead to the erroneous classification of similar starburst galaxies at high redshifts, with the kinematic profile appearing smoother and more symmetric (e.g. Figure \ref{fig:mstar_kinemetry});

(4) Even in cases where the LBAs resemble a disk at low redshift, we cannot rule out mergers as the starburst trigger based solely on the gas kinematics. Some disk-like galaxies show clear signs of recent interaction in the optical imaging data (Figure \ref{fig:merger}), and the gas disk might simply be a result of rapid coalescing of a gas-rich merger. Alternatively, a starburst was triggered by a recent infall event, while the underlying disk formed previously in a more gradual fashion;

(5) We have shown that whether a galaxy resembles a rotating disk depends strongly on stellar mass. The relationships between stellar mass and disk-like properties such as observed velocity shear and dynamical masses as inferred from dispersion-dominated gas motions support this conclusion. This has strong implications regarding the prevalence of disks at high redshift, and might indicate many of the stars in the local universe have not formed in disk-like galaxies.

Future prospects to study gas assembly in extreme starbursts at low and high redshift are excellent. As illustrated by our work on LBAs, a joint analysis of both morphologies and gas kinematics at high resolution and sensitivity is absolutely essential for deriving an unambiguous picture of the dynamical state of these systems. An important step toward achieving this goal at high redshift will be provided by the combination of the existing H$\alpha$ kinematical data with accurate rest-frame optical morphologies that can now be measured with the IR channel on Wide Field Channel 3 aboard HST. Furthermore, ALMA will allow detection of molecular gas in a large number of star-forming galaxies at redshifts $z\sim 2-3$, which should shed more light on the issue of gas-rich mergers. At low redshifts, ALMA will allow high-resolution measurements of molecular gas distribution and kinematics, providing deeper understanding of the conversion of gas into stars. Finally, the upcoming 20- and 30-m class telescopes, which should be operational at the end of the decade, will allow IFU studies of LBGs with higher sensitivity and resolution levels comparable to what is available now to LBAs, while the latter will be resolved at scales of giant molecular clouds, and we will be able to study the physical processes of star formation {\it in situ}. In all cases, the LBA sample studied in this paper offers a unique low redshift dataset useful for contrasting and comparing with starbursts at high redshift.

\acknowledgements

The authors would like to thank Jim Lyke, Al Conrad, Randy Campbell and Hien Tran for invaluable assistance with the laser observations. We also thank the anonymous referee for useful comments regarding dynamical masses. TSG thanks Brant Robertson for useful discussions concerning theoretical modeling of galaxy formation at $z\sim 2$. We would also like to thank those of Hawaiian ancestry for hospitably allowing telescope operations on the summit of Mauna Kea.

\bibliography{OSIRIS}{}
\bibliographystyle{apj}

\nocite{*}

\clearpage
\setlength\headsep{0.8in}

\end{document}